\PassOptionsToPackage{unicode}{hyperref}
\PassOptionsToPackage{hyphens}{url}
\PassOptionsToPackage{dvipsnames,svgnames,x11names}{xcolor}
\documentclass[
  12pt]{article}

\usepackage{siunitx}

\usepackage{amsmath}
\usepackage{amsthm}
\usepackage{amsfonts}
\usepackage{amssymb}
\usepackage{bbm}
\usepackage{bm}
\usepackage{physics}
\usepackage{iftex}
\ifPDFTeX
  \usepackage[T1]{fontenc}
  \usepackage[utf8]{inputenc}
  \usepackage{textcomp} % provide euro and other symbols
\else % if luatex or xetex
  \usepackage{unicode-math}
  \defaultfontfeatures{Scale=MatchLowercase}
  \defaultfontfeatures[\rmfamily]{Ligatures=TeX,Scale=1}
\fi
\usepackage{lmodern}
\usepackage{dirtytalk}
\usepackage{algorithmicx}
\usepackage{algorithm}
\usepackage{algpseudocode}

\usepackage{tikz}
\usetikzlibrary{positioning}
\usetikzlibrary{arrows.meta}

\ifPDFTeX\else  
\fi
\IfFileExists{upquote.sty}{\usepackage{upquote}}{}
\IfFileExists{microtype.sty}{% use microtype if available
  \usepackage[]{microtype}
  \UseMicrotypeSet[protrusion]{basicmath} % disable protrusion for tt fonts
}{}
\makeatletter
\@ifundefined{KOMAClassName}{% if non-KOMA class
  \IfFileExists{parskip.sty}{%
    \usepackage{parskip}
  }{% else
    \setlength{\parindent}{0pt}
    \setlength{\parskip}{6pt plus 2pt minus 1pt}}
}{% if KOMA class
  \KOMAoptions{parskip=half}}
\makeatother
\usepackage{xcolor}
\makeatletter
\ifx\paragraph\undefined\else
  \let\oldparagraph\paragraph
  \renewcommand{\paragraph}{
    \@ifstar
      \xxxParagraphStar
      \xxxParagraphNoStar
  }
  \newcommand{\xxxParagraphStar}[1]{\oldparagraph*{#1}\mbox{}}
  \newcommand{\xxxParagraphNoStar}[1]{\oldparagraph{#1}\mbox{}}
\fi
\ifx\subparagraph\undefined\else
  \let\oldsubparagraph\subparagraph
  \renewcommand{\subparagraph}{
    \@ifstar
      \xxxSubParagraphStar
      \xxxSubParagraphNoStar
  }
  \newcommand{\xxxSubParagraphStar}[1]{\oldsubparagraph*{#1}\mbox{}}
  \newcommand{\xxxSubParagraphNoStar}[1]{\oldsubparagraph{#1}\mbox{}}
\fi
\makeatother

\usepackage{longtable,booktabs,array}
\usepackage{calc} % for calculating minipage widths
\usepackage{etoolbox}
\makeatletter
\patchcmd\longtable{\par}{\if@noskipsec\mbox{}\fi\par}{}{}
\makeatother
\IfFileExists{footnotehyper.sty}{\usepackage{footnotehyper}}{\usepackage{footnote}}
\makesavenoteenv{longtable}
\usepackage{graphicx}
\makeatletter
\def\maxwidth{\ifdim\Gin@nat@width>\linewidth\linewidth\else\Gin@nat@width\fi}
\def\maxheight{\ifdim\Gin@nat@height>\textheight\textheight\else\Gin@nat@height\fi}
\makeatother
\setkeys{Gin}{width=\maxwidth,height=\maxheight,keepaspectratio}
\makeatletter
\def\fps@figure{htbp}
\makeatother

\makeatletter
\@ifpackageloaded{caption}{}{\usepackage{caption}}
\AtBeginDocument{%
\ifdefined\contentsname
  \renewcommand*\contentsname{Table of contents}
\else
  \newcommand\contentsname{Table of contents}
\fi
\ifdefined\listfigurename
  \renewcommand*\listfigurename{List of Figures}
\else
  \newcommand\listfigurename{List of Figures}
\fi
\ifdefined\listtablename
  \renewcommand*\listtablename{List of Tables}
\else
  \newcommand\listtablename{List of Tables}
\fi
\ifdefined\figurename
  \renewcommand*\figurename{Figure}
\else
  \newcommand\figurename{Figure}
\fi
\ifdefined\tablename
  \renewcommand*\tablename{Table}
\else
  \newcommand\tablename{Table}
\fi
}
\@ifpackageloaded{float}{}{\usepackage{float}}
\floatstyle{ruled}
\@ifundefined{c@chapter}{\newfloat{codelisting}{h}{lop}}{\newfloat{codelisting}{h}{lop}[chapter]}
\floatname{codelisting}{Listing}

\makeatother
\makeatletter
\@ifpackageloaded{caption}{}{\usepackage{caption}}
\@ifpackageloaded{subcaption}{}{\usepackage{subcaption}}
\makeatother

\ifLuaTeX
  \usepackage{selnolig}  % disable illegal ligatures
\fi

\usepackage{bookmark}

\IfFileExists{xurl.sty}{\usepackage{xurl}}{} % add URL line breaks if available
\hypersetup{
  pdftitle={The Dynamic Generalized Covariance Measure},
  colorlinks=true,
  linkcolor={blue},
  filecolor={Maroon},
  citecolor={Blue},
  urlcolor={Blue},
  pdfcreator={LaTeX via pandoc}}

\newcommand{\anon}{1}

\usepackage{xpatch}

\usepackage[
maxcitenames=2,
maxbibnames=99,
backend=biber,
style=authoryear-comp,
giveninits=true,
sorting=nyt,
uniquename=false,
uniquelist=false,
sortcites=true
]{biblatex}
\AtBeginRefsection{\GenRefcontextData{sorting=ynt}}
\AtEveryCite{\localrefcontext[sorting=ynt]}

\DeclareNameAlias{sortname}{family-given}
\renewbibmacro{in:}{%
  \ifentrytype{article}{}%
  {\printtext{\bibstring{in}\intitlepunct}}}

\DeclareFieldFormat[article]{title}{\mkbibquote{#1\addcomma}}

\DeclareFieldFormat[article]{number}{\mkbibparens{#1}}

\DeclareFieldFormat[article]{pages}{#1}

\xpatchbibmacro{journal+issuetitle}
  {\setunit*{\addspace}}
  {\setunit*{\addcomma\space}}
  {}{}

\xpatchbibmacro{volume+number+eid}
  {\setunit*{\adddot}}
  {\setunit*{}}
  {}{}
\theoremstyle{plain}% default
\newtheorem{theorem}{Theorem}[section]
\newtheorem{lemma}{Lemma}[section]

\newtheorem{corol}{Corollary}[section]

\newtheorem{definition}{Definition}[section]
\newtheorem{example}{Example}[section]
\newtheorem{assumption}{Assumption}[section]

\begin{document}

\def\spacingset#1{\renewcommand{\baselinestretch}%
{#1}\small\normalsize} \spacingset{1}

%%%%%%%%%%%%%%%%%%%%%%%%%%%%%%%%%%%%%%%%%%%%%%%%%%%%%%%%%%%%%%%%%%%%%%%%%%%%%%

\date{} % no date

\if1\anon
{
  \title{\bf The Dynamic Generalized Covariance Measure for Conditional Independence Testing with Nonstationary Time Series}
  \author{Michael Wieck-Sosa\thanks{Corresponding author: \texttt{mwiecksosa@cmu.edu}}\hspace{.2cm}\\
    Department of Statistics and Data Science\\ Carnegie Mellon University\\
    and \\
    Michel F. C. Haddad \\
    Department of Business Analytics and Applied Economics\\ Queen Mary University of London\\
    and \\
    Aaditya Ramdas\\
    Department of Statistics\\ Stanford University}
  \maketitle
} \fi

\if0\anon
{
  \bigskip
  \bigskip
  \bigskip
  \begin{center}
    {\LARGE\bf The Dynamic Generalized Covariance Measure for Conditional Independence Testing with Nonstationary Time Series}
\end{center}
  \medskip
} \fi

\bigskip
\begin{abstract}
Identifying relationships among stochastic processes is a core objective in many fields, such as economics. While the standard toolkit for multivariate time series analysis has many advantages, it can be difficult to capture nonlinear dynamics using linear vector autoregressive models. This difficulty has motivated the development of methods for causal discovery and variable selection for nonlinear time series, which routinely employ tests for conditional independence. In this paper, we introduce the first framework for conditional independence testing that works with a single realization of a nonstationary nonlinear process. The proposed test is designed to have power against alternatives in which the expected conditional covariance is non-zero for at least some times. We also discuss an approach for gaining power against a broader range of alternatives. The key technical ingredients of our framework are time-varying nonlinear regression, estimation of local long-run covariance matrices of products of error processes, and a distribution-uniform strong Gaussian approximation.

\end{abstract}
%%% Key words/phrases in alphabetical order
\noindent%
\vfill

\newpage

\spacingset{1.8}

\tableofcontents

\begin{refsegment}

\section{Introduction}\label{section:introduction}

A great deal of work has been dedicated to developing tests for conditional independence. That is, testing whether two random vectors $X$ and $Y$ are independent given a third random vector $Z$. For example, there are conditional independence tests based on conditional densities \parencite{su_white_hellinger_2008}, characteristic functions \parencite{su_white_char_ci_test}, likelihood ratios \parencite{su_white_cit_2014}, discretization \parencite{Margaritis_cit_2005,Huang_cit_2010}, permutation \parencite{doran_2014,Sen_2017}, kernels \parencite{Fukumizu2007,kun_zhang_kci_test,assess_granger}, copulas \parencite{copula_cit}, conditional mutual information \parencite{runge_cmi_cit}, and the model-X assumption \parencite{candes_mx,dcrt_ci_liu_2022,katsevich_mx_dr_reconciling_CIT}. Also, there are many conditional independence tests based on regressing $X$ on $Z$ and $Y$ on $Z$, followed by testing for independence between the errors based on the residuals \parencite{Hoyer2009,Peters_2014,Ramsey_CI_ParCorr,flaxman_ci_2015,Zhang_cit_2017,Zhang_cit_2019,Fan_CI_ParCorr}.

Unfortunately, conditional independence tests often fail to control the Type I error in finite samples. The groundbreaking work of \textcite{shah_gcm_2020} proves that conditional independence testing is impossible without making further assumptions. For more discussion of the hardness of conditional independence testing, see Section D in the supplement. 

\textcite{shah_gcm_2020} proposed a conditional independence test for the iid setting based on the \textit{generalized} \textit{covariance} \textit{measure} (GCM), which is a suitably normalized sum of the products of the residuals from the regressions of $X$ on $Z$ and $Y$ on $Z$. Our proposed test can be considered a GCM-type test for the nonstationary nonlinear time series setting. However, moving to this setting introduces several challenges and requires different techniques. Our test shares the advantages of the GCM test, such as rate double robustness, as well as its main limitation of only having power against alternatives with non-zero expected conditional covariance. Thus, our test is not an omnibus conditional independence test in general.

Next, we discuss related work. We then highlight our main contributions and explain the organization of the paper. Section D of the supplement discusses applications of conditional independence testing to variable selection for forecasting and causal discovery for nonstationary time series, as well as connections to Granger causality.

\subsection{Related Work}

Most conditional independence tests lack Type I error control guarantees outside the iid setting. At most, some methods provide guarantees for stationary mixing processes. To the best of our knowledge, only one conditional independence test has been proposed for the setting with one realization of a nonstationary process. 

\textcite{spirtes_var_example} introduce a conditional independence test for nonstationary linear vector autoregressions with iid Gaussian errors. They focus on processes with stochastic trends, so that some components evolve according to a random walk. They consider the case where the characteristic polynomial has unit roots and the first-differenced process is stable. In contrast, we allow linear and nonlinear processes with general nonstationarity, time-varying regression functions, and non-iid non-Gaussian errors. From an economic perspective, our framework models short-term fluctuations as driven by random shocks, and longer-term changes as reflecting shifts in structural mechanisms, e.g., policy and technology.

%%% Note: not mentioning this here because they don't actually pursue the nonstationary setting.
%%% Second, \textcite{flaxman_ci_2015} develop a conditional independence testing framework for non-iid data based on Gaussian process regression. The main idea is to pre-whiten the non-iid data using Gaussian process regression to control for dependencies (e.g., spatial, temporal, or network), which should yield iid residuals. The next step is to test for independence between these residuals using the Hilbert-Schmidt Independence Criterion. The authors state that their framework could be used with nonstationary covariance functions, although this idea was not developed. 

There are several conditional independence tests designed for the setting in which multiple realizations of a stochastic process are available. \textcite{sig_CI_stoch_proc} develop a test for stochastic processes using the signature kernel. \textcite{shah_gcm_hilbert} introduce a test for function-valued random variables. \textcite{cli_test} introduce a framework for testing conditional local independence relationships among point processes.

%%%% Note: not as relevant
% There is a growing literature on unconditional independence testing for nonstationary processes. \textcite{kernel_indep_test_nonstat_ts} develop independence tests based on the Hilbert--Schmidt Independence Criterion (HSIC) \parencite{hsic}, which require multiple realizations of the process. In contrast, the independence tests from \textcite{lsts_indep_testing_beering,loc_stat_indep_test} only require one realization.

\subsection{Our Contributions}\label{subsection:our_contributions}

We summarize the key contributions of the paper here.
\begin{itemize}
    \item We propose the first conditional independence test that can be used with a single realization of a nonstationary nonlinear process. In Theorem~\ref{thm:test_nsts_DR}, we show that our test has asymptotic Type I error control, \textit{uniformly} over a large family of null distributions. 
    \item We provide $\ell_{\infty}$-type and $\ell_2$-type test statistics to achieve high power against sparse and dense alternatives, respectively.    
    \item Our test statistics are based on the products of the residuals from nonlinear regressions of future or past values of the processes $X$ and $Y$ on the covariates $Z$ at time $t$. Hence, our test is doubly robust, in the sense that slower convergence rates of one regression estimator can be compensated for by faster convergence rates of the other estimator. 
    \item   We allow the errors to be nonstationary and temporally dependent, in contrast with other residual-based conditional independence tests that require iid errors. Moreover, we allow the errors to depend on the covariates, whereas many other residual-based tests require covariate-independent errors, thereby ruling out heteroskedasticity.
    \item Our test can be used with any regression estimator for nonstationary time series. However, we provide a guarantee for an instantiation of our test based on the sieve regression estimator from \textcite{zhouzhou_sieve} in Section A of the supplement, which complements prior sieve methods for nonstationary time series \parencite{ar_approx_lsts_xiucai_zhou,pacf_lsts_xiucai_zhou}. 
    \item We introduce a distribution-uniform strong Gaussian approximation in Section C of the supplement, which we use to prove uniformly asymptotic Type I error control. % To do this, we control the nonstationarity and temporal dependence uniformly over collections of distributions for the stochastic nonlinear systems which generate the error processes.
    \item We introduce an unconditional independence test in Section D of the supplement. Thus, our work is the first to supply the two components used in constraint-based causal discovery algorithms for nonstationary nonlinear time series.
\end{itemize}

\paragraph*{Paper Outline.} The rest of the paper is structured as follows. In Section~\ref{section:main_ideas_dgcm_test}, we discuss the main ideas. In Section~\ref{section:CI_test_theory_nsts}, we state a guarantee for our test based on any regression method for nonstationary time series. In Section~\ref{section:simulations}, we report simulation results. In Section~\ref{section:data_analysis}, we illustrate the practicality of our test by applying it to real data. Section~\ref{section:conclusion} concludes. 

The supplement is organized as follows. In Section A, we provide a guarantee for an instantiation of our test based on a sieve estimator in the setting of locally stationary time series. The proofs and technical lemmata are gathered in Section B. Section C presents the distribution-uniform theory.  Section D contains additional discussions.

\section{The Dynamic Generalized Covariance Measure}\label{section:main_ideas_dgcm_test}

In this section, we give a high-level overview of our work. We introduce the notation, main ideas, and algorithm for our \textit{dynamic} \textit{generalized covariance} \textit{measure} (dGCM) test.

\subsection{Setting and Notation}\label{subsection:setting_nsts}

We work in a triangular array framework for high-dimensional nonstationary nonlinear time series. Let $(X_{t,n}, Y_{t,n}, Z_{t,n})_{t\in [n]}$ be the observed sequence of length $n\in\mathbb{N}$, where $[n]=\{1,\ldots,n\}$. We denote $X_n=(X_{t,n})_{t\in [n]}$, $Y_n=(Y_{t,n})_{t\in [n]}$, $Z_n=(Z_{t,n})_{t\in [n]}$, and we denote by $X$, $Y$, $Z$ each process with any length. Let $d_X=d_{X,n}$, $d_Y=d_{Y,n}$, $d_Z=d_{Z,n}$ denote the dimensions, which can grow with $n$. Denote dimension $i \in [d_X]$ of $X_{t,n}$ by $X_{t,n,i}$, dimension $j \in [d_Y]$ of $Y_{t,n}$ by $Y_{t,n,j}$, and dimension $k \in [d_Z]$ of $Z_{t,n}$ by $Z_{t,n,k}$.

Next, we introduce notation for the time-offsets of each dimension of $X_{t,n}$, $Y_{t,n}$, $Z_{t,n}$. Negative time-offsets are called lags, and positive time-offsets are called leads. Let $$A_i \subset \{-n+1,\ldots,n-1\},\text{ }B_j \subset \{-n+1,\ldots,n-1\},\text{ }C_k \subset \{-n+1,\ldots,0\},$$ be the sets of time-offsets of $X_{t,n,i}$, $Y_{t,n,j}$, $Z_{t,n,k}$ under consideration. We require the time-offsets $C_k$ to be non-positive so that the covariates are known at time $t$. The time-offsets should be selected so that there is a sufficient amount of data to conduct the test. 
% reason for allowing for X Y with both lags or both leads: checking for Z being a confounder or collider in a causal discovery framework that allows for dependencies with time-offsets 

Denote the time-offset $a \in A_i$ of $X_{t,n,i}$ by $X_{t,n,i,a}=X_{t+a,n,i}$, the time-offset $b \in B_j$ of $Y_{t,n,j}$ by $Y_{t,n,j,b}=Y_{t+b,n,j}$, and the time-offset $c \in C_k$ of $Z_{t,n,k}$ by $Z_{t,n,k,c}=Z_{t+c,n,k}$. Denote the sets of all time-offsets by $A = \bigcup_{i=1}^{d_X} A_i$, $B = \bigcup_{j=1}^{d_Y} B_j$, $C=\bigcup_{k=1}^{d_Z} C_k$, the largest (signed) time-offsets by $a_{\max}=\max(A)$, $b_{\max}=\max(B)$, $c_{\max}=\max(C)$, and the smallest (signed) time-offsets by $a_{\min}=\min(A)$, $b_{\min}=\min(B)$, $c_{\min}=\min(C)$. Denote the subset of times $$\mathcal{T}_n=\{1-\min(a_{\min},b_{\min},c_{\min},0),\ldots,n-\max(a_{\max},b_{\max},c_{\max},0)\}\subseteq\{1,\ldots,n\},$$ in which all time-offsets of each dimension of $X_{t,n}$, $Y_{t,n}$, $Z_{t,n}$ are actually observed. Denote $\mathbb{T}_n^{-}=\min(\mathcal{T}_n)$, $\mathbb{T}_n^{+}=\max(\mathcal{T}_n)$, and the cardinality $T_n=|\mathcal{T}_n|$. 
%%% Note: If no negative time-offsets (i.e. lags) are used then $\min(a_{\min},b_{\min},c_{\min})=0$, and if no positive time-offsets (i.e. leads) are used then $\max(a_{\max},b_{\max},c_{\max})=0$. 

For all $t\in\mathcal{T}_n$, denote the vectors with all dimensions and time-offsets of interest by
$$\bm{X}_{t,n}=(X_{t,n,i,a})_{i\in[d_X],a\in A_i}, \text{ }\text{ }\bm{Y}_{t,n}=(Y_{t,n,j,b})_{j\in[d_Y],b\in B_j}, \text{ }\text{ }\bm{Z}_{t,n}=(Z_{t,n,k,c})_{k\in[d_Z],c\in C_k},$$ with dimensions $\bm{d_X}=\sum_{i =1}^{d_X} |A_i|$, $\bm{d_Y}=\sum_{j =1}^{d_Y} |B_j|$, $\bm{d_Z}=\sum_{k =1}^{d_Z} |C_k|$, respectively. Denote the entire time series by $\bm{X}_{n} = (\bm{X}_{t,n})_{t\in\mathcal{T}_n}$, $\bm{Y}_{n}=(\bm{Y}_{t,n})_{t\in\mathcal{T}_n}$, and $\bm{Z}_{n}=(\bm{Z}_{t,n})_{t\in\mathcal{T}_n}$. For each $n\in\mathbb{N}$, let $\mathcal{P}_n$ be a collection of distributions, which can change with $n$; see Section~\ref{subsection:observed_proc_nsts}.

We allow the number of time-offsets to grow with $n$, that is, $A_i=A_{i,n}$, $B_j=B_{j,n}$, $C_k=C_{k,n}$ and $A=A_n$, $B=B_n$, $C=C_n$. We require that the largest (in magnitude) time-offset grows at a slower rate than $n$ such that, as $n\xrightarrow[]{}\infty$, we have $\min(a_{\min},b_{\min},c_{\min})/n \xrightarrow[]{}0$ and $\max(a_{\max},b_{\max},c_{\max})/n \xrightarrow[]{} 0$ so that the number of observations $T_n \xrightarrow[]{} \infty$ and $T_n/n \xrightarrow[]{} 1$.

Denote the set containing all dimension/time-offset combinations of interest by $$\mathcal{D}_n \subseteq \{(i,j,a,b):i \in [d_X], j \in [d_Y], a \in A_i, b \in B_j\}.$$  In the most favorable setting, we allow $D_n=|\mathcal{D}_n|=O(T_n^{\frac{1}{8}})$; see Section C in the supplement. 
%%% Note: $D_n$ reflects the intrinsic dimensionality of the problem. 

\subsection{Regression for Nonstationary Time Series}\label{subsection:tv_regr_nsts}

For a fixed sample size $n\in\mathbb{N}$, distribution $P\in\mathcal{P}_n$, time $t \in \mathcal{T}_n$ and dimension/time-offset tuple $m=(i,j,a,b)\in \mathcal{D}_n$, we can always decompose 
\begin{equation}\label{eqn:tv_regr_fncts_nsts}
X_{t,n,i,a}=f_{P,t,n,i,a}(\bm{Z}_{t,n})+\varepsilon_{P,t,n,i,a}, \text{ } Y_{t,n,j,b}=g_{P,t,n,j,b}(\bm{Z}_{t,n})+\xi_{P,t,n,j,b},  \end{equation} where $f_{P,t,n,i,a}(\bm{z})=\mathbb{E}_P(X_{t,n,i,a}|\bm{Z}_{t,n}=\bm{z})$ and $
g_{P,t,n,j,b}(\bm{z})=\mathbb{E}_P(Y_{t,n,j,b}|\bm{Z}_{t,n}=\bm{z})$ are the time-varying regression functions, and $R_{P,t,n,m}=\varepsilon_{P,t,n,i,a}\xi_{P,t,n,j,b}$ denotes the product of these errors. We allow the errors to be temporally dependent and nonstationary; see Section~\ref{section:CI_test_theory_nsts}.
% Intuition for "forecasting" when appealing to infill asymptotics with LSTS: Suppose you are predicting b steps ahead. With infill asymptotics, the "amount of rescaled time" between the time steps decreases since b/n decreases. On the other hand, we let the largest forecasting horizon grow with n also. However, the amount of rescaled time between these time steps will get smaller with n as well "eventually" for each of these new, "longer" forecasting horizons more time steps into the future. Note that the "degree of nonstationarity" with b time steps ahead will get less and less as n grows since b/n shrinks as n increases.

Let $\hat{f}_{t,n,i,a}$, $\hat{g}_{t,n,j,b}$ be estimates of $f_{P,t,n,i,a}$, $g_{P,t,n,j,b}$, based on any regression method. Let 
\begin{equation*}  \hat{\varepsilon}_{t,n,i,a}=X_{t,n,i,a}-\hat{f}_{t,n,i,a}(\bm{Z}_{t,n}),  \text{ }\hat{\xi}_{t,n,j,b}=Y_{t,n,j,b}-\hat{g}_{t,n,j,b}(\bm{Z}_{t,n}),\end{equation*} be the corresponding residuals, and denote the product of these residuals by \begin{equation} \hat{R}_{t,n,m}=\hat{\varepsilon}_{t,n,i,a}  \hat{\xi}_{t,n,j,b}.\label{eqn:prod_residuals_def}\end{equation} Let $\bm{\hat{R}}_{t,n}=(\hat{R}_{t,n,m})_{m\in\mathcal{D}_n}$ be the vector of all residual products at time $t\in\mathcal{T}_n$.

\subsection{Main Ideas and the Algorithm}\label{subsection:test_algo}

We introduce a simulation-assisted testing procedure for the null hypothesis \begin{equation} \bm{X}_{t,n}\!\perp \!\!\! \perp \bm{Y}_{t,n} \mid \bm{Z}_{t,n} \text{ for all times } t \in \mathcal{T}_n,\label{eqn:multivariate_global_null_hypoth_all_times_nsts_simple_notation}\end{equation} which includes all leads and lags of interest.  We give several examples of how this null hypothesis can be used in Section D of the supplement, e.g., in the context of forecasting. 

Denote the time series of residual products by $\bm{\hat{R}}_{n}=(\bm{\hat{R}}_{t,n})_{t\in \mathcal{T}_{n,L}}$, where $\mathcal{T}_{n,L}=\{L_n+\mathbb{T}_n^{-}-1,\ldots,\mathbb{T}_n^{+}-1,\mathbb{T}_n^{+}\}$ is a set of times with cardinality $T_{n,L}=|\mathcal{T}_{n,L}|$ and $L_n\in\mathbb{N}$ is a lag-window size parameter. We discuss how to select $L_n$ in Section D of the supplement.

Our test statistic, which we name the \textit{dynamic generalized covariance measure} (dGCM), is the max $\ell_p$-norm, $p\in [2,\infty]$, achieved by the scaled partial sum process of residual products \begin{equation} S_{n,p}(\bm{\hat{R}}_{n})= \underset{s\in\mathcal{T}_{n,L}}{\max} \text{ } \norm{\frac{1}{\sqrt{T_{n,L}}} \sum_{t\leq s} \bm{\hat{R}}_{t,n}}_{p}.\label{eqn:test_stat_Snp}\end{equation} Section D in the supplement discusses other test statistics (e.g., based on the full sum).

\begin{algorithm}[p] % [ht!]
\caption{The dynamic generalized covariance measure (dGCM) test}\label{algo:dGCM_test}
\begin{algorithmic}[1]
    \State \textbf{Input:} Data $(X_{t,n},Y_{t,n},Z_{t,n})_{t\in [n]}$, regression method for nonstationary time series, test statistic $S_{n,p}(\cdot)$, number of simulations $N^{\mathrm{sim}}$, and significance level $\alpha$
    \State Select hyperparameters for regression method via cross-validation as in Section D
    \For{each time $t\in\mathcal{T}_n$ and dimension/time-offset tuple $m=(i,j,a,b)\in\mathcal{D}_n$}
        \State Obtain the estimates $\hat{f}_{t,n,i,a}$ and $\hat{g}_{t,n,j,b}$ from~\eqref{eqn:tv_regr_fncts_nsts} using the chosen regression method
        \State Calculate the residuals $\hat{\varepsilon}_{t,n,i,a}=X_{t,n,i,a}-\hat{f}_{t,n,i,a}(\bm{Z}_{t,n})$ and $\hat{\xi}_{t,n,j,b}=Y_{t,n,j,b}-\hat{g}_{t,n,j,b}(\bm{Z}_{t,n})$
        \State Calculate the product of the residuals $\hat{R}_{t,n,m}=\hat{\varepsilon}_{t,n,i,a} \hat{\xi}_{t,n,j,b}$ as in~\eqref{eqn:prod_residuals_def}
    \EndFor
    \State Obtain the vector of all residual products $\bm{\hat{R}}_{t,n}=(\hat{R}_{t,n,m})_{m\in\mathcal{D}_n}$ for all times $t\in\mathcal{T}_n$
    \State Select the lag-window size $L_n$ using the minimum volatility method from Section D
    \For{each time $t \in \mathcal{T}_{n,L}$}
        \State Calculate the covariances $\bm{\hat{\Sigma}}_{t,n}^{\bm{R}}=\frac{1}{L_n} \left( \sum_{s=t-L_n+1}^{t} \bm{\hat{R}}_{s,n} \right) \left( \sum_{s=t-L_n+1}^{t} \bm{\hat{R}}_{s,n} \right)^{\top}$
    \EndFor

    \For{each simulation $r=1,\ldots,N^{\mathrm{sim}}$ and time $t \in \mathcal{T}_{n,L}$}
        \State Simulate independent Gaussian random vectors $\bm{\breve{R}}_{t,n}^{(r)} \sim \mathcal{N}(0, \bm{\hat{\Sigma}}_{t,n}^{\bm{R}})$
    \EndFor
    \For{each simulation $r=1,\ldots,N^{\mathrm{sim}}$}
        \State Calculate the test statistic $S_{n,p}(\bm{\breve{R}}_{n}^{(r)})$ from~\eqref{eqn:test_stat_Snp} based on $\bm{\breve{R}}_{n}^{(r)} = (\bm{\breve{R}}_{t,n}^{(r)})_{t \in \mathcal{T}_{n,L}}$
    
    \EndFor
    
    \State Calculate $1 - \alpha$ empirical quantile $\hat{q}_{1-\alpha}^{\text{boot}}$ of simulated test statistics $(S_{n,p}(\bm{\breve{R}}_{n}^{(r)}))_{r\in [N^{\mathrm{sim}}]}$
    
    \State Compute test statistic $S_{n,p}(\bm{\hat{R}}_{n})$ from~\eqref{eqn:test_stat_Snp} with residual products $\bm{\hat{R}}_{n} = (\bm{\hat{R}}_{t,n})_{t \in \mathcal{T}_{n,L}}$

    \If{$S_{n,p}(\bm{\hat{R}}_{n}) > \hat{q}_{1-\alpha}^{\text{boot}}$}
        \State Reject the null hypothesis at the significance level $\alpha$
    \EndIf
 \State \textbf{Output:} Decision to reject or fail to reject the null hypothesis at significance level $\alpha$   
\end{algorithmic}
\end{algorithm}

The main idea of our test is that, under the null hypothesis~\eqref{eqn:multivariate_global_null_hypoth_all_times_nsts_simple_notation}, the products of the residuals will be small, as long as the regression functions are estimated well. Our test statistic~\eqref{eqn:test_stat_Snp} is designed to have power against alternatives in which the covariances of the errors from~\eqref{eqn:tv_regr_fncts_nsts} are non-zero for some times. We do not test for independence between the errors based on the residuals (e.g., using HSIC \parencite{hsic}), because the resulting test can fail to control the Type I error when the errors depend on the covariates; see Example 1 in \textcite{shah_gcm_2020}. This often occurs in economics, e.g., due to heteroskedasticity. In contrast with many other residual-based tests, we allow the errors to depend on the covariates.

Any regression method can be used with our test, as long as modest convergence rate requirements are met. Specifically, the product of the $L^2(P)$ norms of the estimation errors must be about $o(1/\sqrt{T_n})$; see Section~\ref{subsection:test_guarantee_nsts}. Under assumptions about the smoothness of the regression functions, temporal dependence, and nonstationarity, these rates are achieved by kernel \parencite{vogt,wbwu_timevar_npreg, mult_str_wu_2022}, sieve \parencite{zhouzhou_sieve}, and neural network  \parencite{kurisu_NSTS_DNN_2025} estimators. 

As with the GCM test from \textcite{shah_gcm_2020}, our test only has power against alternatives with non-zero expected conditional covariance; see Section D of the supplement for the details. Hence, our test is not, in general, an omnibus test of conditional independence. To gain power against a broader range of alternatives, our test can be applied to transformations of $\bm{X}_{t,n}$ and $\bm{Y}_{t,n}$, as in the discussion following Remark 10 in \textcite{shah_gcm_2020}. To retain our test's Type I error control guarantee, the transformations must be chosen independently of the data used for testing (e.g., using domain knowledge or an independent hold-out dataset). We focus on testing with untransformed data in this paper, and we leave the problem of transformations for future work.

%%% Note: In our simulations in Section~\ref{section:simulations}, we use the sieve method from \textcite{zhouzhou_sieve}. We provide a guarantee for our test with this sieve estimator in Section A of the supplement.

The testing procedure relies on an estimate of the covariance structure of some nonstationary Gaussian process. We postpone the technical details about these covariances until Section~\ref{subsection:test_guarantee_nsts}. 
% To justify this procedure, we introduce a distribution-uniform strong Gaussian approximation in Section C of the supplement. 

\section{Assumptions and the Main Theoretical Result}\label{section:CI_test_theory_nsts} 

%%% Note: Probably too technical if they don't already know what this means. Just keep it at an intuitive level to start.
%%% To do this, we introduce a theoretical framework for high-dimensional nonstationary nonlinear processes. Our framework enables hypothesis testing based on the residuals formed from the predictions of any regression estimator for nonstationary time series. 

In this section, we provide a theoretical guarantee for the dGCM test from Algorithm~\ref{algo:dGCM_test}. We allow the processes to exhibit slow decay in temporal dependence and highly complex forms of nonstationarity, which may be both abrupt and smooth. We control the temporal dependence and nonstationarity of the processes \textit{uniformly} over collections of distributions by employing versions of the functional dependence measure of \textcite{wu_funct_dep_meas} and the total variation-type nonstationarity condition of \textcite{seq_gauss_approx2022}. These distribution-uniform assumptions are needed for the uniform level guarantee for the dGCM test in Theorem~\ref{thm:test_nsts_DR}, which is our main theoretical result.

Our framework nests many classes of stochastic processes. In Section A of the supplement, we show how our framework nests a widely used class of nonstationary processes called locally stationary processes, which exhibit smooth changes over time. In Section D of the supplement, we discuss piecewise locally stationary processes, which exhibit both smooth changes and abrupt change-points. Stationary processes, independent sequences with time-varying distributions, and iid sequences all arise as special cases in our framework.
% , precisely when there is no nonstationarity. Similarly, our framework accommodates temporally independent sequences with time-varying distributions. The iid setting arises when there is neither nonstationarity nor temporal dependence.

%%% Note:
% We discuss how our framework is compatible with even more types of nonstationary processes in Section D of the supplement. Notably, we explain how our framework nests a very general class of nonstationary processes called piecewise locally stationary processes. This class extends the framework for locally stationary processes by permitting both smooth changes and abrupt breakpoints. 
%%% Note: very few people care about cyclostationary processes, so not even going to mention it
%%% We also allow for cyclostationary processes, which are a class of nonstationary processes that exhibit repetition over time. 

%%% Removed: Additionally, we explain how to leverage black-box simulators for $(X,Z)$ by using simulation-and-regression techniques.
%%% Note: If you have a simulator for (X,Z) then you can just consider this the null model where the effect of Y is zero, and then test for goodness-of-fit using *any* discrepancy measure. See e.g. the GoF test from the Robust SBI paper.

\subsection{Observed Processes}\label{subsection:observed_proc_nsts}

We view each dimension of the observed sequence $(X_{t,n},Y_{t,n},Z_{t,n})_{t\in [n]}$ as the outputs of a time-varying nonlinear function that is given a sequence of iid inputs. Let $(\eta_t^X,\eta_t^Y,\eta_t^Z)_{t\in\mathbb{Z}}$ be a sequence of iid random vectors and define $$\mathcal{H}_t^X=(\eta_t^X,\eta_{t-1}^X,\ldots),\text{ }\mathcal{H}_t^Y=(\eta_t^Y,\eta_{t-1}^Y,\ldots),\text{ }\mathcal{H}_t^Z=(\eta_t^Z,\eta_{t-1}^Z,\ldots).$$ Denote the dimensions of $\eta_t^X=\eta_{t,n}^X$, $\eta_t^Y=\eta_{t,n}^Y$, $\eta_t^Z=\eta_{t,n}^Z$, respectively, by $d_{X}^{\eta}=d_{X,n}^{\eta}$, $d_{Y}^{\eta}=d_{Y,n}^{\eta}$, $d_{Z}^{\eta}=d_{Z,n}^{\eta}$, which can change with $n$.
% can change with n because each measurable function Gn (i.e. the time-varying causal mechanism) depends on n 
% defining this way to emphasize that the iid inputs at each time be can dependent (i.e. non-product joint distribution at each time), to ensure this joint distribution of the iid inputs at each time is time-invariant, and to ensure each input is (temporally) independent of all other times of every input sequence. this is important for step 1.2 of the main proof.
% Without loss of generality, we can take each of $\eta_t^X,\eta_t^Y,\eta_t^Z$ to have $\mathrm{Unif}[0,1]$ marginal distributions for each $t\in\mathbb{Z}$.

% Note: we add the subscript n since the measurable functions can change with n. We allow for the dimensions and distributions of X, Y, Z to change with n. 
\begin{assumption}[Causal representations of the observed processes]\label{asmpt_causal_rep_process_nsts} 
Assume that, for each $n\in\mathbb{N}$, $i\in [d_X]$, $j\in [d_Y]$, $k\in [d_Z]$, and $t\in [n]$, we can represent the observed processes as
$$X_{t,n,i}= G_{t,n,i}^X(\mathcal{H}_{t}^X), \text{ } Y_{t,n,j}= G_{t,n,j}^Y(\mathcal{H}_{t}^Y),  \text{ } Z_{t,n,k}=G_{t,n,k}^Z(\mathcal{H}_{t}^Z),$$ where $G_{t,n,i}^{X}(\cdot)$, $G_{t,n,j}^{Y}(\cdot)$, $G_{t,n,k}^{Z}(\cdot)$ are measurable functions from $(\mathbb{R}^{d_{X}^{\eta}})^{\infty}$, $(\mathbb{R}^{d_{Y}^{\eta}})^{\infty}$, $(\mathbb{R}^{d_{Z}^{\eta}})^{\infty}$, respectively, to $\mathbb{R}$ such that $G_{t,n,i}^{X}(\mathcal{H}_s^X)$, $G_{t,n,j}^{Y}(\mathcal{H}_s^Y)$, $G_{t,n,k}^{Z}(\mathcal{H}_s^Z)$ are each well-defined random variables for each $s\in\mathbb{Z}$ and $(G_{t,n,i}^{X}(\mathcal{H}_s^X))_{s\in\mathbb{Z}}$, $(G_{t,n,j}^{Y}(\mathcal{H}_s^Y))_{s\in\mathbb{Z}}$, $(G_{t,n,k}^{Z}(\mathcal{H}_s^Z))_{s\in\mathbb{Z}}$ are each stationary ergodic processes. \end{assumption}

To simplify the notation, we have not defined separate input sequences for each dimension of the observed processes. However, we can define the measurable functions $G_{t,n,i}^{X}(\cdot)$, $G_{t,n,j}^{Y}(\cdot)$, $G_{t,n,k}^{Z}(\cdot)$ and random vectors $\eta_t^X$, $\eta_t^Y$, $\eta_t^Z$ so that each dimension of the observed processes has idiosyncratic inputs. 
Denote $G_{t,n}^X(\mathcal{H}_{t}^X)=(G_{t,n,i}^X(\mathcal{H}_{t}^X))_{i\in [d_X]}$, $G_{t,n}^Y(\mathcal{H}_{t}^Y) = (G_{t,n,j}^Y(\mathcal{H}_{t}^Y))_{j\in [d_Y]}$,  and $G_{t,n}^Z(\mathcal{H}_{t}^Z)= (G_{t,n,k}^Z(\mathcal{H}_{t}^Z))_{k\in [d_Z]}$.
In view of Assumption~\ref{asmpt_causal_rep_process_nsts}, we have the following causal representations for the full observed processes  
\begin{align*}
 X_{t,n} = G_{t,n}^X(\mathcal{H}_{t}^X), \ Y_{t,n}=G_{t,n}^Y(\mathcal{H}_{t}^Y),
 \ Z_{t,n} = G_{t,n}^Z(\mathcal{H}_{t}^Z).
\end{align*} 

For each $n \in \mathbb{N}$, let the measurable space $(\Omega,\mathcal{B})$ be equipped with a family of probability measures $(\mathbb{P}_P)_{P\in\mathcal{P}_n}$ so that the joint distribution of the stochastic nonlinear systems $$(G_{t,n}^X(\mathcal{H}_{s}^X))_{t\in [n],s\in\mathbb{Z}},\text{ } (G_{t,n}^Y(\mathcal{H}_{s}^Y))_{t\in [n],s\in\mathbb{Z}},\text{ } (G_{t,n}^Z(\mathcal{H}_{s}^Z))_{t\in [n],s\in\mathbb{Z}},$$ under $\mathbb{P}_P$ is $P\in\mathcal{P}_n$, where the collection of distributions $\mathcal{P}_n$ can change with $n$. The family of probability measures $(\mathbb{P}_P)_{P\in\mathcal{P}_n}$ is defined with respect to the same measurable space $(\Omega,\mathcal{B})$, but need not have the same dominating measure. For each $n\in\mathbb{N}$, denote the family of probability spaces by $(\Omega,\mathcal{B},\mathbb{P}_P)_{P\in\mathcal{P}_n}$. For a given sample size $n\in\mathbb{N}$ and distribution $P\in\mathcal{P}_n$, let $\mathbb{E}_{P}(\cdot)$ denote the expectation of a random variable with distribution determined by $P$. Let $\mathbb{P}_{P}(E)$ denote the probability of an event $E\in\mathcal{B}$.

In the rest of this section, we state distribution-uniform assumptions with respect to a generic sequence of collections of distributions $(\mathcal{P}_n)_{n\in\mathbb{N}}$. Let $\mathcal{P}_{0,n}^{\mathrm{CI}}$ be a collection of distributions such that the null hypothesis is true, and let $(\mathcal{P}_{0,n}^{\mathrm{CI}})_{n \in \mathbb{N}}$ be a sequence of such collections of distributions. In our main result, which we state as Theorem~\ref{thm:test_nsts_DR}, we will assume that these distribution-uniform assumptions hold for a sequence of collections of distributions $(\mathcal{P}_{0,n}^{\ast})_{n\in\mathbb{N}}$, where $\mathcal{P}_{0,n}^{\ast}\subset\mathcal{P}_{0,n}^{\mathrm{CI}}$ for each $n\in\mathbb{N}$. That is, we make these assumptions for a sequence of subcollections of distributions for which the null hypothesis~\eqref{eqn:multivariate_global_null_hypoth_all_times_nsts_simple_notation} holds. 
% Note: $\mathcal{P}_n \subset \mathrm{Prob}[(\mathbb{R}^{d_X+d_Y+d_Z})^{[n]^3\times\mathbb{Z}^3}]$ is a subset of the set of Borel probability measures on functions from $[n]^3 \times \mathbb{Z}^3$ to $\mathbb{R}^{d_X+d_Y+d_Z}$ i.e. each $\omega$ gives you a different realization of the stochastic nonlinear system for all mechanism times and input times

% Note: Reason for writing the distribution of the stochastic nonlinear system instead of just the observed process is because of the temporal dependence condition (i.e. different time's inputs and same mechanism) and nonstationarity condition (i.e. same inputs and different time's mechanism)

\subsection{Predictions}\label{subsection:prediction_nsts}

Next, we introduce causal representations for the predictors, predictions, and estimation errors. For each $n\in \mathbb{N}$, $t\in\mathcal{T}_n$, $(i,j,a,b)\in\mathcal{D}_n$, let $\eta_{t,n,i,a}^{\text{algo}}$, $\eta_{t,n,j,b}^{\text{algo}}$ be random variables that encode the possible stochasticity of the statistical learning algorithms. If the learning algorithms are not stochastic, then these random variables can be ignored without loss of generality. Going forward, we suppress the dependence of the predictors on $\eta_{t,n,i,a}^{\text{algo}}$, $\eta_{t,n,j,b}^{\text{algo}}$.

Let $\bm{\mathfrak{D}}_{t,n,i,a}^{\hat{f}}$, $\bm{\mathfrak{D}}_{t,n,j,b}^{\hat{g}}$ be the datasets containing the observations used to form the predictors $\hat{f}_{t,n,i,a}$, $\hat{g}_{t,n,j,b}$, and let $\mathcal{H}_{t,a}^{\bm{\mathfrak{D}}^{\hat{f}}}$, $\mathcal{H}_{t,b}^{\bm{\mathfrak{D}}^{\hat{g}}}$ be the corresponding input sequences. We require that the datasets and input sequences are nested, so that for any time $t_1 < t_2$, we have $$\sigma(\mathcal{H}_{t_1,a}^{\bm{\mathfrak{D}}^{\hat{f}}}) \subseteq \sigma(\mathcal{H}_{t_2,a}^{\bm{\mathfrak{D}}^{\hat{f}}}),  \quad \sigma(\mathcal{H}_{t_1,b}^{\bm{\mathfrak{D}}^{\hat{g}}}) \subseteq \sigma(\mathcal{H}_{t_2,b}^{\bm{\mathfrak{D}}^{\hat{g}}}).$$ For example, if only the observations in $\mathcal{T}_n$ up to time $t\in\mathcal{T}_n$ are used to form the predictor $\hat{g}_{t,n,j,b}$, then $\bm{\mathfrak{D}}_{t,n,j,b}^{\hat{g}}=(Y_{s,n,j,b},\bm{Z}_{s,n})_{s\leq t}$ and $\mathcal{H}_{t,b}^{\bm{\mathfrak{D}}^{\hat{g}}}=(\mathcal{H}_{t+b}^Y,\mathcal{H}_{t+c_{\mathrm{max}}}^{Z})$. Similarly, if all of the observations in $\mathcal{T}_n$ are used to form the predictor $\hat{g}_{t,n,j,b}$, then $\bm{\mathfrak{D}}_{t,n,j,b}^{\hat{g}}=(Y_{t,n,j,b},\bm{Z}_{t,n})_{t\in\mathcal{T}_n}$ and $\mathcal{H}_{t,b}^{\bm{\mathfrak{D}}^{\hat{g}}}=(\mathcal{H}_{\mathbb{T}_n^{+}+b}^Y,\mathcal{H}_{\mathbb{T}_n^{+}}^{\bm{Z}})$.

Denote the sets of times corresponding to $\bm{\mathfrak{D}}_{t,n,i,a}^{\hat{f}}$, $\bm{\mathfrak{D}}_{t,n,j,b}^{\hat{g}}$ by $\mathcal{T}_{t,n,i,a}^{\hat{f}}$, $\mathcal{T}_{t,n,j,b}^{\hat{g}}$, respectively, and let $T_{t,n,i,a}^{\hat{f}}=|\mathcal{T}_{t,n,i,a}^{\hat{f}}|$, $T_{t,n,j,b}^{\hat{g}}=|\mathcal{T}_{t,n,j,b}^{\hat{g}}|$ be the cardinalities. For each $n\in \mathbb{N}$, $t\in\mathcal{T}_n$ let $\mathcal{M}(\mathcal{Z},\mathcal{Y})\subseteq \mathcal{Y}^{\mathcal{Z}}$ and $\mathcal{M}(\mathcal{Z},\mathcal{X})\subseteq \mathcal{X}^{\mathcal{Z}}$, where $\mathcal{X}=\mathbb{R}$, $\mathcal{Y}=\mathbb{R}$, and $\mathcal{Z}=\mathbb{R}^{\bm{d_Z}}$. Note that $\bm{d_Z}$ can grow with $n$ as discussed in Section~\ref{subsection:setting_nsts}, although we suppress this in the notation. 

\begin{assumption}[Causal representations of the predictors]\label{asmpt_causal_rep_adaptive_stat_learn_algo_nsts} 

% we stress Borel measurable to emphasize that the composition will be measurable 
For each $n\in \mathbb{N}$, $(i,j,a,b)\in\mathcal{D}_n$, assume that the sequences of statistical learning algorithms $\mathcal{A}_{n,i,a}^{\hat{f}}=(\mathcal{A}_{t,n,i,a}^{\hat{f}})_{t\in\mathcal{T}_n}$, $\mathcal{A}_{n,j,b}^{\hat{g}}=(\mathcal{A}_{t,n,j,b}^{\hat{g}})_{t\in\mathcal{T}_n}$ consist of the Borel measurable functions $\mathcal{A}_{t,n,i,a}^{\hat{f}}:
(\mathcal{Z}\times\mathcal{X})^{T_{t,n,i,a}^{\hat{f}}}\xrightarrow[]{} \mathcal{M}(\mathcal{Z},\mathcal{X})$ and $\mathcal{A}_{t,n,j,b}^{\hat{g}}:(\mathcal{Z}\times\mathcal{Y})^{T_{t,n,j,b}^{\hat{g}}}\xrightarrow[]{} \mathcal{M}(\mathcal{Z},\mathcal{Y})$ defined by $\mathcal{A}_{t,n,i,a}^{\hat{f}}(   
\bm{\mathfrak{D}}_{t,n,i,a}^{\hat{f}})=\hat{f}_{t,n,i,a}$ and $\mathcal{A}_{t,n,j,b}^{\hat{g}}(\bm{\mathfrak{D}}_{t,n,j,b}^{\hat{g}})= \hat{g}_{t,n,j,b}$ such that the predictors have the causal representations
\begin{align*}     \hat{f}_{t,n,i,a}=G_{t,n,i,a}^{\mathcal{A}^{\hat{f}}}(\mathcal{H}_{t,a}^{\bm{\mathfrak{D}}^{\hat{f}}}),     \text{ }     \hat{g}_{t,n,j,b}=G_{t,n,j,b}^{\mathcal{A}^{\hat{g}}}(\mathcal{H}_{t,b}^{\bm{\mathfrak{D}}^{\hat{g}}}), \end{align*}
in view of Assumption~\ref{asmpt_causal_rep_process_nsts}. $G_{t,n,i,a}^{\mathcal{A}^{\hat{f}}}(\cdot)$, $G_{t,n,j,b}^{\mathcal{A}^{\hat{g}}}(\cdot)$ are measurable functions so that $G_{t,n,i,a}^{\mathcal{A}^{\hat{f}}}(\mathcal{H}_{t,a}^{\bm{\mathfrak{D}}^{\hat{f}}})$, $G_{t,n,j,b}^{\mathcal{A}^{\hat{g}}}(\mathcal{H}_{t,b}^{\bm{\mathfrak{D}}^{\hat{g}}})$ are well-defined function-valued random variables.
\end{assumption}

We make the following assumption for the predictions and estimation errors for some sequence of collections of distributions $(\mathcal{P}_n)_{n\in\mathbb{N}}$.

\begin{assumption}[Causal representations of the predictions and estimation errors]\label{asmpt_causal_rep_pred_proc_nsts} 

% we stress Borel measurable to emphasize that the composition will be measurable 
Assume that the predictors $\hat{f}_{t,n,i,a}$, $\hat{g}_{t,n,j,b}$ are Borel measurable functions from $\mathbb{R}^{\bm{d_Z}}$ to $\mathbb{R}$ such that for each $n\in\mathbb{N}$, $t\in\mathcal{T}_n$, $(i,j,a,b)\in\mathcal{D}_n$ we can represent the predictions as
\begin{align*}
    \hat{f}_{t,n,i,a}(\bm{Z}_{t,n})&=G_{t,n,i,a}^{\hat{f}}(\mathcal{H}_{t,a}^{\hat{f}})=[\mathcal{A}_{t,n,i,a}^{\hat{f}}(\bm{\mathfrak{D}}_{t,n,i,a}^{\hat{f}})](\bm{Z}_{t,n}),
    \\
    \hat{g}_{t,n,j,b}(\bm{Z}_{t,n})&=G_{t,n,j,b}^{\hat{g}}(\mathcal{H}_{t,b}^{\hat{g}})=[\mathcal{A}_{t,n,j,b}^{\hat{g}}(\bm{\mathfrak{D}}_{t,n,j,b}^{\hat{g}})](\bm{Z}_{t,n}),
\end{align*} and the estimation errors as
\begin{align*}
\hat{w}_{P,t,n,i,a}^{f}&=G_{P,t,n,i,a}^{\hat{w}^f}(\mathcal{H}_{t,a}^{\hat{f}})=f_{P,t,n,i,a}(\bm{Z}_{t,n})-\hat{f}_{t,n,i,a}(\bm{Z}_{t,n}),
\\ \hat{w}_{P,t,n,j,b}^{g}&=G_{P,t,n,j,b}^{\hat{w}^g}(\mathcal{H}_{t,b}^{\hat{g}})=g_{P,t,n,j,b}(\bm{Z}_{t,n})-\hat{g}_{t,n,j,b}(\bm{Z}_{t,n}),
\end{align*} in view of Assumptions~\ref{asmpt_causal_rep_process_nsts},~\ref{asmpt_causal_rep_adaptive_stat_learn_algo_nsts}, where the input sequences are
$\mathcal{H}_{t,a}^{\hat{f}}=(\mathcal{H}_{t,a}^{\bm{\mathfrak{D}}^{\hat{f}}},\mathcal{H}_{t+c_{\mathrm{max}}}^{Z})$, $\mathcal{H}_{t,b}^{\hat{g}}=(\mathcal{H}_{t,b}^{\bm{\mathfrak{D}}^{\hat{g}}},\mathcal{H}_{t+c_{\mathrm{max}}}^{Z})$, and $G_{t,n,i,a}^{\hat{f}}(\cdot)$, $G_{P,t,n,i,a}^{\hat{w}^f}(\cdot)$, $G_{t,n,j,b}^{\hat{g}}(\cdot)$, $G_{P,t,n,j,b}^{\hat{w}^g}(\cdot)$ are measurable such that $G_{t,n,i,a}^{\hat{f}}(\mathcal{H}_{t,a}^{\hat{f}})$, $G_{t,n,j,b}^{\hat{g}}(\mathcal{H}_{t,b}^{\hat{g}})$, $G_{P,t,n,i,a}^{\hat{w}^f}(\mathcal{H}_{t,a}^{\hat{f}})$, $G_{P,t,n,j,b}^{\hat{w}^g}(\mathcal{H}_{t,b}^{\hat{g}})$ are well-defined random variables. 

Lastly, assume there exists some $q\geq 2$ such that \begin{align*}\underset{n\in\mathbb{N}}{\sup} \ \underset{P\in\mathcal{P}_n}{\sup} \ \underset{t\in\mathcal{T}_n}{\max} \ \underset{(i,j,a,b)\in\mathcal{D}_n}{\max} \mathbb{E}_P(|\hat{w}_{P,t,n,i,a}^{f}|^q)&<\infty, \\ \underset{n\in\mathbb{N}}{\sup} \ \underset{P\in\mathcal{P}_n}{\sup} \ \underset{t\in\mathcal{T}_n}{\max} \ \underset{(i,j,a,b)\in\mathcal{D}_n}{\max} \mathbb{E}_P(|\hat{w}_{P,t,n,j,b}^{g}|^q)&<\infty.\end{align*}
% this is used for more control on the estimation errors in step 1.2.b of the main proof
% in the next section we will show that this is satisfied by the sieve estimator by construction + under mild assumptions on the tv-regr fns
\end{assumption}

% discussion below is needed for martingale arguments in step 1.2.b

% \iffalse
% In view of Assumption~\ref{asmpt_causal_rep_pred_proc_nsts}, we have the following causal representation for all dimensions and time-offsets of the estimation errors \begin{align*} \bm{\hat{w}}_{P,t,n}^{\bm{f}}&=\bm{G}_{P,t,n}^{\bm{\hat{w}}^{\bm{f}}}(\mathcal{H}_{t}^{\bm{\hat{f}}})=(\hat{w}_{P,t,n,i,a}^{f})_{i\in [d_X],a\in A_i},\\ \bm{\hat{w}}_{P,t,n}^{\bm{g}}&=\bm{G}_{P,t,n}^{\bm{\hat{w}}^{\bm{g}}}(\mathcal{H}_{t}^{\bm{\hat{g}}})=(\hat{w}_{P,t,n,j,b}^{g})_{j\in [d_Y],b\in B_j},\end{align*} where $\mathcal{H}_{t}^{\bm{\hat{f}}}=(\mathcal{H}_{t,a}^{\hat{f}})_{a\in A}$ and $\mathcal{H}_{t}^{\bm{\hat{g}}}=(\mathcal{H}_{t,b}^{\hat{g}})_{b\in B}$. 
% \fi

The required convergence rates for the regression estimators are stated in Section~\ref{subsection:test_guarantee_nsts}.

\subsection{Error Processes}\label{subsection:err_proc_nsts}

The most important part of our theoretical framework is the causal representation of the process of error products. Let $(\eta_{t}^{\bm{\varepsilon}},\eta_{t}^{\bm{\xi}})_{t\in\mathbb{Z}}$ be a sequence of iid random vectors, where $\eta_{t}^{\bm{\varepsilon}}=(\eta_{t,a}^{\varepsilon})_{a \in A}$, $\eta_{t}^{\bm{\xi}}=(\eta_{t,b}^{\xi})_{b\in B}$, and each of the $\eta_{t,a}^{\varepsilon}$, $\eta_{t,b}^{\xi}$ are random vectors. For each $a\in A$, $b\in B$, define the input sequences \begin{equation} \mathcal{H}_{t,a}^{\varepsilon}=(\eta_{t,a}^{\varepsilon},\eta_{t-1,a}^{\varepsilon},\ldots), \text{ }\mathcal{H}_{t,b}^{\xi}=(\eta_{t,b}^{\xi},\eta_{t-1,b}^{\xi},\ldots).\label{eqn:input_error} \end{equation} Denote the dimension of $\eta_{t,a}^{\varepsilon}=\eta_{t,a,n}^{\varepsilon}$ by $d_{\varepsilon}^{\eta}=d_{\varepsilon,n}^{\eta}$, and the dimension of $\eta_{t,b}^{\xi}=\eta_{t,b,n}^{\xi}$ by $d_{\xi}^{\eta}=d_{\xi,n}^{\eta}$, which can both change with $n$. Let $\mathcal{H}_{t}^{\bm{\hat{f}}}=(\mathcal{H}_{t,a}^{\hat{f}})_{a\in A}$, $\mathcal{H}_{t}^{\bm{\hat{g}}}=(\mathcal{H}_{t,b}^{\hat{g}})_{b\in B}$, where $\mathcal{H}_{t,b}^{\hat{g}}$, $\mathcal{H}_{t,a}^{\hat{f}}$ are from Assumption~\ref{asmpt_causal_rep_pred_proc_nsts}.

\begin{assumption}[Causal representations of the error processes]\label{asmpt_causal_rep_errors_nsts} 

Assume that for each $n \in \mathbb{N}$, $P\in\mathcal{P}_n$, $(i,j,a,b)\in\mathcal{D}_n$, $t\in\mathcal{T}_n$, we can represent the errors from Section~\ref{subsection:tv_regr_nsts} as 
$$\varepsilon_{P,t,n,i,a}=G_{P,t,n,i,a}^{\varepsilon}(\mathcal{H}_{t,a}^{\varepsilon}),\text{ }\xi_{P,t,n,j,b} = G_{P,t,n,j,b}^{\xi}(\mathcal{H}_{t,b}^{\xi}),$$ % yes, that's right, this martingale difference sequence assumption is for the error conditional on the other predictor's estimation error input sequence
% we don't need  $(\mathcal{H}_{t,a}^{\varepsilon})_{a\in A} \!\perp \!\!\! \perp  \mathcal{H}_{t}^{\bm{\hat{g}}}$, $(\mathcal{H}_{t,b}^{\xi})_{b\in B}\!\perp \!\!\! \perp  \mathcal{H}_{t}^{\bm{\hat{f}}}$
with $\mathbb{E}_P(\varepsilon_{P,t,n,i,a}|\mathcal{H}_{t}^{\bm{\hat{g}}})=0$ and $\mathbb{E}_P(\xi_{P,t,n,j,b}|\mathcal{H}_{t}^{\bm{\hat{f}}})=0$. Assume for all $t'\geq t$ and $h\geq 0$, that
\begin{align*}&\eta_{t-h}^{\bm{\varepsilon}}\perp \!\!\! \perp (\mathcal{H}_{t'}^{\bm{\hat{g}}}, \ldots, \eta_{t-h-2}^{\bm{\varepsilon}}, \eta_{t-h-1}^{\bm{\varepsilon}}, \eta_{t-h+1}^{\bm{\varepsilon}},\eta_{t-h+2}^{\bm{\varepsilon}},\ldots)\mid (\mathcal{H}_{t}^{\bm{\hat{g}}},\eta_{t}^{\bm{\varepsilon}},\ldots,\eta_{t-h+1}^{\bm{\varepsilon}}),\\ & \eta_{t-h}^{\bm{\xi}}\perp \!\!\! \perp (\mathcal{H}_{t'}^{\bm{\hat{f}}}, \ldots,\eta_{t-h-2}^{\bm{\xi}},\eta_{t-h-1}^{\bm{\xi}},\eta_{t-h+1}^{\bm{\xi}},\eta_{t-h+2}^{\bm{\xi}},\ldots) \mid  (\mathcal{H}_{t}^{\bm{\hat{f}}},\eta_{t}^{\bm{\xi}},\ldots,\eta_{t-h+1}^{\bm{\xi}}).\end{align*} $G_{P,t,n,i,a}^{\varepsilon}(\cdot)$, $G_{P,t,n,j,b}^{\xi}(\cdot)$ are measurable functions from $(\mathbb{R}^{d_{\varepsilon}^{\eta}})^{\infty}$, $(\mathbb{R}^{d_{\xi}^{\eta}})^{\infty}$, to $\mathbb{R}$, respectively, so that $G_{P,t,n,i,a}^{\varepsilon}(\mathcal{H}_{s,a}^{\varepsilon})$, $G_{P,t,n,j,b}^{\xi}(\mathcal{H}_{s,b}^{\xi})$ are well-defined random variables for each $s\in\mathbb{Z}$ and $(G_{P,t,n,i,a}^{\varepsilon}(\mathcal{H}_{s,a}^{\varepsilon}))_{s\in\mathbb{Z}}$, $(G_{P,t,n,j,b}^{\xi}(\mathcal{H}_{s,b}^{\xi}))_{s\in\mathbb{Z}}$ are stationary ergodic processes.

\end{assumption}

%%%%% Note: errors can be dependent (e.g., through the conditional variance) on the covariate process, response process, and other response process, but for concrete examples should work it out to ensure all assumptions are satisfied. More generally, the errors can be dependent on the inputs for those processes.
%%%%% Note: we can handle the autoregressive setting if the predictors for prediction at time t are estimated using data up to time t. otherwise, if we trained using all of the data, then the forecast error would be correlated with the future data since that same forecast error is what drives the process in the autoregressive setting... for an AR(1) process, the forecast error will be a MA(h-1) process, see page 39 or so of https://users.ssc.wisc.edu/~bhansen/cbc/cbc2.pdf .... also see the similar result for nonlinear nonstationary setting from Proposition 2 in Wang, Xiaoqian, and Rob J. Hyndman. "Online conformal inference for multi-step time series forecasting." \textit{arXiv preprint arXiv:2410.13115} (2024).  
% None of this dynamic retraining is necessary if you have exogenous time series.
% computationally efficient online estimation procedure for time-varying regression functions can be done with one-way kernels as in Yousuf, Kashif, and Serena Ng. "Boosting high dimensional predictive regressions with time varying parameters." \textit{Journal of Econometrics} 224, no. 1 (2021): 60-87. , and eventually we will develop an online estimation procedure for the sieve estimator from \textcite{zhouzhou_sieve}. 

Again, note that we have not defined the input sequences for the error processes separately for each dimension. The measurable functions $G_{P,t,n,i,a}^{\varepsilon}(\cdot)$, $G_{P,t,n,j,b}^{\xi}(\cdot)$ and random vectors $\eta_{t,a}^{\varepsilon},\eta_{t,b}^{\xi}$ can be defined so that each dimension of the error processes has idiosyncratic inputs.

Denote $\bm{G}_{P,t,n}^{\bm{\varepsilon}}(\mathcal{H}_t^{\bm{\varepsilon}})=(G_{P,t,n,i,a}^{\varepsilon}(\mathcal{H}_{t,a}^{\varepsilon}))_{i\in [d_X], a\in A_i}$ and $\bm{G}_{P,t,n}^{\bm{\xi}}(\mathcal{H}_t^{\bm{\xi}})=(G_{P,t,n,j,b}^{\xi}(\mathcal{H}_{t,b}^{\xi}))_{j\in [d_Y], b\in B_j}$. In view of the causal representations of the univariate error processes, we have the following causal representations for the high-dimensional error processes 
\begin{align*} \bm{\varepsilon}_{P,t,n}=\bm{G}_{P,t,n}^{\bm{\varepsilon}}(\mathcal{H}_t^{\bm{\varepsilon}}), \ 
\bm{\xi}_{P,t,n}=\bm{G}_{P,t,n}^{\bm{\xi}}(\mathcal{H}_t^{\bm{\xi}}).
\end{align*} Here, $\mathcal{H}_t^{\bm{\varepsilon}}=(\eta_{t}^{\bm{\varepsilon}},\eta_{t-1}^{\bm{\varepsilon}},\ldots)$, $\mathcal{H}_t^{\bm{\xi}}=(\eta_{t}^{\bm{\xi}},\eta_{t-1}^{\bm{\xi}},\ldots)$ with $\eta_{t}^{\bm{\varepsilon}}=(\eta_{t,a}^{\varepsilon})_{a \in A}$, $\eta_{t}^{\bm{\xi}}=(\eta_{t,b}^{\xi})_{b\in B}$ for each $t\in\mathbb{Z}$. Similarly, for each dimension/time-offset tuple $m =(i,j,a,b)\in \mathcal{D}_n$ the error products can be represented as $R_{P,t,n,m}=G_{P,t,n,m}^{R}(\mathcal{H}_{t,m}^{R})=G_{P,t,n,i,a}^{\varepsilon}(\mathcal{H}_{t,a}^{\varepsilon})G_{P,t,n,j,b}^{\xi}(\mathcal{H}_{t,b}^{\xi})$, where $\mathcal{H}_{t,m}^{R}=(\eta_{t,m}^{R},\eta_{t-1,m}^{R},\ldots)$ with $\eta_{t,m}^{R}=(\eta_{t,a}^{\varepsilon},\eta_{t,b}^{\xi})$ for each $t\in\mathbb{Z}$. Finally, denoting $\bm{G}_{P,t,n}^{\bm{R}}(\mathcal{H}_t^{\bm{R}})=(G_{P,t,n,m}^R(\mathcal{H}_{t,m}^R))_{m=(i,j,a,b)\in\mathcal{D}_n}$, we can represent all of the error products as  $$\bm{R}_{P,t,n}=\bm{G}_{P,t,n}^{\bm{R}}(\mathcal{H}_t^{\bm{R}}),$$ where $\mathcal{H}_t^{\bm{R}}=(\eta_{t}^{\bm{R}},\eta_{t-1}^{\bm{R}},\ldots)$ and $\eta_{t}^{\bm{R}}=(\eta_{t}^{\bm{\varepsilon}},\eta_{t}^{\bm{\xi}})$ for each $t\in\mathbb{Z}$. For a fixed $P\in\mathcal{P}_n$, $t\in \mathcal{T}_n$, and $n \in \mathbb{N}$, $\bm{G}_{P,t,n}^{\bm{R}}(\mathcal{H}_s^{\bm{R}})$ is a well-defined high-dimensional random vector for each $s\in\mathbb{Z}$ and $(\bm{G}_{P,t,n}^{\bm{R}}(\mathcal{H}_s^{\bm{R}}))_{s\in\mathbb{Z}}$ is a high-dimensional stationary ergodic $\mathbb{R}^{D_n}$-valued process. 

\subsection{Assumptions for Dependence and Nonstationarity}\label{subsection:dependence_nonstationarity_nsts}

We impose mild assumptions on the rate of decay in temporal dependence and the degree of nonstationarity of the error processes. Crucially, these assumptions are stated in a distribution-uniform manner, which is essential for applying the strong Gaussian approximation. This will be further elaborated upon in Section~\ref{subsection:test_guarantee_nsts}.

We quantify temporal dependence using the functional dependence measure of \textcite{wu_funct_dep_meas}, which is related to mixing conditions \parencite{Hill2025_physical_dependence_mixingale,Heinrichs2026_physical_dependence_mixing}. Let $(\tilde{\eta}_{t}^{\bm{\varepsilon}},\tilde{\eta}_{t}^{\bm{\xi}})_{t\in\mathbb{Z}}$ be an iid copy of $(\eta_{t}^{\bm{\varepsilon}},\eta_{t}^{\bm{\xi}})_{t\in\mathbb{Z}}$, where $\tilde{\eta}_{t}^{\bm{\varepsilon}}=(\tilde{\eta}_{t,a}^{\varepsilon})_{a\in A}$ and $\tilde{\eta}_{t}^{\bm{\xi}}=(\tilde{\eta}_{t,b}^{\xi})_{b\in B}$. Denote the set of tuples of error processes, dimensions, and time-offsets by   
$$\mathcal{E}=\{(\varepsilon,i,a):i\in [d_X],a\in A_i\}\cup\{(\xi,j,b):j\in [d_Y],b\in B_j\}.$$ For any tuple $(e,l,d) \in \mathcal{E}$ corresponding to a well-defined combination of an error process, dimension, and time-offset, define $$\tilde{\mathcal{H}}_{t,d,h}^{e}=(\eta_{t,d}^{e},\ldots,\eta_{t-h+1,d}^{e},\tilde{\eta}_{t-h,d}^{e},\eta_{t-h-1,d}^{e},\ldots),$$ to be $\mathcal{H}_{t,d}^{e}$ with the input $\eta_{t-h,d}^{e}$ replaced with the iid copy $\tilde{\eta}_{t-h,d}^{e}$. To be clear, each $\tilde{\eta}_{t-h,d}^{e}$ is obtained from the component indexed by $d$ in the iid copy $\tilde{\eta}_{t-h}^{\bm{e}}$ of the random vector $\eta_{t-h}^{\bm{e}}$. Similarly, define $\tilde{\mathcal{H}}_{t,m,h}^R$ as $\mathcal{H}_{t,m}^R$ with input $\eta_{t-h,m}^R$ replaced with the iid copy $\tilde{\eta}_{t-h,m}^R$ for $m=(i,j,a,b)\in\mathcal{D}_n$, and define $\tilde{\mathcal{H}}_{t,h}^{\bm{R}}$ as $\mathcal{H}_{t}^{\bm{R}}$ with input $\eta_{t-h}^{\bm{R}}$ replaced with the iid copy $\tilde{\eta}_{t-h}^{\bm{R}}$.

Let $\mathcal{H}_{t}^{\bm{\hat{\kappa}}}$ be the inputs for the other target's predictor, i.e., $(\bm{e},\bm{\hat{\kappa}})\in\{(\bm{\varepsilon},\bm{\hat{g}}),(\bm{\xi},\bm{\hat{f}})\}$. Define $$\bar{\mathcal{H}}_{t,d,h}^{e}=(\eta_{t,d}^{e},\ldots,\eta_{t-h+1,d}^{e},\bar{\eta}_{t-h,d}^{e},\eta_{t-h-1,d}^{e},\ldots),$$ as $\mathcal{H}_{t,d}^{e}$ with $\eta_{t-h,d}^{e}$ replaced with the component indexed by $d$, $\bar{\eta}_{t-h,d}^{e}$, in the $\sigma(\eta_{t}^{\bm{e}},\ldots,\eta_{t-h+1}^{\bm{e}},\mathcal{H}_{t}^{\bm{\hat{\kappa}}})$-conditionally iid copy $\bar{\eta}_{t-h}^{\bm{e}}$ of $\eta_{t-h}^{\bm{e}}$, where $\mathcal{L}(\bar{\eta}_{t-h}^{\bm{e}}| \eta_{t}^{\bm{e}},\ldots,\eta_{t-h+1}^{\bm{e}},\mathcal{H}_{t}^{\bm{\hat{\kappa}}})$ is equal to $\mathcal{L}(\eta_{t-h}^{\bm{e}}|\eta_{t}^{\bm{e}},\ldots,\eta_{t-h+1}^{\bm{e}},\mathcal{H}_{t}^{\bm{\hat{\kappa}}})$ almost surely, and $\bar{\eta}_{t-h}^{\bm{e}}$ is conditionally independent of $(\eta_{t-h}^{\bm{e}},\eta_{t-h-1}^{\bm{e}},\ldots)$ given $(\eta_{t}^{\bm{e}},\ldots,\eta_{t-h+1}^{\bm{e}},\mathcal{H}_{t}^{\bm{\hat{\kappa}}})$. We suppress the dependence of $\bar{\eta}_{t-h}^{\bm{e}}$ on the distribution $P$, as well as $t$ and $h$ separately, to simplify the notation.

%%%%% We actually do not need to assume dependence this for the original process for the strong Gaussiann approximation -- only the error processes. This means, we can indeed handle very long range dependence. The original processes still must admit a causal representation but that alone is not a strong assumption at all. Of course, we still need to verify that the regression estimator convergence rates are fast enough.
   
%%%%% Since we are already assuming control for the L-infinity norm and L-infinity version of the functional dependence measure, it is not needed to separately define the control for L-q norms and L-q functional dependence measure. 

\begin{definition}[Functional dependence measure]\label{def_funct_dep_nsts} 

% L-infinity version of FDM to keep estimation errors in terms of second moment 
% for step 1.2
We define the following measures of temporal dependence for each $n\in\mathbb{N}$, $P\in\mathcal{P}_n$, and $t\in\mathcal{T}_n$. First, define the $L^{\infty}(P)$ version of the functional dependence measure for $G_{P,t,n,l,d}^e(\mathcal{H}_{t,d}^e)$, $(e,l,d)\in\mathcal{E}$, with $h\in\mathbb{N}_0$ as  $$\theta_{P,t,n,l,d}^{e,\infty}(h)= \inf\{K \geq 0 : \mathbb{P}_P(|G_{P,t,n,l,d}^e(\mathcal{H}_{t,d}^e)-G_{P,t,n,l,d}^e(\tilde{\mathcal{H}}_{t,d,h}^e)|>K)=0\},$$ and the corresponding conditional version as
$$\bar{\theta}_{P,t,n,l,d}^{e,\infty}(h)= \inf\{K \geq 0 : \mathbb{P}_P(|G_{P,t,n,l,d}^e(\mathcal{H}_{t,d}^e)-G_{P,t,n,l,d}^e(\bar{\mathcal{H}}_{t,d,h}^e)|>K)=0\}.$$

% for strong Gaussian approximation
Second, define the functional dependence measures for the processes of error products $G_{P,t,n,m}^R(\mathcal{H}_{t,m}^R)$ for each $m=(i,j,a,b)\in\mathcal{D}_n$ with $h \in \mathbb{N}_0$, and some $q\geq 1$ as $$\theta_{P,t,n,m}^R(h,q) = [\mathbb{E}_P(|G_{P,t,n,m}^R(\mathcal{H}_{t,m}^R)-G_{P,t,n,m}^R(\tilde{\mathcal{H}}_{t,m,h}^R)|^{q})]^{1/q},$$ and for the vector-valued process $\bm{G}_{P,t,n}^{\bm{R}}(\mathcal{H}_{t}^{\bm{R}})$ with $h \in \mathbb{N}_0$, and some $q\geq 1$, $r\geq 1$ as $$\theta_{P,t,n}^{\bm{R}}(h,q,r) = [\mathbb{E}_P(||\bm{G}_{P,t,n}^{\bm{R}}(\mathcal{H}_{t}^{\bm{R}})-\bm{G}_{P,t,n}^{\bm{R}}(\tilde{\mathcal{H}}_{t,h}^{\bm{R}})||_r^{q})]^{1/q}.$$
\end{definition}

For some sequence of collections of distributions $(\mathcal{P}_n)_{n\in\mathbb{N}}$, we make the following assumption about the temporal dependence. We only require the relatively mild assumption that it decays polynomially, rather than geometrically. Note that we will often write the time of the input sequence as $0$ when it does not matter due to stationarity. 

% Reason we need L infinity moment and L infinity FDM
% For L infinity FDM need for step 1.2 so we only need second moment on the convergence rate for tv np regression estimators…. 
% For L infinity moment on error need for step 1.2 for the bound on the moment of error * PE so only need for second moment on the second moment on estimation error (so regression function if prediction is bounded as in sieve)

\begin{assumption}[Distribution-uniform decay of temporal dependence]\label{asmpt_funct_dep_nsts}

% L infinity and β > 1 required for Step 1.2 to keep the PE requirements in terms of L2 norm
% Proposition 4.3 of \textcite{seq_gauss_approx2022} requires q > 4, β > 2, for the product of errors, so we COULD alternatively assume each error has β > 4 so that the requirements can satisfied for product of errors by Holder's inequality, but INSTEAD we will just assume separately for product of errors (see below)
% only need beta > 1 for partial FDM 
% dont directly need to use bound on Lp norm for each individual error (let alone L-infinity norm)  \underset{P\in\mathcal{P}_n}{\sup}||G_{P,t,n,l,d}^e(\mathcal{H}_{0,d}^e)||_{L^{\infty}(P)} \leq \bar{\Theta}^{\infty},\text{ }\text{ }\text{ } 
% however, we do need the L-infinity PDM for Step 1.2 of main theorem
% note that for Sieve dGCM we do need Lp for p>2 
Assume there exist $\bar{\Theta}^{\infty} >0$, $\bar{\beta}^{\infty} > 1$ such that for all $(e,l,d)\in\mathcal{E}$, we have, for $h\geq 0$, \begin{align*}\underset{n\in\mathbb{N}}{\sup} \ \underset{P\in\mathcal{P}_{n}}{\sup} \ \underset{t\in\mathcal{T}_n}{\max} \  ||G_{P,t,n,l,d}^e(\mathcal{H}_{0,d}^e)||_{L^{\infty}(P)} &\leq \bar{\Theta}^{\infty}, \\ \underset{n\in\mathbb{N}}{\sup} \ \underset{P\in\mathcal{P}_{n}}{\sup} \ \underset{t\in\mathcal{T}_n}{\max} \  \max(\theta_{P,t,n,l,d}^{e,\infty}(h),\bar{\theta}_{P,t,n,l,d}^{e,\infty}(h)) &\leq \bar{\Theta}^{\infty} \cdot (h\lor 1)^{-\bar{\beta}^{\infty}}.\end{align*}

% Proposition 4.3 of \textcite{seq_gauss_approx2022} requires q > 4, β > 2, for the product of errors
% however we will require β > 3 so we can use the limiting rates for growth of dimension
For additional control in terms of the product of errors alone, also assume there exist $\bar{\Theta}^R>0$, $\bar{\beta}^R>3$, $\bar{q}^R>4$, such that we have, for $h\geq 0$, \begin{align*}\underset{n\in\mathbb{N}}{\sup} \ \underset{P\in\mathcal{P}_n}{\sup} \ \underset{t\in\mathcal{T}_n}{\max} \ \underset{m\in\mathcal{D}_n}{\max}[\mathbb{E}_P(|G_{P,t,n,m}^R(\mathcal{H}_{0,m}^R)|^{\bar{q}^R})]^{1/\bar{q}^R} &\leq \bar{\Theta}^R,\\ \underset{n\in\mathbb{N}}{\sup} \ \underset{P\in\mathcal{P}_n}{\sup} \ \underset{t\in\mathcal{T}_n}{\max} \ \underset{m\in\mathcal{D}_n}{\max}\theta_{P,t,n,m}^R(h,\bar{q}^R)&\leq \bar{\Theta}^R \cdot (h\lor 1)^{-\bar{\beta}^R}.\end{align*}

 % PDM on the product of errors is so that the strong Gaussian approximation can be used. Note that if we assumed the L-infinity norm of each of the errors were bounded then the original assumptions for each of the errors would yield a bound for the product of errors
 
\end{assumption}

A few remarks are in order. First, the constants in Assumption~\ref{asmpt_funct_dep_nsts} do not depend on $n$. Second, the assumptions on the individual error processes can be weakened; see Section D of the supplement. Third, by Jensen's inequality, for each $n\in\mathbb{N}$ and $h\geq 0$, we have  \begin{align*} \underset{P\in\mathcal{P}_n}{\sup} \ \underset{t\in\mathcal{T}_n}{\max} \ [\mathbb{E}_P(||\bm{G}_{P,t,n}^{\bm{R}}(\mathcal{H}_{0}^{\bm{R}})||_2^{\bar{q}^R})]^{1/\bar{q}^R} &\leq D_n^{\frac{1}{2}}\bar{\Theta}^R, \\  \ \underset{P\in\mathcal{P}_n}{\sup} \ \underset{t\in\mathcal{T}_n}{\max} \  \theta_{P,t,n}^{\bm{R}}(h,\bar{q}^R,2)&\leq D_n^{\frac{1}{2}}\bar{\Theta}^R \cdot (h\lor 1)^{-\bar{\beta}^R}.\end{align*}

Next, for some sequence of collections of distributions $(\mathcal{P}_n)_{n\in\mathbb{N}}$, we make the following assumption to control the nonstationarity of the process of error products.

\begin{assumption}[Distribution-uniform total variation condition for nonstationarity]\label{asmpt_total_variation_nsts}

Recall $\bar{\Theta}^R$ from Assumption~\ref{asmpt_funct_dep_nsts}. 
Assume for each $n\in\mathbb{N}$, there exists $\bar{\Gamma}_n^R \geq 1$ such that 
$$\underset{P\in\mathcal{P}_n}{\sup}\left(\sum_{t=\mathbb{T}_n^{-}+1}^{\mathbb{T}_n^{+}}  \left(\mathbb{E}_P||\bm{G}_{P,t,n}^{\bm{R}}(\mathcal{H}_{0}^{\bm{R}})-\bm{G}_{P,t-1,n}^{\bm{R}}(\mathcal{H}_{0}^{\bm{R}})||_2^{2}\right)^{1/2}\right) \leq \bar{\Theta}^R \bar{\Gamma}_n^R.$$

\end{assumption}

\subsection{Main Theoretical Result}\label{subsection:test_guarantee_nsts}

We now present the result that justifies the simulation-assisted testing procedure from Algorithm~\ref{algo:dGCM_test}. This result relies on the distribution-uniform strong Gaussian approximation from Section C of the supplement applied to the process of error products. The approximating nonstationary Gaussian process has a time-varying covariance structure that is characterized by local long-run covariances.

\begin{definition}[Local long-run covariance matrices of the process of error products]\label{def_loc_lr_cov_nsts} For each $P\in\mathcal{P}_n$, $t\in \mathcal{T}_n$, $n\in \mathbb{N}$, define the local long-run covariance matrix $\bm{\Sigma}_{P,t,n}^{\bm{R}}\in\mathbb{R}^{D_n\times D_n}$ of the $\mathbb{R}^{D_n}$-valued stationary process $(\bm{G}_{P,t,n}^{\bm{R}}(\mathcal{H}_s^{\bm{R}}))_{s\in\mathbb{Z}}$ by
$$\bm{\Sigma}_{P,t,n}^{\bm{R}}=\sum_{h \in \mathbb{Z}} \mathrm{Cov}_P(\bm{G}_{P,t,n}^{\bm{R}}(\mathcal{H}_0^{\bm{R}}), \bm{G}_{P,t,n}^{\bm{R}}(\mathcal{H}_h^{\bm{R}})).$$
\end{definition}

In view of the Gaussian approximation theory developed in \textcite{seq_gauss_approx2022}, we only require an estimator of the cumulative covariance matrices of the error products $$\sum_{s=\mathbb{T}_n^{-}}^t \bm{\Sigma}_{P,s,n}^{\bm{R}},$$
rather than the local long-run covariance matrices $\bm{\Sigma}_{P,t,n}^{\bm{R}}$ at each time $t$ individually. This is critical for the practical applicability of our method, as estimating individual local long-run covariance matrices can be extremely challenging in practice. We use the estimator
\begin{equation}
\hat{Q}_{t,n}^{\bm{R}} = \sum_{s=L_n+\mathbb{T}_n^{-}-1}^t \frac{1}{L_n} \left( \sum_{r=s-L_n+1}^{s} \bm{\hat{R}}_{r,n} \right)\left( \sum_{r=s-L_n+1}^{s} \bm{\hat{R}}_{r,n} \right)^{\top}, \label{eqn:tv_cov_estimator} \end{equation} for lag-window size $L_n\in\mathbb{N}$. We discuss how to select $L_n$ in Section D of the supplement.

% Vinogradov https://en.wikipedia.org/wiki/Big_O_notation#History_(Bachmann%E2%80%93Landau,_Hardy,_and_Vinogradov_notations)
To account for the estimation errors for the time-varying regression functions and the cumulative covariance matrices, as well as the error for the Gaussian approximation, we introduce offsets $\tau_n\xrightarrow[]{}0$, $\nu_n\xrightarrow[]{}0$ so that $\tau_n=o(\log^{-(1+\delta)}(T_n))$ for some $\delta >0$ and 
\begin{equation}\label{offset_condition_nsts} 
\nu_n \gg \log(T_n)D_n\left[\left(\frac{D_n}{T_n}\right)^{2\xi(\bar{q}^R,\bar{\beta}^R)} + \tau_n^{-2} \left( \varphi_{n,1}+\varphi_{n,2}\right)\right],\end{equation} where $$\varphi_{n,1} = T_n^{-\frac{1}{2}} (\bar{\Gamma}_n^R)^{\frac{1}{2}}L_n^{\frac{1}{4}} + T_n^{-\frac{1}{4}} D_n^{\frac{1}{4}} L_n^{\frac{1}{4}} + L_n^{-\frac{1}{2}} + L_n^{1-\frac{\bar{\beta}^R}{2}} +T_n^{-1},$$ comes from the covariance estimation error, and $$\varphi_{n,2}=\tau_n^{\frac{7}{2}}D_n^{-\frac{5}{4}}+\tau_n^7 D_n^{-\frac{5}{2}},$$ comes from the time-varying regression function estimation errors. Also, the lag-window size $L_n$ from~\eqref{eqn:tv_cov_estimator} must satisfy $L_n \asymp T_n^{\zeta}$ for some $\zeta\in (0,\frac{1}{2})$ so that $\tau_n^{-6}D_n^2 L_n^{-1}=o(1)$ and $\bar{\Gamma}_n^R T_n^{-1} D_n^2 \tau_n^{-6} L_n^{\frac{1}{2}}=o(1)$, where $\bar{\Gamma}_n^R$ is from Assumption~\ref{asmpt_total_variation_nsts}. We additionally require that $$\tau_n^{-1} D_n^{\frac{1}{2}} (\bar{\Gamma}_n^R)^{\frac{1}{2}\frac{\bar{\beta}^R-2}{\bar{\beta}^R-1}} \sqrt{\log(T_{n,L})}\left(\frac{D_n}{T_{n,L}}\right)^{\xi(\bar{q}^R,\bar{\beta}^R)}=o(1),$$ which ensures that the strong Gaussian approximation error is asymptotically negligible. We see that the offsets depend on the number of observations $T_n$ from Section~\ref{subsection:setting_nsts}, the intrinsic dimensionality $D_n$ from Section~\ref{subsection:setting_nsts}, the degree of nonstationarity $\bar{\Gamma}_n^R$ from Assumption~\ref{asmpt_total_variation_nsts}, and the lag-window parameter $L_n$ from~\eqref{eqn:tv_cov_estimator}. $\xi(\bar{q}^R,\bar{\beta}^R)$ is a rate defined in Section C of the supplement that depends on the constants $\bar{\beta}^R$, $\bar{q}^R$ from Assumption~\ref{asmpt_funct_dep_nsts}.

The following result establishes an asymptotic Type I error control guarantee for our testing procedure from Algorithm~\ref{algo:dGCM_test}, which holds approximately in large samples as $n\xrightarrow[]{} \infty$, provided that the previously stated assumptions hold and the regression estimation errors \begin{align}\label{pred_errs} 
\hat{w}_{P,t,n,i,a}^{f}=f_{P,t,n,i,a}(\bm{Z}_{t,n})-\hat{f}_{t,n,i,a}(\bm{Z}_{t,n}),
\text{ }
\hat{w}_{P,t,n,j,b}^{g}=g_{P,t,n,j,b}(\bm{Z}_{t,n})-\hat{g}_{t,n,j,b}(\bm{Z}_{t,n}),\end{align} converge to zero sufficiently fast, in some sense; see Theorem~\ref{thm:test_nsts_DR} for the exact rates. If it were known, we could correctly calibrate our test with the (random) quantile function $\hat{q}$ of $S_{n,p}(\bm{\breve{R}}_{n})$, where $\bm{\breve{R}}_{n}=(\bm{\breve{R}}_{t,n})_{t\in\mathcal{T}_{n,L}}$ and $\bm{\breve{R}}_{t,n} \sim \mathcal{N}(0, \bm{\hat{\Sigma}}_{t,n}^{\bm{R}})$ for all $t\in\mathcal{T}_{n,L}$. In practice, $\hat{q}$ is numerically approximated by conducting a large number of Monte Carlo simulations, and we use $\hat{q}^{\text{boot}}$ from Algorithm~\ref{algo:dGCM_test} in its place.

\begin{theorem}\label{thm:test_nsts_DR}

Suppose Assumptions~\ref{asmpt_causal_rep_process_nsts},~\ref{asmpt_causal_rep_adaptive_stat_learn_algo_nsts},~\ref{asmpt_causal_rep_pred_proc_nsts},~\ref{asmpt_causal_rep_errors_nsts},~\ref{asmpt_funct_dep_nsts},~\ref{asmpt_total_variation_nsts} related to the temporal dependence and nonstationarity hold for the sequence of collections of distributions $(\mathcal{P}_{0,n}^{\ast})_{n\in\mathbb{N}}$, where $\mathcal{P}_{0,n}^{\ast}\subset\mathcal{P}_{0,n}^{\mathrm{CI}}$ for each $n\in\mathbb{N}$. Further, suppose the regression estimation errors satisfy
\begin{align*}
\underset{P\in\mathcal{P}_{0,n}^{\ast}}{\sup}\underset{(i,j,a,b)\in\mathcal{D}_n}{\max}\underset{t\in\mathcal{T}_n}{\max}\text{ }\mathbb{E}_P\left(\left|\hat{w}_{P,t,n,i,a}^{f}\right|^2 \right)^{\frac{1}{2}} \mathbb{E}_P\left(\left|\hat{w}_{P,t,n,j,b}^{g}\right|^2 \right)^{\frac{1}{2}}&= o(T_n^{-\frac{1}{2}} \tau_n^{7} D_n^{-\frac{3}{2}}),\\
 \underset{P\in\mathcal{P}_{0,n}^{\ast}}{\sup}\underset{i\in [d_X], a\in A_i}{\max}\underset{t\in\mathcal{T}_n}{\max}\text{ }  \mathbb{E}_P\left(\left|\hat{w}_{P,t,n,i,a}^{f}\right|^2 \right)^{\frac{1}{2}} &=o(\tau_n^{7} D_n^{-\frac{5}{2}}),\\
 \underset{P\in\mathcal{P}_{0,n}^{\ast}}{\sup}\underset{j\in[d_Y],b\in B_j}{\max}\underset{t\in\mathcal{T}_n}{\max} \text{ }\mathbb{E}_P\left(\left|\hat{w}_{P,t,n,j,b}^{g}\right|^2 \right)^{\frac{1}{2}}&=o( \tau_n^{7} D_n^{-\frac{5}{2}}).
 \end{align*} If the offsets $\tau_n \xrightarrow[]{} 0$ and $\nu_n\xrightarrow[]{} 0$ are chosen such that condition~\eqref{offset_condition_nsts} holds, then we have  $$\underset{n \xrightarrow[]{}\infty}{\limsup}\text{ }\underset{P\in\mathcal{P}_{0,n}^{\ast}}{\sup}\mathbb{P}_P\left(S_{n,p}(\bm{\hat{R}}_{n})  > \hat{q}_{1-\alpha +\nu_n}+\tau_n\right)\leq \alpha.$$ 
\end{theorem}

% Note: we allow for $D_n=O(T_n^{\frac{1}{8}})$ 
% Note: the sieve estimator with Legendre polynomials as basis functions can satisfy this fastest rate of growth since it will be log^4(n)/n^{1/2} for each so log^8(n)/n multiplying

% Note: we just choose alpha' in Algorithm 1, we do not attempt to "choose" the offsets (unknown constants etc etc), just observe the rejection rates as n grows in the simulations with fixed regression complexity and nonstationarity (both smooth for sieve).

The dGCM test possesses a property known as rate double robustness, which means that we place stronger convergence rate requirements on the products of the $L^2(P)$ norms of the estimation errors than on each one individually.

%%% Note: This property can be especially useful in the contexts of causal discovery for time-delayed effects and variable selection in time series forecasting. For instance, a faster convergence rate of a nowcasting model can compensate for a slower convergence rate of a forecasting model.

We demonstrate that the convergence rates required by Theorem~\ref{thm:test_nsts_DR} for estimating the time-varying regression functions can be achieved in Section A of the supplement. Under additional mild assumptions about the temporal dependence and nonstationarity of the time series, ensuring they are approximately stationary over short time segments, the sieve regression estimator from \textcite{zhouzhou_sieve} achieves the required rates. These processes, known as locally stationary processes, are a well-studied class of nonstationary processes that is nested within the general triangular array framework from this section.

\section{Numerical Simulations}\label{section:simulations}

We study the finite-sample performance of the dGCM test. Additional simulation results are reported in the supplement. We emphasize that our method can be used with any regression method for nonstationary time series, e.g., kernel \parencite{vogt,wbwu_timevar_npreg, mult_str_wu_2022}, sieve \parencite{zhouzhou_sieve}, and neural network  \parencite{kurisu_NSTS_DNN_2025} estimators. In the simulations below, we use the sieve method from \textcite{zhouzhou_sieve}. In Section A of the supplement, we prove that our test based on this sieve estimator has uniformly asymptotic Type I error control.

We refer to this instantiation of our dGCM test as the Sieve-dGCM test, which consists of running Algorithm~\ref{algo:dGCM_test} using the sieve method of \textcite{zhouzhou_sieve} to estimate the regression functions. The other aspects of the data generating process (e.g., functions for conditional variance and autocorrelation) need not be estimated. We use $N^{\mathrm{sim}}=5000$ simulations of the nonstationary Gaussian process from Algorithm~\ref{algo:dGCM_test} to approximate the desired quantile of the test statistic. We use Legendre polynomials as the basis functions for the sieve method. In Section D of the supplement, we discuss how to select the number of basis functions for the sieve method with a cross-validation scheme, as well as how to select the lag-window parameter for covariance estimation with a minimum volatility method.

We compare the dGCM test with the generalized covariance measure (GCM) test \parencite{shah_gcm_2020} using a generalized additive model, and the residual prediction test (RPT) \parencite{gof_hdlm_2018,inv_causal_pred_2018} using the Nystr\"{o}m method and a random forest model. We use the implementations from the \texttt{CondIndTests} R package \parencite{inv_causal_pred_2018}. We also examine how the dGCM test performs in the hypothetical scenario in which the time-varying regression functions are estimated perfectly. We refer to this test as Oracle-dGCM, which consists of running Algorithm~\ref{algo:dGCM_test} using the predictions from the true time-varying regression functions. We first generate $100$ realizations of the processes at sample sizes $n\in\{250,500,750, 1000\}$, and then we calculate the empirical rejection rates for each test using the significance level $\alpha=0.05$. To simplify the notation, in this section we suppress the dependence on the distribution $P\in\mathcal{P}_n$.

% Again, we select the lag-window parameter via the minimum volatility method from Section D of the supplement and we use $N^{\mathrm{sim}}=5000$ Monte Carlo simulations. 
% We did not compare with the kernel conditional independence (KCI) test from \textcite{kun_zhang_kci_test} due to its prohibitively high computational cost (its complexity is cubic in the sample size). 

To simplify the exposition, we use Gaussian shocks for the error processes, which are not covered by the theory due to the bounded $L^{\infty}(P)$ norm requirement from Assumption~\ref{asmpt_funct_dep_nsts}. This assumption can be weakened, and is just made to simplify the assumptions; see Section D of the supplement. Instead, one may use truncated Gaussian shocks. Otherwise, our setups adhere to the locally stationary framework described in Section A of the supplement.

% Note: We emphasize that all of our simulation setups adhere to the asymptotic framework of locally stationary processes as described in Section A of the supplement. Within this framework, as $n$ grows we gain more and more observations of each local structure of the nonstationary process. 
% Note: Standard long-run asymptotics for stationary processes emerge as a special case of this infill asymptotic framework when all parameter curves and regression functions are time-invariant. Similarly, the iid setting arises as another special case when there is neither nonstationarity nor temporal dependence. 

% We will report simulation results for other settings and compare with additional conditional independence tests in a separate manuscript; see Section~\ref{subsection:conclusion} for more details. 

% Note: For each of the tvAR(1) processes below, we set the initial condition to be 0.0001 as in Chen, Likai, Ekaterina Smetanina, and Wei Biao Wu. "Estimation of nonstationary nonparametric regression model with multiplicative structure." The Econometrics Journal 25, no. 1 (2022): 176-214. Alternatively, we could use a "buffer" or "burn-in period" of B times with the parameter theta(0) and only apply the tests to the observations B+1,...,B+n to reduce dependence on the initial condition. 

% Note: We emphasize that stationary processes, independent sequences with time-varying distributions (i.e. distribution shift), and iid sequences are all special cases of the frameworks for nonstationary processes from Sections~\ref{section:CI_test_theory_nsts} and~\ref{section:CI_test_SIEVE}. 

We study the setting with $d_X=1$, $d_Y=1$, $d_Z=1$ and no time-offsets, so $A=\{0\}$, $B=\{0\}$, $C=\{0\}$, and $\mathcal{T}_n=[n]$. We test the null hypothesis $X_{t,n} \perp \!\!\! \perp Y_{t,n} \mid  Z_{t,n}$ for all times $t\in \mathcal{T}_n$. % versus the alternative hypothesis   $$X_{t,n}\not\! \perp \!\!\! \perp Y_{t,n} \mid  Z_{t,n} \text{ for all times } t\in \mathcal{T}_n,$$ because we assume that we can restrict the collection of alternative distributions to be those for processes with time-invariant conditional dependencies.

We couple the processes $X$ and $Y$ by using correlated shocks for the error processes. Let the covariate process be a tvAR(1) process (i.e. a time-varying AR(1) process) defined by $$Z_{t,n}=\theta^Z(t/n) Z_{t-1,n}+\eta_t^Z,$$ where the parameter curve $\theta^{Z}:[0,1]\xrightarrow[]{}\mathbb{R}$ is given by $\theta^Z(u) =0.35+0.2\cos(2\pi u)$, and the shocks $(\eta_t^Z)_{t\in \mathcal{T}_n}$ are sampled iid from a standard normal distribution. Let
\begin{align*}
    X_{t,n}= f_K(Z_{t,n},t/n)+\varepsilon_{t,n},\text{ }  Y_{t,n} = g_K(Z_{t,n},t/n)+ \xi_{t,n},\end{align*} where the functions $f_{K},g_{K}:\mathbb{R}\times [0,1]\xrightarrow[]{}\mathbb{R}$ are defined by \begin{align} f_K(z,u)&= (0.5+0.25\cos(2\pi u) )\exp(-z^2)\sin(Kz),\label{eqn:f_tvreg_corrshock} \\ g_K(z,u)&=(0.3+0.15\sin(\pi u) ) \exp(-z^2)\cos(Kz),\label{eqn:g_tvreg_corrshock}\end{align} with regression complexity parameter $K\in \{1,2,3,4\}$. Define the error processes as \begin{align*}
    \varepsilon_{t,n}&=\sigma^{\varepsilon}(Z_{t,n},t/n)\varepsilon_{t,n}',\text{ }\varepsilon_{t,n}' = \theta^{\varepsilon}(t/n) \varepsilon_{t-1,n}' +  \eta_{t}^{\varepsilon}, \\ \xi_{t,n}&= \sigma^{\xi}(Z_{t,n},t/n)\xi_{t,n}',\text{ } \xi_{t,n}' = \theta^{\xi}(t/n)  \xi_{t-1,n}'+  \eta_{t}^{\xi},
\end{align*}
where the parameter curves $\theta^{\varepsilon},\theta^{\xi}:[0,1]\xrightarrow[]{} \mathbb{R}$ are given by $$\theta^{\varepsilon}(u) =0.4 +0.2\sin(\pi u), \text{ } \theta^{\xi}(u) = 0.5+0.25\sin(2\pi u),$$ and where the functions $\sigma^{\varepsilon},\sigma^{\xi}:\mathbb{R}\times [0,1]\xrightarrow[]{} \mathbb{R}$ are given by \begin{align*}
\sigma^{\varepsilon}(z,u)&=0.2+(0.5+0.25\sin(2\pi u))\left(\frac{\exp(-5z)}{1+\exp(-5z)}\right),\\ \sigma^{\xi}(z,u)&=0.5+(0.4+0.2\cos(2\pi u))\exp(-z^2)\sin(z).\end{align*} The shocks $(\eta_{t}^{\varepsilon},\eta_{t}^{\xi})_{t\in \mathcal{T}_n}$ are sampled iid from a centered bivariate normal distribution with unit variances and correlation $\rho \in \{0,0.3,0.6,0.9\}$,
$$\begin{bmatrix} \eta_{t}^{\varepsilon} \\ \eta_{t}^{\xi} \end{bmatrix} \sim \mathcal{N} \left(  \begin{bmatrix} 0 \\ 0 \end{bmatrix},  \begin{bmatrix} 1 & \rho \\ \rho & 1 \end{bmatrix} \right), \text{ } t\in \mathcal{T}_n.$$ The null hypothesis is true when $\rho = 0$, and the alternative hypothesis is true when $\rho \neq 0$.

% For the hypothetical scenario in which the time-varying regression functions are perfectly estimated, we use the same tvAR(1) error processes with correlated shocks as in the previously described setup. 
% Note: For these \say{oracle} simulations, the residual products will be $\sigma^{\varepsilon}(Z_{t,n},t/n)\varepsilon_{t,n} \sigma^{\xi}(Z_{t,n},t/n)\xi_{t,n}$. 

\begin{figure}[ht!]  %The figure environment for floating and captioning    
\centering % Centers the image
\includegraphics[width=1\textwidth, keepaspectratio]{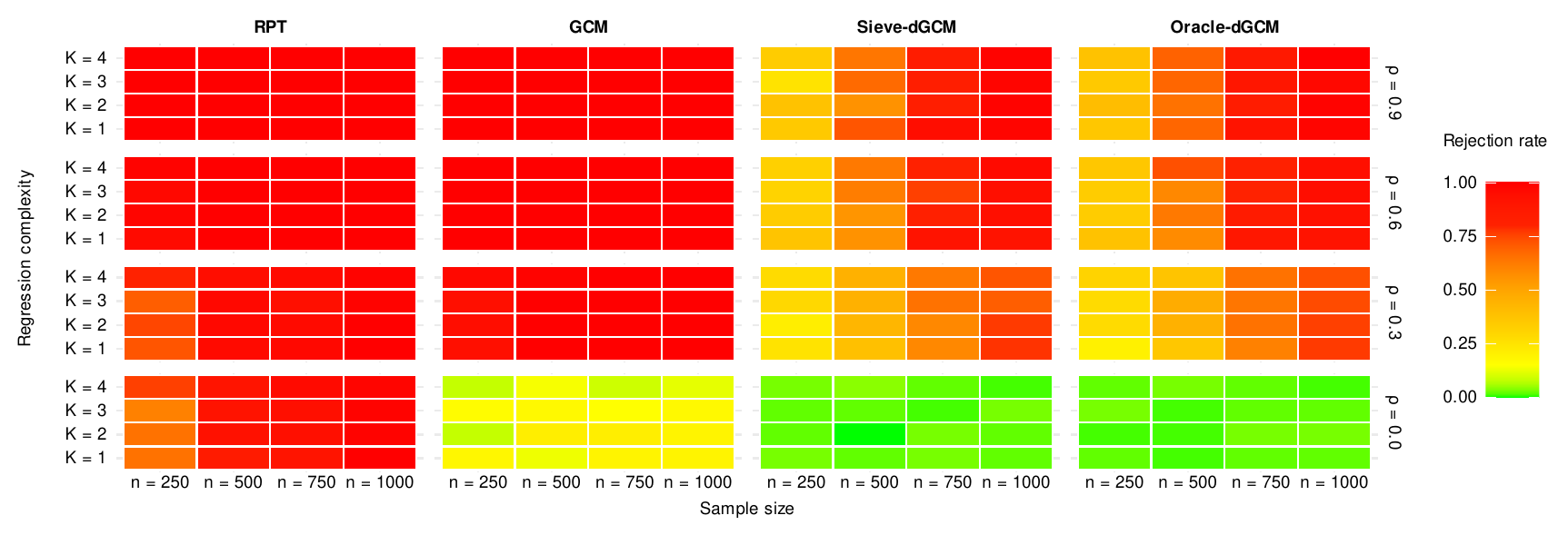}\caption{Our test holds the level even with fairly small sample sizes, and gains power as we increase the correlation and sample size. The other tests fail to hold the level.}
\label{figure:corrshock_heatmap}
\end{figure}

\begin{figure}[ht!]  %The figure environment for floating and captioning    
\centering % Centers the image
\includegraphics[width=1\textwidth, keepaspectratio]{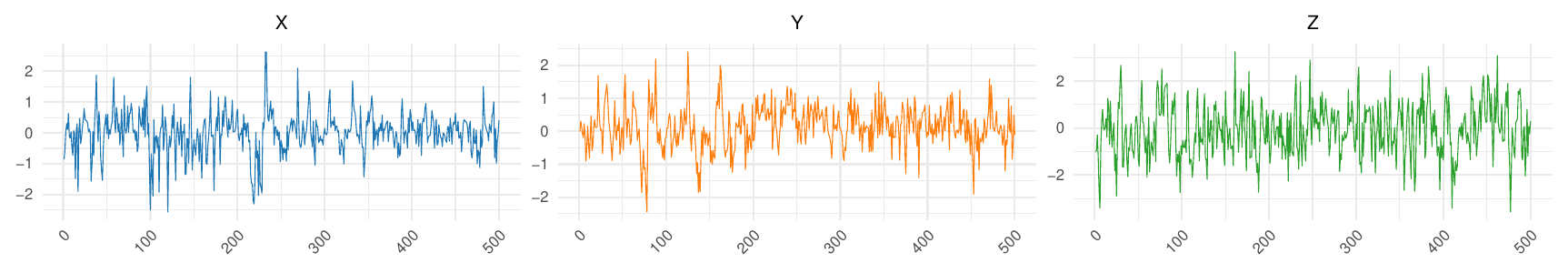}\caption{One realization from the null distribution with $\rho=0$, $K=1$, $n=500$.}
\label{figure:XYZtsplot_corrshock_n_500_k_1_corr_0_r_22}
\end{figure}

\begin{figure}[ht!]  %The figure environment for floating and captioning    
\centering % Centers the image
\includegraphics[width=1\textwidth, keepaspectratio]{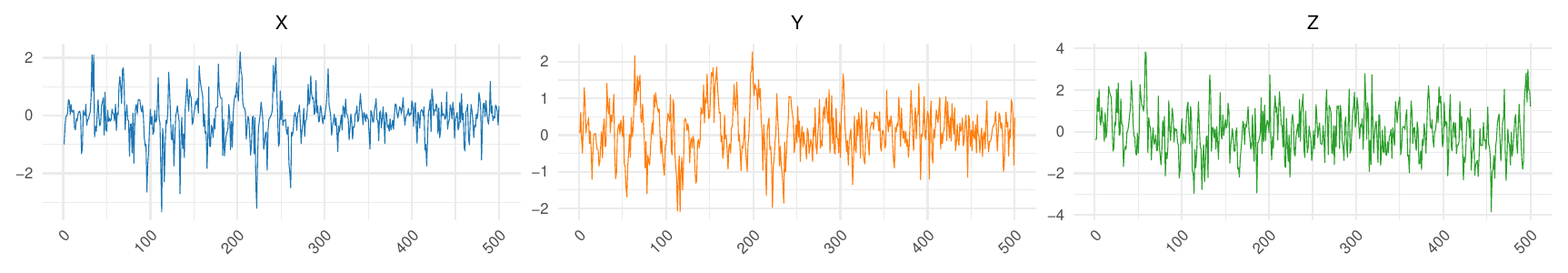}\caption{One realization from the alternative distribution with $\rho=0.9$, $K=1$, $n=500$.}
\label{figure:XYZtsplot_corrshock_n_500_k_1_corr_9_r_88}
\end{figure}

\begin{figure}[ht!]  %The figure environment for floating and captioning    
\centering % Centers the image
\includegraphics[width=1\textwidth, keepaspectratio]{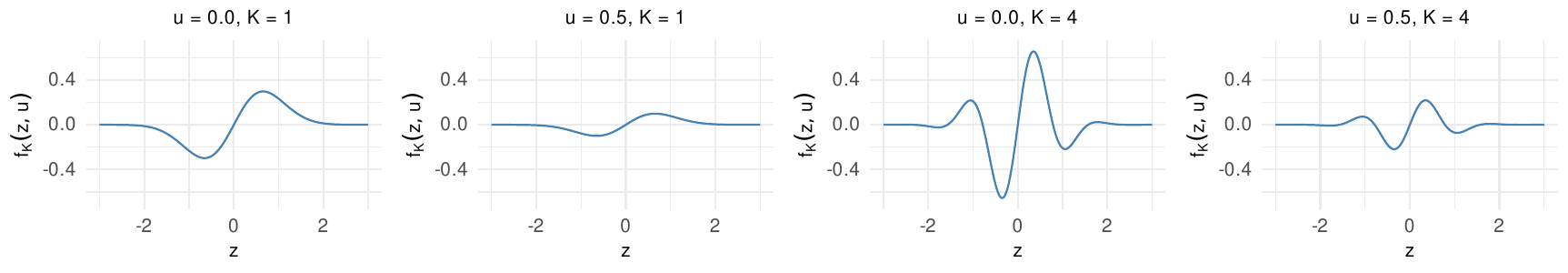}\caption{The time-varying regression function $f_K(z,u)$ from \eqref{eqn:f_tvreg_corrshock} at different $u$ and $K$.}
\label{figure:f_tvreg_function}
\end{figure}

\begin{figure}[ht!]  %The figure environment for floating and captioning    
\centering % Centers the image
\includegraphics[width=1\textwidth, keepaspectratio]{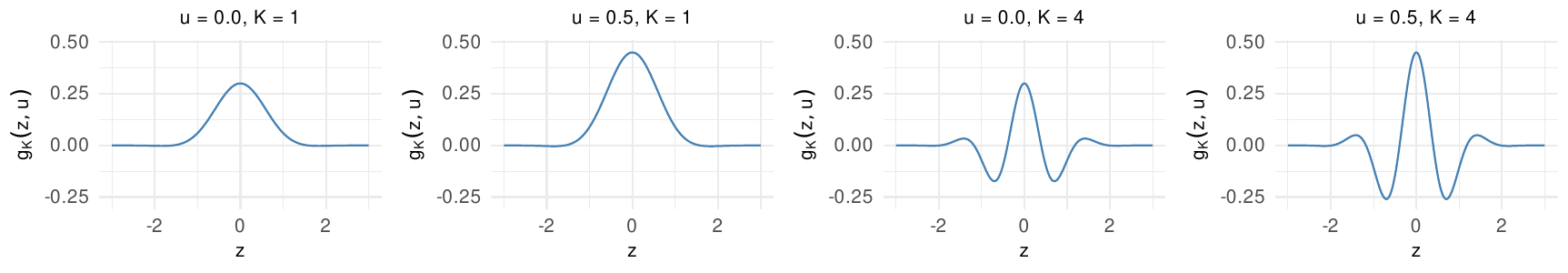}\caption{The time-varying regression function $g_K(z,u)$ from \eqref{eqn:g_tvreg_corrshock} at different $u$ and $K$.}
\label{figure:g_tvreg_function}\end{figure}

\section{Real Data Application}\label{section:data_analysis}

We investigate how stock markets in the United States, United Kingdom, Hong Kong, and Japan are linked. The dataset consists of daily log returns based on the adjusted closing prices of the S\&P 500, FTSE 100, Hang Seng, and Nikkei 225 from January 2022 to March 2025. To deal with holidays observed by each exchange, we linearly interpolate based on the daily log returns from before and after the holiday. This yields $n=845$ observations. Denote the time series of daily log returns by $\texttt{S\&P}$, $\texttt{FTSE}$, $\texttt{HangSeng}$, and $\texttt{Nikkei}$.

For each pair of distinct stock indices $X,Y\in\{\texttt{S\&P}, \texttt{FTSE}, \texttt{HangSeng}, \texttt{Nikkei}\}$, we test  $$X_{t,n} \perp \!\!\! \perp Y_{t,n} \text{ for all times } t =1,2,\ldots,n.$$ 
For each triplet of distinct stock indices $X,Y,Z\in\{\texttt{S\&P}, \texttt{FTSE}, \texttt{HangSeng}, \texttt{Nikkei}\}$, we test $$X_{t,n} \perp \!\!\! \perp Y_{t,n} \mid Z_{t,n} \text{ for all times } t =1,2,\ldots,n.$$ 
Additionally, informed by the related literature \parencite{stock_inderdependencies_2003,stock_inderdependencies_1991}, for each stock market index $X\in\{\texttt{FTSE}, \texttt{HangSeng}, \texttt{Nikkei}\}$,  $Y=\texttt{S\&P}$, we test the null hypothesis 
$$X_{t,n} \perp \!\!\! \perp Y_{t-1,n} \text{ for all times } t =2,3,\ldots,n,$$ and for each pair of distinct indices $X,Y\in\{\texttt{FTSE}, \texttt{HangSeng}, \texttt{Nikkei}\}$,  $Z=\texttt{S\&P}$, we test $$X_{t,n} \perp \!\!\! \perp Y_{t,n} \mid Z_{t-1,n} \text{ for all times } t =2,3,\ldots,n.$$
For each null hypothesis of conditional independence, we apply the Sieve-dGCM test using the setup described in Section~\ref{section:simulations}. For each null hypothesis of unconditional independence, we use the independence test introduced in Section D of the supplement. This test consists of running Algorithm~\ref{algo:dGCM_test} using the sieve estimator to estimate the unconditional means at each time. Otherwise, the setup is identical to that of the Sieve-dGCM test used in Section~\ref{section:simulations}.

We present our main findings based on Benjamini-Hochberg (BH) adjusted p-values \parencite{Benjamini_Hochberg}. Full results for each test are reported in Section D of the supplement. To interpret our findings, we make use of the known trading hours across global stock exchanges. On each day, the exchanges close in the following order from earliest to latest: Tokyo Stock Exchange, Hong Kong Stock Exchange, London Stock Exchange, and New York Stock Exchange. 

We retain the null hypotheses that \texttt{S\&P(t)} is independent of \texttt{Nikkei(t)} (BH-adjusted p-value $0.074$) and that \texttt{FTSE(t)} is independent of \texttt{Nikkei(t)} (BH-adjusted p-value $0.055$). All other null hypotheses of independence are rejected at the significance level $\alpha=0.05$. These results are broadly consistent with prior findings, which suggest that the U.S. and U.K. markets are more strongly linked to one another than to Asia-Pacific markets \parencite{stock_inderdependencies_2003}. This is also consistent with known trading hour effects: European and Asia-Pacific markets react to U.S. market movements with a one-day delay \parencite{stock_inderdependencies_1991}.

The conditional independence tests reveal more structure. We retain the null hypothesis that \texttt{S\&P(t)} is independent of \texttt{HangSeng(t)} given \texttt{FTSE(t)} (BH-adjusted p-value $0.797$). However, we reject the null that \texttt{S\&P(t)} is independent of \texttt{FTSE(t)} given \texttt{HangSeng(t)} (BH-adjusted p-value $0.005$). This suggests that the U.K. market incorporates same-day closing price information from the Hong Kong market, but not vice versa, due to the differences in trading hours. Also, we reject the null hypotheses that \texttt{FTSE(t)} is independent of \texttt{HangSeng(t)} given \texttt{S\&P(t-1)} (BH-adjusted p-value $0.011$), and that \texttt{Nikkei(t)} is independent of \texttt{HangSeng(t)} given \texttt{S\&P(t-1)} (BH-adjusted p-value $0.030$). This suggests direct links between the U.K. and Hong Kong markets and within Asia-Pacific markets, rather than purely reactive co-movements driven by prior-day U.S. market activity.

%%%% Expanding this analysis to include more stock indices and macroeconomic time series, as well as exploring various leads and lags, could yield further insights. Additionally, applying a causal discovery algorithm for nonstationary nonlinear time series (see Section~\ref{subsection:note_cd_nsts}) may reveal interesting global macroeconomic dynamics. For example, \textcite{us_japan_link_2023} employ a causal discovery algorithm to uncover time-delayed causal links among foreign exchange, stock, and bond markets. It would also be possible to gain power by using our multivariate test with the excess log returns of the individual stocks in each stock market index. These extensions are beyond the scope of this paper, but it would be of interest to revisit this analysis in the coming years.

\section{Conclusion}\label{section:conclusion}

The dGCM test shows promise for detecting conditional dependencies among nonstationary nonlinear time series while controlling the Type I error in finite samples. Specifically, we find that the Sieve-dGCM test can hold the level if the sample size is large enough to reliably estimate the time-varying regression functions. Exhaustive comparisons of the dGCM test with other conditional independence tests for time series will be reported in a separate manuscript, as many of these tests lack off-the-shelf implementations. 

%%% We plan to investigate how the dGCM test performs in many settings with varying dimensionality and different forms of nonlinearity, nonstationarity, and temporal correlation. 

\section*{Disclosure Statement}\label{disclosure-statement}

The authors report there are no competing interests to declare.

\section*{Data Availability Statement}\label{data-availability-statement}

The stock return data are available at \url{https://figshare.com/s/93d8176a7194af7a17a8}.

\section*{Supplementary Material}\label{supplementary-material}

The supplementary material contains the proofs, distribution-uniform theory, additional simulations, and details for sieve estimation, independence testing, and parameter selection.

\printbibliography[segment=\therefsegment,
                   title={References}]

%\printbibliography[heading=subbibliography,title={References}]
\end{refsegment}

%%%%%%%%%%%%%%%%%%%%%%%%%%%%%%%%%%%%%%%%%%%%%%%%%%%%%%%%%%%%%%%%%%%%%%%%%%%%%%

\newpage

\appendix 
\begin{refsegment}

\phantomsection\label{section:supplementary-material} \bigskip

\begin{center}

{\large\bf SUPPLEMENT TO \say{The Dynamic Generalized Covariance Measure for Conditional Independence Testing with Nonstationary Time Series}}

\end{center}

\renewcommand{\thesection}{\Alph{section}}

%\begin{description}
%\item[Title:]
%Brief description. (file type)
%\item[R-package for MYNEW routine:]
%R-package MYNEW containing code to perform the diagnostic methods described in the article. The package also contains all datasets used as examples in the article. (GNU zipped tar file) \item[HIV data set:] Data set used in the illustration of MYNEW method in Section XX (.txt file). 

%\end{description}

The supplement is organized as follows. Section~\ref{appendix:CI_test_SIEVE} contains the details for the dGCM test using a sieve time-varying regression estimator. Section~\ref{appendix:proofs} contains proofs of the main theoretical results. Section~\ref{appendix:du_theory} contains the distribution-uniform theory. Section~\ref{appendix:extensions_and_additional_discussion} contains additional discussions including parameter selection, independence testing, and extensions.

\section{dGCM Test with Sieve Time-Varying Regression}\label{appendix:CI_test_SIEVE}

The purpose of this section is to demonstrate that the convergence rates required by Theorem 3.1 for estimating the time-varying regression functions can be achieved. To show this, we consider an instantiation of the dynamic generalized covariance measure (dGCM) test based on the sieve time-varying nonlinear regression estimator from \textcite{zhouzhou_sieve}. We refer to this instantiation of the dGCM test as the Sieve-dGCM test.

\subsection{Setting and Notation}\label{subsection:setting_SIEVE}

We follow \textcite{dahlhaus1997} in rescaling time to the unit interval $t/n\in [0,1]$, so that infill asymptotics can be used to study nonstationary processes. In this setting, the sample size $n$ no longer corresponds to getting information about the future. Instead, as $n$ increases we get more observations about each local structure of the nonstationary process. \textcite{zhou_wu_2009_quantile_loc_stat} introduced the framework for representing locally stationary processes as nonlinear functions of iid inputs as in \textcite{wu_funct_dep_meas}. There is some recent work connecting the functional dependence measure to more familiar mixing conditions \parencite{Hill2025_physical_dependence_mixingale,Heinrichs2026_physical_dependence_mixing}.

We use the same notation as in Section 2.1, with the only difference being that we fix the number of dimensions $d_Z$ and time-offsets $C_k$ for each dimension $k\in [d_Z]$. We still allow the number of dimensions $d_X=d_{X,n}$, $d_Y=d_{Y,n}$ and time-offsets $A_i$, $B_j$ for each $i\in [d_X]$, $j\in [d_Y]$ to grow with $n$. Define $A$, $B$, $C$ as the collection of all time-offsets as in Section 2.1, where $A=A_n$, $B=B_n$ and $C$ is fixed. We emphasize that there is no inherent necessity for fixing the number of dimensions $d_Z$ and time-offsets $C_k$. The reason for this is that we want to leverage the existing theoretical results for the sieve estimator from \textcite{zhouzhou_sieve}. Future investigations can study the performance of the sieve estimator in the high-dimensional setting, so that we can allow the number of dimensions $d_Z$ and time-offsets $C_k$ to grow with $n$.

We will still use the notation $\mathcal{T}_n$ for the subset of original times at which all time-offsets of each dimension of $X_{t,n}$, $Y_{t,n}$, and $Z_{t,n}$ are actually observed,  
$$\mathcal{T}_n=\{1-\min(a_{\min},b_{\min},c_{\min},0),\ldots,n-\max(a_{\max},b_{\max},c_{\max},0)\}\subseteq\{1,\ldots,n\}.$$ Also, we will still denote $T_n=|\mathcal{T}_n|$, $\mathbb{T}_n^{-}=\min(\mathcal{T}_n)$, and $\mathbb{T}_n^{+}=\max(\mathcal{T}_n)$. 
%%% Note: we do not need \mathcal{U}_n or \mathbb{U}_n^{-} or \mathbb{U}_n^{+}. We can have lags be well defined on all of u\in [0,1] even at 0 and 1. think of a parameter curve where the coefficient is defined at 0 and 1. the optimal forecasts are will well defined at these times, even though we never observe the process exactly at time 0 or 1 because we have excluded the beginning and end of the original time 1,...,n to be 1-minLag,...,n-maxLead but asymptotically we will approach observations near 0 and 1 so it is... "asymptotically optimal" forecasting
%%% Note: also even though we don't have observations at 1/n and n/n we do have them at 1/n - minLag/n and 1 - maxLead/n so they are still both going down to 0 and 1 just like it would for the typical sampling scheme with 1/n and n/n...
%%% Similarly, denote the corresponding interval of rescaled times over which all time-offsets are well-defined by $$\mathcal{U}_n=\left[-\frac{\min(a_{\min},b_{\min},c_{\min},0)}{n}, 1 -\frac{\max(a_{\max},b_{\max},c_{\max},0)}{n}\right]\subset [0,1],$$ and denote $\mathbb{U}_n^{-}=\min(\mathcal{U}_n)$ and $\mathbb{U}_n^{+}=\max(\mathcal{U}_n)$. 

Recall the index set containing the dimensions and time-offsets of interest 
$$\mathcal{D}_n \subseteq \{(i,j,a,b):i \in [d_X], j \in [d_Y], a \in A_i, b \in B_j\},$$ where $A_i$ and $B_j$ are the time-offsets for dimensions $i\in[d_X]$ and $j\in[d_Y]$, respectively. Again, we will often refer to the dimension/time-offset tuple by $m=(i,j,a,b)\in\mathcal{D}_n$ to lighten the notation. Denote the cardinality by $D_n=|\mathcal{D}_n|$, which can grow with $n$.

\subsection{Locally Stationary Observed Processes}\label{subsection:observed_proc_SIEVE}

Next, we introduce the causal representation of locally stationary processes, which is most similar to \textcite{zhouzhou_sieve} and Example 3 in \textcite{seq_gauss_approx2022}. This representation is different than the previous causal representation from Assumption 3.1, because we now assume that the stochastic nonlinear system is well-defined for all rescaled times. For the following assumption, let $$\mathcal{H}_t^X=(\eta_t^X,\eta_{t-1}^X,\ldots),\text{ }\mathcal{H}_t^Y=(\eta_t^Y,\eta_{t-1}^Y,\ldots),\text{ }\mathcal{H}_t^Z=(\eta_t^Z,\eta_{t-1}^Z,\ldots),$$ where $(\eta_t^X,\eta_t^Y,\eta_t^Z)_{t\in\mathbb{Z}}$ is an iid sequence of random vectors. Denote the dimensions of $\eta_t^X=\eta_{t,n}^X$, $\eta_t^Y=\eta_{t,n}^Y$, $\eta_t^Z=\eta_{t,n}^Z$, respectively, by $d_{X}^{\eta}=d_{X,n}^{\eta}$, $d_{Y}^{\eta}=d_{Y,n}^{\eta}$, $d_{Z}^{\eta}=d_{Z,n}^{\eta}$, which can change with $n$.

% can change with n because each measurable function Gn depends on n 
% defining this way to emphasize that the iid inputs at each time can be dependent (i.e. non-product joint distribution at each time), to ensure this joint distribution of the iid inputs at each time is time-invariant, and to ensure each input is (temporally) independent of all other times of every input sequence. this is important for step 1.2 of the main proof.
% Without loss of generality, we can take each of $\eta_t^X,\eta_t^Y,\eta_t^Z$ to have $\mathrm{Unif}[0,1]$ marginal distributions for each $t\in\mathbb{Z}$.

% Note: we add the subscript n since the measurable functions can change with n. We allow for the dimensions and distributions of X, Y, Z to change with n. 
\begin{assumption}[Causal representations of the observed processes]\label{asmpt_causal_rep_process_SIEVE} 
Assume, for each $n\in\mathbb{N}$, $i\in [d_X]$, $j\in [d_Y]$, $k\in [d_Z]$, and $t\in [n]$, we can represent the observed processes as $$X_{t,n,i}= \tilde{G}_{n,i}^X(t/n,\mathcal{H}_{t}^X), \text{ } Y_{t,n,j}= \tilde{G}_{n,j}^Y(t/n,\mathcal{H}_{t}^Y),  \text{ } Z_{t,n,k}=\tilde{G}_{n,k}^Z(t/n,\mathcal{H}_{t}^Z),$$ where the systems are defined for all $u\in [0,1]$ by $$\tilde{X}_{t,n,i}(u)= \tilde{G}_{n,i}^X(u,\mathcal{H}_{t}^X),  \text{ }\tilde{Y}_{t,n,j}(u)= \tilde{G}_{n,j}^Y(u,\mathcal{H}_{t}^Y),  \text{ }\tilde{Z}_{t,n,k}(u)= \tilde{G}_{n,k}^Z(u,\mathcal{H}_{t}^Z).$$ %% Not quite% %% so that we have $X_{t,n,i}=\tilde{X}_{t,n,i}(t/n)$, $Y_{t,n,j}=\tilde{Y}_{t,n,j}(t/n)$, $Z_{t,n,k}=\tilde{Z}_{t,n,k}(t/n)$.

Further, assume $\tilde{G}_{n,i}^{X}(u,\cdot)$, $\tilde{G}_{n,j}^{Y}(u,\cdot)$, $\tilde{G}_{n,k}^{Z}(u,\cdot)$ are measurable functions from $(\mathbb{R}^{d_{X}^{\eta}})^{\infty}$, $(\mathbb{R}^{d_{Y}^{\eta}})^{\infty}$, $(\mathbb{R}^{d_{Z}^{\eta}})^{\infty}$, respectively, to $\mathbb{R}$ such that $\tilde{G}_{n,i}^{X}(u,\mathcal{H}_s^X)$, $\tilde{G}_{n,j}^{Y}(u,\mathcal{H}_s^Y)$, $\tilde{G}_{n,k}^{Z}(u,\mathcal{H}_s^Z)$ are each well-defined random variables for each $s\in\mathbb{Z}$ and $(\tilde{G}_{n,i}^{X}(u,\mathcal{H}_s^X))_{s\in\mathbb{Z}}$, $(\tilde{G}_{n,j}^{Y}(u,\mathcal{H}_s^Y))_{s\in\mathbb{Z}}$, $(\tilde{G}_{n,k}^{Z}(u,\mathcal{H}_s^Z))_{s\in\mathbb{Z}}$ are each stationary ergodic processes. 
\end{assumption}

As in Section 3.1, we have not defined the input sequences for the observed processes separately for each dimension. However, without loss of generality, we can define the measurable functions $\tilde{G}_{n,i}^{X}(u,\cdot)$, $\tilde{G}_{n,j}^{Y}(u,\cdot)$, $\tilde{G}_{n,k}^{Z}(u,\cdot)$ and the inputs $\eta_t^X$, $\eta_t^Y$, $\eta_t^Z$ so that each dimension of the observed processes can have idiosyncratic inputs.

In light of Assumption~\ref{asmpt_causal_rep_process_SIEVE}, we have the following causal representations for all dimensions with no time-offsets:
\begin{align*}
\tilde{X}_{t,n}(u) &= \tilde{G}_n^X(u,\mathcal{H}_{t}^X)=(\tilde{G}_{n,i}^X(u,\mathcal{H}_{t}^X))_{i\in [d_{X}]},\\
\tilde{Y}_{t,n}(u) &= \tilde{G}_n^Y(u,\mathcal{H}_{t}^Y)=(\tilde{G}_{n,j}^Y(u,\mathcal{H}_{t}^Y))_{j\in [d_{Y}]},\\
\tilde{Z}_{t,n}(u) &= \tilde{G}_n^Z(u,\mathcal{H}_{t}^Z)=(\tilde{G}_{n,k}^Z(u,\mathcal{H}_{t}^Z))_{k\in [d_Z]},\end{align*}  so that we have $X_{t,n}=\tilde{X}_{t,n}(t/n)$, $Y_{t,n}=\tilde{Y}_{t,n}(t/n)$, $Z_{t,n}=\tilde{Z}_{t,n}(t/n)$.

%%% for time-offsets, it will also shift the rescaled time, so we don't observe all of the stuff for the lags but that's fine, only at time \mathcal{T}_n...
%%% define as in Seq Gauss Approx paper https://arxiv.org/pdf/2203.03237 
For each $n\in\mathbb{N}$, we have causal representations for dimensions $i\in [d_X]$, $j\in [d_Y]$, $k\in [d_Z]$ with time-offsets $a\in A_i$, $b\in B_j$, $c\in C_k$ \begin{align*}    \tilde{X}_{t,n,i,a}(u) = \tilde{G}_{n,i,a}^X(u,\mathcal{H}_{t,a}^X), \ \tilde{Y}_{t,n,j,b}(u) = \tilde{G}_{n,j,b}^Y(u,\mathcal{H}_{t,b}^Y), \ \tilde{Z}_{t,n,k,c}(u) = \tilde{G}_{n,k,c}^Z(u,\mathcal{H}_{t,c}^Z),
%= \tilde{G}_{n,i}^X(u+a/n,\mathcal{H}_{t+a}^X), when this is well-defined i.e. when u in \mathcal{U}_n = \mathcal{T}_n / n
%= \tilde{G}_{n,j}^Y(u,\mathcal{H}_{t+b}^Y), when this is well-defined i.e. when u in \mathcal{U}_n = \mathcal{T}_n / n
%= \tilde{G}_{n,k}^Z(u,\mathcal{H}_{t+c}^Z),, when this is well-defined i.e. when u in \mathcal{U}_n = \mathcal{T}_n / n
\end{align*} where $\mathcal{H}_{t,a}^X=(\eta_{t+a}^X,\eta_{t-1+a}^X,\ldots)$, $\mathcal{H}_{t,b}^Y=(\eta_{t+b}^Y,\eta_{t-1+b}^Y,\ldots)$, and $\mathcal{H}_{t,c}^Z=(\eta_{t+c}^Z,\eta_{t-1+c}^Z,\ldots)$, so that we have $X_{t,n,i,a}=\tilde{X}_{t,n,i,a}(t/n)$, $Y_{t,n,j,b}=\tilde{Y}_{t,n,j,b}(t/n)$, $Z_{t,n,k,c}=\tilde{Z}_{t,n,k,c}(t/n)$ for each time $t\in\mathcal{T}_n$ (i.e., such that the time-offsets are well-defined) for each dimension of the observed sequence with time-offset. We can then write the causal representation of the vectors with all dimensions and time-offsets as \begin{align*}\bm{\tilde{X}}_{t,n}(u)&=\bm{\tilde{G}}_n^{\bm{X}}(u,\mathcal{H}_t^{\bm{X}})=(\tilde{G}_{n,i,a}^X(u,\mathcal{H}_{t,a}^X))_{i\in[d_X],a\in A_i},\\\bm{\tilde{Y}}_{t,n}(u)&=\bm{\tilde{G}}_n^{\bm{Y}}(u,\mathcal{H}_t^{\bm{Y}})=(\tilde{G}_{n,j,b}^Y(u,\mathcal{H}_{t,b}^Y))_{j\in[d_Y],b\in B_j},\\\bm{\tilde{Z}}_{t,n}(u)&=\bm{\tilde{G}}_n^{\bm{Z}}(u,\mathcal{H}_t^{\bm{Z}})=(\tilde{G}_{n,k,c}^Z(u,\mathcal{H}_{t,c}^Z))_{k\in[d_Z],c\in C_k},\end{align*} where $\mathcal{H}_t^{\bm{X}}=(\eta_t^{\bm{X}},\eta_{t-1}^{\bm{X}},\ldots)$, $\mathcal{H}_t^{\bm{Y}}=(\eta_{t}^{\bm{Y}},\eta_{t-1}^{\bm{Y}},\ldots)$, $\mathcal{H}_t^{\bm{Z}}=(\eta_{t}^{\bm{Z}},\eta_{t-1}^{\bm{Z}},\ldots)$, and $\eta_t^{\bm{X}}=\eta_{t+a_{\max}}^X$, $\eta_t^{\bm{Y}}=\eta_{t+b_{\max}}^Y$, $\eta_t^{\bm{Z}}=\eta_{t+c_{\max}}^Z$, so that we have $\bm{X}_{t,n}=\bm{\tilde{X}}_{t,n}(t/n)$, $\bm{Y}_{t,n}=\bm{\tilde{Y}}_{t,n}(t/n)$, $\bm{Z}_{t,n}=\bm{\tilde{Z}}_{t,n}(t/n)$ for each time $t\in\mathcal{T}_n$ (i.e., such that the time-offsets are well-defined) including all dimensions and time-offsets. Note that we only observe the process at times $\mathcal{T}_n$ and corresponding rescaled times $\mathcal{T}_n /n$, though we may define the mappings on this extended rescaled time domain.

For each $n \in \mathbb{N}$, let the measurable space $(\Omega,\mathcal{B})$ be equipped with a family of probability measures $(\mathbb{P}_P)_{P \in \mathcal{P}_n}$ so that the joint distribution of the stochastic nonlinear systems 
$$(\tilde{G}_n^X(u,\mathcal{H}_{t}^X))_{u\in [0,1],t\in\mathbb{Z}},\text{ }(\tilde{G}_n^Y(u,\mathcal{H}_{t}^Y))_{u\in [0,1],t\in\mathbb{Z}},\text{ } (\tilde{G}_n^Z(u,\mathcal{H}_{t}^Z))_{u\in [0,1],t\in\mathbb{Z}},$$ under $\mathbb{P}_P$ is $P\in\mathcal{P}_n$, where the collection of distributions $\mathcal{P}_n$ can change with $n$. The family of probability measures $(\mathbb{P}_P)_{P \in \mathcal{P}_n}$ is defined with respect to the same measurable space $(\Omega,\mathcal{B})$, but need not have the same dominating measure. 
% $\mathcal{P}_n \subset \mathrm{Prob}[(\mathbb{R}^{d_X+d_Y+d_Z})^{[0,1]^3\times\mathbb{Z}^3}]$ is a subset of the set of Borel probability measures on functions from $[0,1]^3 \times \mathbb{Z}^3$ to $\mathbb{R}^{d_X+d_Y+d_Z}$ i.e. each \omega gives you a different realization of the stochastic nonlinear system for all mechanism times and input times

We use the same null hypotheses of conditional independence as those in Section 2.3. Again, for each $n\in\mathbb{N}$, we denote the collection of distributions such that the null hypothesis is true by $\mathcal{P}_{0,n}^{\mathrm{CI}}$. % In the locally stationary setting, the null hypothesis \begin{equation} X_{t,n,i,a}\!\perp \!\!\! \perp Y_{t,n,j,b} \mid \bm{Z}_{t,n} \text{ for all } t\in\mathcal{T}_n, \text{ for all } (i,j,a,b)\in\mathcal{D}_n, \label{eqn:global_null_hypoth_all_times_SIEVE}\end{equation}  can be written equivalently as\begin{align*} &\tilde{X}_{t,n,i,a}(t/n)\!\perp \!\!\! \perp \tilde{Y}_{t,n,j,b}(t/n) \mid \bm{\tilde{Z}}_{t,n}(t/n) \text{ for all } t\in\mathcal{T}_n, \text{ for all } (i,j,a,b)\in\mathcal{D}_n,\end{align*} where $\mathcal{D}_n$ only contains a single dimension/time-offset tuple in the univariate setting. 

We will state more assumptions in the rest of this section for a generic sequence of collections of distributions $(\mathcal{P}_n)_{n\in\mathbb{N}}$. In Theorem~\ref{thm:test_SIEVE_DR}, we will assume that these conditions hold for the sequence of collections of distributions $(\mathcal{P}_{0,n}^{\ast})_{n\in\mathbb{N}}$, where $\mathcal{P}_{0,n}^{\ast}\subset\mathcal{P}_{0,n}^{\mathrm{CI}}$ for each $n\in\mathbb{N}$. Note that we make stronger assumptions in this section than in Section 3 to ensure that the sieve estimators satisfy the convergence rate requirements of Theorem 3.1.

% Reason for writing the distribution of the stochastic nonlinear system instead of just the observed process is because of the temporal dependence condition (i.e. different time's inputs and same mechanism) and nonstationarity condition (i.e. same inputs and different time's mechanism)

\subsection{Sieve Time-Varying Nonlinear Regression Estimator}\label{subsection:tv_regr_SIEVE}

For a given sample size $n\in\mathbb{N}$, distribution $P \in \mathcal{P}_n$, time $t \in \mathcal{T}_n$, and dimension/time-offset tuple $(i,j,a,b)\in \mathcal{D}_n$, we consider the time-varying nonlinear regression model 
\begin{equation}\nonumber
X_{t,n,i,a}=f_{P,n,i,a}(t/n,\bm{Z}_{t,n})+\varepsilon_{P,t,n,i,a}, \text{ } Y_{t,n,j,b}=g_{P,n,j,b}(t/n,\bm{Z}_{t,n})+\xi_{P,t,n,j,b}, \label{tv_regr_fncts_SIEVE} \end{equation} where  $f_{P,n,i,a}(u,\bm{z})$ and $g_{P,n,j,b}(u,\bm{z})$ are smooth functions of rescaled time $u$ and covariate values $\bm{z}$ with $f_{P,n,i,a}(t/n,\bm{z})=\mathbb{E}_P(X_{t,n,i,a}|\bm{Z}_{t,n}=\bm{z})$ and $g_{P,n,j,b}(t/n,\bm{z})=\mathbb{E}_P(Y_{t,n,j,b}|\bm{Z}_{t,n}=\bm{z})$. We emphasize that the functions $f_{P,n,i,a}(u,\bm{z})$ and $g_{P,n,j,b}(u,\bm{z})$ depend on rescaled time $u$ rather than \say{real time} $t$, as in the literature on nonparametric regression for locally stationary processes \parencite{vogt,wbwu_timevar_npreg,boosting_loc_stat_serena_ng,mult_str_wu_2022,zhouzhou_sieve}. For $m=(i,j,a,b)\in\mathcal{D}_n$, denote the error products at time $t$ by $$R_{P,t,n,m}=\varepsilon_{P,t,n,i,a}\xi_{P,t,n,j,b},$$ and the corresponding residual products by $$\hat{R}_{t,n,m}=\hat{\varepsilon}_{t,n,i,a}  \hat{\xi}_{t,n,j,b},$$ where $\hat{\varepsilon}_{t,n,i,a}=X_{t,n,i,a}-\hat{f}_{t,n,i,a}(t/n,\bm{Z}_{t,n})$ and $\hat{\xi}_{t,n,j,b}=Y_{t,n,j,b}-\hat{g}_{t,n,j,b}(t/n,\bm{Z}_{t,n})$.

% Intuition for "forecasting" when appealing to infill asymptotics with LSTS: Suppose you are predicting b steps ahead. With infill asymptotics, the "amount of rescaled time" between the time steps decreases since b/n decreases. On the other hand, we let the largest forecasting horizon grow with n also. However, the amount of rescaled time between these time steps will get smaller with n as well "eventually" for each of these new, "longer" forecasting horizons more time steps into the future. Note that the "degree of nonstationarity" with b time steps ahead will get less and less as n grows since b/n shrinks as n increases.

The estimates $\hat{f}_{t,n,i,a}$ and $\hat{g}_{t,n,j,b}$ of the functions $f_{P,n,i,a}$ and $g_{P,n,j,b}$ are formed using the sieve time-varying regression estimator introduced below. Let $\bm{\hat{R}}_{t,n}=(\hat{R}_{t,n,m})_{m\in\mathcal{D}_n}$ be the high-dimensional vector process containing the residual products for all dimension/time-offset combinations in $\mathcal{D}_n$. The observed processes $X$, $Y$, $Z$ and error processes $\epsilon$, $\xi$ can all be locally stationary processes; see Section~\ref{subsection:dependence_nonstationarity_SIEVE} for the details.
% The subscript $t$ in $\hat{f}_{t,n,i,a}$ and $\hat{g}_{t,n,j,b}$ is to indicate that we allow for sequential estimation, which will be discussed in Remark~\ref{remark_sequential_sieve_est}.

For some sequence of collections of distributions $(\mathcal{P}_n)_{n\in\mathbb{N}}$, we make the following assumption.

% need this to satisfy asmpt_causal_rep_pred_proc_nsts along with the bounded predictions of sieve estimator (i.e. continuous function on closed interval is bounded). in particular, this is used for more control on the estimation errors in step 1.2 of the main proof. 
\begin{assumption}[Additive form and regularity]\label{asmpt_tv_regr_Lq_SIEVE}

For each sample size $n\in\mathbb{N}$, distribution $P \in \mathcal{P}_n$, rescaled time $u \in [0,1]$, and dimension/time-offset tuple $(i,j,a,b)\in \mathcal{D}_n$, assume that \begin{align*}
f_{P,n,i,a}(u,\bm{z}) &= \sum_{k\in [d_Z]}\sum_{c\in C_k} f_{P,n,i,a,k,c}(u,z_{k,c}),\\
g_{P,n,j,b}(u,\bm{z})&= \sum_{k\in [d_Z]}\sum_{c\in C_k} g_{P,n,j,b,k,c}(u, z_{k,c}),\end{align*} where $f_{P,n,i,a,k,c}:[0,1]\times \mathbb{R}\xrightarrow[]{}\mathbb{R}$ and $g_{P,n,j,b,k,c}:[0,1]\times \mathbb{R}\xrightarrow[]{}\mathbb{R}$ are time-varying partial response functions, so that we have \begin{align*}
\mathbb{E}_P(X_{t,n,i,a}|\bm{Z}_{t,n}=\bm{z}) &= \sum_{k\in [d_Z]}\sum_{c\in C_k} f_{P,n,i,a,k,c}(t/n, z_{k,c}),\\
\mathbb{E}_P(Y_{t,n,j,b}|\bm{Z}_{t,n}=\bm{z}) &= \sum_{k\in [d_Z]}\sum_{c\in C_k} g_{P,n,j,b,k,c} (t/n, z_{k,c}),\end{align*} for each time $t \in \mathcal{T}_n$.

Further, assume there exists some $q\geq 2$, such that we have  
\begin{align*} \underset{n\in\mathbb{N}}{\sup} \ \underset{P\in\mathcal{P}_n}{\sup} \ \underset{u \in [0,1]}{\sup} \ \underset{t \in \mathcal{T}_n}{\max} \ \underset{i\in [d_X]}{\max} \ \underset{a\in A_i}{\max} \ \underset{k\in [d_Z]}{\max} \ \underset{c\in C_k}{\max} \ \mathbb{E}_P(|f_{P,n,i,a,k,c}(u,Z_{t,n,k,c})|^q)&<\infty,\\ \underset{n\in\mathbb{N}}{\sup} \ \underset{P\in\mathcal{P}_n}{\sup} \ \underset{u \in [0,1]}{\sup} \ \underset{t \in \mathcal{T}_n}{\max} \ \underset{j\in [d_Y]}{\max} \ \underset{b\in B_j}{\max}  \ \underset{k\in [d_Z]}{\max} \ \underset{c\in C_k}{\max}  \ \mathbb{E}_P(|g_{P,n,j,b,k,c}(u,Z_{t,n,k,c})|^q)&<\infty.\end{align*}

\end{assumption}

To fix ideas, we use the algebraic mapping $h:[-1,1]\xrightarrow[]{}\overline{\mathbb{R}}$ from Example 3.1 in \textcite{zhouzhou_sieve} with positive scaling factor $s=1$,
$$h(\tilde{z})=\begin{cases} -\infty,\text{ }\text{ }\text{ }\text{ }\text{ } &\tilde{z}=-1, \\
\frac{\tilde{z}}{\sqrt{1-\tilde{z}^2}},\text{ }\text{ }\text{ } &\tilde{z}\in (-1,1), \\ \infty, \text{ }\text{ }\text{ }\text{ }\text{ }\text{ }\text{ } &\tilde{z}=1. \end{cases}$$
See the discussion preceding Definition 3.1 in \textcite{zhouzhou_sieve} for additional details. For some sequence of collections of distributions $(\mathcal{P}_n)_{n\in\mathbb{N}}$, for each $n\in\mathbb{N}$, $i\in [d_X]$, $a\in A_i$, $j\in [d_Y]$, $b\in B_j$, $k\in [d_Z]$, $c\in C_k$, and $P\in\mathcal{P}_n$, we relate the time-varying partial response functions $f_{P,n,i,a,k,c}:[0,1]\times \mathbb{R}\xrightarrow[]{}\mathbb{R}$ and $g_{P,n,j,b,k,c}:[0,1]\times \mathbb{R}\xrightarrow[]{}\mathbb{R}$ to $\tilde{f}_{P,n,i,a,k,c}: [0,1]\times [0,1] \xrightarrow[]{}\mathbb{R}$ and $\tilde{g}_{P,n,j,b,k,c}: [0,1]\times [0,1] \xrightarrow[]{}\mathbb{R}$, respectively, where
\begin{align*}
\tilde{f}_{P,n,i,a,k,c}(u^{\ast},z^{\ast})&= f_{P,n,i,a,k,c}(u^{\ast},h(2z^{\ast}-1)),\\
\tilde{g}_{P,n,j,b,k,c}(u^{\ast},z^{\ast})&= g_{P,n,j,b,k,c}(u^{\ast},h(2z^{\ast}-1)). 
\end{align*}

For some sequence of collections of distributions $(\mathcal{P}_n)_{n\in\mathbb{N}}$, we make the following assumption. This assumption sets the smoothness level in Assumption 3.1 from \textcite{zhouzhou_sieve} to $\mathrm{m}_1=\infty$ and $\mathrm{m}_2=\infty$.

\begin{assumption}[Smoothness]\label{asmpt_smooth_regr_fn_SIEVE}

For each $n\in\mathbb{N}$, $i\in [d_X]$, $a\in A_i$, $j\in [d_Y]$, $b\in B_j$, $k\in [d_Z]$, $c\in C_k$, and $P\in\mathcal{P}_n$, assume that for each fixed $u^{\ast}\in [0,1]$ we have
$$\tilde{f}_{P,n,i,a,k,c}(u^{\ast},\cdot)\in C^{\infty}([0,1]), \text{ }\tilde{g}_{P,n,j,b,k,c}(u^{\ast},\cdot)\in C^{\infty}([0,1]),$$ and for each fixed $z^{\ast}\in [0,1]$ we have 
    $$\tilde{f}_{P,n,i,a,k,c}(\cdot,z^{\ast})\in C^{\infty}([0,1]), \text{ }\tilde{g}_{P,n,j,b,k,c}(\cdot,z^{\ast})\in C^{\infty}([0,1]),$$
      where $C^{\infty}([0,1])$ denotes the space of functions on $[0,1]$ that are infinitely differentiable. 
\end{assumption}

%\begin{align*}   f_{P,n,i,a,k,c}(u,z)&\approx \sum_{\ell_1=1}^{\tilde{c}_n} \sum_{\ell_2=1}^{\tilde{d}_n} \beta_{P,n,i,a,k,c,\ell_1,\ell_2}^f b_{\ell_1,\ell_2}(u,z),  \\ g_{P,n,j,b,k,c}(u,z)&\approx\sum_{\ell_1=1}^{\tilde{c}_n} \sum_{\ell_2=1}^{\tilde{d}_n} \beta_{P,n,j,b,k,c,\ell_1,\ell_2}^g b_{\ell_1,\ell_2}(u,z),\end{align*} where $\{b_{\ell_1,\ell_2}(u,z)\}=\{\phi_{\ell_1}(u)\varphi_{\ell_2}(z)\}$ are basis functions and $\{\beta_{P,n,i,a,k,c,\ell_1,\ell_2}^f\}$, $\{\beta_{P,n,j,b,k,c,\ell_1,\ell_2}^g\}$ 

If Assumption~\ref{asmpt_smooth_regr_fn_SIEVE} holds, then by Theorem 3.1 of \textcite{zhouzhou_sieve} we can approximate the time-varying partial response functions by a linear combination of basis functions, and the coefficients can be estimated with OLS. The numbers of basis functions for time and for the covariate values, denoted by $\tilde{c}_n$ and $\tilde{d}_n$, respectively, are chosen to increase with the sample size $n$ at some rate. To fix ideas, we will use Legendre polynomials as the basis functions for both the theoretical analysis in this section and the numerical simulations in Section 4. Specifically, for each $\ell_1 \in [\tilde{c}_n]$ and $\ell_2\in[\tilde{d}_n]$, let the basis functions for time $\{\phi_{\ell_1}(u)\}$ and the covariate values $\{\varphi_{\ell_2}(z)\}$ be \textit{mapped} Legendre polynomials as in Example C.2 and Section 3.1.1 of \textcite{zhouzhou_sieve}. It is straightforward to replace the Legendre polynomials used in our theoretical analysis and simulations with trigonometric polynomials, wavelets, or other Jacobi polynomials.

Next, we introduce the sieve estimators for the time-varying regression functions. Although we do not discuss this topic in detail here, we point interested readers to further discussions of asymptotically optimal forecasting for locally stationary processes \parencite{ar_approx_lsts_xiucai_zhou,inv_cov_krampe_ssr_2024,cui_zhou_2025}.

%\begin{remark}[Sequential sieve estimation]\label{remark_sequential_sieve_est} Our formulation of the sieve estimator from \textcite{zhouzhou_sieve} accommodates sequential estimation, in the sense that the predictors for a particular rescaled time are only constructed using the information available up to that rescaled time. We emphasize that sequential estimation is not required for all settings, particularly when certain exogeneity conditions hold. The need for sequential estimation in some settings is due to the martingale difference sequence condition imposed on the error processes in Assumption 3.4 (cf. Assumption~\ref{asmpt_causal_rep_errors_SIEVE}), which becomes relevant when using our test for variable selection for forecasting and causal inference for time-delayed effects. Note that the same convergence rates are attained whether or not sequential estimation is used, due to the infill asymptotic framework of locally stationary processes. That is, because more observations for each local structure become available as $n$ grows.   \end{remark}

%%%%%%% More details %%%%%%%
%%%% EXAMPLE 1 (forecasting with contemporaneous covariates) If data up to time t is used to form the predictor at time t, then the sieve estimator's original theoretical guarantees apply since time t/n will be the last time in interval [start,t/n] and the convergence rates are time-uniform. 

Recall the following notation from Section 3.2. Let $\bm{\mathfrak{D}}_{t,n,i,a}^{\hat{f}}$, $\bm{\mathfrak{D}}_{t,n,j,b}^{\hat{g}}$ be the datasets used to form the estimators $\hat{f}_{t,n,i,a}(t/n,\cdot)$, $\hat{g}_{t,n,j,b}(t/n,\cdot)$ of the time-varying regression functions at rescaled time $t/n\in [0,1]$, let $\mathcal{H}_{t,a}^{\bm{\mathfrak{D}}^{\hat{f}}}$, $\mathcal{H}_{t,b}^{\bm{\mathfrak{D}}^{\hat{g}}}$ be the corresponding input sequences, and let $\mathcal{T}_{t,n,i,a}^{\hat{f}}$, $\mathcal{T}_{t,n,j,b}^{\hat{g}}$ be the corresponding sets of times with $T_{t,n,i,a}^{\hat{f}}=|\mathcal{T}_{t,n,i,a}^{\hat{f}}|$, $T_{t,n,j,b}^{\hat{g}}=|\mathcal{T}_{t,n,j,b}^{\hat{g}}|$. Note that each of the estimators $\hat{f}_{t,n,i,a,k,c}(t/n,\cdot)$, $\hat{g}_{t,n,j,b,k,c}(t/n,\cdot)$ of the corresponding time-varying partial response functions at rescaled time $t/n\in [0,1]$ may have different numbers of basis functions. Without confusion, we will write the numbers of basis functions as $\tilde{c}_n$ and $\tilde{d}_n$ instead of $\tilde{c}_{t,n,i,a,k,c}^{\hat{f}}$, $\tilde{c}_{t,n,j,b,k,c}^{\hat{g}}$ and $\tilde{d}_{t,n,i,a,k,c}^{\hat{f}}$, $\tilde{d}_{t,n,j,b,k,c}^{\hat{g}}$ to simplify the presentation below.

% The $(s,p)$-th entries of $\bm{\bar{\mathrm{Z}}}_{t,n,i,a}$ and $\bm{\bar{\mathrm{Z}}}_{t,n,j,b}$ are $$\bm{\bar{\mathrm{Z}}}_{t,n,i,a}^{(s,p)}=\phi_{\ell_{1,p}}(t_s/n)\varphi_{\ell_{2,p}}(Z_{t_s,n,k_p,c_p}),$$ $$\bm{\bar{\mathrm{Z}}}_{t,n,j,b}^{(s,p)}=\phi_{\ell_{1,p}}(t_s/n)\varphi_{\ell_{2,p}}(Z_{t_s,n,k_p,c_p}),$$ where we use mappings for the rows $s\mapsto t_s\in\mathcal{T}_{t,n,i,a}^{\hat{f}}$ and $s\mapsto t_s\in\mathcal{T}_{t,n,j,b}^{\hat{g}}$ which maintain the order of time (i.e. $t_{s_1}<t_{s_2}$ if $s_1<s_2$), and some mappings for the columns $p\mapsto (k_p,c_p,\ell_{1,p},\ell_{2,p})$ which determine orderings for the dimension/time-offset/basis-index combinations, where $k_p\in [d_Z]$, $c_p\in C_{k_p}$, $\ell_{1,p} \in [\tilde{c}_n]$, $\ell_{2,p}\in[\tilde{d}_n]$. 

% \in \mathbb{R}^{T_{t,n,i,a}^{\hat{f}} \times \bm{d_Z}\tilde{c}_n\tilde{d}_n}\in \mathbb{R}^{T_{t,n,j,b}^{\hat{g}} \times \bm{d_Z}\tilde{c}_n\tilde{d}_n}
For some fixed $n\in\mathbb{N}$, $t\in\mathcal{T}_n$, and $(i,j,a,b)\in\mathcal{D}_n$, denote the design matrices by $\bm{\bar{\mathrm{Z}}}_{t,n,i,a}$ and $\bm{\bar{\mathrm{Z}}}_{t,n,j,b}$, which are constructed as in \textcite{zhouzhou_sieve}, so that each row corresponds to one time and each column corresponds to one dimension/time-offset combination with a particular basis-index combination, i.e.\ there are no duplicate columns. For each $n\in\mathbb{N}$, $P\in\mathcal{P}_n$, $i\in [d_X]$, $a\in A_i$, $j\in [d_Y]$, $b\in B_j$ the (time-invariant) coefficient vectors % \begin{align*} \bm{\beta}_{P,n,i,a}^f &=(\beta_{P,n,i,a,k,c,\ell_1,\ell_2}^f)_{k,c,\ell_1,\ell_2}^{\top}\in\mathbb{R}^{\bm{d_Z}\tilde{c}_n\tilde{d}_n},\\ \bm{\beta}_{P,n,j,b}^g &=(\beta_{P,n,j,b,k,c,\ell_1,\ell_2}^g)_{k,c,\ell_1,\ell_2}^{\top}\in\mathbb{R}^{\bm{d_Z}\tilde{c}_n\tilde{d}_n},\end{align*}
have the following OLS estimators \begin{align*} \hat{\bm{\beta}}_{t,n,i,a}^f &=(\bm{\bar{\mathrm{Z}}}_{t,n,i,a}^{\top}\bm{\bar{\mathrm{Z}}}_{t,n,i,a})^{-1} \bm{\bar{\mathrm{Z}}}_{t,n,i,a}^{\top} \bm{\bar{\mathrm{X}}}_{t,n,i,a},%=(\hat{\beta}_{t,n,i,a,k,c,\ell_1,\ell_2}^f)_{k,c,\ell_1,\ell_2}^{\top}\in\mathbb{R}^{\bm{d_Z}\tilde{c}_n\tilde{d}_n}
\\
\hat{\bm{\beta}}_{t,n,j,b}^g &=(\bm{\bar{\mathrm{Z}}}_{t,n,j,b}^{\top}\bm{\bar{\mathrm{Z}}}_{t,n,j,b})^{-1} \bm{\bar{\mathrm{Z}}}_{t,n,j,b}^{\top} \bm{\bar{\mathrm{Y}}}_{t,n,j,b},%=(\hat{\beta}_{t,n,j,b,k,c,\ell_1,\ell_2}^g)_{k,c,\ell_1,\ell_2}^{\top}\in\mathbb{R}^{\bm{d_Z}\tilde{c}_n\tilde{d}_n}
\end{align*}
as in \textcite{zhouzhou_sieve}, where $$\bm{\bar{\mathrm{X}}}_{t,n,i,a}=(X_{s,n,i,a})^{\top}_{s\in\mathcal{T}_{t,n,i,a}^{\hat{f}}}\in\mathbb{R}^{T_{t,n,i,a}^{\hat{f}}},\text{ }\bm{\bar{\mathrm{Y}}}_{t,n,j,b}=(Y_{s,n,j,b})^{\top}_{s\in\mathcal{T}_{t,n,j,b}^{\hat{g}}}\in\mathbb{R}^{T_{t,n,j,b}^{\hat{g}}}.$$ Finally, the estimators of the time-varying regression functions $f_{P,n,i,a}(t/n,\cdot)$ and $g_{P,n,j,b}(t/n,\cdot)$ at rescaled time $t/n\in [0,1]$ are given by \begin{align*}
\hat{f}_{t,n,i,a}(t/n,\cdot) &= \sum_{k\in [d_Z]}\sum_{c\in C_k} \hat{f}_{t,n,i,a,k,c}(t/n,\cdot),\\
\hat{g}_{t,n,j,b}(t/n,\cdot)&= \sum_{k\in [d_Z]}\sum_{c\in C_k} \hat{g}_{t,n,j,b,k,c}(t/n,\cdot),\end{align*} where the estimators of the time-varying partial response functions $f_{P,n,i,a,k,c}(t/n,\cdot)$ and $g_{P,n,j,b,k,c}(t/n,\cdot)$ at rescaled time $t/n\in [0,1]$ are given by \begin{align*} \hat{f}_{t,n,i,a,k,c}(t/n,\cdot)&= \sum_{\ell_1=1}^{\tilde{c}_n} \sum_{\ell_2=1}^{\tilde{d}_n} \hat{\beta}_{t,n,i,a,k,c,\ell_1,\ell_2}^f b_{\ell_1,\ell_2}(t/n,\cdot), \\
\hat{g}_{t,n,j,b,k,c}(t/n,\cdot)&=\sum_{\ell_1=1}^{\tilde{c}_n} \sum_{\ell_2=1}^{\tilde{d}_n} \hat{\beta}_{t,n,j,b,k,c,\ell_1,\ell_2}^g b_{\ell_1,\ell_2}(t/n,\cdot).\end{align*}

Although we only discuss the sieve estimator here, we emphasize that any black-box time-varying regression estimator can be used \parencite{vogt,wbwu_timevar_npreg,boosting_loc_stat_serena_ng,mult_str_wu_2022,zhouzhou_sieve}. 
% To use kernel smoothing estimators for sequential estimation, we can use one-sided temporal kernels so that observations after rescaled time $t/n$ receive a weight of zero. This is practically important because \say{local} nonparametric estimators are naturally far more computationally efficient for sequential estimation than \say{global} nonparametric estimators in the absence of efficient online estimation procedures for the latter.

\subsection{Locally Stationary Error Processes}\label{subsection:err_proc_SIEVE}

We will now introduce the causal representations of the locally stationary error processes from Section~\ref{subsection:tv_regr_SIEVE}. Let $(\eta_{t}^{\bm{\varepsilon}},\eta_{t}^{\bm{\xi}})_{t\in\mathbb{Z}}$ be a sequence of iid random vectors, where $\eta_{t}^{\bm{\varepsilon}}=(\eta_{t,a}^{\varepsilon})_{a \in A}$, $\eta_{t}^{\bm{\xi}}=(\eta_{t,b}^{\xi})_{b\in B}$, and each of the $\eta_{t,a}^{\varepsilon}$, $\eta_{t,b}^{\xi}$ are random vectors. For each $a\in A$, $b\in B$, define the input sequences $$\mathcal{H}_{t,a}^{\varepsilon}=(\eta_{t,a}^{\varepsilon},\eta_{t-1,a}^{\varepsilon},\ldots), \text{ }\mathcal{H}_{t,b}^{\xi}=(\eta_{t,b}^{\xi},\eta_{t-1,b}^{\xi},\ldots).$$ Denote the dimension of $\eta_{t,a}^{\varepsilon}=\eta_{t,a,n}^{\varepsilon}$ by $d_{\varepsilon}^{\eta}=d_{\varepsilon,n}^{\eta}$ and the dimension of $\eta_{t,b}^{\xi}=\eta_{t,b,n}^{\xi}$ by $d_{\xi}^{\eta}=d_{\xi,n}^{\eta}$, both of which can change with $n$. For the next assumption, let $\mathcal{H}_{t}^{\bm{\hat{f}}}=(\mathcal{H}_{t,a}^{\hat{f}})_{a\in A}$, $\mathcal{H}_{t}^{\bm{\hat{g}}}=(\mathcal{H}_{t,b}^{\hat{g}})_{b\in B}$ and $\mathcal{H}_{t,a}^{\hat{f}}=(\mathcal{H}_{t,a}^{\bm{\mathfrak{D}}^{\hat{f}}},\mathcal{H}_{t}^{\bm{Z}})$, $\mathcal{H}_{t,b}^{\hat{g}}=(\mathcal{H}_{t,b}^{\bm{\mathfrak{D}}^{\hat{g}}},\mathcal{H}_{t}^{\bm{Z}})$, where the input sequences $\mathcal{H}_{t,a}^{\bm{\mathfrak{D}}^{\hat{f}}}$, $\mathcal{H}_{t,b}^{\bm{\mathfrak{D}}^{\hat{g}}}$ were defined in Section~\ref{subsection:tv_regr_SIEVE} and $\mathcal{H}_{t}^{\bm{Z}}$ was defined in Section~\ref{subsection:observed_proc_SIEVE}.

\begin{assumption}[Causal representations of the error processes]\label{asmpt_causal_rep_errors_SIEVE} 

Assume that for each $n \in \mathbb{N}$, $P \in \mathcal{P}_n$, $(i,j,a,b)\in\mathcal{D}_n$, $t\in\mathcal{T}_n$, the error processes from Section~\ref{subsection:tv_regr_SIEVE} can be represented as 
$$\varepsilon_{P,t,n,i,a} = \tilde{G}_{P,n,i,a}^{\varepsilon}(t/n,\mathcal{H}_{t,a}^{\varepsilon}),\text{ }\xi_{P,t,n,j,b} = \tilde{G}_{P,n,j,b}^{\xi}(t/n,\mathcal{H}_{t,b}^{\xi}),$$ % yes, that's right, this martingale difference sequence assumption is for the error conditional on the other predictor's estimation error input sequence
% we dont need  $(\mathcal{H}_{t,a}^{\varepsilon})_{a\in A} \!\perp \!\!\! \perp  \mathcal{H}_{t}^{\bm{\hat{g}}}$ and $(\mathcal{H}_{t,b}^{\xi})_{b\in B}\!\perp \!\!\! \perp  \mathcal{H}_{t}^{\bm{\hat{f}}}$, 
with $\mathbb{E}_P(\varepsilon_{P,t,n,i,a}|\mathcal{H}_{t}^{\bm{\hat{g}}})=0$ and  $\mathbb{E}_P(\xi_{P,t,n,j,b}|\mathcal{H}_{t}^{\bm{\hat{f}}})=0$. Assume for all $t'\geq t$ and $h\geq 0$, that
\begin{align*}&\eta_{t-h}^{\bm{\varepsilon}}\perp \!\!\! \perp (\mathcal{H}_{t'}^{\bm{\hat{g}}}, \ldots,\eta_{t-h-2}^{\bm{\varepsilon}},\eta_{t-h-1}^{\bm{\varepsilon}},\eta_{t-h+1}^{\bm{\varepsilon}},\eta_{t-h+2}^{\bm{\varepsilon}},\ldots)\mid (\mathcal{H}_{t}^{\bm{\hat{g}}},\eta_{t}^{\bm{\varepsilon}},\ldots,\eta_{t-h+1}^{\bm{\varepsilon}}),\\ & \eta_{t-h}^{\bm{\xi}}\perp \!\!\! \perp (\mathcal{H}_{t'}^{\bm{\hat{f}}}, \ldots,\eta_{t-h-2}^{\bm{\xi}},\eta_{t-h-1}^{\bm{\xi}},\eta_{t-h+1}^{\bm{\xi}},\eta_{t-h+2}^{\bm{\xi}},\ldots) \mid  (\mathcal{H}_{t}^{\bm{\hat{f}}},\eta_{t}^{\bm{\xi}},\ldots,\eta_{t-h+1}^{\bm{\xi}}).\end{align*} The causal representations $$\tilde{\varepsilon}_{P,t,n,i,a}(u)=\tilde{G}_{P,n,i,a}^{\varepsilon} (u,\mathcal{H}_{t,a}^{\varepsilon} ),\text{ } \tilde{\xi}_{P,t,n,j,b}(u)=\tilde{G}_{P,n,j,b}^{\xi} (u,\mathcal{H}_{t,b}^{\xi} ),$$ are defined at all $u\in [0,1]$, so that we have $\varepsilon_{P,t,n,i,a}=\tilde{\varepsilon}_{P,t,n,i,a}(t/n)$, $\xi_{P,t,n,j,b}=\tilde{\xi}_{P,t,n,j,b}(t/n)$. $\tilde{G}_{P,n,i,a}^{\varepsilon}(u,\cdot)$ and $\tilde{G}_{P,n,j,b}^{\xi}(u,\cdot)$ are measurable functions from $(\mathbb{R}^{d_{\varepsilon}^{\eta}})^{\infty}$ and $(\mathbb{R}^{d_{\xi}^{\eta}})^{\infty}$, respectively, to $\mathbb{R}$ so that $\tilde{G}_{P,n,i,a}^{\varepsilon}(u,\mathcal{H}_{s,a}^{\varepsilon})$, $\tilde{G}_{P,n,j,b}^{\xi}(u,\mathcal{H}_{s,b}^{\xi})$ are well-defined random variables for each $s\in\mathbb{Z}$ and $(\tilde{G}_{P,n,i,a}^{\varepsilon}(u,\mathcal{H}_{s,a}^{\varepsilon}))_{s\in\mathbb{Z}}$, $(\tilde{G}_{P,n,j,b}^{\xi}(u,\mathcal{H}_{s,b}^{\xi}))_{s\in\mathbb{Z}}$  are stationary ergodic processes.
\end{assumption}

As in Section 3.3, we have not defined the input sequences for the error processes separately for each dimension, because without loss of generality we may define the measurable functions $\tilde{G}_{P,n,i,a}^{\varepsilon}(u,\cdot)$,  $\tilde{G}_{P,n,j,b}^{\xi}(u,\cdot)$ and inputs $\eta_{t,a}^{\varepsilon},\eta_{t,b}^{\xi}$ so that each dimension of the error processes has idiosyncratic inputs.

Using the causal representations of the univariate error processes, we have the following causal representations of the vector-valued error processes 
\begin{align*} \bm{\tilde{\varepsilon}}_{P,t,n}(u)&=\bm{\tilde{G}}_{P,n}^{\bm{\varepsilon}}(u,\mathcal{H}_{t}^{\bm{\varepsilon}})=(\tilde{G}_{P,n,i,a}^{\varepsilon}(u,\mathcal{H}_{t,a}^{\varepsilon}))_{i\in [d_X], a\in A_i},\\
 \bm{\tilde{\xi}}_{P,t,n}(u)&=\bm{\tilde{G}}_{P,n}^{\bm{\xi}}(u,\mathcal{H}_t^{\bm{\xi}})=(\tilde{G}_{P,n,j,b}^{\xi}(u,\mathcal{H}_{t,b}^{\xi}))_{j\in [d_Y], b\in B_j},
\end{align*}
so that we have $\bm{\varepsilon}_{P,t,n}=\bm{\tilde{\varepsilon}}_{P,t,n}(t/n)$, $\bm{\xi}_{P,t,n}=\bm{\tilde{\xi}}_{P,t,n}(t/n)$, where $\mathcal{H}_t^{\bm{\varepsilon}}=(\eta_{t}^{\bm{\varepsilon}},\eta_{t-1}^{\bm{\varepsilon}},\ldots)$, $\mathcal{H}_t^{\bm{\xi}}=(\eta_{t}^{\bm{\xi}},\eta_{t-1}^{\bm{\xi}},\ldots)$ with $\eta_{t}^{\bm{\varepsilon}}=(\eta_{t,a}^{\varepsilon})_{a\in A}$, $\eta_{t}^{\bm{\xi}}=(\eta_{t,b}^{\xi})_{b\in B}$ for each $t\in\mathbb{Z}$. Similarly, for each $m=(i,j,a,b) \in \mathcal{D}_n$ the error products can be represented as \begin{align*}
\tilde{R}_{P,t,n,m}(u)=\tilde{G}_{P,n,m}^{R}(u,\mathcal{H}_{t,m}^{R})=\tilde{G}_{P,n,i,a}^{\varepsilon}(u,\mathcal{H}_{t,a}^{\varepsilon})\tilde{G}_{P,n,j,b}^{\xi}(u,\mathcal{H}_{t,b}^{\xi}),\end{align*} so that we have $R_{P,t,n,m}=\tilde{R}_{P,t,n,m}(t/n)$, where $\mathcal{H}_{t,m}^{R}=(\eta_{t,m}^{R},\eta_{t-1,m}^{R},\ldots)$ with $\eta_{t,m}^{R}=(\eta_{t,a}^{\varepsilon},\eta_{t,b}^{\xi})$ for each $t\in\mathbb{Z}$. Also, we have the following causal representation of the nonstationary $\mathbb{R}^{D_n}$-valued process of all the products of errors  \begin{align*}
    \bm{\tilde{R}}_{P,n,t}(u)=\bm{\tilde{G}}_{P,n}^{\bm{R}}(u,\mathcal{H}_{t}^{\bm{R}})=(\tilde{G}_{P,n,m}^{R}(u,\mathcal{H}_{t,m}^{R}) )_{m\in\mathcal{D}_n},
\end{align*} so that we have $\bm{R}_{P,t,n}=\bm{\tilde{R}}_{P,n,t}(t/n)$, where $\mathcal{H}_t^{\bm{R}}=(\eta_{t}^{\bm{R}},\eta_{t-1}^{\bm{R}},\ldots)$ and $\eta_{t}^{\bm{R}}=(\eta_{t}^{\bm{\varepsilon}},\eta_{t}^{\bm{\xi}})$ for each $t\in\mathbb{Z}$. We emphasize that for a fixed $P\in\mathcal{P}_n$, $u\in [0,1]$, and $n \in \mathbb{N}$, we have that $\bm{\tilde{G}}_{P,n}^{\bm{R}}(u,\mathcal{H}_s^{\bm{R}})$ is a well-defined random vector for each $s\in\mathbb{Z}$ and $(\bm{\tilde{G}}_{P,n}^{\bm{R}}(u,\mathcal{H}_s^{\bm{R}}))_{s\in\mathbb{Z}}$ is a stationary ergodic $\mathbb{R}^{D_n}$-valued process.

\subsection{Assumptions on Dependence and Nonstationarity}\label{subsection:dependence_nonstationarity_SIEVE}

We impose assumptions on the rate of decay in temporal dependence and the degree of nonstationarity of the observed processes and error processes. We emphasize that the assumptions here are strictly stronger than those in Section 3.4. We impose these stronger assumptions to guarantee that the sieve time-varying nonlinear regression estimator achieves the convergence rates required by Theorem 3.1. Note that the assumptions here require that the nonstationary processes evolve \say{smoothly} in time, which excludes nonstationary processes with abrupt changes. We do this mainly to simplify the presentation, and we discuss extensions to nonstationary processes with both smooth and abrupt changes in Section~\ref{subsection:piecewise_locally_stationary}.

Denote the set of well-defined tuples of observed processes, dimensions, and time-offsets by  $$\mathcal{W}=\{(X,i,a):i\in [d_X],a\in A_i\}\cup\{(Y,j,b):j\in [d_Y],b\in B_j\}\cup\{(Z,k,c):k\in [d_Z],c\in C_k\},$$ so that we may conveniently refer to such well-defined combinations by $(W,l,d) \in \mathcal{W}$. Also, denote the set of well-defined tuples of error processes, dimensions, and time-offsets by   
$$\mathcal{E}=\{(\varepsilon,i,a):i\in [d_X],a\in A_i\}\cup\{(\xi,j,b):j\in [d_Y],b\in B_j\},$$ so that we may write $(e,l,d) \in \mathcal{E}$ to refer to any such combination.

Again, we quantify temporal dependence via the functional dependence measure of \textcite{wu_funct_dep_meas}. Let $(\tilde{\eta}_t^X,\tilde{\eta}_t^Y,\tilde{\eta}_t^Z)_{t\in\mathbb{Z}}$ be an iid copy of $(\eta_t^X,\eta_t^Y,\eta_t^Z)_{t\in\mathbb{Z}}$. Going forward, the inputs with the tilde are from $(\tilde{\eta}_t^X,\tilde{\eta}_t^Y,\tilde{\eta}_t^Z)_{t\in\mathbb{Z}}$. Similarly, let $(\tilde{\eta}_{t}^{\bm{\varepsilon}},\tilde{\eta}_{t}^{\bm{\xi}})_{t\in\mathbb{Z}}$ be an iid copy of $(\eta_{t}^{\bm{\varepsilon}},\eta_{t}^{\bm{\xi}})_{t\in\mathbb{Z}}$, where $\tilde{\eta}_{t}^{\bm{\varepsilon}}=(\tilde{\eta}_{t,a}^{\varepsilon})_{a\in A}$ and $\tilde{\eta}_{t}^{\bm{\xi}}=(\tilde{\eta}_{t,b}^{\xi})_{b\in B}$.

For any tuple $(W,l,d) \in \mathcal{W}$ corresponding to a well-defined combination of an observed process, dimension, and time-offset, define $$\tilde{\mathcal{H}}_{t,d,h}^{W}=(\eta_{t+d}^{W},\ldots,\eta_{t-h+1+d}^{W},\tilde{\eta}_{t-h+d}^{W},\eta_{t-h-1+d}^{W},\ldots),$$ to be $\mathcal{H}_{t,d}^{W}$ with the input $\eta_{t-h+d}^{W}$ replaced with the iid copy $\tilde{\eta}_{t-h+d}^{W}$. For example, for $i\in [d_X]$, $a\in A_i$, we have that $\tilde{\eta}_{t-h+a}^X$ is the copy of the input $\eta_{t-h+a}^{X}$ in the input sequence $\mathcal{H}_{t,a}^{X}$ used in the causal representation of $X_{t,n,i,a}$. Analogously, for $\bm{W}\in\{\bm{X},\bm{Y},\bm{Z}\}$ define $\tilde{\mathcal{H}}_{t,h}^{\bm{W}}$ as $\mathcal{H}_{t}^{\bm{W}}$ with the input $\eta_{t-h}^{\bm{W}}$ replaced with the iid copy $\tilde{\eta}_{t-h}^{\bm{W}}$ as in Section~\ref{subsection:observed_proc_SIEVE}.

For any tuple $(e,l,d) \in \mathcal{E}$ corresponding to a well-defined combination of an error process, dimension, and time-offset, define $$\tilde{\mathcal{H}}_{t,d,h}^{e}=(\eta_{t,d}^{e},\ldots,\eta_{t-h+1,d}^{e},\tilde{\eta}_{t-h,d}^{e},\eta_{t-h-1,d}^{e},\ldots),$$ to be $\mathcal{H}_{t,d}^{e}$ with the input $\eta_{t-h,d}^{e}$ replaced with the iid copy $\tilde{\eta}_{t-h,d}^{e}$, where each $\tilde{\eta}_{t-h,d}^{e}$ is the component indexed by $d$ in the iid copy $\tilde{\eta}_{t-h}^{\bm{e}}$ of the random vector $\eta_{t-h}^{\bm{e}}$. Similarly, for $\bm{e}\in\{\bm{\varepsilon},\bm{\xi}\}$ define $\tilde{\mathcal{H}}_{t,h}^{\bm{e}}$ as $\mathcal{H}_{t}^{\bm{e}}$ with the input $\eta_{t-h}^{\bm{e}}$ replaced with the iid copy $\tilde{\eta}_{t-h}^{\bm{e}}$ as in Section~\ref{subsection:err_proc_SIEVE}. Also, for the product of errors define $\tilde{\mathcal{H}}_{t,m,h}^R$ as $\mathcal{H}_{t,m}^R$ with the input $\eta_{t-h,m}^R$ replaced with the iid copy $\tilde{\eta}_{t-h,m}^R$ for $m=(i,j,a,b)\in\mathcal{D}_n$, and define $\tilde{\mathcal{H}}_{t,h}^{\bm{R}}$ as $\mathcal{H}_{t}^{\bm{R}}$ with the input $\eta_{t-h}^{\bm{R}}$ replaced with the iid copy $\tilde{\eta}_{t-h}^{\bm{R}}$ as in Section~\ref{subsection:err_proc_SIEVE}.

Let $\mathcal{H}_{t}^{\bm{\hat{\kappa}}}$ be the input for the other target's predictor, i.e., $(\bm{e},\bm{\hat{\kappa}})\in\{(\bm{\varepsilon},\bm{\hat{g}}),(\bm{\xi},\bm{\hat{f}})\}$. Define $$\bar{\mathcal{H}}_{t,d,h}^{e}=(\eta_{t,d}^{e},\ldots,\eta_{t-h+1,d}^{e},\bar{\eta}_{t-h,d}^{e},\eta_{t-h-1,d}^{e},\ldots),$$ as $\mathcal{H}_{t,d}^{e}$ with $\eta_{t-h,d}^{e}$ replaced with the component indexed by $d$, $\bar{\eta}_{t-h,d}^{e}$, in the $\sigma(\eta_{t}^{\bm{e}},\ldots,\eta_{t-h+1}^{\bm{e}},\mathcal{H}_{t}^{\bm{\hat{\kappa}}})$-conditionally iid copy $\bar{\eta}_{t-h}^{\bm{e}}$ of $\eta_{t-h}^{\bm{e}}$, where $\mathcal{L}(\bar{\eta}_{t-h}^{\bm{e}}| \eta_{t}^{\bm{e}},\ldots,\eta_{t-h+1}^{\bm{e}},\mathcal{H}_{t}^{\bm{\hat{\kappa}}})$ is equal to $\mathcal{L}(\eta_{t-h}^{\bm{e}}|\eta_{t}^{\bm{e}},\ldots,\eta_{t-h+1}^{\bm{e}},\mathcal{H}_{t}^{\bm{\hat{\kappa}}})$ almost surely, and $\bar{\eta}_{t-h}^{\bm{e}}$ is conditionally independent of $(\eta_{t-h}^{\bm{e}},\eta_{t-h-1}^{\bm{e}},\ldots)$ given $(\eta_{t}^{\bm{e}},\ldots,\eta_{t-h+1}^{\bm{e}},\mathcal{H}_{t}^{\bm{\hat{\kappa}}})$. We suppress the dependence of $\bar{\eta}_{t-h}^{\bm{e}}$ on the distribution $P$, as well as $t$ and $h$ separately, to simplify the notation.

Next, we define the functional dependence measures of the processes.

\begin{definition}[Functional dependence measures]\label{def_funct_dep_SIEVE}

% q > 2 required for def 2.1 and asmpt 2.1 in sieve paper
% q \geq 2 required for local long-run covariances of $\bm{w}_{t,n}^{\varphi(Z)}$, $\bm{w}_{P,t,n,i,a}^{\varphi(Z),\varepsilon}$, $\bm{w}_{P,t,n,j,b}^{\varphi(Z),\xi}$ to be well-defined by Prop 5.4 from \textcite{seq_gauss_approx2022} paper 
We define the following measures of temporal dependence for each $n\in\mathbb{N}$, $P\in\mathcal{P}_n$, $u\in [0,1]$, $t\in\mathcal{T}_n$. First, define the functional dependence measures of the observed processes $\tilde{G}_{n,l,d}^W(u,\mathcal{H}_{t,d}^W)$ for each $(W,l,d)\in\mathcal{W}$ with $h\in\mathbb{N}_0$, and some $q\geq 1$ as  $$\theta_{P,u,t,n,l,d}^W(h,q)= [\mathbb{E}_P(|\tilde{G}_{n,l,d}^W(u,\mathcal{H}_{t,d}^W)-\tilde{G}_{n,l,d}^W(u,\tilde{\mathcal{H}}_{t,d,h}^W)|^q)]^{1/q},$$
 and for the vector-valued process $\bm{\tilde{G}}_{n}^{\bm{W}}(u,\mathcal{H}_{t}^{\bm{W}})$ for each $\bm{W}\in\{\bm{X},\bm{Y},\bm{Z}\}$ with $h\in\mathbb{N}_0$, and some $q \geq 1$, $r\geq 1$ as $$\theta_{P,u,t,n}^{\bm{W}}(h,q,r)= [\mathbb{E}_P(||\bm{\tilde{G}}_{n}^{\bm{W}}(u,\mathcal{H}_{t}^{\bm{W}})-\bm{\tilde{G}}_{n}^{\bm{W}}(u,\tilde{\mathcal{H}}_{t,h}^{\bm{W}})||_r^q)]^{1/q}.$$

% L^infinity required for step 1.2 of proof to keep estimation errors in terms of second moment
% partial version for step 1.2
% q > 2 required for def 2.1 and asmpt 2.1 in sieve paper
% q \geq 2 required for local long-run covariances of $\bm{w}_{t,n}^{\varphi(Z)}$, $\bm{w}_{P,t,n,i,a}^{\varphi(Z),\varepsilon}$, $\bm{w}_{P,t,n,j,b}^{\varphi(Z),\xi}$ to be well-defined by Prop 5.4 from \textcite{seq_gauss_approx2022} paper 
Second, define the $L^{\infty}(P)$ versions of the functional dependence measures of the error processes $\tilde{G}_{P,n,l,d}^e(u,\mathcal{H}_{t,d}^e)$ for each $(e,l,d)\in\mathcal{E}$ with $h\in\mathbb{N}_0$ as
$$\theta_{P,u,t,n,l,d}^{e,\infty}(h)= \inf\{K \geq 0 : \mathbb{P}_P(|\tilde{G}_{P,n,l,d}^e(u,\mathcal{H}_{t,d}^e)-\tilde{G}_{P,n,l,d}^e(u,\tilde{\mathcal{H}}_{t,d,h}^e)|>K)=0\},$$ and the corresponding conditional version as $$\bar{\theta}_{P,u,t,n,l,d}^{e,\infty}(h)= \inf\{K \geq 0 : \mathbb{P}_P(|\tilde{G}_{P,n,l,d}^e(u,\mathcal{H}_{t,d}^e)-\tilde{G}_{P,n,l,d}^e(u,\bar{\mathcal{H}}_{t,d,h}^e)|>K)=0\}.$$ Similarly, for the vector-valued process $\bm{\tilde{G}}_{P,n}^{\bm{e}}(u,\mathcal{H}_{t}^{\bm{e}})$ for each $\bm{e}\in\{\bm{\varepsilon}, \bm{\xi}\}$ with $h\in\mathbb{N}_0$, and some $r\geq 1$, define $$\theta_{P,u,t,n}^{\bm{e},\infty}(h,r)= \inf\{K \geq 0 : \mathbb{P}_P(||\bm{\tilde{G}}_{P,n}^{\bm{e}}(u,\mathcal{H}_{t}^{\bm{e}})-\bm{\tilde{G}}_{P,n}^{\bm{e}}(u,\tilde{\mathcal{H}}_{t,h}^{\bm{e}})||_r>K)=0\}.$$

% Euclidean vector norm with q > 4 required for step 2 of proof SGA
Third, define the functional dependence measures of the processes of error products $\tilde{G}_{P,n,m}^R(u,\mathcal{H}_{t,m}^R)$ for each $m=(i,j,a,b)\in\mathcal{D}_n$ with $h\in\mathbb{N}_0$, and some $q\geq 1$ as $$\theta_{P,u,t,n,m}^R(h,q) = [\mathbb{E}_P(|\tilde{G}_{P,n,m}^R(u,\mathcal{H}_{t,m}^R)-\tilde{G}_{P,n,m}^R(u,\tilde{\mathcal{H}}_{t,m,h}^R)|^{q})]^{1/q},$$ and for the vector-valued process $\bm{\tilde{G}}_{P,n}^{\bm{R}}(u,\mathcal{H}_{t}^{\bm{R}})$ with $h\in\mathbb{N}_0$, and some $q\geq 1$, $r\geq 1$ as $$\theta_{P,u,t,n}^{\bm{R}}(h,q,r) = [\mathbb{E}_P(||\bm{\tilde{G}}_{P,n}^{\bm{R}}(u,\mathcal{H}_{t}^{\bm{R}})-\bm{\tilde{G}}_{P,n}^{\bm{R}}(u,\tilde{\mathcal{H}}_{t,h}^{\bm{R}})||_r^{q})]^{1/q}.$$

\end{definition}

Next, we introduce an assumption imposing a uniform polynomial decay of the temporal dependence. Note that we will often write the time as $0$ when the time of the input sequence does not matter because of stationarity. For some sequence of collections of distributions $(\mathcal{P}_n)_{n\in\mathbb{N}}$, we make the following assumption.

\begin{assumption}[Distribution-uniform decay of temporal dependence]\label{asmpt_funct_dep_SIEVE}

% q > 2 required for def 2.1 and asmpt 2.1 in sieve paper, but only requires beta>1 due to remark 3.1 after theorem 3.2 in sieve paper with the Legendre polynomials. See Lemma C.3 in Appendix C of sieve paper.
% q \geq 2 and \beta>2 required for local long-run covariances of $\bm{w}_{t,n}^{\varphi(Z)}$, $\bm{w}_{P,t,n,i,a}^{\varphi(Z),\varepsilon}$, $\bm{w}_{P,t,n,j,b}^{\varphi(Z),\xi}$ to be well-defined by Prop 5.4 from \textcite{seq_gauss_approx2022} paper 
% since we are assuming L infinity norm of errors bounded, we will use Holders (L infinity version) to bound in terms of L2 moment and FDM, so this is why we require q > 2 instead of q>4
% however, to simplify things later on, we will assume q>4 so that the stochastic lipschitz condition for the product of basis and error has q>2...
Assume that there exist $\bar{\Theta} >0$, $\bar{\beta} >2$, $\bar{q} >4$, such that for all observed processes $(W,l,d) \in \mathcal{W}$, it holds that, for $h\geq 0$, \begin{align*} \underset{n\in\mathbb{N}}{\sup} \ \underset{P\in\mathcal{P}_n}{\sup} \ \underset{u\in [0,1]}{\sup} [\mathbb{E}_P(|\tilde{G}_{n,l,d}^W(u,\mathcal{H}_{0,d}^W)|^{\bar{q}})]^{1/\bar{q}} &\leq \bar{\Theta}, \\  \underset{n\in\mathbb{N}}{\sup} \ \underset{P\in\mathcal{P}_n}{\sup} \ \underset{u\in [0,1]}{\sup} \theta_{P,u,0,n,l,d}^W(h,\bar{q}) &\leq \bar{\Theta} \cdot (h\lor 1)^{-\bar{\beta}}.\end{align*}

% to do: remove, dont need anymore as we are assuming L infinity norm
% and for the error processes $(e,l,d)\in\mathcal{E}$, it holds that $$\underset{P\in\mathcal{P}_n}{\sup}[\mathbb{E}_P(|\tilde{G}_{P,n,l,d}^e(u,\mathcal{H}_{0,d}^e)|^{\bar{q}})]^{1/\bar{q}} \leq \bar{\Theta}.$$

% L infinity and β > 1 required for Step 1.2 to keep the PE requirements in terms of L2 norm
% q \geq 2 and \beta>2 required for local long-run covariances of $\bm{w}_{t,n}^{\varphi(Z)}$, $\bm{w}_{P,t,n,i,a}^{\varphi(Z),\varepsilon}$, $\bm{w}_{P,t,n,j,b}^{\varphi(Z),\xi}$ to be well-defined by Prop 5.4 from \textcite{seq_gauss_approx2022} paper 
% for these reasons, we will assume beta^inf > 2 
% note that Proposition 4.3 of \textcite{seq_gauss_approx2022} requires q > 4, β > 2, for the *product of errors*, so we could alternatively assume each error has β > 4 so that the requirements can be satisfied for product of errors by Holder's inequality... but we will just assume separately for product of errors (see below)
% dont need L-infinity norm bound, just need Lp for p>2 for each error for sieve (see above)
Also, assume that there exist $\bar{\Theta}^{\infty} >0$, $\bar{\beta}^{\infty} >2$, such that for all error processes $(e,l,d)\in\mathcal{E}$, it holds that, for $h\geq 0$,
\begin{align*}\underset{n\in\mathbb{N}}{\sup} \ \underset{P\in\mathcal{P}_n}{\sup} \ \underset{u\in [0,1]}{\sup} ||G_{P,n,l,d}^e(u,\mathcal{H}_{0,d}^e)||_{L^{\infty}(P)} &\leq \bar{\Theta}^{\infty},\\   \underset{n\in\mathbb{N}}{\sup} \ \underset{P\in\mathcal{P}_n}{\sup} \ \underset{u\in [0,1]}{\sup} \ \underset{t\in\mathcal{T}_n}{\max} \  \max(\theta_{P,u,t,n,l,d}^{e,\infty}(h),\bar{\theta}_{P,u,t,n,l,d}^{e,\infty}(h)) &\leq \bar{\Theta}^{\infty} \cdot (h\lor 1)^{-\bar{\beta}^{\infty}}.\end{align*}

% Proposition 4.3 of \textcite{seq_gauss_approx2022} requires q > 4, β > 2, for the product of errors
% sieve does not require anything more 
% we will require β>3 for limiting rates on dim growth to simplify
For additional control in terms of the product of errors alone, also assume that there exist $\bar{\Theta}^R>0$, $\bar{\beta}^R>3$, $\bar{q}^R>4$, such that we have, for $h\geq 0$, \begin{align*} \underset{n\in\mathbb{N}}{\sup} \ \underset{P\in\mathcal{P}_n}{\sup} \ \underset{u\in [0,1]}{\sup} \ \underset{m\in\mathcal{D}_n}{\max} \ [\mathbb{E}_P(|\tilde{G}_{P,n,m}^R(u,\mathcal{H}_{0,m}^R)|^{\bar{q}^R})]^{1/\bar{q}^R} &\leq \bar{\Theta}^R, \\ \underset{n\in\mathbb{N}}{\sup} \ \underset{P\in\mathcal{P}_n}{\sup} \ \underset{u\in [0,1]}{\sup} \ \underset{m\in\mathcal{D}_n}{\max} \ \theta_{P,u,0,n,m}^R(h,\bar{q}^R)&\leq \bar{\Theta}^R \cdot (h\lor 1)^{-\bar{\beta}^R}.\end{align*} 
% PDM on the product of errors is so that the strong Gaussian approximation can be used

\end{assumption}

% The requirement that $\bar{\beta}>1$ is due to the sieve estimator with Legendre polynomials as the basis functions and choosing the numbers of basis functions to satisfy $\tilde{c}_n=\tilde{d}_n=O(\log(T_n))$. See Theorem 3.2, Remark 3.1, Example C.2, and Lemma C.3 in \textcite{zhouzhou_sieve}. Note that this is an assumption required by the sieve estimator, whereas in the general framework from Section 3.4 we did not make any assumptions about the dependence of the observed processes themselves.

In view of Assumption~\ref{asmpt_funct_dep_SIEVE}, we have the following bounds on the functional dependence measures of the corresponding vector-valued processes for each $n \in \mathbb{N}$ by Jensen's inequality. For the vector-valued process of error products we have, for $h\geq 0$,
\begin{align*}\underset{P\in\mathcal{P}_n}{\sup} \ \underset{u\in [0,1]}{\sup} \ [\mathbb{E}_P(||\bm{\tilde{G}}_{P,n}^{\bm{R}}(u,\mathcal{H}_{0}^{\bm{R}})||_2^{\bar{q}^R})]^{1/\bar{q}^R} & \leq D_n^{\frac{1}{2}}\bar{\Theta}^R, \\ \underset{P\in\mathcal{P}_n}{\sup} \ \underset{u\in [0,1]}{\sup} \  \theta_{P,u,0,n}^{\bm{R}}(h,\bar{q}^R,2) &\leq D_n^{\frac{1}{2}} \bar{\Theta}^R \cdot (h\lor 1)^{-\bar{\beta}^R}.\end{align*} % Proposition 4.3 of \textcite{seq_gauss_approx2022} requires q > 4, β > 2, for the vector of product of errors
% sieve does not require anything more
 Also, for each of the vector-valued observed processes $\bm{W} \in \{\bm{X},\bm{Y},\bm{Z}\}$, we have, for $h\geq 0$,
   \begin{align*} \underset{P\in\mathcal{P}_n}{\sup} \ \underset{u\in [0,1]}{\sup} \ [\mathbb{E}_P(||\bm{\tilde{G}}_n^{\bm{W}}(u,\mathcal{H}_0^{\bm{W}})||_2^{\bar{q}})]^{1/\bar{q}} &\leq \bm{d_W}^{\frac{1}{2}}\bar{\Theta}, \\ \underset{P\in\mathcal{P}_n}{\sup} \ \underset{u\in [0,1]}{\sup} \ \theta_{P,u,0,n}^{\bm{W}}(h,\bar{q},2) &\leq \bm{d_W}^{\frac{1}{2}}\bar{\Theta} \cdot (h\lor 1)^{-\bar{\beta}},\end{align*} % for the discussion below about the product of error and basis function for z 
   where we denote the dimension of the vector-valued observed process by $\bm{d_W}$. Lastly, for each of the vector-valued error processes $\bm{e} \in \{\bm{\varepsilon},\bm{\xi}\}$, we have, for $h\geq 0$, 
     \begin{align*}\underset{P\in\mathcal{P}_n}{\sup} \ \underset{u\in [0,1]}{\sup} \ \left|\left|||\bm{\tilde{G}}_{P,n}^{\bm{e}}(u,\mathcal{H}_{0}^{\bm{e}})||_2\right|\right|_{L^{\infty}(P)} &\leq \bm{d_e}^{\frac{1}{2}}\bar{\Theta}^{\infty},\\ \underset{P\in\mathcal{P}_n}{\sup} \ \underset{u\in [0,1]}{\sup} \ \theta_{P,u,0,n}^{\bm{e},\infty}(h,2) &\leq \bm{d_e}^{\frac{1}{2}}\bar{\Theta}^{\infty} \cdot (h\lor 1)^{-\bar{\beta}^{\infty}},\end{align*}% for the discussion below about the product of error and basis function for z 
     where we denote the dimension of the vector-valued error process by $\bm{d_e}$.

% this is why we assumed q \geq 2 and \beta>2 for FDM of processes and errors, so that the local long-run covariances of $\bm{w}_{t,n}^{\varphi(Z)}$, $\bm{w}_{P,t,n,i,a}^{\varphi(Z),\varepsilon}$, $\bm{w}_{P,t,n,j,b}^{\varphi(Z),\xi}$ are well-defined by Prop 5.4 from \textcite{seq_gauss_approx2022} paper 
Next, we discuss an additional regularity condition required by the sieve estimator that is analogous to Lemma 3.1 in \textcite{zhouzhou_sieve}. Recall the set of basis functions $\{\varphi_{\ell_2}(z)\}$ from Section~\ref{subsection:tv_regr_SIEVE}. For each $n\in\mathbb{N}$, $P\in\mathcal{P}_n$, $t\in\mathcal{T}_n$, $i\in [d_X]$, $a\in A_i$, $j\in [d_Y]$, $b\in B_j$ let 
\begin{align*}
\bm{w}_{t,n}^{\varphi(Z)}&=(\varphi_{\ell_2}(Z_{t,n,k,c}))_{k\in[d_Z], c\in C_k, 2 \leq \ell_2 \leq \tilde{d}_n},\\
\bm{w}_{P,t,n,i,a}^{\varphi(Z),\varepsilon}&=(\varphi_{\ell_2}(Z_{t,n,k,c})\varepsilon_{P,t,n,i,a})_{k\in[d_Z], c\in C_k, 1 \leq \ell_2 \leq \tilde{d}_n},\\
\bm{w}_{P,t,n,j,b}^{\varphi(Z),\xi}&=(\varphi_{\ell_2}(Z_{t,n,k,c})\xi_{P,t,n,j,b})_{k\in[d_Z], c\in C_k, 1 \leq \ell_2 \leq \tilde{d}_n},\end{align*} so that there is no constant column in $\bm{w}_{t,n}^{\varphi(Z)}$. As in Section 3.2 in \textcite{zhouzhou_sieve}, the processes $\bm{w}_{t,n}^{\varphi(Z)}$, $\bm{w}_{P,t,n,i,a}^{\varphi(Z),\varepsilon}$, and $\bm{w}_{P,t,n,j,b}^{\varphi(Z),\xi}$ all have causal representations 
\begin{align*}\bm{w}_{t,n}^{\varphi(Z)}&=\bm{\tilde{G}}_n^{\bm{w}^{\varphi(Z)}}(t/n,\mathcal{H}_t^{\bm{w}^{\varphi(Z)}}),\\ \bm{w}_{P,t,n,i,a}^{\varphi(Z),\varepsilon}&=\bm{\tilde{G}}_{P,n,i,a}^{\bm{w}^{\varphi(Z),\varepsilon}}(t/n,\mathcal{H}_{t,a}^{\bm{w}^{\varphi(Z),\varepsilon}}),\\
\bm{w}_{P,t,n,j,b}^{\varphi(Z),\xi}&=\bm{\tilde{G}}_{P,n,j,b}^{\bm{w}^{\varphi(Z),\xi}}(t/n,\mathcal{H}_{t,b}^{\bm{w}^{\varphi(Z),\xi}}),\end{align*}
where 
\begin{align*}
\mathcal{H}_t^{\bm{w}^{\varphi(Z)}}&=(\eta_t^{\bm{w}^{\varphi(Z)}},\eta_{t-1}^{\bm{w}^{\varphi(Z)}},\ldots),\\ \mathcal{H}_{t,a}^{\bm{w}^{\varphi(Z),\varepsilon}}&=(\eta_{t,a}^{\bm{w}^{\varphi(Z),\varepsilon}},\eta_{t-1,a}^{\bm{w}^{\varphi(Z),\varepsilon}},\ldots),\\ \mathcal{H}_{t,b}^{\bm{w}^{\varphi(Z),\xi}}&=(\eta_{t,b}^{\bm{w}^{\varphi(Z),\xi}},\eta_{t-1,b}^{\bm{w}^{\varphi(Z),\xi}},\ldots),\end{align*} with $\eta_t^{\bm{w}^{\varphi(Z)}}=\eta_{t+c_{\max}}^{Z}$, $\eta_{t,a}^{\bm{w}^{\varphi(Z),\varepsilon}}=(\eta_{t+c_{\max}}^{Z},\eta_{t,a}^{\varepsilon})$, and $\eta_{t,b}^{\bm{w}^{\varphi(Z),\xi}}=(\eta_{t+c_{\max}}^{Z},\eta_{t,b}^{\xi})$. 
Define the functional dependence measures of the vector-valued processes $\bm{w}_{t,n}^{\varphi(Z)}$, $\bm{w}_{P,t,n,i,a}^{\varphi(Z),\varepsilon}$, $\bm{w}_{P,t,n,j,b}^{\varphi(Z),\xi}$, by
\begin{align*}
\theta_{P,u,t,n}^{\bm{w}^{\varphi(Z)}}(h,q,2)&= [\mathbb{E}_P(||\bm{\tilde{G}}_{n}^{\bm{w}^{\varphi(Z)}}(u,\mathcal{H}_{t}^{\bm{w}^{\varphi(Z)}})-\bm{\tilde{G}}_{n}^{\bm{w}^{\varphi(Z)}}(u,\tilde{\mathcal{H}}_{t,h}^{\bm{w}^{\varphi(Z)}})||_2^q)]^{1/q},\\
\theta_{P,u,t,n,i,a}^{\bm{w}^{\varphi(Z),\varepsilon}}(h,q,2)&= [\mathbb{E}_P(||\bm{\tilde{G}}_{P,n,i,a}^{\bm{w}^{\varphi(Z),\varepsilon}}(u,\mathcal{H}_{t,a}^{\bm{w}^{\varphi(Z),\varepsilon}})-\bm{\tilde{G}}_{P,n,i,a}^{\bm{w}^{\varphi(Z),\varepsilon}}(u,\tilde{\mathcal{H}}_{t,a,h}^{\bm{w}^{\varphi(Z),\varepsilon}})||_2^q)]^{1/q},\\
\theta_{P,u,t,n,j,b}^{\bm{w}^{\varphi(Z),\xi}}(h,q,2)&= [\mathbb{E}_P(||\bm{\tilde{G}}_{P,n,j,b}^{\bm{w}^{\varphi(Z),\xi}}(u,\mathcal{H}_{t,b}^{\bm{w}^{\varphi(Z),\xi}})-\bm{\tilde{G}}_{P,n,j,b}^{\bm{w}^{\varphi(Z),\xi}}(u,\tilde{\mathcal{H}}_{t,b,h}^{\bm{w}^{\varphi(Z),\xi}})||_2^q)]^{1/q}.\end{align*}

% note that we have q>4 instead of q>2 due to using L infinity Holders
Recall $\bar{\Theta}$, $\bar{\Theta}^{\infty}$, $\bar{\beta}$, $\bar{\beta}^{\infty}$, and $\bar{q}$ from Assumption~\ref{asmpt_funct_dep_SIEVE}. By the arguments from Lemma 3.1 from \textcite{zhouzhou_sieve}, for each $n\in\mathbb{N}$, the processes $\bm{w}_{P,t,n,i,a}^{\varphi(Z),\varepsilon}$, $\bm{w}_{P,t,n,j,b}^{\varphi(Z),\xi}$ of dimension $\bm{d_{w,e}}$ satisfy % (i.e. adding and subtracting cross-terms, the triangle inequality, the distributive property, and H\"{o}lder's inequality)
\begin{align*}
\underset{P\in\mathcal{P}_n}{\sup} \ \underset{u\in [0,1]}{\sup} \ [\mathbb{E}_P(||\bm{\tilde{G}}_{P,n,i,a}^{\bm{w}^{\varphi(Z),\varepsilon}}(u,\mathcal{H}_{t,a}^{\bm{w}^{\varphi(Z),\varepsilon}})||_2^{\tilde{q}})]^{1/\tilde{q}} &\leq \bm{d_{w,e}}^{\frac{1}{2}}\tilde{\Theta},\\ \underset{P\in\mathcal{P}_n}{\sup} \ \underset{u\in [0,1]}{\sup} \  \theta_{P,u,t,n,i,a}^{\bm{w}^{\varphi(Z),\varepsilon}}(h,\tilde{q},2)&\leq \bm{d_{w,e}}^{\frac{1}{2}}\tilde{\Theta} \cdot (h\lor 1)^{-\tilde{\beta}},\\
\underset{P\in\mathcal{P}_n}{\sup} \ \underset{u\in [0,1]}{\sup} \ [\mathbb{E}_P(||\bm{\tilde{G}}_{P,n,j,b}^{\bm{w}^{\varphi(Z),\xi}}(u,\mathcal{H}_{t,b}^{\bm{w}^{\varphi(Z),\xi}})||_2^{\tilde{q}})]^{1/\tilde{q}} &\leq \bm{d_{w,e}}^{\frac{1}{2}}\tilde{\Theta},\\  \underset{P\in\mathcal{P}_n}{\sup} \ \underset{u\in [0,1]}{\sup} \  \theta_{P,u,t,n,j,b}^{\bm{w}^{\varphi(Z),\xi}}(h,\tilde{q},2)&\leq \bm{d_{w,e}}^{\frac{1}{2}}\tilde{\Theta} \cdot (h\lor 1)^{-\tilde{\beta}},\end{align*}
for $h\geq 0$, where $\tilde{q}=\bar{q}>4$ with $\tilde{\beta}=\min(\bar{\beta},\bar{\beta}^{\infty})>2$ and $\tilde{\Theta}=2K_n(\max(\bar{\Theta}^{\infty},\bar{\Theta}))^2>0$ where $K_n>0$ is due to the growing number of basis functions. % ie the smoothness and boundedness for the two terms, respectively, then just upper bound by whichever is greater... 
Similarly, the vector-valued process $\bm{w}_{t,n}^{\varphi(Z)}$ of dimension $\bm{d_{w}}$ satisfies
\begin{align*}\underset{n\in\mathbb{N}}{\sup} \ \underset{P\in\mathcal{P}_n}{\sup} \ \underset{u\in [0,1]}{\sup} \  [\mathbb{E}_P(||\bm{\tilde{G}}_{n}^{\bm{w}^{\varphi(Z)}}(u,\mathcal{H}_{t}^{\bm{w}^{\varphi(Z)}})||_2^{\bar{q}})]^{1/\bar{q}} &\leq \bm{d_{w}}^{\frac{1}{2}}\bar{\Theta}K_n',\\ \underset{n\in\mathbb{N}}{\sup} \ \underset{P\in\mathcal{P}_n}{\sup} \ \underset{u\in [0,1]}{\sup} \  \theta_{P,u,t,n}^{\bm{w}^{\varphi(Z)}}(h,\bar{q},2)&\leq \bm{d_{w}}^{\frac{1}{2}}\bar{\Theta} K_n' \cdot (h\lor 1)^{-\bar{\beta}},\end{align*}
for $h\geq 0$, where $K_n'>0$ is due to the growing number of basis functions. % ie the smoothness of the basis functions
% note that we have q>4 instead of q>2 due to using L infinity Holders

For Theorem 3.1, we only require that the total variation of the causal mechanism of the process of error products can be bounded distribution-uniformly. However, the sieve estimator requires the stronger assumption that the causal mechanisms of the observed processes and error processes are stochastic Lipschitz functions of rescaled time. We impose the following regularity conditions to control the nonstationarity uniformly over a sequence of collections of distributions $(\mathcal{P}_n)_{n\in\mathbb{N}}$.

\begin{assumption}[Distribution-uniform stochastic Lipschitz condition for nonstationarity]\label{asmpt_stoch_lip_SIEVE}

% q>2 required by sieve asmpt 2.1 for lemma 3.1
% only need q>2 for observed process, but need q>4 for error process to ensure that the product of error and sieve basis of z has q>2 using L-infinity Holder's for the first term and L-2 holders for the second term. this is also why we need q>4 for moment for observed process. 
% to simplify everything, we will just have q>4 for moment and for stochastic lipschitz...
For each $n\in\mathbb{N}$, $(W,l,d)\in \mathcal{W}$, $(e,l,d)\in \mathcal{E}$, and $t\in\mathbb{Z}$, we assume that $\tilde{G}_{n,l,d}^W(\cdot, \mathcal{H}_{t,d}^W)$ and $\tilde{G}_{P,n,l,d}^e(\cdot, \mathcal{H}_{t,d}^e)$ are stochastic Lipschitz functions of rescaled time $u \in [0,1]$. Recall $\bar{\Theta} >0$, $\bar{q} >4$ from Assumption~\ref{asmpt_funct_dep_SIEVE}. Assume that there exists a constant $\bar{L}>0$ such that for all $u,v \in [0,1]$, $(W,l,d) \in\mathcal{W}$, and $(e,l,d) \in\mathcal{E}$, it holds that \begin{align*} \underset{n\in\mathbb{N}}{\sup} \ \underset{P\in\mathcal{P}_n}{\sup} \ [\mathbb{E}_P(|\tilde{G}_{n,l,d}^W(u,\mathcal{H}_{0,d}^W) - \tilde{G}_{n,l,d}^W(v,\mathcal{H}_{0,d}^W)|^{\bar{q}})]^{1/\bar{q}} &\leq \bar{L} \bar{\Theta}  |u - v|,\\ \underset{n\in\mathbb{N}}{\sup} \ \underset{P\in\mathcal{P}_n}{\sup} \   [\mathbb{E}_P(|\tilde{G}_{P,n,l,d}^e(u,\mathcal{H}_{0,d}^e) - \tilde{G}_{P,n,l,d}^e(v,\mathcal{H}_{0,d}^e)|^{\bar{q}})]^{1/\bar{q}} &\leq \bar{L}\bar{\Theta} |u - v|.\end{align*}

\end{assumption}

% used for satisfying asmpt_total_variation_nsts which is used step 2 of the main proof in this paper for applying thm 3.2 in \textcite{seq_gauss_approx2022}. note that this is basically prop 4.3 from seq22, and G.2 only requires q>2. similarly, for thm 3.1 we only require q>2 for this, which we get for free from using L-infinity version of Holder's inequality and assuming q>2 for the stochastic lipschitz for error processes individually from asmpt_stoch_lip_SIEVE and using the L-infinity norm bound on error processes from asmpt_funct_dep_SIEVE
% however for the reasons explained above we assume q>4 stoch lipschitz on each error and thus we get q>4 for prod error stoch lipschitz for free by using L infinity holders due to L infinity norm on each error
In view of Assumption~\ref{asmpt_stoch_lip_SIEVE}, there exist $\tilde{L}^{R}=\bar{L}>0$, $\tilde{q}^R=\bar{q} > 4$, $\tilde{\Theta}^R=2(\max(\bar{\Theta}^{\infty},\bar{\Theta}))^2$ such that for all $u,v\in [0,1]$, we have $$\underset{n\in\mathbb{N}}{\sup} \ \underset{P\in\mathcal{P}_n}{\sup} \  \underset{m\in\mathcal{D}_n}{\max} \  [\mathbb{E}_P(|\tilde{G}_{P,n,m}^R(u,\mathcal{H}_{0,m}^R) - \tilde{G}_{P,n,m}^R(v,\mathcal{H}_{0,m}^R)|^{\tilde{q}^R})]^{1/\tilde{q}^R} \leq \tilde{L}^{R}\tilde{\Theta}^R |u - v|.$$ This follows from adding and subtracting cross-terms, the triangle inequality, the distributive property, H\"{o}lder's inequality, and applying the moment bounds and stochastic Lipschitz conditions for the individual error processes from Assumptions~\ref{asmpt_funct_dep_SIEVE} and~\ref{asmpt_stoch_lip_SIEVE}. It is easy to verify that Assumption 3.6 is satisfied under this stronger condition on the nonstationarity.
% stochastic Lipschitz is enough to imply the total variation-type condition of Assumption ~\ref{asmpt_total_variation_nsts}, i.e. just apply this bound at each time "increment" 

% the prod of basis function and error is q/2 > 2 because of the second term in the triangle inequality with stochastic lipschitz for error and moment for process, both are in terms of L4 since used holders (ie since dont have L infinity norm or L infinity stoch lipschitz)  
Also, using the same arguments as Lemma 3.1 from \textcite{zhouzhou_sieve}, the individual dimensions of the vector-valued processes $\bm{w}_{P,t,n,i,a}^{\varphi(Z),\varepsilon}$, and $\bm{w}_{P,t,n,j,b}^{\varphi(Z),\xi}$ can be shown to satisfy this stochastic Lipschitz condition for moment $\bar{q}/2>2$ with Lipschitz constant $2 K_n \bar{L} (\max(\bar{\Theta}^{\infty},\bar{\Theta}))^2>0$. Similarly, the individual dimensions of the vector-valued process $\bm{w}_{t,n}^{\varphi(Z)}$ can be shown to satisfy this stochastic Lipschitz condition for moment $\bar{q}>4$ with Lipschitz constant $K_n'\bar{L}\bar{\Theta}>0$.

\subsection{Assumptions on Local Long-Run Covariances}\label{subsection:tv_covariances_SIEVE}

To ensure fast convergence rates by the sieve estimator, we require the following assumptions on the local long-run covariance matrices. Note that these assumptions are not made in Section 3.

\begin{definition}[Local long-run covariance matrices of error products]\label{def_local_LR_cov_SIEVE} For each $n\in \mathbb{N}$, $P\in \mathcal{P}_n$, $u\in [0,1]$, define the local long-run covariance matrix $\bm{\tilde{\Sigma}}_{P,n}^{\bm{R}}(u)\in\mathbb{R}^{D_n\times D_n}$ for the $\mathbb{R}^{D_n}$-valued stationary process $(\bm{\tilde{G}}_{P,n}^{\bm{R}}(u,\mathcal{H}_t^{\bm{R}}))_{t\in\mathbb{Z}}$ by
$$\bm{\tilde{\Sigma}}_{P,n}^{\bm{R}}(u)=\sum_{h \in \mathbb{Z}} \mathrm{Cov}_P(\bm{\tilde{G}}_{P,n}^{\bm{R}}(u,\mathcal{H}_0^{\bm{R}}), \bm{\tilde{G}}_{P,n}^{\bm{R}}(u,\mathcal{H}_h^{\bm{R}})).$$
\end{definition}

By Lemma~\ref{lma:du_prop54}, the local long-run covariance matrices of $\bm{w}_{t,n}^{\varphi(Z)}$, $\bm{w}_{P,t,n,i,a}^{\varphi(Z),\varepsilon}$, $\bm{w}_{P,t,n,j,b}^{\varphi(Z),\xi}$ are well-defined in view of the discussion following Assumption~\ref{asmpt_funct_dep_SIEVE}. Now, we will define the local long-run and integrated long-run covariance matrices of these processes as in Section 3.2 of \textcite{zhouzhou_sieve}.

\begin{definition}[Local long-run and integrated long-run covariance matrices]\label{def_integrated_LR_cov_SIEVE} 
% \in\mathbb{R}^{\bm{d_Z}\tilde{d}_n \times \bm{d_Z}\tilde{d}_n }
% $\mathbb{R}^{\bm{d_Z}\tilde{d}_n }$-valued
For each $n\in \mathbb{N}$, $P\in \mathcal{P}_n$, $u\in [0,1]$, $i\in [d_X]$, $a\in A_i$, $j\in [d_Y]$, $b\in B_j$, define the local long-run covariance matrices $\bm{\tilde{\Sigma}}_{P,n}^{\bm{w}^{\varphi(Z)}}(u)$, $\bm{\tilde{\Sigma}}_{P,n,i,a}^{\bm{w}^{\varphi(Z),\varepsilon}}(u)$, $\bm{\tilde{\Sigma}}_{P,n,j,b}^{\bm{w}^{\varphi(Z),\xi}}(u)$ for the stationary processes $(\bm{\tilde{G}}_n^{\bm{w}^{\varphi(Z)}}(u,\mathcal{H}_t^{\bm{w}^{\varphi(Z)}}))_{t\in\mathbb{Z}}$, $(\bm{\tilde{G}}_{P,n,i,a}^{\bm{w}^{\varphi(Z),\varepsilon}}(u,\mathcal{H}_{t,a}^{\bm{w}^{\varphi(Z),\varepsilon}}))_{t\in\mathbb{Z}}$, $(\bm{\tilde{G}}_{P,n,j,b}^{\bm{w}^{\varphi(Z),\xi}}(u,\mathcal{H}_{t,b}^{\bm{w}^{\varphi(Z),\xi}}))_{t\in\mathbb{Z}}$, respectively, by
\begin{align*}    
\bm{\tilde{\Sigma}}_{P,n}^{\bm{w}^{\varphi(Z)}}(u)&=\sum_{h \in \mathbb{Z}} \mathrm{Cov}_P(\bm{\tilde{G}}_n^{\bm{w}^{\varphi(Z)}}(u,\mathcal{H}_0^{\bm{w}^{\varphi(Z)}}), \bm{\tilde{G}}_n^{\bm{w}^{\varphi(Z)}}(u,\mathcal{H}_h^{\bm{w}^{\varphi(Z)}})),\\
\bm{\tilde{\Sigma}}_{P,n,i,a}^{\bm{w}^{\varphi(Z),\varepsilon}}(u)&=\sum_{h \in \mathbb{Z}} \mathrm{Cov}_P(\bm{\tilde{G}}_{P,n,i,a}^{\bm{w}^{\varphi(Z),\varepsilon}}(u,\mathcal{H}_{0,a}^{\bm{w}^{\varphi(Z),\varepsilon}}), \bm{\tilde{G}}_{P,n,i,a}^{\bm{w}^{\varphi(Z),\varepsilon}}(u,\mathcal{H}_{h,a}^{\bm{w}^{\varphi(Z),\varepsilon}})),\\
\bm{\tilde{\Sigma}}_{P,n,j,b}^{\bm{w}^{\varphi(Z),\xi}}(u)&=\sum_{h \in \mathbb{Z}} \mathrm{Cov}_P(\bm{\tilde{G}}_{P,n,j,b}^{\bm{w}^{\varphi(Z),\xi}}(u,\mathcal{H}_{0,b}^{\bm{w}^{\varphi(Z),\xi}}), \bm{\tilde{G}}_{P,n,j,b}^{\bm{w}^{\varphi(Z),\xi}}(u,\mathcal{H}_{h,b}^{\bm{w}^{\varphi(Z),\xi}})).\end{align*}
Next, for each $n\in \mathbb{N}$, $P\in \mathcal{P}_n$, $i\in [d_X]$, $a\in A_i$, $j\in [d_Y]$, $b\in B_j$, define the corresponding integrated long-run covariance matrices $\bm{\tilde{\Sigma}}_{P,n}^{\bm{w}^{\varphi(Z)}}$, $\bm{\tilde{\Sigma}}_{P,n,i,a}^{\bm{w}^{\varphi(Z),\varepsilon}}$, $\bm{\tilde{\Sigma}}_{P,n,j,b}^{\bm{w}^{\varphi(Z),\xi}}$ by % \in\mathbb{R}^{\bm{d_Z}\tilde{c}_n \tilde{d}_n\times \bm{d_Z}\tilde{c}_n \tilde{d}_n}
\begin{align*}
\bm{\tilde{\Sigma}}_{P,n}^{\bm{w}^{\varphi(Z)}}&=\int_0^1\bm{\tilde{\Sigma}}_{P,n}^{\bm{w}^{\varphi(Z)}}(u)\otimes(\bm{\phi}(u)\bm{\phi}^{\top}(u))du,\\
\bm{\tilde{\Sigma}}_{P,n,i,a}^{\bm{w}^{\varphi(Z),\varepsilon}}&=\int_0^1\bm{\tilde{\Sigma}}_{P,n,i,a}^{\bm{w}^{\varphi(Z),\varepsilon}}(u)\otimes(\bm{\phi}(u)\bm{\phi}^{\top}(u))du,\\
\bm{\tilde{\Sigma}}_{P,n,j,b}^{\bm{w}^{\varphi(Z),\xi}}&=\int_0^1\bm{\tilde{\Sigma}}_{P,n,j,b}^{\bm{w}^{\varphi(Z),\xi}}(u)\otimes(\bm{\phi}(u)\bm{\phi}^{\top}(u))du,\end{align*}
where $\bm{\phi}(u)=(\phi_1(u),\ldots,\phi_{\tilde{c}_n}(u))^{\top}$. % tensor product (Kronecker product)

\end{definition}

We require that, for some sequence of collections of distributions $(\mathcal{P}_n)_{n\in\mathbb{N}}$, there exists a distribution-uniform lower bound on the eigenvalues of the integrated long-run covariance matrices. This assumption is analogous to Assumption 3.2 from \textcite{zhouzhou_sieve}.

\begin{assumption}[Eigenvalue condition for integrated long-run covariance matrices]\label{asmpt_eigenvalues_integrated_LR_cov_SIEVE}

Recall $\bm{\tilde{\Sigma}}_{P,n}^{\bm{w}^{\varphi(Z)}}$, $\bm{\tilde{\Sigma}}_{P,n,i,a}^{\bm{w}^{\varphi(Z),\varepsilon}}$, $\bm{\tilde{\Sigma}}_{P,n,j,b}^{\bm{w}^{\varphi(Z),\xi}}$ from Definition~\ref{def_integrated_LR_cov_SIEVE}. Assume that there exists a universal constant $K^{\mathrm{eig}} >0$ such that for all $n\in \mathbb{N}$, $i\in [d_X]$, $a\in A_i$, $j\in [d_Y]$, $b\in B_j$, we have 
$$\underset{P\in\mathcal{P}_n}{\inf}\min(\lambda_{\min}(\bm{\tilde{\Sigma}}_{P,n}^{\bm{w}^{\varphi(Z)}}),\lambda_{\min}( \bm{\tilde{\Sigma}}_{P,n,i,a}^{\bm{w}^{\varphi(Z),\varepsilon}}), \lambda_{\min}(\bm{\tilde{\Sigma}}_{P,n,j,b}^{\bm{w}^{\varphi(Z),\xi}}))\geq K^{\mathrm{eig}},$$
where $\lambda_{\min}(\cdot)$ is the smallest eigenvalue of the given matrix.    
\end{assumption}

Again, we emphasize that the locally stationary framework in this section fits into the more general triangular array framework from Section 3. Hence, we can use the same cumulative covariance estimator $\hat{Q}_{t,n}^{\bm{R}}$ from Section 3.5 for the cumulative covariance matrices $\sum_{s=\mathbb{T}_n^{-}}^t\bm{\Sigma}_{P,s,n}^{\bm{R}}$, where $\bm{\Sigma}_{P,s,n}^{\bm{R}}=\bm{\tilde{\Sigma}}_{P,n}^{\bm{R}}(s/n)$ denotes the local long-run covariance matrix at time $s\in\mathcal{T}_n$.

\subsection{Theoretical Result for Sieve-dGCM}\label{subsection:test_all_times_hypoth_SIEVE}

The main result of this section is that the Sieve-dGCM test, implemented by running Algorithm 1 with the predictions from the sieve estimator, will have uniformly asymptotic Type I error control under the previously stated assumptions.
% because (1) the previously stated assumptions imply the assumptions of Theorem 3.1
% and (2) under these strictly stronger assumptions the sieve time-varying regression estimator will achieve the convergence rates required by Theorem 3.1
% Note that the offsets $\tau_n$, $\nu_n$ are chosen in the same way as in Section 3.5 such that condition~\eqref{offset_condition_nsts} holds.

\begin{theorem}\label{thm:test_SIEVE_DR}

Suppose that Assumptions~\ref{asmpt_causal_rep_process_SIEVE},~\ref{asmpt_tv_regr_Lq_SIEVE},~\ref{asmpt_smooth_regr_fn_SIEVE},~\ref{asmpt_causal_rep_errors_SIEVE},~\ref{asmpt_funct_dep_SIEVE},~\ref{asmpt_stoch_lip_SIEVE},~\ref{asmpt_eigenvalues_integrated_LR_cov_SIEVE} all hold for the sequence of collections of distributions $(\mathcal{P}_{0,n}^{\ast})_{n\in\mathbb{N}}$, where $\mathcal{P}_{0,n}^{\ast}\subset\mathcal{P}_{0,n}^{\mathrm{CI}}$ for each $n\in\mathbb{N}$. Further, suppose that we use the sieve time-varying regression estimator from Section~\ref{subsection:tv_regr_SIEVE} with the basis functions $\{\phi_{\ell_1}(u)\}$, $\{\varphi_{\ell_2}(z)\}$ chosen to be mapped Legendre polynomials, where the numbers of basis functions are chosen to satisfy  $\tilde{c}_n=O(\log(T_n))$, $\tilde{d}_n=O(\log(T_n))$. Then Assumptions 3.1, 3.2, 3.3, 3.4, 3.5, 3.6 hold for $(\mathcal{P}_{0,n}^{\ast})_{n\in\mathbb{N}}$, and the sieve estimators will achieve the convergence rates required by Theorem 3.1.
% Note: That is, if the offsets $\tau_n \xrightarrow[]{} 0$ and $\nu_n\xrightarrow[]{} 0$ are chosen such that condition~\eqref{offset_condition_nsts} holds, then we have  $$\underset{n \xrightarrow[]{}\infty}{\mathrm{lim\text{ }sup}}\text{ }\underset{P\in\mathcal{P}_{0,n}^{\ast}}{\sup}\mathbb{P}_P\left(S_{n,p}(\bm{\hat{R}}_{n})  > \hat{q}_{1-\alpha +\nu_n}+\tau_n\right)\leq \alpha.$$
    
\end{theorem}

% we allow for $D_n=O(T_n^{\frac{1}{6}})$  
% the sieve estimator with Legendre polynomials as basis functions can satisfy this fastest rate of growth since it will be log^4(n)/n^{1/2} for each so log^8(n)/n multiplying

Throughout this section, we have used Legendre polynomials as the basis functions. In our simulations, we investigate the finite-sample performance of the Sieve-dGCM test using Legendre basis functions. We emphasize that the Legendre polynomials in our theoretical analysis and simulations can easily be substituted with trigonometric polynomials, wavelets, or other Jacobi polynomials.

\section{Proofs of Main Theoretical Results}\label{appendix:proofs}

Let $(V_{P,n})_{P\in\mathcal{P}_n, n\in\mathbb{N}}$ be a sequence of families of random variables with distributions determined by $P\in\mathcal{P}_n$ for some $n\in\mathbb{N}$, for some sequence of collections of distributions $(\mathcal{P}_n)_{n\in\mathbb{N}}$ which will be made clear from the context. We write $V_{P,n}=o_{\mathcal{P}}(1)$ to mean that for all $\epsilon >0$, we have $$\underset{P\in\mathcal{P}_n}{\sup}\mathbb{P}_P(|V_{P,n}|>\epsilon)\xrightarrow[]{} 0.$$ Also, by $V_{P,n}=O_{\mathcal{P}}(1)$ we mean that for all $\epsilon >0$, there exists a constant $K>0$ and $N\in\mathbb{N}$ such that for all $n > N$, we have $$\underset{P\in\mathcal{P}_n}{\sup}\mathbb{P}_P(|V_{P,n}|>K)<\epsilon.$$ Let $(W_{P,n})_{P\in\mathcal{P}_n, n\in\mathbb{N}}$ be another sequence of families of random variables. By $V_{P,n}=o_{\mathcal{P}}(W_{P,n})$ we mean $V_{P,n}=W_{P,n}U_{P,n}$ and $U_{P,n}=o_{\mathcal{P}}(1)$, and by $V_{P,n}=O_{\mathcal{P}}(W_{P,n})$ we mean $V_{P,n}=W_{P,n}U_{P,n}$ and $U_{P,n}=O_{\mathcal{P}}(1)$. Crucially, when we write $o_{\mathcal{P}}(\cdot)$ and $O_{\mathcal{P}}(\cdot)$ in the proofs below, we will always be doing so with reference to the sequence of collections of distributions $(\mathcal{P}_{0,n}^{\ast})_{n\in\mathbb{N}}$ defined in the corresponding theorem's statement.

Next, we introduce some notation. We will denote the three bias terms by
\begin{align*}
\bm{\hat{w}}_{P,t,n}^{\bm{f},\bm{g}}&=(\hat{w}_{P,t,n,m}^{f,g})_{m\in\mathcal{D}_n}=(\hat{w}_{P,t,n,i,a}^{f} \hat{w}_{P,t,n,j,b}^{g})_{m\in\mathcal{D}_n},\\
\bm{\hat{w}}_{P,t,n}^{\bm{g},\bm{\varepsilon}}&=( \hat{w}_{P,t,n,m}^{g,\varepsilon})_{m\in\mathcal{D}_n}=( \hat{w}_{P,t,n,j,b}^{g}\varepsilon_{P,t,n,i,a})_{m\in\mathcal{D}_n},\\
\bm{\hat{w}}_{P,t,n}^{\bm{f},\bm{\xi}}&=(\hat{w}_{P,t,n,m}^{f, \xi})_{m\in\mathcal{D}_n}=(\hat{w}_{P,t,n,i,a}^{f} \xi_{P,t,n,j,b})_{m\in\mathcal{D}_n},\end{align*} where $m=(i,j,a,b)\in\mathcal{D}_n$. Also, denote \begin{align*} \bm{\hat{w}}_{P,n}^{\bm{f},\bm{g}}&=(\bm{\hat{w}}_{P,t,n}^{\bm{f},\bm{g}})_{t\in\mathcal{T}_{n,L}},\\ \bm{\hat{w}}_{P,n}^{\bm{g},\bm{\varepsilon}}&=(\bm{\hat{w}}_{P,t,n}^{\bm{g},\bm{\varepsilon}})_{t\in\mathcal{T}_{n,L}},\\ \bm{\hat{w}}_{P,n}^{\bm{f},\bm{\xi}}&=(\bm{\hat{w}}_{P,t,n}^{\bm{f},\bm{\xi}})_{t\in\mathcal{T}_{n,L}}.\end{align*}

\subsection{Proof of Theorem 3.1}\label{subsection:proof_of_test_thm_nsts}

\textbf{Step 1 (Bias Terms):} We decompose the products of residuals into the products of errors and the three bias terms, and then apply the triangle inequality and subadditivity, which yields \begin{align*}
&\underset{P\in\mathcal{P}_{0,n}^{\ast}}{\sup}\mathbb{P}_P(S_{n,p}(\bm{\hat{R}}_{n}) > \hat{q}_{1-\alpha +\nu_n}+\tau_n) 
\\&
\leq \underset{P\in\mathcal{P}_{0,n}^{\ast}}{\sup}\mathbb{P}_P(S_{n,p}(\bm{R}_{P,n}) > \hat{q}_{1-\alpha +\nu_n}+\frac{\tau_n}{2}) 
\\& + \underset{P\in\mathcal{P}_{0,n}^{\ast}}{\sup}\mathbb{P}_P(S_{n,p}(\bm{\hat{w}}_{P,n}^{\bm{f},\bm{g}}) > \frac{\tau_n}{6}) 
\\&
+ \underset{P\in\mathcal{P}_{0,n}^{\ast}}{\sup}\mathbb{P}_P(S_{n,p}(\bm{\hat{w}}_{P,n}^{\bm{g},\bm{\varepsilon}}) > \frac{\tau_n}{6})
\\&
+ \underset{P\in\mathcal{P}_{0,n}^{\ast}}{\sup}\mathbb{P}_P(S_{n,p}(\bm{\hat{w}}_{P,n}^{\bm{f},\bm{\xi}}) > \frac{\tau_n}{6}).\end{align*} We will handle each of the three bias terms separately.

\textbf{Step 1.1:} Observe that for any $\delta >0$, we have\begin{align*}    &\underset{P\in\mathcal{P}_{0,n}^{\ast}}{\sup}\mathbb{P}_P(\tau_n^{-1}S_{n,p}(\bm{\hat{w}}_{P,n}^{\bm{f},\bm{g}}) > \delta)    
\\ &\overset{(1)}{\leq} 
\delta^{-1}\tau_n^{-1}T_{n,L}^{-\frac{1}{2}}\underset{P\in\mathcal{P}_{0,n}^{\ast}}{\sup}\mathbb{E}_P\left(\underset{s \in \mathcal{T}_{n,L}}{\max}\left|\left|\sum_{t\leq s} \bm{\hat{w}}_{P,t,n}^{\bm{f},\bm{g}} \right|\right|_2 \right)
\\&\overset{(2)}{\leq}  
\delta^{-1} \tau_n^{-1} T_{n,L}^{-\frac{1}{2}} D_n \underset{P\in\mathcal{P}_{0,n}^{\ast}}{\sup}\underset{(i,j,a,b)\in\mathcal{D}_n}{\max}  \mathbb{E}_P\left(\sum_{t\in\mathcal{T}_{n,L}} |\hat{w}_{P,t,n,i,a}^{f}|| \hat{w}_{P,t,n,j,b}^{g}|\right)
\\&\overset{(3)}{\leq}  
\delta^{-1} \tau_n^{-1} T_{n,L}^{\frac{1}{2}} D_n \underset{P\in\mathcal{P}_{0,n}^{\ast}}{\sup}\underset{(i,j,a,b)\in\mathcal{D}_n}{\max} \underset{t\in\mathcal{T}_n}{\max} \ \mathbb{E}_P\left( |\hat{w}_{P,t,n,i,a}^{f}|^2\right)^{\frac{1}{2}}\mathbb{E}_P\left(| \hat{w}_{P,t,n,j,b}^{g}|^2\right)^{\frac{1}{2}}
\\&\overset{(4)}{=}
o(1),
\end{align*} where the previous lines follow by (1) Markov's inequality and $\ell_p$-norm inequalities, (2) the triangle inequality, $\ell_p$-norm inequalities, linearity of expectation, (3) linearity of expectation and the Cauchy-Schwarz inequality, (4) the convergence rate requirements for the time-varying regression estimators.

\textbf{Step 1.2:} Observe that for any $\delta>0$, we have
\begin{align*}
&\underset{P\in\mathcal{P}_{0,n}^{\ast}}{\sup}\mathbb{P}_P(\tau_n^{-1}S_{n,p}(\bm{\hat{w}}_{P,n}^{\bm{g},\bm{\varepsilon}}) > \delta) 
\\&
\overset{(1)}{\leq} 
\underset{P\in\mathcal{P}_{0,n}^{\ast}}{\sup}\mathbb{P}_P\left(\tau_n^{-2}\underset{s \in \mathcal{T}_{n,L}}{\max}\left|\left|\frac{1}{\sqrt{T_{n,L}}} \sum_{t\leq s} \bm{\hat{w}}_{P,t,n}^{\bm{g},\bm{\varepsilon}} \right|\right|_2^2  \geq \delta^2 \right)
\\&
\overset{(2)}{\leq}
\delta^{-2} \tau_n^{-2} T_{n,L}^{-1} \underset{P\in\mathcal{P}_{0,n}^{\ast}}{\sup}\mathbb{E}_P\left(\underset{s \in \mathcal{T}_{n,L}}{\max}\left|\left| \sum_{t\leq s} \bm{\hat{w}}_{P,t,n}^{\bm{g},\bm{\varepsilon}} \right|\right|_2^2 \right)
\\&
\overset{(3)}{\leq}
\delta^{-2} \tau_n^{-2} T_{n,L}^{-1} 
(\bar{K} T_{n,L}^{\frac{1}{2}} D_n^{\frac{1}{2}} \underset{P\in\mathcal{P}_{0,n}^{\ast}}{\sup} \underset{t\in \mathcal{T}_{n,L}}{\max}\text{ }\underset{j\in [d_Y]}{\max}\text{ }\underset{b\in B_j}{\max}\text{ }\mathbb{E}_P(|\hat{w}_{P,t,n,j,b}^{g}|^2)^{\frac{1}{2}})^2
\\&
\overset{(4)}{\leq}
\delta^{-2} \tau_n^{-2}  
\bar{K}^{2} D_n \underset{P\in\mathcal{P}_{0,n}^{\ast}}{\sup}\underset{t\in \mathcal{T}_{n,L}}{\max}\text{ }\underset{j\in [d_Y]}{\max}\text{ }\underset{b\in B_j}{\max}\text{ }\mathbb{E}_P(|\hat{w}_{P,t,n,j,b}^{g}|^2)
\\&
\overset{(5)}{=}
o(1),
\end{align*}
where the previous lines follow by (1) the assumption about the form of the test statistic and squaring, (2) Markov's inequality and linearity of expectation, (3) for some constant $\bar{K}>0$ by the arguments below, (4) simplifying the expression, and (5) the convergence rate requirements for the time-varying regression estimator. The following arguments show (3). 

Our arguments to show (3) use constructions that build on those used in the proof of Theorem 3.2 in \textcite{seq_gauss_approx2022}, which themselves build on the techniques from Theorem 1 in \textcite{liu_2013}. For instance, our constructions use a conditional version of the functional dependence measure of \textcite{wu_funct_dep_meas} and Doob's conditional independence property (Theorem 8.9 of \textcite{kallenberg_FOMP_ed3}). Our techniques may be of independent interest for future work involving conditional independence for time series.
% plus, we fill in the rigorous details, which are usually ommitted in other papers...

For each $t\in\mathcal{T}_{n,L}$ and $h\in\mathbb{N}_0$, let $$\mathcal{F}_{t,h}^{\bm{\hat{g}},\bm{\varepsilon}}=\sigma(\eta_{t}^{\bm{\varepsilon}},\eta_{t-1}^{\bm{\varepsilon}},\ldots,\eta_{t-h}^{\bm{\varepsilon}},\mathcal{H}_{t}^{\bm{\hat{g}}}),$$ where the input $\eta_{t}^{\bm{\varepsilon}}$ is from Section 3.3 and the input sequence $\mathcal{H}_{t}^{\bm{\hat{g}}}$ is defined following Assumption 3.3. For each $n\in\mathbb{N}$, $P\in\mathcal{P}_{0,n}^{\ast}$, $t\in\mathcal{T}_{n,L}$, and $h\in\mathbb{N}_0$ let  
\begin{align*}
\bm{\hat{S}}_{P,t,n,h}^{\bm{g},\bm{\varepsilon}}&=\sum_{k\leq t} \bm{\hat{w}}_{P,k,n,h}^{\bm{g},\bm{\varepsilon}},\\
\bm{\hat{w}}_{P,t,n,h}^{\bm{g},\bm{\varepsilon}}&=\mathbb{E}_P(\bm{\hat{w}}_{P,t,n}^{\bm{g},\bm{\varepsilon}} | \mathcal{F}_{t,h}^{\bm{\hat{g}},\bm{\varepsilon}}),\\
\bm{\hat{w}}_{P,t,n,-1}^{\bm{g},\bm{\varepsilon}}&=\mathbb{E}_P(\bm{\hat{w}}_{P,t,n}^{\bm{g},\bm{\varepsilon}}| \mathcal{H}_{t}^{\bm{\hat{g}}})=\bm{0},\end{align*}
almost surely, because for each $n\in\mathbb{N}$, $P\in\mathcal{P}_{0,n}^{\ast}$, $(i,j,a,b)\in\mathcal{D}_n$, $t\in\mathcal{T}_{n,L}$ we have 
\begin{equation}
\mathbb{E}_P(\hat{w}_{P,t,n,j,b}^{g}\varepsilon_{P,t,n,i,a}|\mathcal{H}_{t}^{\bm{\hat{g}}})=\hat{w}_{P,t,n,j,b}^{g}\mathbb{E}_P(\varepsilon_{P,t,n,i,a}|\mathcal{H}_{t}^{\bm{\hat{g}}})=0, \label{eqn:pred_err_mble_and_err_ci_argument}
\end{equation} almost surely, by Assumptions 3.3 and 3.4. For each $n\in\mathbb{N}$, $P\in\mathcal{P}_{0,n}^{\ast}$, $t\in\mathcal{T}_{n,L}$, and $h\in\mathbb{N}_0$ we have 
\begin{equation}\label{eqn:finite_PE}
    \mathbb{E}_P(||\bm{\hat{w}}_{P,t,n,h}^{\bm{g},\bm{\varepsilon}}||_2^2)<\infty,
\end{equation}
by linearity of expectation, the contraction property of conditional expectation, Assumption 3.3, and Assumption 3.5. For each $n\in\mathbb{N}$, $P\in\mathcal{P}_{0,n}^{\ast}$, $t\in\mathcal{T}_{n,L}$, and $h\in\mathbb{N}_0$, by the tower property we have
$$\mathbb{E}_P(\bm{\hat{w}}_{P,t,n,h+1}^{\bm{g},\bm{\varepsilon}}|\mathcal{F}_{t,h}^{\bm{\hat{g}},\bm{\varepsilon}})=\bm{\hat{w}}_{P,t,n,h}^{\bm{g},\bm{\varepsilon}},$$ almost surely. Hence, for each $n\in\mathbb{N}$, $P\in\mathcal{P}_{0,n}^{\ast}$, and $t\in\mathcal{T}_{n,L}$, $(\bm{\hat{w}}_{P,t,n,h}^{\bm{g},\bm{\varepsilon}})_{h=0}^{\infty}$ is a martingale with respect to the filtration $(\mathcal{F}_{t,h}^{\bm{\hat{g}},\bm{\varepsilon}})_{h=0}^{\infty}$. The martingale convergence theorem (Theorem 1.5 of \textcite{pisier2016}) ensures that for each $n\in\mathbb{N}$, $P\in\mathcal{P}_{0,n}^{\ast}$, $t\in\mathcal{T}_{n,L}$ there exists some random vector $$\bm{\tilde{w}}_{P,t,n}^{\bm{g},\bm{\varepsilon}}:=\bm{\hat{w}}_{P,t,n,\infty}^{\bm{g},\bm{\varepsilon}}=\mathbb{E}_P(\bm{\hat{w}}_{P,t,n}^{\bm{g},\bm{\varepsilon}} | \mathcal{F}_{t,\infty}^{\bm{\hat{g}},\bm{\varepsilon}}),\quad \mathcal{F}_{t,\infty}^{\bm{\hat{g}},\bm{\varepsilon}}=\sigma\left(\bigcup_{h\geq 0} \mathcal{F}_{t,h}^{\bm{\hat{g}},\bm{\varepsilon}}\right),$$ such that $\mathbb{E}_P||\bm{\tilde{w}}_{P,t,n}^{\bm{g},\bm{\varepsilon}}-\bm{\hat{w}}_{P,t,n,h}^{\bm{g},\bm{\varepsilon}}||_2^2\xrightarrow[]{}0$ as $h\xrightarrow[]{}\infty$. The measurability of $\bm{\hat{w}}_{P,t,n}^{\bm{g},\bm{\varepsilon}}$ with respect to $\mathcal{F}_{t,\infty}^{\bm{\hat{g}},\bm{\varepsilon}}$, due to its causal representation $\bm{G}_{P,t,n}^{\bm{\hat{w}}^{\bm{g},\bm{\varepsilon}}}$ in view of Assumptions 3.1, 3.2, 3.3, 3.4, ensures that $\bm{\tilde{w}}_{P,t,n}^{\bm{g},\bm{\varepsilon}}=\bm{\hat{w}}_{P,t,n}^{\bm{g},\bm{\varepsilon}}$ almost surely. Thus, for each $t\in\mathcal{T}_{n,L}$ we have 
\begin{equation}
\bm{\hat{S}}_{P,t,n}^{\bm{g},\bm{\varepsilon}}=\sum_{k\leq t} \bm{\hat{w}}_{P,k,n}^{\bm{g},\bm{\varepsilon}}=\sum_{h=0}^{\infty}(\bm{\hat{S}}_{P,t,n,h}^{\bm{g},\bm{\varepsilon}}-\bm{\hat{S}}_{P,t,n,h-1}^{\bm{g},\bm{\varepsilon}}),\label{eqn:telescoping_fdm}\end{equation}
by telescoping. For each $n\in\mathbb{N}$, $P\in\mathcal{P}_{0,n}^{\ast}$, and $h\in\mathbb{N}_0$, $$(\bm{\hat{w}}_{P,\mathbb{T}_n^{+}-k,n,h}^{\bm{g},\bm{\varepsilon}}-\bm{\hat{w}}_{P,\mathbb{T}_n^{+}-k,n,h-1}^{\bm{g},\bm{\varepsilon}})_{k=0}^{\mathbb{T}_n^{+}-\mathbb{T}_n^{-}-L_n+1},$$ are martingale differences with respect to the filtration $(\mathcal{G}_{\mathbb{T}_n^{+},k,h}^{\bm{\hat{g}},\bm{\varepsilon}})_{k=0}^{\mathbb{T}_n^{+}-\mathbb{T}_n^{-}-L_n+1}$, where $$\mathcal{G}_{\mathbb{T}_n^{+},k,h}^{\bm{\hat{g}},\bm{\varepsilon}}=\sigma(\mathcal{H}_{\mathbb{T}_n^{+}}^{\bm{\hat{g}}},\eta_{\mathbb{T}_n^{+}-k-h}^{\bm{\varepsilon}},\eta_{\mathbb{T}_n^{+}-k-h+1}^{\bm{\varepsilon}}\ldots),\quad k\geq 0,
% not subtracting by k for \mathcal{H}_{\mathbb{T}_n^{+}-k}^{\bm{\hat{w}}^{\bm{g}}}, so that it is indeed a nested sequence of sigma-algebras i.e. filtration and so we can use Doob's later  
% ie X inputs AFTER Tn-k-h, and all Y and Z inputs up to time Tn+b, Tn+cmax 
% note the difference with 
% $$\mathcal{F}_{t,h}^{\bm{\hat{g}},\bm{\varepsilon}}$$
% which is X inputs t,..,t-h, and all Y and Z inputs up to time Tn+bmax, Tn+cmax 
$$ with $\mathcal{G}_{\mathbb{T}_n^{+},-1,h}^{\bm{\hat{g}},\bm{\varepsilon}}=\sigma(\mathcal{H}_{\mathbb{T}_n^{+}}^{\bm{\hat{g}}},\eta_{\mathbb{T}_n^{+}+1-h}^{\bm{\varepsilon}},\eta_{\mathbb{T}_n^{+}+1-h+1}^{\bm{\varepsilon}}\ldots)$, because for any $n\in\mathbb{N}$, $P\in\mathcal{P}_{0,n}^{\ast}$, $h\in\mathbb{N}_0$ and $k=0,1,\ldots,$ we have
\begin{align*}
    & \mathbb{E}_P(\bm{\hat{w}}_{P,\mathbb{T}_n^{+}-k,n,h}^{\bm{g},\bm{\varepsilon}}-\bm{\hat{w}}_{P,\mathbb{T}_n^{+}-k,n,h-1}^{\bm{g},\bm{\varepsilon}}|\mathcal{G}_{\mathbb{T}_n^{+},k-1,h}^{\bm{\hat{g}},\bm{\varepsilon}})
    \\&
    = \mathbb{E}_P(\mathbb{E}_P(\bm{\hat{w}}_{P,\mathbb{T}_n^{+}-k,n}^{\bm{g},\bm{\varepsilon}} | \mathcal{F}_{\mathbb{T}_n^{+}-k,h}^{\bm{\hat{g}},\bm{\varepsilon}})|\mathcal{G}_{\mathbb{T}_n^{+},k-1,h}^{\bm{\hat{g}},\bm{\varepsilon}}) 
    \\&
    -  \mathbb{E}_P(\mathbb{E}_P(\bm{\hat{w}}_{P,\mathbb{T}_n^{+}-k,n}^{\bm{g},\bm{\varepsilon}} | \mathcal{F}_{\mathbb{T}_n^{+}-k,h-1}^{\bm{\hat{g}},\bm{\varepsilon}})|\mathcal{G}_{\mathbb{T}_n^{+},k-1,h}^{\bm{\hat{g}},\bm{\varepsilon}})
    \\&
    =\bm{0},
\end{align*} almost surely, because for each $n\in\mathbb{N}$, $P\in\mathcal{P}_{0,n}^{\ast}$, $(i,j,a,b)\in\mathcal{D}_n$, $h\in\mathbb{N}_0$ and $k=0,1,\ldots,$ we have
\begin{align*}    
&
\mathbb{E}_P(\mathbb{E}_P(\hat{w}_{P,\mathbb{T}_n^{+}-k,n,j,b}^{g}\varepsilon_{P,\mathbb{T}_n^{+}-k,n,i,a} |\mathcal{H}_{\mathbb{T}_n^{+}-k}^{\bm{\hat{g}}}, \eta_{\mathbb{T}_n^{+}-k-h}^{\bm{\varepsilon}},\eta_{\mathbb{T}_n^{+}-k-h+1}^{\bm{\varepsilon}},\ldots,\eta_{\mathbb{T}_n^{+}-k}^{\bm{\varepsilon}}) 
\\&
|\mathcal{H}_{\mathbb{T}_n^{+}}^{\bm{\hat{g}}},\eta_{\mathbb{T}_n^{+}-k-h+1}^{\bm{\varepsilon}},\eta_{\mathbb{T}_n^{+}-k-h+2}^{\bm{\varepsilon}},\ldots)  
\\&     
-  \mathbb{E}_P(\mathbb{E}_P(\hat{w}_{P,\mathbb{T}_n^{+}-k,n,j,b}^{g}\varepsilon_{P,\mathbb{T}_n^{+}-k,n,i,a} | \mathcal{H}_{\mathbb{T}_n^{+}-k}^{\bm{\hat{g}}},\eta_{\mathbb{T}_n^{+}-k-h+1}^{\bm{\varepsilon}},\eta_{\mathbb{T}_n^{+}-k-h+2}^{\bm{\varepsilon}},\ldots,\eta_{\mathbb{T}_n^{+}-k}^{\bm{\varepsilon}})    
\\&
|\mathcal{H}_{\mathbb{T}_n^{+}}^{\bm{\hat{g}}},\eta_{\mathbb{T}_n^{+}-k-h+1}^{\bm{\varepsilon}},\eta_{\mathbb{T}_n^{+}-k-h+2}^{\bm{\varepsilon}},\ldots)
\\&
\overset{(1)}{=}
\hat{w}_{P,\mathbb{T}_n^{+}-k,n,j,b}^{g}\mathbb{E}_P(\mathbb{E}_P(\varepsilon_{P,\mathbb{T}_n^{+}-k,n,i,a} | \mathcal{H}_{\mathbb{T}_n^{+}-k}^{\bm{\hat{g}}},\eta_{\mathbb{T}_n^{+}-k-h}^{\bm{\varepsilon}},\eta_{\mathbb{T}_n^{+}-k-h+1}^{\bm{\varepsilon}},\ldots,\eta_{\mathbb{T}_n^{+}-k}^{\bm{\varepsilon}}) 
\\&
|\mathcal{H}_{\mathbb{T}_n^{+}}^{\bm{\hat{g}}},\eta_{\mathbb{T}_n^{+}-k-h+1}^{\bm{\varepsilon}},\eta_{\mathbb{T}_n^{+}-k-h+2}^{\bm{\varepsilon}},\ldots)     
\\& 
- \hat{w}_{P,\mathbb{T}_n^{+}-k,n,j,b}^{g} \mathbb{E}_P(\mathbb{E}_P(\varepsilon_{P,\mathbb{T}_n^{+}-k,n,i,a} | \mathcal{H}_{\mathbb{T}_n^{+}-k}^{\bm{\hat{g}}},\eta_{\mathbb{T}_n^{+}-k-h+1}^{\bm{\varepsilon}},\eta_{\mathbb{T}_n^{+}-k-h+2}^{\bm{\varepsilon}},\ldots,\eta_{\mathbb{T}_n^{+}-k}^{\bm{\varepsilon}})    
\\&
|\mathcal{H}_{\mathbb{T}_n^{+}}^{\bm{\hat{g}}},\eta_{\mathbb{T}_n^{+}-k-h+1}^{\bm{\varepsilon}},\eta_{\mathbb{T}_n^{+}-k-h+2}^{\bm{\varepsilon}},\ldots)
\\&
\overset{(2)}{=}
\hat{w}_{P,\mathbb{T}_n^{+}-k,n,j,b}^{g}\mathbb{E}_P(\varepsilon_{P,\mathbb{T}_n^{+}-k,n,i,a} | \mathcal{H}_{\mathbb{T}_n^{+}-k}^{\bm{\hat{g}}},\eta_{\mathbb{T}_n^{+}-k-h+1}^{\bm{\varepsilon}},\eta_{\mathbb{T}_n^{+}-k-h+2}^{\bm{\varepsilon}},\ldots,\eta_{\mathbb{T}_n^{+}-k}^{\bm{\varepsilon}})    
\\& 
- \hat{w}_{P,\mathbb{T}_n^{+}-k,n,j,b}^{g} \mathbb{E}_P(\varepsilon_{P,\mathbb{T}_n^{+}-k,n,i,a} | \mathcal{H}_{\mathbb{T}_n^{+}-k}^{\bm{\hat{g}}},\eta_{\mathbb{T}_n^{+}-k-h+1}^{\bm{\varepsilon}},\eta_{\mathbb{T}_n^{+}-k-h+2}^{\bm{\varepsilon}},\ldots,\eta_{\mathbb{T}_n^{+}-k}^{\bm{\varepsilon}})    
\\&
\overset{(3)}{=}
0,
\end{align*} % for (2), 
% the sigma-algebra for the inner conditional expectation is: $\mathcal{F} = \sigma(\mathcal{H}_{\mathbb{T}_n^{+}-k}^{\mathbf{\hat{g}}},\eta_{\mathbb{T}_n^{+}-k-h}^{\mathbf{\varepsilon}},\eta_{\mathbb{T}_n^{+}-k-h+1}^{\mathbf{\varepsilon}},\ldots,\eta_{\mathbb{T}_n^{+}-k}^{\mathbf{\varepsilon}})$ 
% the sigma-algebra for the outer conditional expectation is: \sigma(\mathcal{H}_{\mathbb{T}_n^{+}}^{\mathbf{\hat{g}}},\eta_{\mathbb{T}_n^{+}-k-h+1}^{\mathbf{\varepsilon}},\eta_{\mathbb{T}_n^{+}-k-h+2}^{\mathbf{\varepsilon}},\ldots,\eta_{\mathbb{T}_n^{+}-k}^{\mathbf{\varepsilon}})
% the intersection sigma-algebra is: \sigma(\mathcal{H}_{\mathbb{T}_n^{+}-k}^{\mathbf{\hat{g}}},\eta_{\mathbb{T}_n^{+}-k-h+1}^{\mathbf{\varepsilon}},\eta_{\mathbb{T}_n^{+}-k-h+2}^{\mathbf{\varepsilon}},\ldots,\eta_{\mathbb{T}_n^{+}-k}^{\mathbf{\varepsilon}})
% indeed the "early" error noise input $\eta_{\mathbb{T}_n^{+}-k-h+1}^{\mathbf{\varepsilon}}$ is independent of the "later other predictor" noise inputs for $\bm{\hat{g}}$ at times \mathbb{T}_n^{+}-k+1,...,\mathbb{T}_n^{+} given the shared information that both sigma-algebras have (i.e. the intersection sigma-algebra)
almost surely, by (1) Assumption 3.3, (2) Doob's conditional independence property (Theorem 8.9 of \textcite{kallenberg_FOMP_ed3}) applied to the first nested conditional expectations (where the conditional independence relationships from Assumption 3.4 are used), and for the second nested conditional expectations use the measurability of the inner conditional expectation with respect to the $\sigma$-algebra in the outer conditional expectation, and (3) subtraction. Also, $\mathbb{E}_P(||\bm{\hat{w}}_{P,\mathbb{T}_n^{+}-k,n,h}^{\bm{g},\bm{\varepsilon}}-\bm{\hat{w}}_{P,\mathbb{T}_n^{+}-k,n,h-1}^{\bm{g},\bm{\varepsilon}}||_2^2)<\infty$ by the triangle inequality, squaring, linearity of expectation, the Cauchy-Schwarz inequality, and the same arguments as~\eqref{eqn:finite_PE} (i.e. linearity of expectation, the contraction property of conditional expectation, and Assumption 3.3). 
% more details for (1): the $\sigma(\mathcal{H}_{\mathbb{T}_n^{+},b}^{Y},\mathcal{H}_{\mathbb{T}_n^{+}}^{\bm{Z}})$-measurability of $\hat{w}_{P,\mathbb{T}_n^{+}-k,n,j,b}^{g}$ in view of its causal representation from Assumption 3.3 and the conditional independence (see e.g., https://math.stackexchange.com/questions/365310/conditional-expectation-on-more-than-one-sigma-algebra ) of $\varepsilon_{P,\mathbb{T}_n^{+}-k,n,i,a}$ and $\sigma(\mathcal{H}_{\mathbb{T}_n^{+},b}^{Y},\eta_{\mathbb{T}_n^{+}-k+1}^{\bm{Z}},\ldots,\eta_{\mathbb{T}_n^{+}}^{\bm{Z}})$ given $\sigma(\mathcal{H}_{\mathbb{T}_n^{+}}^{\bm{Z}},\eta_{\mathbb{T}_n^{+}-k-h,a}^{X},\eta_{\mathbb{T}_n^{+}-k-h+1,a}^{X},\ldots,\eta_{\mathbb{T}_n^{+}-k,a}^{X})$ and $\sigma(\mathcal{H}_{\mathbb{T}_n^{+}}^{\bm{Z}},\eta_{\mathbb{T}_n^{+}-k-h+1,a}^{X},\eta_{\mathbb{T}_n^{+}-k-h+2,a}^{X},\ldots,\eta_{\mathbb{T}_n^{+}-k,a}^{X})$ for the first and second terms, respectively, in view of the causal representation of $\varepsilon_{P,\mathbb{T}_n^{+}-k,n,i,a}$ from Assumption 3.4, and then conditional independence for the outer conditional expectation

Next, observe that for each $n\in\mathbb{N}$ and $h\in\mathbb{N}_0$, we have
\begin{align*}
    &
    \underset{P\in\mathcal{P}_{0,n}^{\ast}}{\sup}(\mathbb{E}_P\text{ }\underset{t\in\mathcal{T}_{n,L}}{\max} ||\bm{\hat{S}}_{P,t,n,h}^{\bm{g},\bm{\varepsilon}}-\bm{\hat{S}}_{P,t,n,h-1}^{\bm{g},\bm{\varepsilon}}||_2^2)^{\frac{1}{2}}
    \\& 
    \overset{(1)}{\leq} 
    \underset{P\in\mathcal{P}_{0,n}^{\ast}}{\sup}(\mathbb{E}_P\text{ }||\bm{\hat{S}}_{P,\mathbb{T}_n^{+},n,h}^{\bm{g},\bm{\varepsilon}}-\bm{\hat{S}}_{P,\mathbb{T}_n^{+},n,h-1}^{\bm{g},\bm{\varepsilon}}||_2^2)^{\frac{1}{2}}
    \\&+
    \underset{P\in\mathcal{P}_{0,n}^{\ast}}{\sup}(\mathbb{E}_P\text{ } \underset{t\in\mathcal{T}_{n,L}}{\max}||(\bm{\hat{S}}_{P,\mathbb{T}_n^{+},n,h}^{\bm{g},\bm{\varepsilon}}-\bm{\hat{S}}_{P,\mathbb{T}_n^{+},n,h-1}^{\bm{g},\bm{\varepsilon}})-(\bm{\hat{S}}_{P,t,n,h}^{\bm{g},\bm{\varepsilon}}-\bm{\hat{S}}_{P,t,n,h-1}^{\bm{g},\bm{\varepsilon}})||_2^2)^{\frac{1}{2}}
    \\& 
       \overset{(2)}{\leq} 
       \underset{P\in\mathcal{P}_{0,n}^{\ast}}{\sup}(\mathbb{E}_P\text{ }||\bm{\hat{S}}_{P,\mathbb{T}_n^{+},n,h}^{\bm{g},\bm{\varepsilon}}-\bm{\hat{S}}_{P,\mathbb{T}_n^{+},n,h-1}^{\bm{g},\bm{\varepsilon}}||_2^2)^{\frac{1}{2}}
    \\&+
    \underset{P\in\mathcal{P}_{0,n}^{\ast}}{\sup}\left(\mathbb{E}_P\text{ } \underset{\ell=0,\ldots,\mathbb{T}_n^{+}-\mathbb{T}_n^{-}-L_n+1}{\max}\left|\left|\sum_{k=0}^{\ell}(\bm{\hat{w}}_{P,\mathbb{T}_n^{+}-k,n,h}^{\bm{g},\bm{\varepsilon}}-\bm{\hat{w}}_{P,\mathbb{T}_n^{+}-k,n,h-1}^{\bm{g},\bm{\varepsilon}})\right|\right|_2^2\right)^{\frac{1}{2}}
\end{align*} by (1) adding and subtracting $\bm{\hat{S}}_{P,\mathbb{T}_n^{+},n,h}^{\bm{g},\bm{\varepsilon}}-\bm{\hat{S}}_{P,\mathbb{T}_n^{+},n,h-1}^{\bm{g},\bm{\varepsilon}}$ and the triangle inequality, (2) including the \say{last} term in this reversed partial sum and rewriting as the corresponding martingale. Continuing on from (2), we have \begin{align*}
    &
       \underset{P\in\mathcal{P}_{0,n}^{\ast}}{\sup}(\mathbb{E}_P\text{ }||\bm{\hat{S}}_{P,\mathbb{T}_n^{+},n,h}^{\bm{g},\bm{\varepsilon}}-\bm{\hat{S}}_{P,\mathbb{T}_n^{+},n,h-1}^{\bm{g},\bm{\varepsilon}}||_2^2)^{\frac{1}{2}}
    \\&+
    \underset{P\in\mathcal{P}_{0,n}^{\ast}}{\sup}\left(\mathbb{E}_P\text{ } \underset{\ell=0,\ldots,\mathbb{T}_n^{+}-\mathbb{T}_n^{-}-L_n+1}{\max}\left|\left|\sum_{k=0}^{\ell}(\bm{\hat{w}}_{P,\mathbb{T}_n^{+}-k,n,h}^{\bm{g},\bm{\varepsilon}}-\bm{\hat{w}}_{P,\mathbb{T}_n^{+}-k,n,h-1}^{\bm{g},\bm{\varepsilon}})\right|\right|_2^2\right)^{\frac{1}{2}}
       \\& 
       \overset{(3)}{\leq} 
       3 \underset{P\in\mathcal{P}_{0,n}^{\ast}}{\sup}(\mathbb{E}_P\text{ }||\bm{\hat{S}}_{P,\mathbb{T}_n^{+},n,h}^{\bm{g},\bm{\varepsilon}}-\bm{\hat{S}}_{P,\mathbb{T}_n^{+},n,h-1}^{\bm{g},\bm{\varepsilon}}||_2^2)^{\frac{1}{2}}
       \\& 
       \overset{(4)}{\leq}  K \underset{P\in\mathcal{P}_{0,n}^{\ast}}{\sup}\left(\sum_{t\in\mathcal{T}_{n,L}}\mathbb{E}_P\text{ }||\bm{\hat{w}}_{P,t,n,h}^{\bm{g},\bm{\varepsilon}}-\bm{\hat{w}}_{P,t,n,h-1}^{\bm{g},\bm{\varepsilon}}||_2^2\right)^{\frac{1}{2}}, % Note: for (1) we also equated norm of negative, (2) \textcite{seq_gauss_approx2022} Thm 3.2 proof skips this minor step (their result is still correct) but we include it for the sake of clarity -- also, compare with Liu 2013 Thm 1 proof, (3) use vector-valued Doob's Lp maximal inequality on second term and the upper bound has a factor 2 so adding together both terms yields the factor of 3, which is recovered as a special case from the general Banach setting, (4) upper bound by max of partial sums ||full sum|| \leq max||psum|| just to clarify before applying the Lemma
\end{align*} by (3) Doob's maximal inequality (see e.g., Theorem 1.9 of \textcite{pisier2016}), and (4) upper bounding by max of partial sums and applying Lemma~\ref{lma:du_thm56} with the finite constant $K/3>0$. Hence, we have the inequality
\begin{align} & \underset{P\in\mathcal{P}_{0,n}^{\ast}}{\sup}(\mathbb{E}_P\text{ }\underset{t\in\mathcal{T}_{n,L}}{\max} ||\bm{\hat{S}}_{P,t,n,h}^{\bm{g},\bm{\varepsilon}}-\bm{\hat{S}}_{P,t,n,h-1}^{\bm{g},\bm{\varepsilon}}||_2^2)^{\frac{1}{2}}\label{eqn:PE_ineq1} \\ & \leq K \underset{P\in\mathcal{P}_{0,n}^{\ast}}{\sup}\left(\sum_{t\in\mathcal{T}_{n,L}}\mathbb{E}_P\text{ }||\bm{\hat{w}}_{P,t,n,h}^{\bm{g},\bm{\varepsilon}}-\bm{\hat{w}}_{P,t,n,h-1}^{\bm{g},\bm{\varepsilon}}||_2^2\right)^{\frac{1}{2}}. \nonumber
\end{align}

Observe that for $h=1,2,\ldots$, we have
\begin{align*}   &   \mathbb{E}_P\text{ }||\bm{\hat{w}}_{P,t,n,h}^{\bm{g},\bm{\varepsilon}}-\bm{\hat{w}}_{P,t,n,h-1}^{\bm{g},\bm{\varepsilon}}||_2^2    
\\&    
\overset{(1)}{=}
\sum_{m=(i,j,a,b)\in\mathcal{D}_n} \mathbb{E}_P(|\mathbb{E}_P(\hat{w}_{P,t,n,j,b}^{g}\varepsilon_{P,t,n,i,a}|\eta_{t}^{\bm{\varepsilon}},\ldots,\eta_{t-h}^{\bm{\varepsilon}},\mathcal{H}_{t}^{\bm{\hat{g}}})
\\&
-\mathbb{E}_P(\hat{w}_{P,t,n,j,b}^{g}\varepsilon_{P,t,n,i,a}|\eta_{t}^{\bm{\varepsilon}},\ldots,\eta_{t-h+1}^{\bm{\varepsilon}},\mathcal{H}_{t}^{\bm{\hat{g}}})|^2)  
\\&   
\overset{(2)}{=}
\sum_{m=(i,j,a,b)\in\mathcal{D}_n} \mathbb{E}_P(|\hat{w}_{P,t,n,j,b}^{g}(\mathbb{E}_P(\varepsilon_{P,t,n,i,a}|\eta_{t}^{\bm{\varepsilon}},\ldots,\eta_{t-h}^{\bm{\varepsilon}},\mathcal{H}_{t}^{\bm{\hat{g}}})
\\&
-\mathbb{E}_P(\varepsilon_{P,t,n,i,a}|\eta_{t}^{\bm{\varepsilon}},\ldots,\eta_{t-h+1}^{\bm{\varepsilon}},\mathcal{H}_{t}^{\bm{\hat{g}}}))|^2) 
\\&    
\overset{(3)}{=}\sum_{m=(i,j,a,b)\in\mathcal{D}_n} \mathbb{E}_P(|\hat{w}_{P,t,n,j,b}^{g}\mathbb{E}_P[(\mathbb{E}_P(\varepsilon_{P,t,n,i,a}|\eta_{t}^{\bm{\varepsilon}},\ldots,\eta_{t-h}^{\bm{\varepsilon}},\mathcal{H}_{t}^{\bm{\hat{g}}})
\\&
-\mathbb{E}_P(\varepsilon_{P,t,n,i,a}|\eta_{t}^{\bm{\varepsilon}},\ldots,\eta_{t-h+1}^{\bm{\varepsilon}},\mathcal{H}_{t}^{\bm{\hat{g}}}))|\eta_{t}^{\bm{\varepsilon}},\ldots,\eta_{t-h}^{\bm{\varepsilon}},\mathcal{H}_{t}^{\bm{\hat{g}}}]|^2) % measurability of the conditional expectations and linearity of conditional expectation  
\\&    
\overset{(4)}{=}\sum_{m=(i,j,a,b)\in\mathcal{D}_n} \mathbb{E}_P(|\hat{w}_{P,t,n,j,b}^{g}\mathbb{E}_P[(\mathbb{E}_P(G_{P,t,n,i,a}^{\varepsilon}(\mathcal{H}_{t,a}^{\varepsilon})|\eta_{t}^{\bm{\varepsilon}},\ldots,\eta_{t-h}^{\bm{\varepsilon}},\mathcal{H}_{t}^{\bm{\hat{g}}})
\\&
-\mathbb{E}_P(G_{P,t,n,i,a}^{\varepsilon}(\mathcal{H}_{t,a}^{\varepsilon})|\eta_{t}^{\bm{\varepsilon}},\ldots,\eta_{t-h+1}^{\bm{\varepsilon}},\mathcal{H}_{t}^{\bm{\hat{g}}}))|\eta_{t}^{\bm{\varepsilon}},\ldots,\eta_{t-h}^{\bm{\varepsilon}},\mathcal{H}_{t}^{\bm{\hat{g}}}]|^2) % measurability of the conditional expectations and linearity of conditional expectation  
\\ &
\overset{(5)}{=}\sum_{m=(i,j,a,b)\in\mathcal{D}_n} \mathbb{E}_P(|\hat{w}_{P,t,n,j,b}^{g}\mathbb{E}_P[(\mathbb{E}_P(G_{P,t,n,i,a}^{\varepsilon}(\mathcal{H}_{t,a}^{\varepsilon})|\eta_{t}^{\bm{\varepsilon}},\ldots,\eta_{t-h}^{\bm{\varepsilon}},\mathcal{H}_{t}^{\bm{\hat{g}}})
\\&
-\mathbb{E}_P(G_{P,t,n,i,a}^{\varepsilon}(\bar{\mathcal{H}}_{t,a,h}^{\varepsilon})|\eta_{t}^{\bm{\varepsilon}},\ldots,\eta_{t-h+1}^{\bm{\varepsilon}},\mathcal{H}_{t}^{\bm{\hat{g}}}))|\eta_{t}^{\bm{\varepsilon}},\ldots,\eta_{t-h}^{\bm{\varepsilon}},\mathcal{H}_{t}^{\bm{\hat{g}}}]|^2), % measurability of the conditional expectations and linearity of conditional expectation. 
% Note: the error inputs can be correlated with the covariate inputs and other response inputs. We just require the error inputs to have decaying conditional functional dependence measure given the inputs for the other predictor. 
%%%%% Note: do NOT replace with the entire past inputs, just replace the input at time t-h. as in page 6 of seq22 paper https://arxiv.org/pdf/2203.03237 you can indeed upper bound the replacement of all past innovations. indeed, that's exactly what you're doing below, so there is no need to do that here...
\end{align*} by (1) rewriting the expression, (2) the causal representation from Assumption 3.3, (3) measurability of the conditional expectations and the linearity property of conditional expectation, (4) the causal representation from Assumption 3.4, and (5) replacing $\eta_{t-h,a}^{\varepsilon}$ in $\mathcal{H}_{t,a}^{\varepsilon}$ with the $\sigma(\eta_{t}^{\bm{\varepsilon}},\ldots,\eta_{t-h+1}^{\bm{\varepsilon}},\mathcal{H}_{t}^{\bm{\hat{g}}})$-conditionally iid copy $\bar{\eta}_{t-h,a}^{\varepsilon}$ yielding $\bar{\mathcal{H}}_{t,a,h}^{\varepsilon}$. %%% Note: the error inputs can include the inputs for the covariates (e.g. heteroskedasticity). nevertheless, even when we condition on the noise inputs for the other predictor up to time t, i.e., $\mathcal{H}_{t}^{\bm{\hat{g}}}$, that will not make the noise inputs for epsilonn correlated across time (i.e. no collider), because they cause the covariate noise inputs at different times and those are temporally indepenndent... hence we do NOT need to condition on the recent noise inputs not just $\mathcal{H}_{t}^{\bm{\hat{g}}}$...  so there wont be a collider problem
% nevertheless it's still possible to be independent but conditionally dependent despite there not being a collider (pathological examples, but still), so just to be safe define as conditional on the inputs for the other predictor and the "recent" innovations
Continuing on from (5), \begin{align*} &\overset{(6)}{=}  \sum_{m=(i,j,a,b)\in\mathcal{D}_n} \mathbb{E}_P(|\hat{w}_{P,t,n,j,b}^{g}\mathbb{E}_P[(\mathbb{E}_P(G_{P,t,n,i,a}^{\varepsilon}(\mathcal{H}_{t,a}^{\varepsilon})|\eta_{t}^{\bm{\varepsilon}},\ldots,\eta_{t-h}^{\bm{\varepsilon}},\mathcal{H}_{t}^{\bm{\hat{g}}})
\\&
-\mathbb{E}_P(G_{P,t,n,i,a}^{\varepsilon}(\bar{\mathcal{H}}_{t,a,h}^{\varepsilon})|\eta_{t}^{\bm{\varepsilon}},\ldots,\eta_{t-h}^{\bm{\varepsilon}},\mathcal{H}_{t}^{\bm{\hat{g}}}))|\eta_{t}^{\bm{\varepsilon}},\ldots,\eta_{t-h}^{\bm{\varepsilon}},\mathcal{H}_{t}^{\bm{\hat{g}}}]|^2) % measurability of the conditional expectations and linearity of conditional expectation  
\\&    
\overset{(7)}{=}\sum_{m=(i,j,a,b)\in\mathcal{D}_n} \mathbb{E}_P(|\hat{w}_{P,t,n,j,b}^{g}\mathbb{E}_P[G_{P,t,n,i,a}^{\varepsilon}(\mathcal{H}_{t,a}^{\varepsilon})
-G_{P,t,n,i,a}^{\varepsilon}(\bar{\mathcal{H}}_{t,a,h}^{\varepsilon})|\eta_{t}^{\bm{\varepsilon}},\ldots,\eta_{t-h}^{\bm{\varepsilon}},\mathcal{H}_{t}^{\bm{\hat{g}}}]|^2)
\\&
\overset{(8)}{\leq} D_n  \underset{(i,j,a,b)\in \mathcal{D}_n}{\max}\mathbb{E}_P(|\hat{w}_{P,t,n,j,b}^{g}|^2)  (\bar{\theta}_{P,t,n,i,a}^{\varepsilon,\infty}(h))^2 
\\&
\overset{(9)}{\leq}D_n  \underset{(i,j,a,b)\in \mathcal{D}_n}{\max}\mathbb{E}_P(|\hat{w}_{P,t,n,j,b}^{g}|^2)  
(\bar{\Theta}^{\infty}(h\lor 1)^{-\bar{\beta}^{\infty}})^2,
\end{align*} by (6) conditioning on $\eta_{t-h}^{\bm{\varepsilon}}$ in the second inner conditional expectation, which follows from the conditional independence of $\eta_{t-h}^{\bm{\varepsilon}}$ and Doob's conditional independence property (Theorem 8.9 of \textcite{kallenberg_FOMP_ed3}), (7) measurability and linearity of the conditional expectations, and (8) H\"{o}lder's inequality, contraction property of conditional expectation, rewriting as the conditional functional dependence measure from Definition 3.1, and upper bounding by the sum by $D_n$ times the maximum over the dimension/time-offset combinations in $\mathcal{D}_n$, and (9) the upper bound on the $L^{\infty}(P)$ functional dependence measure from Assumption 3.5. Similarly, for $h=0$, we have
 \begin{align*}   &   \mathbb{E}_P\text{ }||\bm{\hat{w}}_{P,t,n,0}^{\bm{g},\bm{\varepsilon}}-\bm{\hat{w}}_{P,t,n,-1}^{\bm{g},\bm{\varepsilon}}||_2^2
\\&    
\overset{(1)}{=}
\mathbb{E}_P\text{ }||\bm{\hat{w}}_{P,t,n,0}^{\bm{g},\bm{\varepsilon}}||_2^2
\\&
\overset{(2)}{=}
\sum_{m=(i,j,a,b)\in\mathcal{D}_n} \mathbb{E}_P(|\mathbb{E}_P(\hat{w}_{P,t,n,j,b}^{g}\varepsilon_{P,t,n,i,a}|\eta_{t}^{\bm{\varepsilon}},\mathcal{H}_{t}^{\bm{\hat{g}}})|^2)
\\&
\overset{(3)}{=}
\sum_{m=(i,j,a,b)\in\mathcal{D}_n} \mathbb{E}_P(|\hat{w}_{P,t,n,j,b}^{g}\mathbb{E}_P(\varepsilon_{P,t,n,i,a}|\eta_{t}^{\bm{\varepsilon}},\mathcal{H}_{t}^{\bm{\hat{g}}})|^2)
\\&
\overset{(4)}{\leq}
D_n  \underset{(i,j,a,b)\in \mathcal{D}_n}{\max}\mathbb{E}_P(|\hat{w}_{P,t,n,j,b}^{g}|^2)  
(\bar{\Theta}^{\infty})^2,
\end{align*} because (1) $\bm{\hat{w}}_{P,t,n,-1}^{\bm{g},\bm{\varepsilon}}=0$, (2) rewriting the expression, (3) Assumption 3.3, and (4) H\"{o}lder's inequality, contraction property of conditional expectation, applying the upper bound on the $L^{\infty}(P)$ norm from Assumption 3.5, and upper bounding by the sum by $D_n$ times the maximum over the dimension/time-offset combinations in $\mathcal{D}_n$. Hence, for all $h\in\mathbb{N}_0$ we have \begin{equation}    \mathbb{E}_P\text{ }||\bm{\hat{w}}_{P,t,n,h}^{\bm{g},\bm{\varepsilon}}-\bm{\hat{w}}_{P,t,n,h-1}^{\bm{g},\bm{\varepsilon}}||_2^2 \leq D_n  \underset{(i,j,a,b)\in \mathcal{D}_n}{\max}\mathbb{E}_P(|\hat{w}_{P,t,n,j,b}^{g}|^2)  
(\bar{\Theta}^{\infty}(h\lor 1)^{-\bar{\beta}^{\infty}})^2.  \label{eqn:PE_ineq2}  \end{equation}

% more detail for (2): by Assumption 3.3 the estimation error w^g is measurable wrt sigma-algebra generated by the iputs for Y and Z up to time mathbb{T}^+, the sigma-algebra generated by epsilon and the inputs for X and Z up to time t is independet of the all of inputs for Y and the inputs for Z up after time t and hence by basic properties of conditional expectation (see e.g., https://math.stackexchange.com/questions/365310/conditional-expectation-on-more-than-one-sigma-algebra ) we have that the error is independent of the inputs for Y and inputs for Z after time t and hence we can simplify the conditional expectation

%the causal representation and measurability of $\hat{w}_{P,t,n,j,b}^g$ by Assumption, measurability of $\hat{w}_{P,t,n,j,b}^g$, by distributing,
%by the tower property of conditional expectation, where $\tilde{\mathcal{H}}_{t,t-h}^{\bm{\hat{w}^{g,\varepsilon}}}=(\tilde{\mathcal{H}}_{t,t-h}^{\bm{\varepsilon}},\mathcal{H}_{\mathbb{T}_n^{+}}^{\bm{Y}},\mathcal{H}_{\mathbb{T}_n^{+}}^{\bm{Z}})$ using the iid copy $\tilde{\eta}_{t-h}^{\bm{\varepsilon}}$ of $\eta_{t-h}^{\bm{\varepsilon}}$ in the $(t-h)$-th position in the input sequence, linearity of expectation, the contraction property of conditional expectation, and H\"{o}lder's inequality.

Summing over $h\in\mathbb{N}_0$, we have
\begin{align*}
    &
    \underset{P\in\mathcal{P}_{0,n}^{\ast}}{\sup}(\mathbb{E}_P\text{ }\underset{t\in\mathcal{T}_{n,L}}{\max} ||\bm{\hat{S}}_{P,t,n}^{\bm{g},\bm{\varepsilon}}||_2^2)^{\frac{1}{2}}
    \\& 
    \overset{(1)}{\leq}
    \sum_{h=0}^{\infty}\underset{P\in\mathcal{P}_{0,n}^{\ast}}{\sup}(\mathbb{E}_P\text{ }\underset{t\in\mathcal{T}_{n,L}}{\max} ||\bm{\hat{S}}_{P,t,n,h}^{\bm{g},\bm{\varepsilon}}-\bm{\hat{S}}_{P,t,n,h-1}^{\bm{g},\bm{\varepsilon}}||_2^2)^{\frac{1}{2}}
    \\& 
    \overset{(2)}{\leq}
 \sum_{h=0}^{\infty}   K \underset{P\in\mathcal{P}_{0,n}^{\ast}}{\sup}\left(\sum_{t\in\mathcal{T}_{n,L}}\mathbb{E}_P\text{ }||\bm{\hat{w}}_{P,t,n,h}^{\bm{g},\bm{\varepsilon}}-\bm{\hat{w}}_{P,t,n,h-1}^{\bm{g},\bm{\varepsilon}}||_2^2\right)^{\frac{1}{2}} 
 \\& 
 \overset{(3)}{\leq}
 \sum_{h=0}^{\infty}   K \underset{P\in\mathcal{P}_{0,n}^{\ast}}{\sup}\left(\sum_{t\in\mathcal{T}_{n,L}} D_n \underset{j\in [d_Y]}{\max}\text{ } \underset{b\in B_j}{\max}\text{ }\mathbb{E}_P(|\hat{w}_{P,t,n,j,b}^{g}|^2)(\bar{\Theta}^{\infty}(h\lor 1)^{-\bar{\beta}^{\infty}})^2 \right)^{\frac{1}{2}},
\end{align*} by (1) the telescoping argument from~\eqref{eqn:telescoping_fdm} and the triangle inequality, (2) applying the inequality~\eqref{eqn:PE_ineq1}, and (3) applying the inequality~\eqref{eqn:PE_ineq2}. Continuing on from line (3), we have \begin{align*} & \sum_{h=0}^{\infty}   K \underset{P\in\mathcal{P}_{0,n}^{\ast}}{\sup}\left(\sum_{t\in\mathcal{T}_{n,L}} D_n \underset{j\in [d_Y]}{\max}\text{ }\underset{b\in B_j}{\max}\text{ }\mathbb{E}_P(|\hat{w}_{P,t,n,j,b}^{g}|^2)(\bar{\Theta}^{\infty}(h\lor 1)^{-\bar{\beta}^{\infty}})^2 \right)^{\frac{1}{2}}
  \\& 
    \overset{(4)}{\leq}
 \sum_{h=0}^{\infty}   K \underset{P\in\mathcal{P}_{0,n}^{\ast}}{\sup}(T_{n,L} D_n \underset{t\in \mathcal{T}_{n,L}}{\max}\underset{j\in [d_Y]}{\max}\text{ }\underset{b\in B_j}{\max}\text{ }\mathbb{E}_P(|\hat{w}_{P,t,n,j,b}^{g}|^2)(\bar{\Theta}^{\infty}(h\lor 1)^{-\bar{\beta}^{\infty}})^2 )^{\frac{1}{2}}
 \\& 
  \overset{(5)}{\leq}
\bar{\Theta}^{\infty} K T_{n,L}^{\frac{1}{2}} D_n^{\frac{1}{2}} \underset{P\in\mathcal{P}_{0,n}^{\ast}}{\sup} \underset{t\in \mathcal{T}_{n,L}}{\max}\text{ }\underset{j\in [d_Y]}{\max}\text{ }\underset{b\in B_j}{\max}\text{ }\mathbb{E}_P(|\hat{w}_{P,t,n,j,b}^{g}|^2)^{\frac{1}{2}} \sum_{h=0}^{\infty}    (h\lor 1)^{-\bar{\beta}^{\infty}}
\\&
\overset{(6)}{\leq}
  \bar{\Theta}^{\infty} \bar{K}^{\infty} K T_{n,L}^{\frac{1}{2}} D_n^{\frac{1}{2}} \underset{P\in\mathcal{P}_{0,n}^{\ast}}{\sup} \underset{t\in \mathcal{T}_{n,L}}{\max}\text{ }\underset{j\in [d_Y]}{\max}\text{ }\underset{b\in B_j}{\max}\text{ }\mathbb{E}_P(|\hat{w}_{P,t,n,j,b}^{g}|^2)^{\frac{1}{2}} 
\\&
    \overset{(7)}{\leq}
\bar{K} T_{n,L}^{\frac{1}{2}} D_n^{\frac{1}{2}} \underset{P\in\mathcal{P}_{0,n}^{\ast}}{\sup} \underset{t\in \mathcal{T}_{n,L}}{\max}\text{ }\underset{j\in [d_Y]}{\max}\text{ }\underset{b\in B_j}{\max}\text{ }\mathbb{E}_P(|\hat{w}_{P,t,n,j,b}^{g}|^2)^{\frac{1}{2}},
\end{align*} by (4) upper bounding each term by the maximum over time $t$, (5) simplifying the expression, (6) writing $\bar{K}^{\infty}=\sum_{h=0}^{\infty}(h\lor 1)^{-\bar{\beta}^{\infty}}<\infty$ since $\bar{\beta}^{\infty}>1$ by Assumption 3.5, and (7) grouping together the positive constants into the positive constant $\bar{K}$.

\textbf{Step 1.3:} The same arguments as Step 1.2 (i.e. exchanging $g,\varepsilon$ with $f,\xi$) can be used to show that for $n\in\mathbb{N}$ and $\delta >0$ we have
$$\underset{P\in\mathcal{P}_{0,n}^{\ast}}{\sup}\mathbb{P}_P(\tau_n^{-1}S_{n,p}(\bm{\hat{w}}_{P,n}^{\bm{f},\bm{\xi}}) > \delta)=o(1).$$

\textbf{Step 2 (Strong Gaussian Approximation):} Next, we turn to the products of errors $(\bm{R}_{P,t,n})_{t\in\mathcal{T}_{n,L}}$. Denote the Gaussian random vectors associated with the strong Gaussian approximation of the product of errors by $\bm{R^\dagger}_{t,n}\sim\mathcal{N}(0,\bm{\Sigma}_{P,t,n}^{\bm{R}})$ for ${t\in\mathcal{T}_{n,L}}$. Observe that
\begin{align*}
&\underset{P\in\mathcal{P}_{0,n}^{\ast}}{\sup}\mathbb{P}_P(S_{n,p}(\bm{R}_{P,n}) > \hat{q}_{1-\alpha +\nu_n}+\frac{\tau_n}{2})
\\& \overset{(1)}{\leq}
\underset{P\in\mathcal{P}_{0,n}^{\ast}}{\sup}\mathbb{P}_P(S_{n,p}(\bm{R^\dagger}_{n}) > \hat{q}_{1-\alpha +\nu_n}+\frac{\tau_n}{4})
\\&+
\underset{P\in\mathcal{P}_{0,n}^{\ast}}{\sup}\mathbb{P}_P\left(\underset{s\in\mathcal{T}_{n,L}}{\max}\left|\left|\frac{1}{\sqrt{T_{n,L}}}\sum_{t\leq s}(\bm{R}_{P,t,n}-\bm{R^\dagger}_{t,n})\right|\right|_2 > \frac{\tau_n}{4}\right)
\\&\overset{(2)}{\leq}
\underset{P\in\mathcal{P}_{0,n}^{\ast}}{\sup}\mathbb{P}_P(S_{n,p}(\bm{R^\dagger}_{n}) > \hat{q}_{1-\alpha +\nu_n}+\frac{\tau_n}{4})
\\&+ 4 \tau_n^{-1} \underset{P\in\mathcal{P}_{0,n}^{\ast}}{\sup}\mathbb{E}_P\left(\underset{s\in\mathcal{T}_{n,L}}{\max}\left|\left|\frac{1}{\sqrt{T_{n,L}}}\sum_{t\leq s}(\bm{R}_{P,t,n}-\bm{R^\dagger}_{t,n})\right|\right|_2 \right)
\\&\overset{(3)}{\leq} 
\underset{P\in\mathcal{P}_{0,n}^{\ast}}{\sup}\mathbb{P}_P(S_{n,p}(\bm{R^\dagger}_{n}) > \hat{q}_{1-\alpha +\nu_n}+\frac{\tau_n}{4})
\\&+ 4 \tau_n^{-1} K D_n^{\frac{1}{2}}\bar{\Theta}^R (\bar{\Gamma}_n^R)^{\frac{1}{2}\frac{\bar{\beta}^R-2}{\bar{\beta}^R-1}} \sqrt{\log(T_{n,L})}\left(\frac{D_n}{T_{n,L}}\right)^{\xi(\bar{q}^R,\bar{\beta}^R)},
\end{align*} where (1) follows from the triangle inequality, subadditivity, and the assumption about the form of the test statistic, (2) follows by Markov's inequality, and (3) follows by the distribution-uniform strong Gaussian approximation for high-dimensional nonstationary processes from Lemma~\ref{lma:du_sga}. By subadditivity and monotonicity, we have
\begin{align*}
&\underset{P\in\mathcal{P}_{0,n}^{\ast}}{\sup}\mathbb{P}_P(S_{n,p}(\bm{R^\dagger}_{n}) > \hat{q}_{1-\alpha +\nu_n}+\frac{\tau_n}{4})
\\&
\leq 
\underset{P\in\mathcal{P}_{0,n}^{\ast}}{\sup}\mathbb{P}_P(S_{n,p}(\bm{R^\dagger}_{n}) > q_{1-\alpha})
\\& 
+\underset{P\in\mathcal{P}_{0,n}^{\ast}}{\sup}\mathbb{P}_P(q_{1-\alpha} > \hat{q}_{1-\alpha +\nu_n}+\frac{\tau_n}{4}) 
\\&
= 
\alpha +\underset{P\in\mathcal{P}_{0,n}^{\ast}}{\sup}\mathbb{P}_P(q_{1-\alpha} > \hat{q}_{1-\alpha +\nu_n}+\frac{\tau_n}{4}).
\end{align*}

\textbf{Step 3 (Covariance Approximation):} Now, we focus on upper bounding $$\underset{P\in\mathcal{P}_{0,n}^{\ast}}{\sup}\mathbb{P}_P(q_{1-\alpha} > \hat{q}_{1-\alpha +\nu_n}+\frac{\tau_n}{4}).$$

\textbf{Step 3.1:} Let us reflect on the implications of Proposition 4.2 of \textcite{seq_gauss_approx2022}, which is the distribution-pointwise version of Lemma~\ref{lma:du_stat_inf_cov_est}. Proposition 4.2 states that for each $n\in\mathbb{N}$ and $P\in\mathcal{P}_{0,n}^{\ast}$, for some cumulative covariances $(\bar{Q}_{P,t,n}^{\bm{R}})_{t\in\mathcal{T}_{n,L}}$, there exist \textit{independent} Gaussian random vectors $\bm{\bar{R}}_{t,n} \sim \mathcal{N}(0,\bm{\bar{\Sigma}}_{P,t,n}^{\bm{R}})$ for ${t\in\mathcal{T}_{n,L}}$ with $\bm{\bar{\Sigma}}_{P,t,n}^{\bm{R}}=\bar{Q}_{P,t,n}^{\bm{R}}-\bar{Q}_{P,t-1,n}^{\bm{R}}$ that are coupled with the Gaussian random vectors from the strong Gaussian approximation of the product of errors $\bm{R^\dagger}_{t,n}\sim\mathcal{N}(0,\bm{\Sigma}_{P,t,n}^{\bm{R}})$ for ${t\in\mathcal{T}_{n,L}}$, such that
$$\mathbb{E}_P\text{ }\underset{k\in\mathcal{T}_{n,L}}{\max}\left|\left|\sum_{t\leq k} \bm{R^\dagger}_{t,n} - \sum_{t\leq k} \bm{\bar{R}}_{t,n}  \right|\right|_2^2 \leq K \text{ }\log(T_{n,L}) \text{ } [\sqrt{T_{n,L}\bar{\delta}_{P,n} \rho_{P,n}}+\rho_{P,n}]=\bar{\Delta}_{P,n},$$ where $$\bar{\delta}_{P,n}=\underset{k\in\mathcal{T}_{n,L}}{\max}\left|\left|\sum_{t\leq k} \bm{\Sigma}_{P,t,n}^{\bm{R}}-\sum_{t \leq k} \bm{\bar{\Sigma}}_{P,t,n}^{\bm{R}} \right|\right|_{\mathrm{tr}}$$ and $$\rho_{P,n}=\underset{t\in \mathcal{T}_{n,L}}{\max}||\bm{\Sigma}_{P,t,n}^{\bm{R}}||_{\mathrm{tr}}.$$

Let $\bm{\bar{R}}_{n}=(\bm{\bar{R}}_{t,n})_{t\in\mathcal{T}_{n,L}}$ and denote the $(1-\alpha)$ quantile of $S_{n,p}(\bm{\bar{R}}_{n})$ by $\bar{q}_{1-\alpha}$. For each $n\in\mathbb{N}$ and $P\in\mathcal{P}_{0,n}^{\ast}$, we have \begin{align*}
& \mathbb{P}_P(S_{n,p}(\bm{R^\dagger}_{n}) > \bar{q}_{1-\alpha +\nu_n}+\frac{\tau_n}{4})
\\&
\overset{(1)}{\leq} 
 \mathbb{P}_P(S_{n,p}(\bm{\bar{R}}_{n}) > \bar{q}_{1-\alpha +\nu_n})
\\&
+  \mathbb{P}_P\left(\underset{s\in\mathcal{T}_{n,L}}{\max}\left|\left|\frac{1}{\sqrt{T_{n,L}}}\sum_{t\leq s}(\bm{R^\dagger}_{t,n}-\bm{\bar{R}}_{t,n})\right|\right|_2 > \frac{\tau_n}{4}\right)
\\&
\overset{(2)}{=} 
 \mathbb{P}_P(S_{n,p}(\bm{\bar{R}}_{n}) > \bar{q}_{1-\alpha +\nu_n})
\\&
+  \mathbb{P}_P\left(\underset{s\in\mathcal{T}_{n,L}}{\max}\left|\left|\frac{1}{\sqrt{T_{n,L}}}\sum_{t\leq s}(\bm{R^\dagger}_{t,n}-\bm{\bar{R}}_{t,n})\right|\right|_2^2 > \frac{\tau_n^2}{16}\right)
\\&
\overset{(3)}{\leq}
 \mathbb{P}_P(S_{n,p}(\bm{\bar{R}}_{n}) > \bar{q}_{1-\alpha +\nu_n})
\\&
+ 16\tau_n^{-2}T_{n,L}^{-1} \mathbb{E}_P\left(\underset{s\in\mathcal{T}_{n,L}}{\max}\left|\left|\sum_{t\leq s}(\bm{R^\dagger}_{t,n}-\bm{\bar{R}}_{t,n})\right|\right|_2^2 \right)
\\&
\overset{(4)}{\leq} 
(\alpha -\nu_n)+ 16\tau_n^{-2} \bar{\Delta}_{P,n} T_{n,L}^{-1}\overset{(5)}{=}\alpha +\left[16\tau_n^{-2} \bar{\Delta}_{P,n} T_{n,L}^{-1}-\nu_n \right],
\end{align*} where the previous lines follow by (1) the triangle inequality, subadditivity, the assumption about the form of the test statistic, (2) squaring, (3) Markov's inequality, (4) Proposition 4.2 from \textcite{seq_gauss_approx2022}, and (5) rearranging terms. We see that if $$\left[16\tau_n^{-2} \bar{\Delta}_{P,n} T_{n,L}^{-1}-\nu_n \right]<0,$$ then $$\mathbb{P}_P(S_{n,p}(\bm{R^\dagger}_{n}) > \bar{q}_{1-\alpha +\nu_n}+\frac{\tau_n}{4})<\alpha,$$ which implies that $\bar{q}_{1-\alpha +\nu_n}+\frac{\tau_n}{4}$ is greater than $q_{1-\alpha}^{\dagger}$, the $(1-\alpha)$ quantile of $S_{n,p}(\bm{R^\dagger}_{n})$. Hence, if $$q_{1-\alpha}^{\dagger} \geq \bar{q}_{1-\alpha +\nu_n}+\frac{\tau_n}{4},$$ then $$\left[16\tau_n^{-2} \bar{\Delta}_{P,n} T_{n,L}^{-1}-\nu_n \right] \geq 0,$$ or equivalently $$\bar{\Delta}_{P,n} \geq \frac{1}{16}T_{n,L}\nu_n \tau_n^{2}.$$

\textbf{Step 3.2:} Now, we apply this idea with the cumulative covariance of the residual products. By the implication stated at the end of Step 3.1 and monotonicity, we have
\begin{align*}
&\underset{P\in\mathcal{P}_{0,n}^{\ast}}{\sup}\mathbb{P}_P(q_{1-\alpha} > \hat{q}_{1-\alpha +\nu_n}+\frac{\tau_n}{4})
\\&
\leq
\underset{P\in\mathcal{P}_{0,n}^{\ast}}{\sup}\mathbb{P}_P(\hat{\Delta}_{P,n} \geq \frac{1}{16}T_{n,L}\nu_n \tau_n^{2}),
\end{align*} where we have replaced $\bar{\Delta}_{P,n}$, $\bar{\delta}_{P,n}$ with $\hat{\Delta}_{P,n}$, $\hat{\delta}_{P,n}$ which are defined by  \begin{align*}\hat{\Delta}_{P,n} &= K \text{ }\log(T_{n,L}) \text{ } [\sqrt{T_{n,L}\hat{\delta}_{P,n} \rho_{P,n}}+\rho_{P,n}],\\ \hat{\delta}_{P,n}&=\underset{k\in\mathcal{T}_{n,L}}{\max}\left|\left|\sum_{t\leq k} \bm{\Sigma}_{P,t,n}^{\bm{R}}-\hat{Q}_{k,n}^{\bm{R}} \right|\right|_{\mathrm{tr}},\\ \rho_{P,n}&=\underset{t\in\mathcal{T}_{n,L}}{\max}||\bm{\Sigma}_{P,t,n}^{\bm{R}}||_{\mathrm{tr}}.\end{align*} Thus, if we can find $\varphi_n$ such that $\hat{\Delta}_{P,n}=O_{\mathcal{P}}(\varphi_n)$ and if we select the offsets so that $\nu_n \tau_n^2 \gg T_{n,L}^{-1} \varphi_n$, or equivalently $ \nu_n \gg \tau_n^{-2} T_{n,L}^{-1} \varphi_n$, then we will have  
% Vinogradov https://en.wikipedia.org/wiki/Big_O_notation#History_(Bachmann%E2%80%93Landau,_Hardy,_and_Vinogradov_notations)
%or equivalently $ \tau_n \gg \nu_n^{-\frac{1}{2}} T_{n,L}^{-\frac{1}{2}} \varphi_n^{\frac{1}{2}}$, then we will have  
$$\underset{P\in\mathcal{P}_{0,n}^{\ast}}{\sup}\mathbb{P}_P(\hat{\Delta}_{P,n} \geq \frac{1}{16}T_{n,L}\nu_n \tau_n^{2})=o(1).$$ By Lemma~\ref{lma:du_prop54} and Assumption 3.5, we have
$$\underset{P\in\mathcal{P}_{0,n}^{\ast}}{\sup}\rho_{P,n} \leq K_{\rho} D_n (\bar{\Theta}^R)^2,$$
for some constant $K_{\rho}>0$, so we obtain $\hat{\Delta}_{P,n}=O_{\mathcal{P}}\left(\varphi_n\right)$ with $$\varphi_n=\log(T_{n,L})D_n\left[T_{n,L}^{\frac{1}{2}}D_n^{-\frac{1}{2}}(r_{n,1}^{\delta}+r_{n,2}^{\delta})^{\frac{1}{2}} + 1\right],$$
where $$\hat{\delta}_{P,n}=O_{\mathcal{P}}(r_{n,1}^{\delta}+r_{n,2}^{\delta}),$$ for some rates $r_{n,1}^{\delta}$, $r_{n,2}^{\delta}$ that remain to be derived.

Plugging $\varphi_n$ into the offset condition $\nu_n \gg \tau_n^{-2} T_{n,L}^{-1} \varphi_n$ that we wish to satisfy, if we have
% Vinogradov https://en.wikipedia.org/wiki/Big_O_notation#History_(Bachmann%E2%80%93Landau,_Hardy,_and_Vinogradov_notations)
% or $\nu_n \tau_n^2 \gg T_{n,L}^{-1} \varphi_n$
% or $ \tau_n \gg \nu_n^{-\frac{1}{2}} T_{n,L}^{-\frac{1}{2}} \varphi_n^{\frac{1}{2}}$ 
$$\nu_n \gg \log(T_{n,L})D_n(\tau_n^{-2} (T_{n,L}^{-\frac{1}{2}}D_n^{-\frac{1}{2}}(r_{n,1}^{\delta}+r_{n,2}^{\delta})^{\frac{1}{2}} + T_{n,L}^{-1})),$$
% Vinogradov https://en.wikipedia.org/wiki/Big_O_notation#History_(Bachmann%E2%80%93Landau,_Hardy,_and_Vinogradov_notations)
% $$\tau_n \gg  \sqrt{\text{log}(T_{n,L})}D_n^{\frac{1}{2}}\left(\nu_n^{-\frac{1}{2}}\left( T_{n,L}^{-\frac{1}{4}}D_n^{-\frac{1}{4}}\hat{\delta}_n^{\frac{1}{4}}+T_{n,L}^{-\frac{1}{2}}\right)\right),$$
then  
$$\underset{P\in\mathcal{P}_{0,n}^{\ast}}{\sup}\mathbb{P}_P(\hat{\Delta}_{P,n} \geq \frac{1}{16}T_{n,L}\nu_n \tau_n^{2})=o(1).$$

\textbf{Step 3.3:} It remains to analyze $\hat{\delta}_{P,n}$ and the rates $r_{n,1}^{\delta}$, $r_{n,2}^{\delta}$. For $P\in\mathcal{P}_n$ and $t\in\mathcal{T}_{n,L}$, define
\begin{equation}
Q_{P,t,n}^{\bm{R}} = \sum_{s=L_n+\mathbb{T}_n^{-}-1}^t \frac{1}{L_n} \left( \sum_{r=s-L_n+1}^{s} \bm{R}_{P,r,n} \right)\left( \sum_{r=s-L_n+1}^{s} \bm{R}_{P,r,n} \right)^{\top}, \label{eqn:tv_cov_estimator_with_errors_not_resid} \end{equation} which is the cumulative covariance estimator $\hat{Q}_{t,n}^{\bm{R}}$ from Section 3.5 with the residual products substituted by the error products. By the triangle inequality, we have
\begin{align*}
\hat{\delta}_{P,n}
&=
\underset{k\in\mathcal{T}_{n,L}}{\max}\left|\left|\sum_{t\leq k} \bm{\Sigma}_{P,t,n}^{\bm{R}}-\hat{Q}_{k,n}^{\bm{R}} \right|\right|_{\mathrm{tr}}
\\&\leq 
\text{ }\underset{k\in\mathcal{T}_{n,L}}{\max}\left|\left|\sum_{t\leq k} \bm{\Sigma}_{P,t,n}^{\bm{R}}-Q_{P,k,n}^{\bm{R}} \right|\right|_{\mathrm{tr}}
\\&+
\text{ }\underset{k\in\mathcal{T}_{n,L}}{\max}|| \hat{Q}_{k,n}^{\bm{R}} - Q_{P,k,n}^{\bm{R}}  ||_{\mathrm{tr}}.\end{align*} By Lemma~\ref{lma:du_cumul_cov_estimator}, Assumption 3.5, and Assumption 3.6, the covariance estimation error when using the error products instead of the residual products can be bounded as
\begin{align*}
&\underset{P\in\mathcal{P}_{0,n}^{\ast}}{\sup}\mathbb{E}_P\left(\text{ }\underset{k\in\mathcal{T}_{n,L}}{\max}\left|\left|\sum_{t\leq k} \bm{\Sigma}_{P,t,n}^{\bm{R}}-Q_{P,k,n}^{\bm{R}} \right|\right|_{\mathrm{tr}}\right)
\\&\leq
K  (\bar{\Theta}^R)^2 D_n(\bar{\Gamma}_n^R  L_n^{\frac{1}{2}} + T_{n,L}^{\frac{1}{2}} D_n^{\frac{1}{2}} L_n^{\frac{1}{2}} + T_{n,L} L_n^{-1} + T_{n,L} L_n^{2-\bar{\beta}^R})
\\&
=O(r_{n,1}^{\delta}),
\end{align*} where $$r_{n,1}^{\delta}= D_n( \bar{\Gamma}_n^R  L_n^{\frac{1}{2}} + T_{n,L}^{\frac{1}{2}} D_n^{\frac{1}{2}} L_n^{\frac{1}{2}} + T_{n,L} L_n^{-1} + T_{n,L} L_n^{2-\bar{\beta}^R}).$$ Next, we must handle the estimation errors due to using the residual products instead of the error products. For any $\epsilon >0$, we have
\begin{align*}
&\underset{P\in\mathcal{P}_{0,n}^{\ast}}{\sup}\mathbb{E}_P(\text{ } \underset{k\in\mathcal{T}_{n,L}}{\max}|| \hat{Q}_{k,n}^{\bm{R}} - Q_{P,k,n}^{\bm{R}} ||_{\mathrm{tr}}\land \epsilon)
\\&\overset{(1)}{=} 
\underset{P\in\mathcal{P}_{0,n}^{\ast}}{\sup}\mathbb{E}_P\left(\text{ }\underset{k\in\mathcal{T}_{n,L}}{\max}\left|\left|\frac{1}{L_n} \sum_{r=L_n+\mathbb{T}_n^{-}-1}^k  \left[\left( \sum_{s=r-L_n+1}^{r} \bm{\hat{R}}_{s,n} \right)^{\otimes 2}
-\left( \sum_{s=r-L_n+1}^{r} \bm{R}_{P,s,n} \right)^{\otimes 2}\right]  \right|\right|_{\mathrm{tr}}\land \epsilon\right)
\\&\overset{(2)}{\leq}
 \underset{P\in\mathcal{P}_{0,n}^{\ast}}{\sup}\mathbb{E}_P\left(\text{ }\left[\frac{1}{L_n}\sum_{r\in\mathcal{T}_{n,L}} 
\left|\left| \left( \sum_{s=r-L_n+1}^{r} \bm{\hat{R}}_{s,n} \right)^{\otimes 2}  - \left( \sum_{s=r-L_n+1}^{r} \bm{R}_{P,s,n} \right)^{\otimes 2} \right|\right|_{\mathrm{tr}}\right]\land \epsilon\right)
\\&\overset{(3)}{\leq}
 \underset{P\in\mathcal{P}_{0,n}^{\ast}}{\sup}\mathbb{E}_P\left(\text{ }\left[\frac{2}{L_n}\sum_{r\in\mathcal{T}_{n,L}}  \left( \left|\left|  \sum_{s=r-L_n+1}^{r} \left(\bm{\hat{R}}_{s,n} - \bm{R}_{P,s,n} \right) \right|\right|_2 \left|\left|  \sum_{s=r-L_n+1}^{r} \bm{R}_{P,s,n} \right|\right|_2 \right.\right.\right. 
\\&+ \left.\left.\left. \left|\left|  \sum_{s=r-L_n+1}^{r} \left(\bm{\hat{R}}_{s,n} - \bm{R}_{P,s,n} \right) \right|\right|_2^2\right)\right]\land\epsilon\right),\end{align*} where (1) is from the definitions of $Q_{P,k,n}^{\bm{R}}$, $\hat{Q}_{k,n}^{\bm{R}}$, (2) is from the triangle inequality, and (3) is from the following outer product inequality for vectors $\hat{v},v \in \mathbb{R}^d$  
\begin{align*}
&||\hat{v}\hat{v}^{\top}-vv^{\top}||_{\text{tr}}
\\&\overset{(1)}{=}
||(\hat{v}-v)v^{\top} + v(\hat{v}-v)^{\top} + (\hat{v}-v)(\hat{v}-v)^{\top} ||_{\text{tr}}
\\&\overset{(2)}{\leq}
2||(\hat{v}-v)v^{\top}||_{\text{tr}} + ||(\hat{v}-v)(\hat{v}-v)^{\top} ||_{\text{tr}}
\\&\overset{(3)}{=}
2||\hat{v}-v||_2 ||v||_2 + ||\hat{v}-v||_2^2, 
\end{align*} where (1) follows from adding and subtracting terms, (2) follows from the triangle inequality, and (3) follows by the properties of outer products and the definition of the trace norm. For any $r\in\mathcal{T}_{n,L}$, we have the following decomposition into the three bias terms from Step 1 by the triangle inequality
\begin{align*}&\left|\left|  \sum_{s=r-L_n+1}^{r} \left(\bm{\hat{R}}_{s,n} - \bm{R}_{P,s,n} \right) \right|\right|_2 \\&\leq \left|\left|  \sum_{s=r-L_n+1}^{r}\bm{\hat{w}}_{P,s,n}^{\bm{f},\bm{g}}\right|\right|_2+\left|\left|  \sum_{s=r-L_n+1}^{r}  \bm{\hat{w}}_{P,s,n}^{\bm{g},\bm{\varepsilon}}\right|\right|_2+\left|\left|  \sum_{s=r-L_n+1}^{r} \bm{\hat{w}}_{P,s,n}^{\bm{f},\bm{\xi}} \right|\right|_2.\end{align*} Observe that for any $\delta >0$ and any $r\in\mathcal{T}_{n,L}$, we have\begin{align*}    &
\underset{P\in\mathcal{P}_{0,n}^{\ast}}{\sup}\mathbb{P}_P\left(\left|\left|  \sum_{s=r-L_n+1}^{r}\bm{\hat{w}}_{P,s,n}^{\bm{f},\bm{g}}\right|\right|_2 > \delta L_n^{\frac{1}{2}} \tau_n^{7} D_n^{-2} \right)    
\\ &\leq \delta^{-1}L_n^{-\frac{1}{2}} \tau_n^{-7} D_n^{2}\underset{P\in\mathcal{P}_{0,n}^{\ast}}{\sup}\mathbb{E}_P\left(\left|\left|  \sum_{s=r-L_n+1}^{r}\bm{\hat{w}}_{P,s,n}^{\bm{f},\bm{g}}\right|\right|_2 \right)
\\& \leq \delta^{-1}L_n^{\frac{1}{2}} \tau_n^{-7} D_n^{3}\underset{P\in\mathcal{P}_{0,n}^{\ast}}{\sup}\underset{(i,j,a,b)\in\mathcal{D}_n}{\max}\underset{t\in\mathcal{T}_n}{\max} \ \mathbb{E}_P( |\hat{w}_{P,t,n,i,a}^{f}|^2)^{\frac{1}{2}}\mathbb{E}_P(| \hat{w}_{P,t,n,j,b}^{g}|^2)^{\frac{1}{2}}\\&=o(1),\end{align*} using the same arguments as Step 1.1 replacing $T_{n,L}$ with $L_n$, and noting that the fastest $D_n$ can grow is $D_n=O(T_n^{\frac{1}{8}})$. % which corresponds to a lag-window size of $L_n=O(T_n^{\frac{1}{3-\delta'}})$ for any $\delta'>0$. % hence the rate assumption stated in the theorem is sufficient to handle even the "fastest case" 
Next, for any $\delta >0$ and any $r\in\mathcal{T}_{n,L}$, we have
\begin{align*}
 & 
 \underset{P\in\mathcal{P}_{0,n}^{\ast}}{\sup}\mathbb{P}_P\left(\left|\left|  \sum_{s=r-L_n+1}^{r}  \bm{\hat{w}}_{P,s,n}^{\bm{g},\bm{\varepsilon}}\right|\right|_2 > \delta L_n^{\frac{1}{2}}D_n^{-2}\tau_n^{7}\right)
 \\& = 
 \underset{P\in\mathcal{P}_{0,n}^{\ast}}{\sup}\mathbb{P}_P\left(\left|\left|  \sum_{s=r-L_n+1}^{r}  \bm{\hat{w}}_{P,s,n}^{\bm{g},\bm{\varepsilon}}\right|\right|_2^2 > \delta^2 L_n D_n^{-4}\tau_n^{14}\right)
 \\& \leq
\delta^{-2}L_n^{-1} D_n^{4}\tau_n^{-14}  \underset{P\in\mathcal{P}_{0,n}^{\ast}}{\sup}\mathbb{E}_P\left(\left|\left|  \sum_{s=r-L_n+1}^{r}  \bm{\hat{w}}_{P,s,n}^{\bm{g},\bm{\varepsilon}}\right|\right|_2^2 \right)
 \\& \leq
\delta^{-2} D_n^{5}\tau_n^{-14}  \bar{K}^{2}  \underset{P\in\mathcal{P}_{0,n}^{\ast}}{\sup}\underset{t\in \mathcal{T}_{n,L}}{\max}\text{ }\underset{j\in [d_Y]}{\max}\text{ }\underset{b\in B_j}{\max}\text{ }\mathbb{E}_P(|\hat{w}_{P,t,n,j,b}^{g}|^2)
\\& = o(1),
\end{align*} for some $\bar{K}>0$ using the same arguments as Step 1.2 replacing $T_{n,L}$ with $L_n$. The same arguments as Step 1.2 (i.e. exchanging $g,\varepsilon$ with $f,\xi$) can be used to show that $$\left|\left|  \sum_{s=r-L_n+1}^{r} \bm{\hat{w}}_{P,s,n}^{\bm{f},\bm{\xi}} \right|\right|_2=o_{\mathcal{P}}( L_n^{\frac{1}{2}}D_n^{-2}\tau_n^{7}).$$ Hence, for any $r\in\mathcal{T}_{n,L}$ we have $$\left|\left|  \sum_{s=r-L_n+1}^{r} \left(\bm{\hat{R}}_{s,n} - \bm{R}_{P,s,n} \right) \right|\right|_2=o_{\mathcal{P}}( L_n^{\frac{1}{2}}D_n^{-2}\tau_n^{7}).$$ By Lemma~\ref{lma:du_rosenthal}, we have for all $r\in\mathcal{T}_{n,L}$ that
\begin{align*}
& \underset{P\in\mathcal{P}_{0,n}^{\ast}}{\sup}\mathbb{P}_P\left( \left|\left|  \sum_{s=r-L_n+1}^{r} \bm{R}_{P,s,n} \right|\right|_2 > L_n^{\frac{1}{2}}D_n^{\frac{1}{2}}\epsilon \right)
\\&=
\underset{P\in\mathcal{P}_{0,n}^{\ast}}{\sup}\mathbb{P}_P\left( \left|\left|  \sum_{s=r-L_n+1}^{r} \bm{R}_{P,s,n} \right|\right|_2^2 > L_nD_n \epsilon^2 \right)
\\& \leq
L_n^{-1}D_n^{-1} \epsilon^{-2} \underset{P\in\mathcal{P}_{0,n}^{\ast}}{\sup} \mathbb{E}_P\left( \left|\left|   \sum_{s=r-L_n+1}^{r} \bm{R}_{P,s,n} \right|\right|_2^2\right)
\\&\leq 
L_n^{-1}D_n^{-1} \epsilon^{-2} (2  L_n^{\frac{1}{2}} D_n^{\frac{1}{2}} \bar{\Theta}^R K \sum_{h=1}^{\infty} h^{-\bar{\beta}^R})^2
\\&= 
 \epsilon^{-2} (2 \bar{\Theta}^R K C^{\beta})^2,
\end{align*} where $C^{\beta}=\sum_{h=1}^{\infty} h^{-\bar{\beta}^R}<\infty$ since $\bar{\beta}^R>1$ by Assumption 3.5, so that
$$\left|\left|  \sum_{s=r-L_n+1}^{r} \bm{R}_{P,s,n} \right|\right|_2 = O_{\mathcal{P}}(L_n^{\frac{1}{2}}D_n^{\frac{1}{2}}).$$

Therefore, by Markov's inequality, bounded convergence (Lemma~\ref{lma:bounded_convergence}), and noting that the previous statements hold for all times in $\mathcal{T}_{n,L}$, we have $$\underset{k\in\mathcal{T}_{n,L}}{\max}|| \hat{Q}_{k,n}^{\bm{R}}-Q_{P,k,n}^{\bm{R}} ||_{\mathrm{tr}}=O_{\mathcal{P}}(r_{n,2}^{\delta}),$$
where $$r_{n,2}^{\delta}=T_{n,L} \tau_n^7 D_n^{-\frac{3}{2}}+T_{n,L} D_n^{-4}\tau_n^{14}.$$ % original Prob(>) equals Prob(>) with min eps eps, Markov's, E() with min eps as above, then use sum of the rates for the Op() from before since which implies Op() with min eps has same rate since it is greater than min eps, and since the little op() with min eps is obviously bounded by the eps so the E() with min eps is little o(1) by bounded convergence lemma. thus the original Prob(>) is little o(1). the sum of two covariance approx terms are thus O(1) as desired. all of this is used to show that delta hat is big Op() of the sum of the teo rates for the covariance approximation terms. %Explicitly calculating the rate:
%$L_n^{-1}(T_{n,L}-L_n)[(L_n^{\frac{1}{2}}D_n^{-2}\tau_n^{7})(L_n^{\frac{1}{2}}D_n^{\frac{1}{2}})+(L_n^{\frac{1}{2}}D_n^{-2}\tau_n^{7})^2]$
%$=L_n^{-1}(T_{n,L}-L_n)[L_nD_n^{-3/2}\tau_n^{7}+L_n D_n^{-4}\tau_n^{14}]$
%$=L_n^{-1}T_{n,L}[L_nD_n^{-3/2}\tau_n^{7}+L_n D_n^{-4}\tau_n^{14}]$
%$=T_{n,L} D_n^{-3/2}\tau_n^{7}+T_{n,L} D_n^{-4}\tau_n^{14}$
Putting it all together, by the triangle inequality, Markov's inequality, and the covariance approximation results above, we have that  
$$\hat{\delta}_{P,n}=O_{\mathcal{P}}(r_{n,1}^{\delta}+r_{n,2}^{\delta}),$$
where
% from cov approx
\begin{align*} r_{n,1}^{\delta}&= D_n( \bar{\Gamma}_n^R  L_n^{\frac{1}{2}} + T_{n,L}^{\frac{1}{2}} D_n^{\frac{1}{2}} L_n^{\frac{1}{2}} + T_{n,L} L_n^{-1} + T_{n,L} L_n^{2-\bar{\beta}^R}),% from residuals
\\ r_{n,2}^{\delta}&=T_{n,L} \tau_n^7 D_n^{-\frac{3}{2}}+T_{n,L} D_n^{-4}\tau_n^{14}.% matches rate from \textcite{seq_gauss_approx2022} 
\end{align*}

%%% Note: this is just one proof approach. we don't actually require bounded converge, it's just for convenience so we can directly use the expectation inequalities and write the term with sup_P expectation with min eps then use bounded convergence. alternatively, we can just show second term is op() instead of using expectation min eps and bounded convergence, but either way is fine.

% Vinogradov https://en.wikipedia.org/wiki/Big_O_notation#History_(Bachmann%E2%80%93Landau,_Hardy,_and_Vinogradov_notations)
Next, recall the offset condition $$\nu_n \gg \log(T_{n,L})D_n(\tau_n^{-2} (T_{n,L}^{-\frac{1}{2}}D_n^{-\frac{1}{2}}(r_{n,1}^{\delta}+r_{n,2}^{\delta})^{\frac{1}{2}} + T_{n,L}^{-1})).$$ Observe that
\begin{align*}
&T_{n,L}^{-\frac{1}{2}}D_n^{-\frac{1}{2}} (r_{n,1}^{\delta})^{\frac{1}{2}} +T_{n,L}^{-1}
\\& \leq
T_{n,L}^{-\frac{1}{2}}D_n^{-\frac{1}{2}}(D_n^{\frac{1}{2}}((\bar{\Gamma}_n^R)^{\frac{1}{2}}  L_n^{\frac{1}{4}} + T_{n,L}^{\frac{1}{4}} D_n^{\frac{1}{4}} L_n^{\frac{1}{4}} + T_{n,L}^{\frac{1}{2}} L_n^{-\frac{1}{2}} + T_{n,L}^{\frac{1}{2}} L_n^{1-\frac{\bar{\beta}^R}{2}}))+T_{n,L}^{-1}
\\& =
T_{n,L}^{-\frac{1}{2}} (\bar{\Gamma}_n^R)^{\frac{1}{2}}L_n^{\frac{1}{4}} + T_{n,L}^{-\frac{1}{4}} D_n^{\frac{1}{4}} L_n^{\frac{1}{4}} + L_n^{-\frac{1}{2}} + L_n^{1-\frac{\bar{\beta}^R}{2}} +T_{n,L}^{-1}
 \\&= \varphi_{n,1},
\end{align*} which comes from the covariance estimation error. Also, we have 
\begin{align*}
&T_{n,L}^{-\frac{1}{2}}D_n^{-\frac{1}{2}}(r_{n,2}^{\delta})^{\frac{1}{2}}
\\&\leq
T_{n,L}^{-\frac{1}{2}}D_n^{-\frac{1}{2}}(T_{n,L}^{\frac{1}{2}} \tau_n^{\frac{7}{2}} D_n^{-\frac{3}{4}}+T_{n,L}^{\frac{1}{2}} \tau_n^7D_n^{-2} ) 
\\&= \tau_n^{\frac{7}{2}}D_n^{-\frac{5}{4}}+\tau_n^7 D_n^{-\frac{5}{2}}
\\&=
\varphi_{n,2},
\end{align*} which comes from the estimation errors since we use the residual products instead of the error products. The assumption on the offset condition from Section 3.5 implies that
$$\nu_n \gg \log(T_{n,L})D_n\left(\tau_n^{-2} \left( \varphi_{n,1}+\varphi_{n,2}\right)\right),$$ and therefore   
$$\underset{P\in\mathcal{P}_{0,n}^{\ast}}{\sup}\mathbb{P}_P(\hat{\Delta}_{P,n} \geq \frac{1}{16}T_{n,L}\nu_n \tau_n^{2})=o(1).$$ Combining the results from Step 1, Step 2, and Step 3, we have
$$\underset{n \xrightarrow[]{}\infty}{\limsup}\underset{P\in\mathcal{P}_{0,n}^{\ast}}{\sup}\mathbb{P}_P(S_{n,p}(\bm{\hat{R}}_{n}) > \hat{q}_{1-\alpha +\nu_n}+\tau_n)\leq \alpha,$$ which completes the proof. 

% no QED symbol for proofs with their own section
% \hfill $\qedsymbol$

\subsection{Proof of Theorem~\ref{thm:test_SIEVE_DR}}\label{subsection:proof_of_test_thm_SIEVE}

It suffices to establish the following two points. First, that the assumptions of Theorem~\ref{thm:test_SIEVE_DR} imply those of Theorem 3.1. Second, that the sieve time-varying regression estimators, under the setup of Theorem~\ref{thm:test_SIEVE_DR}, satisfy the convergence rate requirements of Theorem 3.1.
% i.e. with the basis functions $\{\phi_{\ell_1}(u)\}$, $\{\varphi_{\ell_2}(z)\}$ chosen to be mapped Legendre polynomials and the numbers of basis functions chosen to satisfy $\tilde{c}_n=O(\log(T_n))$, $\tilde{d}_n=O(\log(T_n))$. 

By using the following notation, we see that Assumption~\ref{asmpt_causal_rep_process_SIEVE} implies Assumption 3.1 and Assumption~\ref{asmpt_causal_rep_errors_SIEVE} implies Assumption 3.4. Note that the causal representations for the observed processes and error processes from Assumptions~\ref{asmpt_causal_rep_process_SIEVE} and~\ref{asmpt_causal_rep_errors_SIEVE} are defined for all rescaled times, and therefore for all $\{t/n\}_{t\in\mathcal{T}_n}\subset [0,1]$ in particular. For a generic high-dimensional locally stationary observed process $W \in \{X,Y,Z\}$ and any time $t$, sample size $n$, dimension $l$, and time-offset $d$, we write
$$G_{t,n}^W(\cdot)=\tilde{G}_n^W(t/n,\cdot),\text{ } G_{t,n,l}^W(\cdot)=\tilde{G}_{n,l}^W(t/n,\cdot),\text{ } G_{t,n,l,d}^W(\cdot)=\tilde{G}_{n,l,d}^W(t/n,\cdot),$$
to respectively denote the causal representations of all dimensions of the process $W$, dimension $l$ of the process $W$, and dimension $l$ of the process $W$ with time-offset $d$. For a generic high-dimensional locally stationary error process $e\in \{\varepsilon,\xi\}$ and any distribution $P$, time $t$, sample size $n$, dimension $l$, and time-offset $d$, we denote
$$G_{P,t,n}^e(\cdot)=\tilde{G}_{P,n}^e(t/n,\cdot),\text{ }G_{P,t,n,l}^e(\cdot)=\tilde{G}_{P,n,l}^e(t/n,\cdot), \text{ } G_{P,t,n,l,d}^e(\cdot)=\tilde{G}_{P,n,l,d}^e(t/n,\cdot),$$ to respectively denote the causal representations of all dimensions of the error process $e$, dimension $l$ of the error process $e$, and dimension $l$ of the error process $e$ with time-offset $d$. The causal representations of the error products are defined similarly. Using this notation, we see that Assumption~\ref{asmpt_funct_dep_SIEVE} implies Assumption 3.5 and Assumption~\ref{asmpt_stoch_lip_SIEVE} implies Assumption 3.6. Specifically, Assumption 3.6 is satisfied with $\bar{\Gamma}_n^R=D_n^{\frac{1}{2}}$ by using linearity of expectation and directly applying the stochastic Lipschitz condition for the product of errors from the discussion below Assumption~\ref{asmpt_stoch_lip_SIEVE} to each term in the sum.

It remains to show that Assumptions 3.2 and 3.3 are implied. To see this, let us consider the following notation. For any distribution $P$, time $t$, sample size $n$, dimensions $i$, $j$, and time-offsets $a$, $b$, we write
\begin{align*}
     f_{P,t,n,i,a}(\cdot)&=f_{P,n,i,a}(t/n,\cdot), \text{ }\hat{f}_{t,n,i,a}(\cdot)=\hat{f}_{t,n,i,a}(t/n,\cdot), 
    \\ g_{P,t,n,j,b}(\cdot)&=g_{P,n,j,b}(t/n,\cdot),\text{ }\hat{g}_{t,n,j,b}(\cdot)=\hat{g}_{t,n,j,b}(t/n,\cdot),
\end{align*} to denote the time-varying regression functions and the corresponding sieve estimators from Section~\ref{subsection:tv_regr_SIEVE} using the notation of Section 2.2. For all times $t\in\mathcal{T}_n$, the algorithms used to construct the sieve estimators from Section~\ref{subsection:tv_regr_SIEVE} for rescaled time $t/n\in [0,1]$ are Borel measurable functions of the datasets $\bm{\mathfrak{D}}_{t,n,i,a}^{\hat{f}}$ and $\bm{\mathfrak{D}}_{t,n,j,b}^{\hat{g}}$. The measurability of the causal mechanisms of the observed processes from Assumption~\ref{asmpt_causal_rep_process_SIEVE} ensures that these sieve estimators have the causal representations $G_{t,n,i,a}^{\mathcal{A}^{\hat{f}}}(\mathcal{H}_{t,a}^{\bm{\mathfrak{D}}^{\hat{f}}})$ and $G_{t,n,j,b}^{\mathcal{A}^{\hat{g}}}(\mathcal{H}_{t,b}^{\bm{\mathfrak{D}}^{\hat{g}}})$ from Assumption 3.2.

Further, note that the sieve estimators are Borel measurable functions from $\mathbb{R}^{\bm{d_Z}}$ to $\mathbb{R}$. The measurability of the causal mechanisms of the covariate processes from Assumption~\ref{asmpt_causal_rep_process_SIEVE} ensures that the sieve estimator's predictions have the causal representations $G_{t,n,i,a}^{\hat{f}}(\mathcal{H}_{t,a}^{\hat{f}})$ and $G_{t,n,j,b}^{\hat{g}}(\mathcal{H}_{t,b}^{\hat{g}})$ from Assumption 3.3. Similarly, note that the Borel measurability of the conditional expectations $f_{P,t,n,i,a}$ and $g_{P,t,n,j,b}$ ensures that the sieve estimator's estimation errors are Borel measurable functions from $\mathbb{R}^{\bm{d_Z}}$ to $\mathbb{R}$. Again, by the measurability of the causal mechanisms of the covariate processes from Assumption~\ref{asmpt_causal_rep_process_SIEVE}, the estimation errors are ensured to have the causal representations $G_{P,t,n,i,a}^{\hat{w}^f}(\mathcal{H}_{t,a}^{\hat{f}})$ and $G_{P,t,n,j,b}^{\hat{w}^g}(\mathcal{H}_{t,b}^{\hat{g}})$ from Assumption 3.3. In view of the additive form of the time-varying regression functions from Assumption~\ref{asmpt_tv_regr_Lq_SIEVE}, Theorem 3.2 in \textcite{zhouzhou_sieve} implies that there exists a $q\geq 2$ (namely, $q=2$) such that the estimation errors satisfy 
\begin{align*}&\underset{n\in\mathbb{N}}{\sup} \ \underset{P\in\mathcal{P}_{0,n}^{\ast}}{\sup} \ \underset{t\in\mathcal{T}_n}{\sup} \ \underset{(i,j,a,b)\in\mathcal{D}_n}{\sup} \mathbb{E}_P(|\hat{w}_{P,t,n,i,a}^{f}|^q)<\infty, \\ & \underset{n\in\mathbb{N}}{\sup} \ \underset{P\in\mathcal{P}_{0,n}^{\ast}}{\sup} \ \underset{t\in\mathcal{T}_n}{\sup} \ \underset{(i,j,a,b)\in\mathcal{D}_n}{\sup} \mathbb{E}_P(|\hat{w}_{P,t,n,j,b}^{g}|^q)<\infty.\end{align*}
Hence, the sieve estimator's predictions and estimation errors meet all the conditions required by Assumption 3.3.
% Note: the each of the Legendre polynomials are continuous functions on a closed interval and hence bounded, which ensures that the predictions are bounded since it is merely the finite sum of all possible products of these Legendre polynomials multiplied by the estimated coefficients multiplied

The distribution-uniform assumptions of Theorem~\ref{thm:test_SIEVE_DR} imply that the distribution-pointwise assumptions of Theorem 3.2 in \textcite{zhouzhou_sieve} hold for each distribution in the collection. Specifically, for each $n\in\mathbb{N}$ and $P\in\mathcal{P}_{0,n}^{\ast}$, Assumption~\ref{asmpt_tv_regr_Lq_SIEVE} implies the additive form of the time-varying regression functions in \textcite{zhouzhou_sieve}, Assumptions~\ref{asmpt_causal_rep_process_SIEVE},~\ref{asmpt_causal_rep_errors_SIEVE},~\ref{asmpt_stoch_lip_SIEVE} imply Assumption 2.1 in \textcite{zhouzhou_sieve}, Assumption~\ref{asmpt_funct_dep_SIEVE} implies Assumption 2.2 in \textcite{zhouzhou_sieve}, Assumption~\ref{asmpt_smooth_regr_fn_SIEVE} implies Assumption 3.1 in \textcite{zhouzhou_sieve}, and Assumption~\ref{asmpt_eigenvalues_integrated_LR_cov_SIEVE} implies Assumption 3.2 in \textcite{zhouzhou_sieve}. Next, we consider the additional regularity condition required by Theorem 3.2 in \textcite{zhouzhou_sieve} involving the rate of decay in temporal dependence and the rate of growth of the largest sup-norm of the basis functions for time.

% additional regularity assumption in Theorem 3.2 in \textcite{zhouzhou_sieve}
Recall the numbers of observations $T_{t,n,i,a}^{\hat{f}}$, $T_{t,n,j,b}^{\hat{g}}$ in the datasets $\bm{\mathfrak{D}}_{t,n,i,a}^{\hat{f}}$, $\bm{\mathfrak{D}}_{t,n,j,b}^{\hat{g}}$ used to construct the sieve estimators $\hat{f}_{t,n,i,a}(t/n,\cdot)$, $\hat{g}_{t,n,j,b}(t/n,\cdot)$ of the time-varying regression functions at rescaled time $t/n\in [0,1]$. Also, recall the numbers of basis functions $\tilde{c}_n$, $\tilde{d}_n$ from Section~\ref{subsection:tv_regr_SIEVE}. As previously noted in Section~\ref{subsection:tv_regr_SIEVE}, we simplified the notation for the numbers of basis functions $\{\phi_{\ell_1}(u)\}$, $\{\varphi_{\ell_2}(z)\}$ for the estimators $\hat{f}_{t,n,i,a,k,c}(t/n,\cdot)$ and $\hat{g}_{t,n,j,b,k,c}(t/n,\cdot)$ of the time-varying partial response functions at rescaled time $t/n\in [0,1]$ from $\tilde{c}_{t,n,i,a,k,c}^{\hat{f}}$, $\tilde{d}_{t,n,i,a,k,c}^{\hat{f}}$ and $\tilde{c}_{t,n,j,b,k,c}^{\hat{g}}$, $\tilde{d}_{t,n,j,b,k,c}^{\hat{g}}$ to $\tilde{c}_n$, $\tilde{d}_n$. We will now require the full notation for the numbers of basis functions.

For the convergence rate guarantees from Theorem 3.2 in \textcite{zhouzhou_sieve} to be applicable in our setting, we must have
\begin{align*}
\tilde{c}_{t,n,i,a,k,c}^{\hat{f}} \tilde{d}_{t,n,i,a,k,c}^{\hat{f}} \left(\frac{1}{\sqrt{T_{t,n,i,a}^{\hat{f}}}}+ \frac{(T_{t,n,i,a}^{\hat{f}})^{\frac{2}{\min(\bar{\beta},\bar{\beta}^{\infty})+1}}}{T_{t,n,i,a}^{\hat{f}}} \right) \underset{\ell_1\in [\tilde{c}_{t,n,i,a,k,c}^{\hat{f}}]}{\sup}\underset{u\in [0,1]}{\sup}|\phi_{\ell_1}(u)|^2 &=o(1),
\\
\tilde{c}_{t,n,j,b,k,c}^{\hat{g}}\tilde{d}_{t,n,j,b,k,c}^{\hat{g}}\left(\frac{1}{\sqrt{T_{t,n,j,b}^{\hat{g}}}}+ \frac{(T_{t,n,j,b}^{\hat{g}})^{\frac{2}{\min(\bar{\beta},\bar{\beta}^{\infty})+1}}}{T_{t,n,j,b}^{\hat{g}}} \right) \underset{\ell_1\in [\tilde{c}_{t,n,j,b,k,c}^{\hat{g}}]}{\sup}\underset{u\in [0,1]}{\sup}|\phi_{\ell_1}(u)|^2 &=o(1),\end{align*}
for each time $t\in\mathcal{T}_n$ and combination of dimensions $i\in [d_X]$, $j\in [d_Y]$, $k\in [d_Z]$ and time-offsets $a\in A_i$, $b\in B_j$, $c\in C_k$. This condition is satisfied for the following reasons. First, we have  $$\underset{\ell_1\in [\tilde{c}_{t,n,i,a,k,c}^{\hat{f}}]}{\sup}\underset{u\in [0,1]}{\sup}|\phi_{\ell_1}(u)|^2 \lesssim (\tilde{c}_{t,n,i,a,k,c}^{\hat{f}})^2,\text{ } \underset{\ell_1\in [\tilde{c}_{t,n,j,b,k,c}^{\hat{g}}]}{\sup}\underset{u\in [0,1]}{\sup}|\phi_{\ell_1}(u)|^2 \lesssim (\tilde{c}_{t,n,j,b,k,c}^{\hat{g}})^2,$$ because the basis functions are chosen to be mapped Legendre polynomials; see Appendix C in \textcite{zhouzhou_sieve} and Section 3 in \textcite{bcck2015}. For more information about sieve estimators and other basis functions, see \parencite{newey1997,huang1998,xchen2007,ding_zhou_2020,pacf_lsts_xiucai_zhou}. Second, because we have chosen the numbers of basis functions to be $O(\log(T_n))$ in the setup of Theorem~\ref{thm:test_SIEVE_DR}. Third, because the constants $\bar{\beta}$, $\bar{\beta}^{\infty}$ from Assumption~\ref{asmpt_funct_dep_SIEVE} are both greater than $2$. Fourth, because every $T_{t,n,i,a}^{\hat{f}}\asymp T_n$ and $T_{t,n,j,b}^{\hat{g}}\asymp T_n$.
% regardless of whether the sieve estimators are fit once based on all the data or sequentially as in Remark~\ref{remark_sequential_sieve_est} (i.e.\ only using data up to a particular rescaled time $u\in [0,1]$). To be clear, this is due to the infill asymptotic framework of locally stationary processes, so that more and more observations related to each local structure of the process are available as $n$ grows, but we only consider the observations before rescaled time $u\in [0,1]$. 

Therefore, the main inequality in the proof of Theorem 3.2 in \textcite{zhouzhou_sieve} holds for each $P\in\mathcal{P}_{0,n}^{\ast}$ and $n\in\mathbb{N}$ because all of the theorem's assumptions are satisfied under the stronger assumptions of Theorem~\ref{thm:test_SIEVE_DR}. Moreover, for each $n\in\mathbb{N}$, the supremum over $P\in\mathcal{P}_{0,n}^{\ast}$ of the final upper bound for the main inequality in the proof of Theorem 3.2 in \textcite{zhouzhou_sieve} is finite under the distribution-uniform assumptions of Theorem~\ref{thm:test_SIEVE_DR}. Thus, by basic properties of the supremum, the main inequality in the proof of Theorem 3.2 in \textcite{zhouzhou_sieve} holds with a supremum over $P\in\mathcal{P}_{0,n}^{\ast}$ for each $n\in\mathbb{N}$. In view of the notational changes described in Remark 3.2 in \textcite{zhouzhou_sieve}, the same steps in the proof of Theorem 3.2 in \textcite{zhouzhou_sieve} imply that this distribution-uniform inequality also holds in the general regression setting with time-offsets. We do not repeat the proof of Theorem 3.2 in \textcite{zhouzhou_sieve} here, as the only changes are in the notation. Putting it all together, the estimation errors of the sieve estimators with the setup of Theorem~\ref{thm:test_SIEVE_DR} satisfy 
\begin{align*} \underset{P\in\mathcal{P}_{0,n}^{\ast}}{\sup}\underset{i\in [d_X], a\in A_i}{\max}\underset{t\in\mathcal{T}_n}{\max}\text{ }  \mathbb{E}_P\left(\left|\hat{w}_{P,t,n,i,a}^{f}\right|^2 \right)^{\frac{1}{2}} &=o(T_n^{-\frac{1}{2+\delta}}\log^3(T_n)),\\ \underset{P\in\mathcal{P}_{0,n}^{\ast}}{\sup}\underset{j\in[d_Y],b\in B_j}{\max}\underset{t\in\mathcal{T}_n}{\max} \text{ }\mathbb{E}_P\left(\left|\hat{w}_{P,t,n,j,b}^{g}\right|^2 \right)^{\frac{1}{2}}&=o(T_n^{-\frac{1}{2+\delta}}\log^3(T_n)),\end{align*}
for any $\delta >0$. Since, in the most favorable setting, the fastest $D_n$ can grow is $D_n=O(T_n^{\frac{1}{8}})$, the convergence rates required by Theorem 3.1 are achieved by the sieve estimators with the setup of Theorem~\ref{thm:test_SIEVE_DR}. This completes the proof.

\section{Distribution-Uniform Theory}\label{appendix:du_theory}

In this section, we state distribution-uniform versions of the results from \textcite{seq_gauss_approx2022}. All of the results in this section can be applied to general triangular array frameworks for high-dimensional nonstationary nonlinear processes, such as locally stationary processes.

\subsection{Prior Work on Distribution-Uniform Inference}\label{subsection:distr_unif_inference_lit_review}

We briefly review prior work on distribution-uniform inference. First, we discuss the conditional independence testing literature. Recently, there has been a great deal of work on distribution-uniform conditional independence testing frameworks due to the hardness result and conditional independence testing framework from \textcite{shah_gcm_2020}. For instance, \textcite{shah_gcm_hilbert} introduced many distribution-uniform convergence results for separable Banach and Hilbert spaces. Recently, \textcite{cli_test} introduced a distribution-uniform \say{conditional local independence} testing framework for the setting in which $n$ realizations of a point process are observed. \textcite{cli_test} also introduce a distribution-uniform version of Rebolledo's martingale central limit theorem \parencite{rebolledo_mart_clt} and extend many distribution-uniform convergence results from \textcite{shah_gcm_hilbert} to metric spaces.

Second, we mention some relevant work from the literature on anytime-valid inference. Recently, \textcite{ian_distr_unif} introduced a distribution-uniform strong (almost-sure) Gaussian approximation for the full sum of iid random variables. The work in \textcite{ian_distr_unif} is motivated by prior work on asymptotic anytime-valid inference from \textcite{ian_asympcs}, in which the authors defined the concept of an \say{asymptotic confidence sequence}. In particular, \textcite{ian_asympcs} introduced asymptotic confidence sequences for iid random variables and a Lindeberg-type asymptotic confidence sequence which can capture time-varying means under martingale dependence.

Third, we discuss other areas in which distribution-uniform inference is studied under different names. There is a vast literature discussing the importance of distribution-uniform inference under the name of \say{honest} or \say{uniform} inference \parencite{Li_Honest,Kasy_Uniformity,tibshirani_rinaldo_tibshirani_wasserman_2018,rinaldo_wasserman_gsell_2019, arun_siva_larry_2023}. Also, there is a plethora of literature on distribution-uniform moment inequality testing \parencite{Imbens_Manski_2004, Romano_Shaikh_2008, andrews_guggenberger_2009,andrews_soares_2010,andrews_barwick_2012,romano_shaikh_wolf_2014}. Most recently, \textcite{unif_str_approx_mixingale2022} developed a distribution-uniform test for general functional inequalities which admits conditional moment inequalities as a special case. In their supplementary appendix, \textcite{unif_str_approx_mixingale2022} introduce a distribution-uniform strong Gaussian approximation for the full sum of a high-dimensional mixingale.

\subsection{Distribution-Uniform Strong Gaussian Approximation}\label{subsection:du_sga}

To begin, let us introduce the setting rigorously. Let $\Omega$ be a sample space, $\mathcal{B}$ the Borel $\sigma$-algebra, and $(\Omega,\mathcal{B})$ a measurable space. Fix $n \in \mathbb{N}$ and define $\mathcal{T}_n=[n]$ throughout this section. Let $(\Omega,\mathcal{B})$ be equipped with a family of probability measures $(\mathbb{P}_P)_{P\in\mathcal{P}_n}$ so that the distribution of the high-dimensional stochastic nonlinear system % $\mathcal{P}_n \subset \mathrm{Prob}[(\mathbb{R}^{d_n})^{\mathcal{T}_n\times\mathbb{Z}}]$ is a subset of the set of Borel probability measures on functions from $\mathcal{T}_n \times \mathbb{Z}$ to $\mathbb{R}^{d_n}$ i.e. each \omega gives you a different realization of the stochastic nonlinear system for all mechanism times and input times
$$(G_{t,n}(\mathcal{H}_s))_{t\in \mathcal{T}_n, s\in \mathbb{Z}},$$ or, in the locally stationary setting % $\mathcal{P}_n \subset \mathrm{Prob}[(\mathbb{R}^{d_n})^{[0,1]\times\mathbb{Z}}]$ is a subset of the set of Borel probability measures on functions from $[0,1] \times \mathbb{Z}$ to $\mathbb{R}^{d_n}$ i.e. each \omega gives you a different realization of the stochastic nonlinear system for all mechanism times and input times
$$(\tilde{G}_{n}(u,\mathcal{H}_s))_{u\in [0,1],s\in\mathbb{Z}},$$
 under $\mathbb{P}_P$ is $P\in\mathcal{P}_n$. Here $\mathcal{H}_t =(\eta_t,\eta_{t-1},\ldots)$, where $(\eta_t)_{t\in\mathbb{Z}}$ is a sequence of iid random vectors with dimension $d^{\eta}=d_n^{\eta}$, and $G_{t,n}:(\mathbb{R}^{d^{\eta}})^{\infty}\xrightarrow[]{}\mathbb{R}^{d_n}$ is a measurable function. For each $t\in \mathcal{T}_n$, $G_{t,n}(\mathcal{H}_s)$ is a well-defined high-dimensional random vector for every $s\in\mathbb{Z}$, and $(G_{t,n}(\mathcal{H}_s))_{s\in\mathbb{Z}}$ is a high-dimensional stationary ergodic process.

For each $n\in\mathbb{N}$, write the $\mathbb{R}^{d_n}$-valued process of interest as $(W_{t,n})_{t\in \mathcal{T}_n}$. We assume that for each $n\in\mathbb{N}$ and $t\in \mathcal{T}_n$, the random vector $W_{t,n}$ has a causal representation; that is, it can be represented as a measurable function of these iid random vectors
$$W_{t,n} = G_{t,n}(\mathcal{H}_t).$$ Similarly, for the causal representations in the locally stationary setting, the measurable function $\tilde{G}_{n}(u,\cdot):(\mathbb{R}^{d^{\eta}})^{\infty}\xrightarrow[]{}\mathbb{R}^{d_n}$ is defined for each rescaled time $u\in [0,1]$, and we assume that  $$W_{t,n} =\tilde{G}_{n}(t/n,\mathcal{H}_t).$$ We can use the results in this section for locally stationary processes by writing $$G_{t,n}(\mathcal{H}_t)=\tilde{G}_{n}(t/n,\mathcal{H}_t).$$

The family of probability measures $(\mathbb{P}_P)_{P\in\mathcal{P}_n}$ is defined with respect to the same measurable space $(\Omega,\mathcal{B})$, but need not have the same dominating measure. Denote a family of probability spaces by $(\Omega,\mathcal{B},\mathbb{P}_P)_{P\in\mathcal{P}_n}$. When we say that the process $(W_{t,n})_{t\in \mathcal{T}_n}$ is defined on the collection of probability spaces $(\Omega,\mathcal{B},\mathbb{P}_P)_{P\in\mathcal{P}_n}$ for some $n\in\mathbb{N}$, we mean that $(W_{t,n})_{t\in \mathcal{T}_n}$ is defined on the probability space $(\Omega, \mathcal{B}, \mathbb{P}_P)$ for each $P\in\mathcal{P}_n$. 
% and a sequence of such families of probability spaces by $((\Omega,\mathcal{B},\mathbb{P}_P)_{P\in\mathcal{P}_n})_{n \in \mathbb{N}}$

Note that the causal representations in this paper use sequences of iid random vectors, whereas the causal representations in \textcite{seq_gauss_approx2022} use sequences of iid $\mathrm{Unif}[0,1]$ random variables. The same arguments used in \textcite{seq_gauss_approx2022} can be applied when using our formulation of the causal representations with iid random vectors. The only reason we write the causal representations in this way is for the sake of clarity. 

In fact, standard results in probability theory imply that the causal representations based on measurable functions of sequences of iid $\mathrm{Unif}[0,1]$ random variables as in \textcite{seq_gauss_approx2022} are already sufficiently general. For example, see \textcite{kallenberg_FOMP_ed3} Lemma 4.21, Lemma 4.22, and the surrounding discussion. More specifically, the causal representations with sequences of iid $\mathrm{Unif}[0,1]$ random variables can express the causal representations with sequences of random vectors by including compositions with additional measurable functions for replicating each of the iid $\mathrm{Unif}[0,1]$ random variables and for doing Rosenblatt transformations as in Section 4 of \textcite{wu_mielniczuk_2010} (i.e.\ procedurally generating the random vector using the corresponding conditional quantile functions).

 Next, we define our measure of temporal dependence for the process. For the following definition, let $(\tilde{\eta}_t)_{t\in\mathbb{Z}}$ be an iid copy of $(\eta_t)_{t\in\mathbb{Z}}$ and denote 
\begin{align*}
\tilde{\mathcal{H}}_{t,j}=(\eta_t,\ldots,\eta_{t-j+1},\tilde{\eta}_{t-j},\eta_{t-j-1},\ldots),\end{align*}
to be $\mathcal{H}_t$ with the $j$-th input in the past $\eta_{t-j}$ replaced with the iid copy $\tilde{\eta}_{t-j}$.

\begin{definition}[Functional dependence measure]\label{funct_dep_measure_du_theory_section}

For $n\in\mathbb{N}$, $P\in\mathcal{P}_n$, $t\in\mathcal{T}_n$, define the functional dependence measure of $W_{t,n}=G_{t,n}(\mathcal{H}_t)$ as $$\theta_{P,t,n}(j,q,r)=(\mathbb{E}_P||G_{t,n}(\mathcal{H}_t)-G_{t,n}(\tilde{\mathcal{H}}_{t,j})||_r^q)^{\frac{1}{q}},$$
with $j\in\mathbb{N}_0$, $q\geq 1$, $r\geq 1$. 
    
\end{definition}

We state the following distribution-uniform assumptions about the temporal dependence and nonstationarity of the process for some collections of distributions $\mathcal{P}_n$ for some $n\in\mathbb{N}$. Note that we write the time of the input sequence as $0$ when it does not matter due to stationarity. 
% assumption is for a fixed n, i.e. not for sequence of collections of distributions as in Sections 3 and 4, since the results below will be applied for a fixed n

\begin{assumption}[Distribution-uniform decay of temporal dependence]\label{DU_G1_dependence_seq2022}
 We assume that there exist $\beta >0$, $q \geq 2$ and a constant $\Theta_n >0$, such that for all times $t\in \mathcal{T}_n$ it holds $$\underset{P\in\mathcal{P}_n}{\sup} \theta_{P,t,n}(j,q,r)\leq \Theta_n \cdot (j\lor 1)^{-\beta},$$
for $j\geq 0$, and that $$\underset{P\in\mathcal{P}_n}{\sup}(\mathbb{E}_P||G_{t,n}(\mathcal{H}_0)||_2^{q})^{1/q} \leq \Theta_n.$$
\end{assumption}

\begin{assumption}[Distribution-uniform total variation condition for nonstationarity]\label{DU_G2_nonstationarity_seq2022}
Recall $\Theta_n$ from Assumption~\ref{DU_G1_dependence_seq2022}. Assume that there exists some $\Gamma_n \geq 1$ such that  $$\underset{P\in\mathcal{P}_n}{\sup}\left(\sum_{t=2}^n (\mathbb{E}_P||G_{t,n}(\mathcal{H}_0)-G_{t-1,n}(\mathcal{H}_0)||_2^2)^{\frac{1}{2}}\right)\leq \Gamma_n \cdot \Theta_n.$$ 

\end{assumption}

Note that the assumptions regarding the temporal dependence and nonstationarity of the process of error products, as stated in Section 3.4, ensure that both Assumptions~\ref{DU_G1_dependence_seq2022} and~\ref{DU_G2_nonstationarity_seq2022} hold for each $n\in\mathbb{N}$. Furthermore, since the assumptions in Section~\ref{subsection:dependence_nonstationarity_SIEVE} are strictly stronger than those in Section 3.4, the results from this section can be applied to the process of error products in both Sections 3 and~\ref{appendix:CI_test_SIEVE}.

Define the two rates
$$
\chi(q,\beta)=
\begin{cases}
    \frac{q-2}{6q-4}, \text{ } &\beta \geq \frac{3}{2}, \\ 
    \frac{(\beta-1)(q-2)}{q(4\beta - 3)-2}, \text{ } &\beta \in (1,\frac{3}{2}),
\end{cases}
$$ and $$
\xi(q,\beta)=
\begin{cases}
    \frac{q-2}{6q-4},\text{ } &\beta \geq 3, \\ 
    \frac{(\beta-2)(q-2)}{(4\beta-6)q-4},\text{ } &\frac{3+\frac{2}{q}}{1+\frac{2}{q}}<\beta<3,\\
    \frac{1}{2}-\frac{1}{\beta}, \text{ } &2<\beta \leq \frac{3+\frac{2}{q}}{1+\frac{2}{q}},
\end{cases} $$ which will appear in the results in this section. In general, the Gaussian approximation allows the dimensions to grow as $d_n=O(n^{\frac{1-\delta}{1+\frac{1}{2\xi(q,\beta)}}})$ for some $\delta >0$. In the limiting case when $\beta \geq 3$ and $q \xrightarrow[]{} \infty$, this corresponds to $d_n=O(n^{\frac{1}{4}-\delta'})$ for some $\delta'>0$.

Let us briefly recall some notation used in the main text. Recall $\bar{\beta}^R>3$, $\bar{q}^R>4$ from Assumption 3.5, as well as the number of times $T_n$ and the number of dimension/time-offset combinations $D_n=|\mathcal{D}_n|$ from Section 2.1. For the dGCM test, we allow $D_n=O(T_n^{r(\bar{q}^R,\bar{\beta}^R)})$ where
\begin{equation}
r(\bar{q}^R,\bar{\beta}^R)=\min\left(\frac{1-\delta}{1+\frac{1}{2\xi(\bar{q}^R,\bar{\beta}^R)}},\frac{1}{8}\right),\label{eqn:rate_dim_growth}\end{equation} for some $\delta >0$.

The limiting factor that leads to this requirement is not from the strong Gaussian approximation, but rather to ensure that the convergence rate requirements can be achieved by the time-varying nonparametric regression estimators. For the same reason, the number of locations $\mathcal{L}_n$ from Section~\ref{subsection:cond_indep_null_hypoth_for_nsts} can grow as $|\mathcal{L}_n|=O(T_n^{r(\bar{q}^R,\bar{\beta}^R)})$ when $D_n=|\mathcal{D}_n|$ is fixed. In general, when both $|\mathcal{D}_n|$ and $|\mathcal{L}_n|$ grow with the sample size, the rate at which each of them grow must be such that $|\mathcal{D}_n||\mathcal{L}_n|=O(T_n^{r(\bar{q}^R,\bar{\beta}^R)})$.

The following result is a distribution-uniform version of the strong Gaussian approximation from Theorem 3.1 in \textcite{seq_gauss_approx2022}. 

\begin{lemma}\label{lma:du_sga} 
For some sample size $n\in\mathbb{N}$ and collection of distributions $\mathcal{P}_n$ for the stochastic nonlinear system $(G_{t,n}(\mathcal{H}_s))_{t\in \mathcal{T}_n,s\in\mathbb{Z}}$, suppose that Assumption~\ref{DU_G1_dependence_seq2022} is satisfied for $\mathcal{P}_n$ with some $q>2$, $\beta >1$ and constant $\Theta_n>0$. Let the $\mathbb{R}^{d_n}$-valued process $(W_{t,n})_{t\in \mathcal{T}_n}$ be defined on the collection of probability spaces $(\Omega,\mathcal{B},\mathbb{P}_P)_{P\in\mathcal{P}_n}$ so that $W_{t,n}=G_{t,n}(\mathcal{H}_t)$ with $\mathbb{E}_P(W_{t,n})=0$ for each time $t\in \mathcal{T}_n$ and distribution $P\in\mathcal{P}_n$. Also, suppose the dimension $d_n < c n$ for some constant $c>0$. Then, on a potentially enriched collection of probability spaces $(\Omega',\mathcal{B}',\mathbb{P}_P')_{P\in\mathcal{P}_n}$, there exist random vectors $(W_{t,n}')_{t\in \mathcal{T}_n}$ with the same distribution as $(W_{t,n})_{t\in \mathcal{T}_n}$ for each $P\in\mathcal{P}_n$, and independent Gaussian random vectors $(V_{t,n}')_{t\in \mathcal{T}_n}$ with $\mathbb{E}_P(V_{t,n}')=0$ for each $t\in \mathcal{T}_n$, $P\in\mathcal{P}_n$, such that $$\underset{P\in\mathcal{P}_n}{\sup}\left(\mathbb{E}_P\text{ }\underset{k \leq n}{\max}\left|\left|\frac{1}{\sqrt{n}}\sum_{t=1}^k (W_{t,n}'-V_{t,n}')\right|\right|_2^2\right)^{\frac{1}{2}} \leq K \Theta_n \sqrt{\log(n)}\left(\frac{d_n}{n}\right)^{\chi(q,\beta)}$$ for some universal constant $K$ depending only on $q$, $c$, and $\beta$.

If $\beta>2$, then the local long-run covariance matrix $\Sigma_{P,t,n}=\sum_{h=-\infty}^{\infty} \mathrm{Cov}_P(G_{t,n}(\mathcal{H}_0),G_{t,n}(\mathcal{H}_h))$ is well-defined for each $t\in \mathcal{T}_n$, $P\in\mathcal{P}_n$ by Lemma~\ref{lma:du_prop54}. If Assumption~\ref{DU_G2_nonstationarity_seq2022} is also satisfied for $\mathcal{P}_n$, then on $(\Omega',\mathcal{B}',\mathbb{P}_P')_{P\in\mathcal{P}_n}$ there exist random vectors $(W_{t,n}')_{t\in \mathcal{T}_n}$ which have the same distribution as $(W_{t,n})_{t\in \mathcal{T}_n}$ for each $P\in\mathcal{P}_n$, and independent Gaussian random vectors $(V_{t,n}^{\ast})_{t\in \mathcal{T}_n}$ where $V_{t,n}^{\ast}\sim\mathcal{N}(0,\Sigma_{P,t,n})$ for each $t\in \mathcal{T}_n$, $P\in\mathcal{P}_n$, such that 
$$\underset{P\in\mathcal{P}_n}{\sup}\left(\mathbb{E}_P\text{ }\underset{k \leq n}{\max}\left|\left|\frac{1}{\sqrt{n}}\sum_{t=1}^k (W_{t,n}'-V_{t,n}^{\ast})\right|\right|_2^2\right)^{\frac{1}{2}} \leq K \Theta_n \Gamma_n^{\frac{1}{2}\frac{\beta-2}{\beta-1}}\sqrt{\log(n)}\left(\frac{d_n}{n}\right)^{\xi(q,\beta)}$$  for some universal constant $K$ depending only on $q$, $c$, and $\beta$.
\end{lemma}
\begin{proof}
Assumptions~\ref{DU_G1_dependence_seq2022} and~\ref{DU_G2_nonstationarity_seq2022} are distribution-uniform versions of conditions (G.1) and (G.2) from \textcite{seq_gauss_approx2022}. Hence, under the assumptions of the Lemma related to Assumptions~\ref{DU_G1_dependence_seq2022} and~\ref{DU_G2_nonstationarity_seq2022}, the distribution-pointwise inequalities from Theorem 3.1 in \textcite{seq_gauss_approx2022} hold for each $P\in\mathcal{P}_n$. Since the suprema over all distributions in the collection $\mathcal{P}_n$ of the upper bounds are finite, the distribution-uniform inequalities from the Lemma hold for $\mathcal{P}_n$ by basic properties of the supremum. \qedhere \end{proof}

Recently, \textcite{wbwu_sga_2024} introduced univariate strong Gaussian approximation results with optimal rates and explicit constructions, building on prior work by \textcite{wbwu_sga_2020}. We emphasize that the distribution-uniform strong Gaussian approximation for high-dimensional nonstationary nonlinear processes from Lemma~\ref{lma:du_sga} does not achieve this optimal rate. However, the convergence rates for the estimation errors of the time-varying regression functions dominate the strong Gaussian approximation rates. Therefore, we do not \say{lose anything} by using Lemma~\ref{lma:du_sga} in our regression-based conditional independence test instead of a distribution-uniform version of the strong Gaussian approximation from \textcite{wbwu_sga_2024}, as our main results would not change in any meaningful way.

The next result is a distribution-uniform version of Theorem 3.2 in \textcite{seq_gauss_approx2022}.

\begin{lemma}\label{lma:du_rosenthal} 
For some sample size $n\in\mathbb{N}$ and collection of distributions $\mathcal{P}_n$ for the stochastic nonlinear system $(G_{t,n}(\mathcal{H}_s))_{t\in \mathcal{T}_n,s\in\mathbb{Z}}$, let the $\mathbb{R}^{d_n}$-valued process $(W_{t,n})_{t\in \mathcal{T}_n}$ be defined on the collection of probability spaces $(\Omega,\mathcal{B},\mathbb{P}_P)_{P\in\mathcal{P}_n}$ so that $W_{t,n}=G_{t,n}(\mathcal{H}_t)$ with $W_{t,n}\in L_q(P)$ and $\theta_{P,t,n}(j,q,r)$ as in Definition~\ref{funct_dep_measure_du_theory_section} for each $P\in\mathcal{P}_n$ and some $2\leq r \leq q < \infty$. There exists a universal constant $K=K(q,r)$ such that
\begin{align*} \underset{P\in\mathcal{P}_n}{\sup}\left(\mathbb{E}_P \text{ }\underset{k\leq n}{\max}\left|\left|\sum_{t=1}^k (W_{t,n}-\mathbb{E}_P(W_{t,n}))\right|\right|_r^q\right)^{\frac{1}{q}} &\leq \underset{P\in\mathcal{P}_n}{\sup} \left( K \text{ }  n^{\frac{1}{2}-\frac{1}{q}} \sum_{j=1}^{\infty} \left(\sum_{t=1}^n \theta_{P,t,n}^q(j,q,r)\right)^{\frac{1}{q}} \right)\\ &\leq \underset{P\in\mathcal{P}_n}{\sup}\left( K \text{ } n^{\frac{1}{2}} \sum_{j=1}^{\infty}  \underset{t\leq n}{\max}\text{ }\theta_{P,t,n}(j,q,r)\right).
\end{align*}

In the special case $r=2$, the inequality may be improved to 
\begin{align*} &\underset{P\in\mathcal{P}_n}{\sup}\left(\mathbb{E}_P \text{ }\underset{k\leq n}{\max}\left|\left|\sum_{t=1}^k (W_{t,n}-\mathbb{E}_P(W_{t,n}))\right|\right|_2^q\right)^{\frac{1}{q}} \\&\leq \underset{P\in\mathcal{P}_n}{\sup}\left( K  \sum_{j=1}^{\infty} (j\land n)^{\frac{1}{2}-\frac{1}{q}} \left(\sum_{t=1}^n \theta_{P,t,n}^q(j,q,2)\right)^{\frac{1}{q}} +  K  \sum_{j=1}^{n} \left(\sum_{t=1}^n\theta_{P,t,n}^2(j,2,2)\right)^{\frac{1}{2}}\right).
\end{align*}
\end{lemma}

\begin{proof}
Under the assumptions of the Lemma, the distribution-pointwise inequalities from Theorem 3.2 in \textcite{seq_gauss_approx2022} hold for each $P\in\mathcal{P}_n$. Since the suprema over all distributions in the collection $\mathcal{P}_n$ of the upper bounds are always finite, the distribution-uniform inequalities from the Lemma hold for $\mathcal{P}_n$ by basic properties of the supremum. \qedhere \end{proof}

\subsection{Distribution-Uniform Feasible Gaussian Approximation}\label{subsection:du_fga}

We introduce distribution-uniform versions of Theorem 4.1 and Proposition 4.2 from \textcite{seq_gauss_approx2022} so that the distribution-uniform strong Gaussian approximation from Section~\ref{subsection:du_sga} can be used for statistical inference. The key is a distribution-uniform cumulative covariance estimator $\hat{Q}_{k,n}$ of the cumulative covariance matrices $\sum_{t=1}^k \Sigma_{P,t,n}$ where $\Sigma_{P,t,n}=\sum_{h=-\infty}^{\infty} \text{Cov}_P(G_{t,n}(\mathcal{H}_0),G_{t,n}(\mathcal{H}_h))$ and $W_{t,n}=G_{t,n}(\mathcal{H}_t)$. We will prove these guarantees for the same estimator from \textcite{seq_gauss_approx2022}, namely $$\hat{Q}_{k,n}=\sum_{r=L_n}^k \frac{1}{L_n}\left(\sum_{s=r-L_n+1}^r W_{s,n}\right)^{\otimes 2}$$ for some window size $L_n \asymp n^{\zeta}$ for some $\zeta\in (0,\frac{1}{2})$.

The next result is a distribution-uniform version of Theorem 4.1 in \textcite{seq_gauss_approx2022}.

\begin{lemma}\label{lma:du_cumul_cov_estimator} 
For some sample size $n\in\mathbb{N}$ and collection of distributions $\mathcal{P}_n$ for the stochastic nonlinear system $(G_{t,n}(\mathcal{H}_s))_{t\in \mathcal{T}_n,s\in\mathbb{Z}}$, let the $\mathbb{R}^{d_n}$-valued mean-zero process $(W_{t,n})_{t\in \mathcal{T}_n}$ be defined on the collection of probability spaces $(\Omega,\mathcal{B},\mathbb{P}_P)_{P\in\mathcal{P}_n}$ so that $W_{t,n}=G_{t,n}(\mathcal{H}_t)$ and Assumptions~\ref{DU_G1_dependence_seq2022} and~\ref{DU_G2_nonstationarity_seq2022} are satisfied for $\mathcal{P}_n$ with $q\geq 4$ and $\beta > 2$. Then $$\underset{P\in\mathcal{P}_n}{\sup}\left( \mathbb{E}_P \text{ }\underset{k=L_n,\ldots,n}{\max} \left|\left| \hat{Q}_{k,n}-\sum_{t=1}^k \Sigma_{P,t,n}\right|\right|_{\mathrm{tr}}\right)\leq K \Theta_n^2 \left(\Gamma_n \sqrt{L_n} + \sqrt{n d_n L_n} + nL_n^{-1} + nL_n^{2-\beta}\right)$$
for some universal constant $K$ depending only on $\beta$ and $q$.
\end{lemma}
\begin{proof}
Assumptions~\ref{DU_G1_dependence_seq2022} and~\ref{DU_G2_nonstationarity_seq2022} are distribution-uniform versions of conditions (G.1) and (G.2) from \textcite{seq_gauss_approx2022}. Hence, under the assumptions of the Lemma related to Assumptions~\ref{DU_G1_dependence_seq2022} and~\ref{DU_G2_nonstationarity_seq2022}, the distribution-pointwise inequalities from Theorem 4.1 in \textcite{seq_gauss_approx2022} hold for each $P\in\mathcal{P}_n$. Since the supremum over all distributions in the collection $\mathcal{P}_n$ of the upper bound is always finite, the distribution-uniform inequality from the Lemma holds for $\mathcal{P}_n$ by basic properties of the supremum. \qedhere \end{proof}

Next is a distribution-uniform version of Proposition 4.2 in \textcite{seq_gauss_approx2022}.

\begin{lemma}\label{lma:du_stat_inf_cov_est} For some sample size $n\in\mathbb{N}$, let $\mathcal{P}_n$ be a collection of distributions for the stochastic nonlinear system $(G_{t,n}(\mathcal{H}_s))_{t\in \mathcal{T}_n,s\in\mathbb{Z}}$. Let $\Sigma_{P,t,n}, \Sigma_{P,t,n}'\in\mathbb{R}^{d_n \times d_n}$ be symmetric, positive definite matrices for each $t\in \mathcal{T}_n$, $P\in\mathcal{P}_n$, and let $(V_{t,n})_{t\in \mathcal{T}_n}$ be independent random vectors defined on the collection of probability spaces $(\Omega,\mathcal{B},\mathbb{P}_P)_{P\in\mathcal{P}_n}$ so that $V_{t,n}\sim\mathcal{N}(0,\Sigma_{P,t,n})$ for each $t\in \mathcal{T}_n$, $P\in\mathcal{P}_n$. Then, on a potentially enriched collection of probability spaces $(\Omega',\mathcal{B}',\mathbb{P}_P')_{P\in\mathcal{P}_n}$, there exist independent random vectors $(V_{t,n}')_{t\in \mathcal{T}_n}$ with $V_{t,n}'\sim\mathcal{N}(0,\Sigma_{P,t,n}')$ for each $t\in \mathcal{T}_n$, $P\in\mathcal{P}_n$ such that 
$$\underset{P\in\mathcal{P}_n}{\sup}\left(\mathbb{E}_P\text{ }\underset{k\leq n}{\max}\left|\left|\sum_{t=1}^k V_{t,n} - \sum_{t=1}^k V_{t,n}' \right|\right|_2^2 \right) \leq \underset{P\in\mathcal{P}_n}{\sup}\left( K \text{ }\log(n) \text{ } [\sqrt{n\delta_{P,n} \rho_{P,n}}+\rho_{P,n}]\right),$$ where
\begin{align*} \delta_{P,n}&=\underset{k\leq n}{\max}\text{ }\left|\left|\sum_{t=1}^k \Sigma_{P,t,n}-\sum_{t=1}^k \Sigma_{P,t,n}'\right|\right|_{\mathrm{tr}},\\
\rho_{P,n}&=\underset{t\leq n}{\max}\text{ }||\Sigma_{P,t,n}||_{\mathrm{tr}}.\end{align*}
\end{lemma}
\begin{proof} 
The distribution-pointwise inequalities from Proposition 4.2 in \textcite{seq_gauss_approx2022} hold for each $P\in\mathcal{P}_n$. Since the supremum over all distributions in the collection $\mathcal{P}_n$ of the upper bound is always finite, the distribution-uniform inequality from the Lemma holds for $\mathcal{P}_n$ by basic properties of the supremum.\qedhere \end{proof}

\subsection{Auxiliary Lemmas}\label{subsection:aux_lemmas}

The following result is a distribution-uniform version of Proposition 5.4 from \textcite{seq_gauss_approx2022}.
\begin{lemma}\label{lma:du_prop54} For some sample size $n\in\mathbb{N}$ and collection of distributions $\mathcal{P}_n$ for the stochastic nonlinear system $(G_{t,n}(\mathcal{H}_s))_{t\in \mathcal{T}_n,s\in\mathbb{Z}}$, let Assumption~\ref{DU_G1_dependence_seq2022} be satisfied for $\mathcal{P}_n$ with some $q\geq 2$, $\beta >0$, and constant $\Theta_n>0$. Denote $$\gamma_{P,t,n,h}=\mathrm{Cov}_P[G_{t,n}(\mathcal{H}_0),G_{t,n}(\mathcal{H}_h)]\in \mathbb{R}^{d_n \times d_n}.$$ Then for all $t\in \mathcal{T}_n$, $h\in\mathbb{Z}$, we have $$\underset{P\in\mathcal{P}_n}{\sup}||\gamma_{P,t,n,h}||_{\mathrm{tr}}\leq \Theta_n^2 \sum_{j=h}^{\infty} j^{-\beta},$$ where $||\cdot||_{\mathrm{tr}}$ denotes the trace norm. Hence, if $\beta>2$, then the long-run covariance matrix $$\gamma_{P,t,n}=\sum_{h=-\infty}^{\infty} \gamma_{P,t,n,h},$$ is well-defined for all $t\in \mathcal{T}_n$, $P\in\mathcal{P}_n$.
\end{lemma}
\begin{proof}
The distribution-pointwise inequality from Proposition 5.4 in \textcite{seq_gauss_approx2022} holds for each $P\in\mathcal{P}_n$. Since the supremum over all distributions in the collection $\mathcal{P}_n$ of the upper bound is always finite, the distribution-uniform inequality from the Lemma holds for $\mathcal{P}_n$ by basic properties of the supremum. \qedhere \end{proof}

The following result is a distribution-uniform version of the Rosenthal inequality from the first part of Theorem 5.6 from \textcite{seq_gauss_approx2022}.

\begin{lemma}\label{lma:du_thm56} For some sample size $n\in\mathbb{N}$ and collection of distributions $\mathcal{P}_n$, let $(M_{t,n})_{t\in \mathcal{T}_n}$ be a $\mathbb{R}^{d_n}$-valued martingale-difference sequence with distribution determined by $P\in\mathcal{P}_n$. For each $2\leq r \leq q < \infty$, there exists a finite factor $C_{q,r}$ such that for any $n,d_n\in\mathbb{N}$, we have 
\begin{align*}
\underset{P\in\mathcal{P}_n}{\sup}\left(\mathbb{E}_P\text{ }\underset{k\leq n}{\max}\left|\left|\sum_{t=1}^k M_{t,n}\right|\right|_r^q\right)^{\frac{1}{q}} &\leq C_{q,r} n^{\frac{1}{2}-\frac{1}{q}}\underset{P\in\mathcal{P}_n}{\sup}\left(\sum_{t=1}^n \mathbb{E}_P ||M_{t,n}||_r^q \right)^{\frac{1}{q}}\\&\leq   C_{q,r} n^{\frac{1}{2}}\underset{P\in\mathcal{P}_n}{\sup}\left(\underset{t\leq n}{\max}\left(\mathbb{E}_P ||M_{t,n}||_r^q \right)^{\frac{1}{q}}\right).
\end{align*}
\end{lemma}
\begin{proof}
The distribution-pointwise inequalities from the first part of Theorem 5.6 in \textcite{seq_gauss_approx2022} hold for each $P\in\mathcal{P}_n$. Since the suprema over all distributions in the collection $\mathcal{P}_n$ of the upper bounds are always finite, the distribution-uniform inequality from the Lemma holds for $\mathcal{P}_n$ by basic properties of the supremum. \qedhere \end{proof}

The following result is similar to the bounded convergence lemma from Lemma 25 in \textcite{shah_gcm_2020}.

\begin{lemma}\label{lma:bounded_convergence} For some sample size $n\in\mathbb{N}$ and collection of distributions $\mathcal{P}_n$, let $X_n$ be a generic real-valued random variable with distribution determined by $P\in\mathcal{P}_n$, where the collection of distributions $\mathcal{P}_n$ can change with $n$. Let $K>0$, and suppose that $|X_n| \leq K$ for all $n\in\mathbb{N}$ and $X_n=o_{\mathcal{P}}(1)$. Then we have $$\underset{P\in\mathcal{P}_n}{\sup}\mathbb{E}_P(|X_n|)=o(1).$$
\end{lemma}
\begin{proof}
For any given $\epsilon>0$, $$|X_n|=|X_n|\mathbbm{1}_{\{|X_n|>\epsilon\}}+|X_n|\mathbbm{1}_{\{|X_n|\leq \epsilon\}}\leq K \mathbbm{1}_{\{|X_n|>\epsilon\}}+\epsilon.$$ By the assumption that $X_n=o_{\mathcal{P}}(1)$, we can find some $N\in \mathbb{N}$ such that $\underset{P\in\mathcal{P}_n}{\sup}\mathbb{P}_P(|X_n|> \epsilon) < \epsilon /K$ for $n \geq N$. Hence, for $n\geq N$ we have $$\underset{P\in\mathcal{P}_n}{\sup}\mathbb{E}_P(|X_n|)\leq K\underset{P\in\mathcal{P}_n}{\sup}\mathbb{P}_P(|X_n|> \epsilon) +\epsilon< 2 \epsilon.$$ Since $\epsilon >0$ was arbitrary, we obtain the desired result. \qedhere \end{proof}

\section{Additional Discussions}\label{appendix:extensions_and_additional_discussion}

 \subsection{Different Null Hypotheses of Conditional Independence}\label{subsection:cond_indep_null_hypoth_for_nsts}

 We present various ways that our test can be used.

 \paragraph*{Univariate Testing.}

Consider forecasting in the univariate setting with $d_X=1$, $d_Y=1$, $\bm{d_Z}\geq1$. 
\begin{example}[Univariate test for time series forecasting]\label{example:univariate_forecasting_ci}
    Suppose we are interested in determining whether our existing forecasting signals $\bm{Z}_{t,n}$ render the current value of a new signal irrelevant for forecasting a target seven time steps ahead. % We have dimensions $i=1$, $j=1$ and time-offsets $a=0$, $b=7$.
    Note that $\bm{Z}_{t,n}$ can consist of current values and lags of each signal. In this case, we would use our univariate test and the null hypothesis 
\begin{equation*} X_{t,n,1,0}\!\perp \!\!\! \perp Y_{t,n,1,7} \mid \bm{Z}_{t,n} \text{ for all } t \in \mathcal{T}_n.\end{equation*}
\end{example} For example, we can individually test for conditional independence between $X_{t,n,i,a}$ and $Y_{t,n,j,b}$ given $\bm{Z}_{t,n}$ for any combination of forecasting horizons $b\in\{7,14,21,28\}$, lags $a\in\{0,-7,-14\}$, targets $j\in\{1,2,3\}$, and signals $i\in\{1,2,3,4,5\}$.

The following null hypothesis of \say{no causal effect at all times} is pertinent to the growing literature on causal inference for time series \parencite{reconstruct_causal_relations_ts_runge,causalinf_temp_runge,ci_ts_runge,causal_assoc_runge,causal_network_runge}, particularly for the setting in which just one realization of a nonstationary process is available. Crucially, several additional causal modeling assumptions would need to be made, e.g., no unobserved confounders. \begin{example}[Univariate test for time series causal inference]\label{example:univariate_causal_ci}    Suppose we are interested in determining whether the current value (time-offset $a=0$) of a continuous treatment (dimension $i=1$) has any causal effect on an outcome of interest (dimension $j=1$) one time step into the future (time-offset $b=1$) after accounting for the confounders $\bm{Z}_{t,n}$. In this case, we would use the univariate version of our test with the null hypothesis \begin{equation*} X_{t,n,1,0}\!\perp \!\!\! \perp Y_{t,n,1,1} \mid \bm{Z}_{t,n} \text{ for all } t \in \mathcal{T}_n.\end{equation*} \end{example}

 \paragraph*{Restricted Alternative Hypothesis.}

Our univariate test is for the null hypothesis \begin{equation} X_{t,n,i,a}\!\perp \!\!\! \perp Y_{t,n,j,b} \mid \bm{Z}_{t,n} \text{ for all } t \in \mathcal{T}_n,\label{eqn:univariate_global_null_hypoth_all_times_nsts}\end{equation}
for a single dimension/time-offset tuple $(i,j,a,b)\in\mathcal{D}_n$. 
The largest possible alternative is the complement of the null, but it may be impossible to detect dependencies that occur momentarily and then disappear.
Thus, we must look for reasonable restrictions; for example, if domain knowledge suggests that we may reasonably restrict the alternative distributions to those for which the conditional dependencies are time-invariant, then we can use the more restricted alternative hypothesis
\begin{equation} X_{t,n,i,a}\not\!\perp\!\!\!\perp Y_{t,n,j,b} \mid \bm{Z}_{t,n} \text{ for all } t \in \mathcal{T}_n.\label{eqn:univariate_altern_hypoth_all_times_nsts}\end{equation} %%% Note: We begin by focusing on time-invariant conditional independence relationships, and we address the time-varying case afterwards. 

 \paragraph*{Gaining Power with Multivariate Testing.}
We can conduct several of these univariate tests and use multiple testing procedures to control the false discovery rate. In high dimensions, we can group together correlated dimensions and neighboring time-offsets (i.e. similar leads and lags) and proceed as in \textcite{Meinshausen_2008} and the literature on hierarchical testing. That is, start by testing for conditional independence at a coarse resolution \begin{equation} X_{t,n,i,a}\!\perp \!\!\! \perp Y_{t,n,j,b} \mid \bm{Z}_{t,n} \text{ for all } t \in \mathcal{T}_n, \text{ for all } (i,j,a,b)\in\mathcal{D}_n,\label{eqn:multivariate_global_null_hypoth_all_times_nsts}\end{equation} using our multivariate test, and continue trying to attribute significance at finer resolutions while using procedures to control the false discovery rate or familywise error rate.
% Note that when using the multivariate test, different alternative hypotheses can be used depending on whether it is reasonable to restrict $\mathcal{P}_n$ to consist of distributions in which the conditional dependencies are dimension-invariant. 

\paragraph*{Gaining Power with Groups of Time Series.} Suppose we have time series data from different countries, companies, or cities, as is common in economics. Write $X_{t,n,i,a}^{\ell}$, $Y_{t,n,j,b}^{\ell}$, $\bm{Z}_{t,n}^{\ell}$ to denote $X_{t,n,i,a}$, $Y_{t,n,j,b}$, $\bm{Z}_{t,n}$ at index $\ell\in\mathcal{L}_n$, where $\mathcal{L}_n$ is an index set (e.g., for different locations). We can gain power by aggregating the time series in $\mathcal{L}_n$ and using our multivariate test with the null hypothesis 
\begin{equation} X_{t,n,i,a}^{\ell}\!\perp \!\!\! \perp Y_{t,n,j,b}^{\ell} \mid \bm{Z}_{t,n}^{\ell} \text{ for all } t \in \mathcal{T}_n,\text{ for all } \ell \in \mathcal{L}_n, \text{ for all }(i,j,a,b)\in\mathcal{D}_n.\label{eqn:multiLocation_univSignal_global_null_hypoth_all_times_nsts}\end{equation} In the best case scenario, the rates at which $|\mathcal{D}_n|$ and $|\mathcal{L}_n|$ grow must be such that $|\mathcal{D}_n||\mathcal{L}_n|=O(T_n^{\frac{1}{8}})$; see Section~\ref{appendix:du_theory} for more details. Note that we suppress the superscript $\ell$ when the number of locations $|\mathcal{L}_n|=1$.

%%% Note: In general, we can use the alternative hypothesis \begin{equation} X_{t,n,i,a}^{\ell}\not\!\perp\!\!\!\perp Y_{t,n,j,b}^{\ell} \mid \bm{Z}_{t,n}^{\ell} \text{ for some } t \in \mathcal{T}_n, \text{ for some } \ell \in \mathcal{L}_n, \text{ for some }(i,j,a,b)\in\mathcal{D}_n,\label{eqn:alt_hypoth_CD} \end{equation} but if we can further restrict the alternative distributions so that the conditional dependencies are time-invariant, location-invariant, and dimension-invariant, then we can use \begin{equation} X_{t,n,i,a}^{\ell}\not\!\perp\!\!\!\perp Y_{t,n,j,b}^{\ell} \mid \bm{Z}_{t,n}^{\ell} \text{ for all } t \in \mathcal{T}_n, \text{ for all } \ell \in \mathcal{L}_n, \text{ for all }(i,j,a,b)\in\mathcal{D}_n, \label{eqn:alt_hypoth_CD_all_times_all_indices} \end{equation} as the alternative hypothesis.  

The following example is about forecasting a group of time series with  $d_X=1$, $d_Y=1$, $\bm{d_Z}\geq1$, and $|\mathcal{L}_n|>1$. In many cases, the multivariate test used in Example~\ref{example:groups_forecasting_ci} will have more power than the univariate test used in Example~\ref{example:univariate_forecasting_ci}. Crucially, the processes at different indices (i.e. $\ell_1,\ell_2 \in \mathcal{L}_n$) can be correlated with one another and have different distributions, which is often the case in economics. 
% Similar hypotheses arise in causal inference with groups of time series (i.e. multivariate analogies of Example~\ref{example:univariate_causal_ci}); see Section 6 of \textcite{cd_groups_runge_2023}. 
 
\begin{example}[Multivariate test for forecasting a group of time series]\label{example:groups_forecasting_ci}
As in Example~\ref{example:univariate_forecasting_ci}, we are interested in forecasting a target seven time steps ahead. We want to determine whether the current value of a new forecasting signal is relevant or not after accounting for the existing forecasting signals. % As before, we have dimensions $i=1$, $j=1$ and time-offsets $a=0$, $b=7$. 
However, now we have access to the same set of forecasting signals and targets for each index $\ell\in\mathcal{L}_n$. In this case, we would use the multivariate version of our test with the null hypothesis \begin{equation*}X_{t,n,1,0}^{\ell}\!\perp \!\!\! \perp Y_{t,n,1,7}^{\ell} \mid \bm{Z}_{t,n}^{\ell} \text{ for all } t \in \mathcal{T}_n, \text{ for all } \ell \in \mathcal{L}_n.\end{equation*}    
\end{example}

In the groups of time series setting with $|\mathcal{L}_n| > 1$, the product of residuals $\hat{R}_{t,n,m}^{\ell}$ at location $\ell \in \mathcal{L}_n$ is formed analogously by only using the time series at location $\ell \in \mathcal{L}_n$. In general, we define $\bm{\hat{R}}_{t,n}=(\hat{R}_{t,n,m}^{\ell})_{m\in\mathcal{D}_n,\ell \in \mathcal{L}_n}$ to be the high-dimensional process containing all the residual products for all dimension/time-offset combinations $\mathcal{D}_n$ from all locations $\mathcal{L}_n$. To simplify the notation, we present our work in the setting with $|\mathcal{L}_n|=1$, and suppress the superscript $\ell$. However, one may extend the methodology and theory to the setting with $|\mathcal{L}_n|>1$ without too much extra effort.

\paragraph*{Time-Varying Structure.} To deal with alternatives having time-varying conditional dependencies, we suggest modeling the conditional dependencies as though they are stable during certain time windows. If the breakpoints separating these time windows are known, then we can simply use our conditional independence test on each of these time windows and use multiple testing procedures to control the false discovery rate. However, this becomes more challenging if the breakpoints are unknown, so we leave this for future work. We discuss how to test for time-varying conditional independence relationships at particular points in time by using the framework of locally stationary processes in Section~\ref{subsection:local_testing}. 
 
\subsection{Parameter Selection for Regression Estimation}\label{subsection:parameter_selection_for_regression}

We introduce a cross-validation approach based on subsampling which can be used for selecting the parameters of \say{global} estimators of time-varying regression functions. Our approach complements the cross-validation procedure suggested in Section 5.1 of \textcite{zhouzhou_sieve}, which is only for parameter selection in the autoregressive forecasting setting. Also, we note that \textcite{dahlahus_cv_2019,dahlahus_adaptation_2023} theoretically investigated cross-validation for locally stationary processes in the context of selecting bandwidths for kernel smoothing estimators (i.e. a \say{local} estimation approach). In contrast, our proposed cross-validation approach is for \say{global} estimators, such as the sieve estimator from Section~\ref{appendix:CI_test_SIEVE}. 
% Note that commonly used approaches for cross-validation in the time series forecasting setting (see e.g., Section 5.10 of \textcite{hyndman_book}) are not applicable in this time-varying regression setting, although we can still use that type of approach in the forecasting setting.
% The approach we present here is designed for the case where the global estimator is fit once on all the data. 
% When using sequential estimation as in Remark~\ref{remark_sequential_sieve_est}, standard approaches for time series cross-validation can be used; see Section 5.10 of \textcite{hyndman_book}.

The main idea of our cross-validation scheme is to create several folds constructed by sampling the original time series at a lower sampling frequency. Specifically, for some buffer $\gamma\in\mathbb{N}_0$ and index $k=1,\ldots,2(\gamma+1)$, the $k$-th fold will consist of the subsampled time series $$\mathcal{T}_{n}^{(k)}=\{\mathbb{T}_n^{-}+k-1+2j(\gamma+1):j=0,1,\ldots,\lfloor\frac{\mathbb{T}_n^{+}-\mathbb{T}_n^{-}-k+1}{2(\gamma+1)}\rfloor\}.$$ For instance, when the buffer $\gamma=0$, we have $\mathcal{T}_{n}^{(1)}=\{\mathbb{T}_n^{-},\mathbb{T}_n^{-}+2,\ldots\}$ and $\mathcal{T}_{n}^{(2)}=\{\mathbb{T}_n^{-}+1,\mathbb{T}_n^{-}+3,\ldots\}$. Similarly, when the buffer $\gamma=1$, we have $\mathcal{T}_{n}^{(1)}=\{\mathbb{T}_n^{-},\mathbb{T}_n^{-}+4,\ldots\}$, $\mathcal{T}_{n}^{(2)}=\{\mathbb{T}_n^{-}+1,\mathbb{T}_n^{-}+5,\ldots\}$, $\mathcal{T}_{n}^{(3)}=\{\mathbb{T}_n^{-}+2,\mathbb{T}_n^{-}+6,\ldots\}$, and $\mathcal{T}_{n}^{(4)}=\{\mathbb{T}_n^{-}+3,\mathbb{T}_n^{-}+7,\ldots\}$. The reason we refer to $\gamma$ as a buffer will be made clear below.

We describe our cross-validation scheme in the context of a basic grid search procedure for pedagogical reasons. For each parameter combination, do the following. For each index $k=1,\ldots,\gamma+1$, use the $k$-th fold $\mathcal{T}_{n}^{(k)}$ to estimate the entire time-varying regression function (i.e. on a suitably fine grid of rescaled times and covariate values) using the \say{global} estimator. Afterward, calculate the residuals based on the observations in the $(k+\gamma+1)$-th fold $\mathcal{T}_{n}^{(k+\gamma+1)}$. By construction, there are $\gamma$ time points between the observations in $\mathcal{T}_{n}^{(k)}$ and $\mathcal{T}_{n}^{(k+\gamma+1)}$. Next, reverse the roles of the folds. That is, for each index $k=1,\ldots,\gamma+1$, estimate the entire time-varying regression function (i.e. on a suitably fine grid) using the $(k+\gamma+1)$-th fold $\mathcal{T}_{n}^{(k+\gamma+1)}$, and then calculate the corresponding residuals based on the observations in the $k$-th fold $\mathcal{T}_{n}^{(k)}$. Finally, for each $k=1,\ldots,2(\gamma+1)$, calculate the mean squared error $\mathrm{MSE}^{(k)}$ based on the residuals in fold $\mathcal{T}_{n}^{(k)}$. Select the parameter combination which yields the lowest average mean squared error $$\overline{\mathrm{MSE}}=\frac{1}{2(\gamma+1)}\sum_{k=1}^{2(\gamma+1)}\mathrm{MSE}^{(k)}.$$

In practice, $\gamma$ should be chosen large enough to account for the temporal dependence, but small enough so that there is enough data to estimate the time-varying regression functions. In our simulations with Sieve-dGCM, we use the buffer $\gamma=1$ and the grid $\{2,4,6,8,10\}\times\{2,4,6,8,10\}$ for the numbers of sieve basis functions for time and for the covariate values. Note that we allow for each regression to have a different number of basis functions. In future work, we will study the statistical properties of this cross-validation procedure as the buffer $\gamma=\gamma_n$ grows with the sample size $n$ using infill asymptotics. For now, this cross-validation approach serves as a practical technique for parameter selection for generic \say{global} estimators of time-varying regression functions, such as the sieve estimator.

\subsection{Parameter Selection for Covariance Estimation}\label{subsection:parameter_selection_for_covariance}

We discuss how to select the lag-window size parameter $L_n$ for the covariance estimator with a version of the minimum volatility method suggested by \textcite{weichi_wu_monotone}. First, select $H\in \mathbb{N}$ candidate lag-window sizes $l_1 < l_2 < \ldots < l_H$. For each index $h=1,\ldots,H$, let $$\bm{\hat{\Sigma}}_{t,n,l_h}=\frac{1}{l_h}\left(\sum_{s=t-l_h+1}^t \bm{\hat{R}}_{s,n}\right)^{\otimes 2}$$ be the lag-window estimate of the local long-run covariance matrix at time $t$ using the candidate lag-window size $l_h \in \mathbb{N}$. Second, calculate the minimum volatility criterion for each $j=1,\ldots,H$, $$\textbf{MV}(j)=\underset{t=\mathbb{T}_n^{-}+l_H,\ldots,\mathbb{T}_n^{+}}{\max} \textbf{se}[(\bm{\hat{\Sigma}}_{t,n,l_h})_{h=1\lor (j-\Delta)}^{H\land (j+\Delta)}],$$
where $\Delta\in \mathbb{N}$ is chosen heuristically to balance robustness and adaptivity, and $$\textbf{se}[(\bm{\hat{\Sigma}}_{t,n,l_h})_{h=h_{1}}^{h_{2}}]=\text{tr}\left[\frac{1}{h_{2}-h_{1}+1}\sum_{h=h_{1}}^{h_{2}}\left(\bm{\hat{\Sigma}}_{t,n,l_h}-\frac{1}{h_{2}-h_{1}+1}\sum_{r=h_{1}}^{h_{2}}\bm{\hat{\Sigma}}_{t,n,l_r}\right)^2\right]^{1/2},$$ with $h_1=1\lor (j-\Delta)$ and $h_2=H\land (j+\Delta)$. Third, select the lag-window size $L_n^{\ast}$ that corresponds to the index $j^{\ast}$ which yields the smallest minimum volatility criterion $$j^{\ast}=\underset{j=1,\ldots,H}{\arg\min}\text{ }\textbf{MV}(j).$$
% The *heuristic* that motivates the minimum volatility is that the eigenvalues of the estimated covariance matrix should be more "stable" around the "optimal" lag-window size parameter

We use the following setup in our simulations. We consider $H=\lfloor n^{1/2}\rfloor$ candidate lag-windows with sizes $l_1=1,l_2=2,\ldots,l_H=\lfloor n^{1/2} \rfloor$. We use $\Delta=12$ so that $25$ consecutive lag-window sizes are typically used in the calculation of the minimum volatility criterion $\textbf{MV}(j)$ for each $j=1,\ldots,H$.

\subsection{No-Free-Lunch in Conditional Independence Testing}\label{subsection:discuss_no_free_lunch}

% Note: conditional independence testing for general discrete-time stochastic processes is also a hard statistical problem \parencite{grangerCausalityExtremes2024}. This is not surprising in view of the original no-free-lunch result from \textcite{shah_gcm_2020}, which already shows the hardness of conditional independence testing in the idealized iid setting. 

\textcite{shah_gcm_2020} introduce a no-free-lunch result which states that if one wants to have a conditional independence test with Type I error control for all absolutely continuous (with respect to the Lebesgue measure) triplets of random vectors $(X,Y,Z)$, then this conditional independence test cannot have power against any alternative hypothesis. To make the conditional independence testing problem feasible, we must consider a smaller subset of the null hypothesis and use domain knowledge to select an appropriate conditional independence test. This hardness result was revisited by \textcite{matey_mmo_ci_test,ilmun_matey_siva_larry_locPermCI_2022}, and was extended to the time series setting by \textcite{grangerCausalityExtremes2024}.

For the GCM test from \textcite{shah_gcm_2020}, the practitioner's domain knowledge is used to select appropriate regression methods for the problem at hand. Crucially, \textcite{shah_gcm_2020} show that the GCM test has asymptotic Type I error control, \textit{uniformly} over a large collection of distributions for which the null hypothesis of conditional independence holds. Since then, numerous tests have been developed which draw inspiration from the original GCM test \parencite{weighted_cov_measure, shah_gcm_hilbert, cli_test,ian_distr_unif,dp_cit_gcm_2024,katsevish_2024_variable_selection,proj_cov}.

To demonstrate the hardness of conditional independence testing in our setting, we consider a variant of a setup from \textcite{shah_gcm_2020}. We work in the setting with $d_X=1$, $d_Y=1$, $d_Z=1$ and no time-offsets, so that $A=\{0\}$, $B=\{0\}$, $C=\{0\}$, and $\mathcal{T}_n=[n]$. We test for the null hypothesis  $$X_{t,n} \perp \!\!\! \perp Y_{t,n} \mid  Z_{t,n} \text{ for all times } t\in \mathcal{T}_n,$$ versus the alternative    
$$X_{t,n}\not\! \perp \!\!\! \perp Y_{t,n} \mid  Z_{t,n} \text{ for all times } t\in \mathcal{T}_n,$$ since we assume we can restrict the collection of alternative distributions to be those for processes with time-invariant conditional dependencies. This setup is challenging because $X$ and $Y$ have the same time-varying regression functions, so similar estimation errors are likely to occur at the same covariate values and rescaled times. When the regression complexity is high relative to the sample size, this induces non-negligible correlation between the residuals even when the error processes are uncorrelated under the null.

Let the covariate process be defined by  $$Z_{t,n}=\theta^Z(t/n)Z_{t-1,n}+\eta_t^Z,$$ where the parameter curve $\theta^Z:[0,1]\xrightarrow[]{} \mathbb{R}$ is defined by $\theta^Z(u)=0.5+0.25\cos(\pi u)$ and the shocks $(\eta_t^Z)_{t\in \mathcal{T}_n}$ are sampled iid from a standard normal distribution. The processes $X$ and $Y$ here are coupled due to $X$ having an additive effect on $Y$ at each time. Let
$$X_{t,n}= f_{K}(Z_{t,n},t/n)+\varepsilon_{t,n}, \text{ } Y_{t,n} = f_{K}(Z_{t,n},t/n)+ \beta X_{t,n}  + \xi_{t,n},$$ with effect size $\beta \in \{0,0.3,0.6,0.9\}$, where the function $f_{K}:\mathbb{R}\times [0,1]\xrightarrow[]{}\mathbb{R}$ is defined by \begin{align*}f_{K}(z,u)&=  (0.4+0.2\sin(2\pi u) )\exp(-z^2)\sin(K z),\end{align*}
with regression complexity parameter $K\in\{1,2,3,4\}$. Let the error processes be given by  \begin{align*}
    \varepsilon_{t,n}&=0.3\varepsilon_{t,n}',\text{ }\varepsilon_{t,n}' = \theta^{e}(t/n)\varepsilon_{t-1,n}' +\eta_t^{\varepsilon}, \\ \xi_{t,n}&=0.3\xi_{t,n}',\text{ }\xi_{t,n}' = \theta^{e}(t/n)  \xi_{t-1,n}'+\eta_t^{\xi},
\end{align*} where the parameter curve $\theta^{e}:[0,1]\xrightarrow[]{}\mathbb{R}$ is given by $\theta^{e}(u)=0.45+0.3\sin(2\pi u)$. The shocks $(\eta_t^{\varepsilon})_{t\in \mathcal{T}_n}$, $(\eta_t^{\xi})_{t\in \mathcal{T}_n}$ are each sampled iid from a standard normal distribution. The null hypothesis is true when the effect size $\beta$ is zero, and the alternative hypothesis is true when it is nonzero.

% For the hypothetical scenario where the time-varying regression functions are estimated perfectly, we use the same tvAR(1) error processes as in the prior setup. 
% Note: For these \say{oracle} simulations, the residual products will be $\sigma^{e}(Z_{t,n},t/n) \varepsilon_{t,n} \sigma^{e}(Z_{t,n},t/n) (\beta \varepsilon_{t,n}  + \xi_{t,n})$. 

\begin{figure}[ht!] %The figure environment for floating and captioning    
\centering % Centers the image
\includegraphics[width=1\textwidth, keepaspectratio]{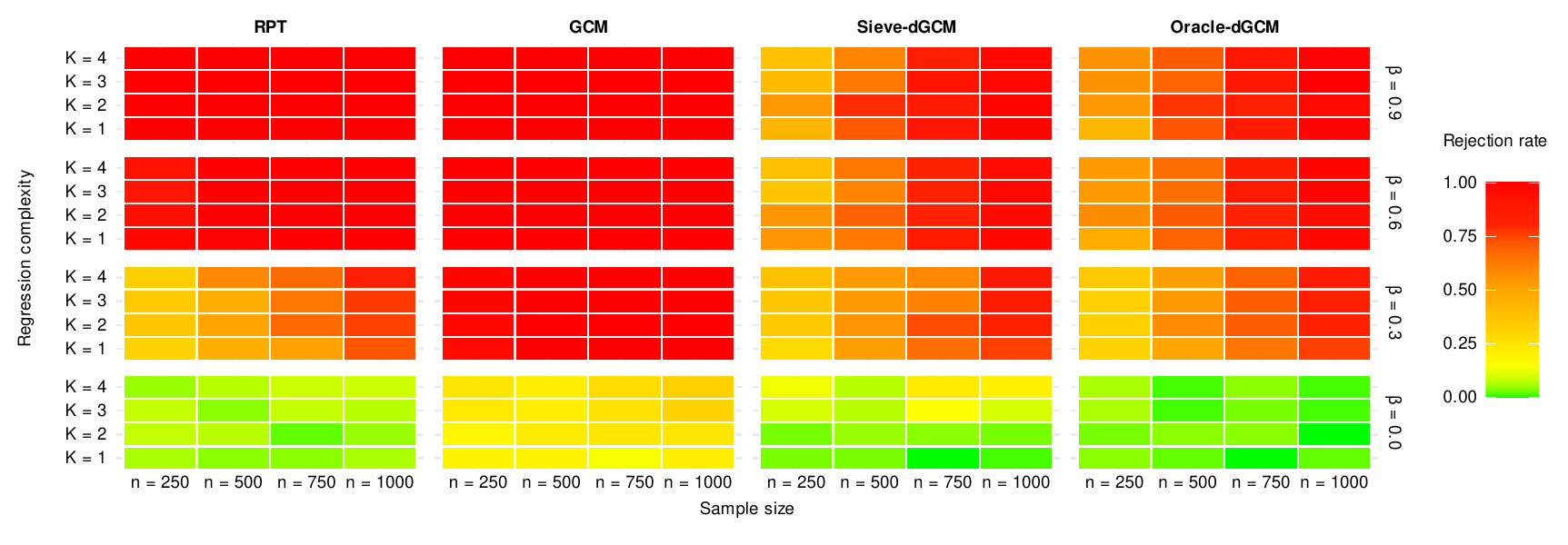}\caption{In this setup with identical time-varying regression functions, the Sieve-dGCM test fails to hold the level when the time-varying regression functions are too complex to reliably estimate at the given sample size. All of the tests can detect conditional dependencies.}
\label{figure:addeffect_heatmap}
\end{figure}

This simulation setup provides an empirical demonstration of the no-free-lunch results from \textcite{shah_gcm_2020,grangerCausalityExtremes2024}. By letting the regression complexity $K_n$ grow with the sample size $n$ at a fairly rapid rate so that $K_{250}=1$, $K_{500}=2$, $K_{750}=3$, $K_{1000}=4$, and so on, we see the Sieve-dGCM test lose control of the Type I error. A similar phenomenon can also be observed with the original GCM test in Section 5 of \textcite{shah_gcm_2020}.

Consider a sequence of null distributions from this simulation setup parametrized by the sequence of regression complexity parameters $K_n=\lfloor  C n^{r} \rfloor$, $n\in\mathbb{N}$, for some $r\geq 1/5$, $C>0$. In this case, the numbers of basis functions for the covariate values must also grow polynomially in $n$, since the $j$-th Legendre polynomial is a degree $j$ polynomial. As a result, the convergence rate of the sieve estimator will be slower than the rate required by Theorem 3.1. See Theorem 3.2 and Appendix C of \textcite{zhouzhou_sieve} for the details about how the convergence rate of the sieve estimator is affected by the growth rate of the numbers of basis functions.

Consequently, the uniform asymptotic Type I error control guarantee we provide for the dGCM test in Theorem 3.1 is not applicable. On the other hand, this guarantee is applicable if we fix $K_n=K$ to some positive constant or have $K_n$ grow slowly in the sample size (e.g., logarithmically). This illustration helps us understand when GCM-type tests can fail to control Type I error, and highlights the transparency of uniform level guarantees.

In light of the no-free-lunch results, we understand that it is impossible to ensure the correct significance level for every null distribution, no matter how large the sample size. That is, there will always be some null distribution under which the Type I error exceeds the significance level. 
% As \textcite{shah_slides_2025} puts it, \say{with great power comes great Type I error.} 

% Note: Although we do not provide a guarantee, the Sieve-dGCM test can still hold the level. For example, in the first simulation setup, the Sieve-dGCM test holds the level for each degree of regression complexity under consideration. However, if the regression complexity continued to increase rapidly relative to the sample size (or if we consider a fixed sample size), then we would expect a loss of Type I error control.
% Note: In Theorem~\ref{thm:test_SIEVE_DR}, we assumed that the numbers of basis functions can be selected so that they grow logarithmically in the sample size.

\subsection{Unconditional Independence Testing}\label{subsection:bonus_indep_test}

The theoretical framework in Section 3 can also be used for unconditional independence testing between multivariate nonstationary nonlinear time series, which corresponds to the special case in which the covariate process is taken to be a constant $Z\equiv K\in\mathbb{R}$. Naturally, this independence test is designed for the practical setting in which only one realization of the process $(X,Y)$ is observed. Recall the variety of null hypotheses of conditional independence from Section~\ref{subsection:cond_indep_null_hypoth_for_nsts}. The independence test is for the unconditional versions of these null hypotheses. Specifically, the univariate test is for the null hypothesis $$X_{t,n,i,a}\!\perp \!\!\! \perp Y_{t,n,j,b} \text{ for all } t\in\mathcal{T}_n,$$ for one dimension/time-offset tuple $(i,j,a,b)\in\mathcal{D}_n$, and the multivariate test is for the null hypothesis
$$X_{t,n,i,a}\!\perp \!\!\! \perp Y_{t,n,j,b} \text{ for all } t\in\mathcal{T}_n,\text{ for all } (i,j,a,b)\in\mathcal{D}_n,$$ or for the \say{groups of time series} setting with multiple locations as in Section~\ref{subsection:cond_indep_null_hypoth_for_nsts}. For each $n\in\mathbb{N}$, let $\mathcal{P}_n$ be a collection of distributions for the processes, and let $\mathcal{P}_{0,n}^{\mathrm{Indep}}$ be a collection of distributions for the processes such that the null hypothesis of independence is true. In contrast, the independence tests for locally stationary time series from \textcite{lsts_indep_testing_beering,loc_stat_indep_test} are for the null hypothesis of independence at a particular rescaled time. Also, \textcite{weichi_wu_network} introduced a framework for inferring correlation curves between piecewise locally stationary time series in the high-dimensional setting.

For a fixed sample size $n\in\mathbb{N}$, distribution $P\in\mathcal{P}_n$, time $t \in \mathcal{T}_n$ and dimension/time-offset tuple $(i,j,a,b)\in \mathcal{D}_n$, we decompose 
\begin{equation}\label{eqn:tv_means_nsts}
X_{t,n,i,a}=\mu^X_{P,t,n,i,a}+\varepsilon_{P,t,n,i,a}, \text{ } Y_{t,n,j,b}=\mu^Y_{P,t,n,j,b}+\xi_{P,t,n,j,b},  \end{equation} where we denote the unconditional means at time $t$ by $$\mu^X_{P,t,n,i,a}=\mathbb{E}_P(X_{t,n,i,a}),\text{ } \mu^Y_{P,t,n,j,b}=\mathbb{E}_P(Y_{t,n,j,b}).$$ The observed processes $X$, $Y$ and error processes $\varepsilon$, $\xi$ can all be nonstationary nonlinear processes as in Section 3 for the general nonstationary time series setting, or as in Section~\ref{appendix:CI_test_SIEVE} for the locally stationary time series setting. For $m=(i,j,a,b)\in \mathcal{D}_n$, denote the products of errors at time $t$ by \begin{equation}\label{prod_err_def_trend} R_{P,t,n,m}=\varepsilon_{P,t,n,i,a}\xi_{P,t,n,j,b}.\end{equation} Denote the high-dimensional $\mathbb{R}^{D_n}$-valued process of all the products of errors by $\bm{R}_{P,t,n}=(R_{P,t,n,m})_{m\in\mathcal{D}_n}.$ Let $\hat{\mu}^X_{t,n,i,a}$, $\hat{\mu}^Y_{t,n,j,b}$ be the estimates of the means $\mu^X_{P,t,n,i,a}$, $\mu^Y_{P,t,n,j,b}$, respectively. Let 
\begin{equation*}  \hat{\varepsilon}_{t,n,i,a}=X_{t,n,i,a}-\hat{\mu}^X_{t,n,i,a},  \text{ }\hat{\xi}_{t,n,j,b}=Y_{t,n,j,b}-\hat{\mu}^Y_{t,n,j,b},\end{equation*} be the residuals, and denote the product of these residuals at time $t$ by \begin{equation} \hat{R}_{t,n,m}=\hat{\varepsilon}_{t,n,i,a}  \hat{\xi}_{t,n,j,b},\label{eqn:prod_residuals_def_indep}\end{equation} for $m=(i,j,a,b)\in \mathcal{D}_n$. Let $\bm{\hat{R}}_{t,n}=(\hat{R}_{t,n,m})_{m\in\mathcal{D}_n}$ be the process containing the residual products for every dimension/time-offset combination in $\mathcal{D}_n$.

Next, we situate the setting for the independence test within the general framework of Section 3. Since we have fixed $\bm{Z}_{t,n}\equiv K\in\mathbb{R}$, we define the functions $f_{P,t,n,i,a}$, $g_{P,t,n,j,b}$ and $\hat{f}_{t,n,i,a}$, $\hat{g}_{t,n,j,b}$ from Assumptions 3.2 and 3.3 so that they map the constant $K\in\mathbb{R}$ to the corresponding unconditional means and estimates. That is, we define $f_{P,t,n,i,a}$, $g_{P,t,n,j,b}:\{K\}\xrightarrow[]{}\mathbb{R}$ so that
$$\mu^X_{P,t,n,i,a}=f_{P,t,n,i,a}(K),\text{ } \mu^Y_{P,t,n,j,b}=g_{P,t,n,j,b}(K),$$ and we define $\hat{f}_{t,n,i,a}$, $\hat{g}_{t,n,j,b}:\{K\}\xrightarrow[]{}\mathbb{R}$ so that $$\hat{\mu}^X_{t,n,i,a}=\hat{f}_{t,n,i,a}(K),\text{ } \hat{\mu}^Y_{t,n,j,b}=\hat{g}_{t,n,j,b}(K).$$  This way, we may write the estimation errors of the time-varying mean estimators as \begin{align}
\hat{w}_{P,t,n,i,a}^{f}&=f_{P,t,n,i,a}(K)-\hat{f}_{t,n,i,a}(K)=\mu^X_{P,t,n,i,a} -\hat{\mu}^X_{t,n,i,a},\label{estim_errs}
\\
\hat{w}_{P,t,n,j,b}^{g}&=g_{P,t,n,j,b}(K)-\hat{g}_{t,n,j,b}(K)=\mu^Y_{P,t,n,j,b}-\hat{\mu}^Y_{t,n,j,b}.\nonumber\end{align}
Also, note that since we have fixed $\bm{Z}_{t,n}\equiv K\in\mathbb{R}$, we have implicitly defined the causal representation of the covariate process from Assumption 3.1 in such a way that it maps to the constant $K\in\mathbb{R}$.
%%% Note: Although this notation is somewhat peculiar, we emphasize that it is unproblematic for our purposes. 
%%% Note: Allows us to avoid making separate assumptions, writing a separate proof, etc for the independence test.

The test is based on the fact that under the null hypothesis of independence, the corresponding combinations of dimensions and time-offsets of $X$, $Y$ will have zero covariance. That is, for every null distribution $P\in\mathcal{P}_{0,n}^{\mathrm{Indep}}$, we have $\mathbb{E}_P(\bm{R}_{P,t,n})=\bm{0}$ for all times $t\in\mathcal{T}_n$. The multivariate independence test is given by Algorithm 1, where we use estimates of the unconditional means at each time rather than estimates of the conditional means at each time. The test statistics $S_{n,p}(\cdot)$, $S_{n}(\cdot)$ and estimators $\bm{\hat{\Sigma}}_{t,n}^{\bm{R}}$, $\hat{\sigma}_{t,n,m}^{R}$ are defined in the same way as Section 2.3, except we use the residuals based on the estimates of the time-varying unconditional means.

The following result establishes the validity of our simulation-assisted testing procedure. Specifically, the result states that the independence test will have uniformly asymptotic Type I error control. We could correctly calibrate our test with the (random) quantile function $\hat{q}$ of $S_{n,p}(\bm{\breve{R}}_{n})$, where $\bm{\breve{R}}_{n}=(\bm{\breve{R}}_{t,n})_{t\in\mathcal{T}_{n,L}}$ and $\bm{\breve{R}}_{t,n} \sim \mathcal{N}(0, \bm{\hat{\Sigma}}_{t,n}^{\bm{R}})$ for all $t\in\mathcal{T}_{n,L}$. However, $\hat{q}$ is not known in practice. Instead, we numerically approximate $\hat{q}$  by conducting many Monte Carlo simulations, and we base our test on $\hat{q}^{\text{boot}}$ from Algorithm 1.

\begin{corol}\label{cor:indep_test_nsts_DR}

Suppose that Assumptions 3.1, 3.2, 3.3, 3.4, 3.5, 3.6 all hold for the sequence of collections of distributions $(\mathcal{P}_{0,n}^{\ast})_{n\in\mathbb{N}}$, where $\mathcal{P}_{0,n}^{\ast}\subset\mathcal{P}_{0,n}^{\mathrm{Indep}}$ for each $n\in\mathbb{N}$, and where we fix the covariate process to be a constant. Further, suppose that 
\begin{align*}
\underset{P\in\mathcal{P}_{0,n}^{\ast}}{\sup}\underset{(i,j,a,b)\in\mathcal{D}_n}{\max}\underset{t\in\mathcal{T}_n}{\max}\text{ }\mathbb{E}_P\left(\left|\hat{w}_{P,t,n,i,a}^{f}\right|^2 \right)^{\frac{1}{2}} \mathbb{E}_P\left(\left|\hat{w}_{P,t,n,j,b}^{g}\right|^2 \right)^{\frac{1}{2}}&= o(T_n^{-\frac{1}{2}} \tau_n^{7} D_n^{-\frac{3}{2}}),\\
 \underset{P\in\mathcal{P}_{0,n}^{\ast}}{\sup}\underset{i\in [d_X], a\in A_i}{\max}\underset{t\in\mathcal{T}_n}{\max}\text{ }  \mathbb{E}_P\left(\left|\hat{w}_{P,t,n,i,a}^{f}\right|^2 \right)^{\frac{1}{2}} &=o(\tau_n^{7} D_n^{-\frac{5}{2}}),\\
 \underset{P\in\mathcal{P}_{0,n}^{\ast}}{\sup}\underset{j\in[d_Y],b\in B_j}{\max}\underset{t\in\mathcal{T}_n}{\max} \text{ }\mathbb{E}_P\left(\left|\hat{w}_{P,t,n,j,b}^{g}\right|^2 \right)^{\frac{1}{2}}&=o( \tau_n^{7} D_n^{-\frac{5}{2}}).
 \end{align*} If the offsets $\tau_n \xrightarrow[]{} 0$ and $\nu_n\xrightarrow[]{} 0$ are chosen such that the offset condition from Section 3.5 holds, then we have  $$\underset{n \xrightarrow[]{}\infty}{\limsup}\text{ }\underset{P\in\mathcal{P}_{0,n}^{\ast}}{\sup}\mathbb{P}_P\left(S_{n,p}(\bm{\hat{R}}_{n})  > \hat{q}_{1-\alpha +\nu_n}+\tau_n\right)\leq \alpha.$$ 
\end{corol}
\begin{proof}    
The proof follows by the same steps as the proof of Theorem 3.1. In this special case with $\bm{Z}_{t,n}\equiv K\in\mathbb{R}$, the $\sigma$-algebra generated by $\bm{Z}_{t,n}$ is the trivial $\sigma$-algebra, so the conditional expectations reduce to the unconditional expectations. Hence, for every null distribution $P\in\mathcal{P}_{0,n}^{\ast}\subset\mathcal{P}_{0,n}^{\mathrm{Indep}}$, the error products have mean zero $\mathbb{E}_P(\bm{R}_{P,t,n})=\bm{0}$ for all times $t\in\mathcal{T}_n$. Therefore, the strong Gaussian approximation from Section~\ref{appendix:du_theory} used in Step 2 in the proof of Theorem 3.1 is applicable. \qedhere \end{proof}

The following result provides a theoretical guarantee for an instantiation of our independence test in which the sieve estimator from Section~\ref{appendix:CI_test_SIEVE} is used to estimate the unconditional means at each time.

\begin{corol}\label{cor:indep_test_SIEVE_DR}

Suppose that Assumptions~\ref{asmpt_causal_rep_process_SIEVE},~\ref{asmpt_tv_regr_Lq_SIEVE},~\ref{asmpt_smooth_regr_fn_SIEVE},~\ref{asmpt_causal_rep_errors_SIEVE},~\ref{asmpt_funct_dep_SIEVE},~\ref{asmpt_stoch_lip_SIEVE} all hold, and that Assumption~\ref{asmpt_eigenvalues_integrated_LR_cov_SIEVE} holds for $\bm{\tilde{\Sigma}}_{P,n,i,a}^{\bm{w}^{\varphi(Z),\varepsilon}}$ and $\bm{\tilde{\Sigma}}_{P,n,j,b}^{\bm{w}^{\varphi(Z),\xi}}$, for the sequence of collections of distributions $(\mathcal{P}_{0,n}^{\ast})_{n\in\mathbb{N}}$, where $\mathcal{P}_{0,n}^{\ast}\subset\mathcal{P}_{0,n}^{\mathrm{Indep}}$ for each $n\in\mathbb{N}$, and where we fix the covariate process to be a constant. Further, suppose that we use the sieve estimator from Section~\ref{subsection:tv_regr_SIEVE} with the basis functions for time $\{\phi_{\ell_1}(u)\}$ chosen to be mapped Legendre polynomials, where the numbers of basis functions are chosen to satisfy $\tilde{c}_n=O(\log(T_n))$. Also, suppose we only use the first Legendre polynomial for the constant covariate value. Then the assumptions of Corollary~\ref{cor:indep_test_nsts_DR} hold for $(\mathcal{P}_{0,n}^{\ast})_{n\in\mathbb{N}}$, and the sieve estimators will achieve the convergence rates required by Corollary~\ref{cor:indep_test_nsts_DR}.
% Note: That is, if the offsets $\tau_n \xrightarrow[]{} 0$ and $\nu_n\xrightarrow[]{} 0$ are chosen such that condition~\eqref{offset_condition_nsts} holds, then we have  $$\underset{n \xrightarrow[]{}\infty}{\mathrm{lim\text{ }sup}}\text{ }\underset{P\in\mathcal{P}_{0,n}^{\ast}}{\sup}\mathbb{P}_P\left(S_{n,p}(\bm{\hat{R}}_{n})  > \hat{q}_{1-\alpha +\nu_n}+\tau_n\right)\leq \alpha.$$
    
\end{corol}
\begin{proof}
The proof follows by the same arguments as the proof of Theorem~\ref{thm:test_SIEVE_DR}. These arguments demonstrate how the general triangular array framework for nonstationary time series from Section 3 nests the locally stationary time series setting from Section~\ref{appendix:CI_test_SIEVE}. In the special case with $\bm{Z}_{t,n}\equiv K\in\mathbb{R}$ and where only the first Legendre polynomial (which maps to $1$) is used for the constant covariate value, the sieve estimator will estimate the time-varying mean function because the $\sigma$-algebra generated by $\bm{Z}_{t,n}$ is the trivial $\sigma$-algebra. \qedhere \end{proof}

Note that the Legendre polynomials can be substituted with trigonometric polynomials, wavelets, or other Jacobi polynomials. For time-varying mean estimators based on kernel smoothing, we refer readers to \textcite{dahlhaus_richter_wu_locally_stationary_general}.

\subsection{Simulations for Unconditional Independence Test}\label{subsection:indep_test_sims}

In this section, we evaluate the independence test from Section~\ref{subsection:bonus_indep_test}, which consists of running Algorithm 1 based on the estimates of the unconditional means at each time from the sieve estimator. We use Legendre polynomials as the basis functions. The numbers of basis functions for time were chosen from the set $\{2,4,6,8,10\}$ using the subsampling cross-validation procedure from Section~\ref{subsection:parameter_selection_for_regression} with the buffer $\gamma = 1$, and the lag-window parameter for covariance estimation was selected via the minimum volatility method from Section~\ref{subsection:parameter_selection_for_covariance}. To approximate the quantile of the test statistic, we used $N^{\mathrm{sim}}=5000$ Monte Carlo simulations.

We compare our independence test (\say{Our Test}) with a robust version of Pearson's correlation test (\say{Correlation Test}) and an independence test of Kolmogorov-Smirnov type (\say{KS Test}) from \textcite{robustest_r_package}. We use the implementations from the \texttt{robusTest} R package \parencite{robustest_r_package}. We also examine how our independence test performs in the hypothetical scenario in which the time-varying means are estimated perfectly. This test (\say{Oracle Test}) consists of running Algorithm 1 using the true time-varying means as the estimates. We first generate $100$ realizations of the processes at sample sizes $n\in\{250,500,750, 1000\}$. Then we calculate the empirical rejection rates for each test using the significance level $\alpha=0.05$. We suppress the dependence on the distribution $P\in\mathcal{P}_n$ to simplify the notation.

We investigate the setting with $d_X=1$, $d_Y=1$, and no time-offsets, so that $A=\{0\}$, $B=\{0\}$, and $\mathcal{T}_n=[n]$. We test for the null hypothesis  $$X_{t,n} \perp \!\!\! \perp Y_{t,n} \text{ for all times } t\in \mathcal{T}_n,$$ versus the alternative hypothesis   
$$X_{t,n}\not\! \perp \!\!\! \perp Y_{t,n} \text{ for all times } t\in \mathcal{T}_n,$$ because we assume that we can restrict the collection of alternative distributions to be those for processes with time-invariant dependencies. We couple the processes $X$ and $Y$ by using correlated shocks for the error processes. Let
\begin{align*}
    X_{t,n}= \mu^X_{\Psi}(t/n)+\varepsilon_{t,n},\text{ }  Y_{t,n} = \mu^Y_{\Psi}(t/n)+ \xi_{t,n},\end{align*} where the time-varying mean functions $\mu^X_{\Psi},\mu^Y_{\Psi}: [0,1]\xrightarrow[]{}\mathbb{R}$ are defined by \begin{align} \mu^X_{\Psi}(u)&= (0.5+0.25\cos(\Psi\pi u) ),\label{eqn:mu_x_trend_corrshock} \\ \mu^Y_{\Psi}(u)&=(0.3+0.15\sin(\Psi\pi u) ) ,\label{eqn:mu_y_trend_corrshock}\end{align} with trend complexity parameter $\Psi \in \{1,2,3,4\}$.

   Define the error processes as \begin{align*}
    \varepsilon_{t,n}&=\sigma^{\varepsilon}(t/n)\varepsilon_{t,n}',\text{ }\varepsilon_{t,n}' = \theta^{\varepsilon}(t/n) \varepsilon_{t-1,n}' +  \eta_{t}^{\varepsilon}, \\ \xi_{t,n}&= \sigma^{\xi}(t/n)\xi_{t,n}',\text{ } \xi_{t,n}' = \theta^{\xi}(t/n)  \xi_{t-1,n}'+  \eta_{t}^{\xi},
\end{align*}
where the parameter curves $\theta^{\varepsilon},\theta^{\xi}:[0,1]\xrightarrow[]{} \mathbb{R}$ are given by $$\theta^{\varepsilon}(u) =0.4 +0.2\sin(\pi u), \text{ } \theta^{\xi}(u) = 0.5+0.25\sin(2\pi u),$$ and where the functions $\sigma^{\varepsilon},\sigma^{\xi}: [0,1]\xrightarrow[]{} \mathbb{R}$ are given by \begin{align*}
\sigma^{\varepsilon}(u)=0.2+(0.5+0.25\sin(2\pi u)),\text{ } \sigma^{\xi}(u)=0.5+(0.4+0.2\cos(2\pi u)).\end{align*} The shocks $(\eta_{t}^{\varepsilon},\eta_{t}^{\xi})_{t\in \mathcal{T}_n}$ are sampled iid from a centered bivariate normal distribution with unit variances and correlation $\rho \in \{0,0.3,0.6,0.9\}$

$$\begin{bmatrix} \eta_{t}^{\varepsilon} \\ \eta_{t}^{\xi} \end{bmatrix} \sim \mathcal{N} \left(  \begin{bmatrix} 0 \\ 0 \end{bmatrix},  \begin{bmatrix} 1 & \rho \\ \rho & 1 \end{bmatrix} \right), \text{ } t\in \mathcal{T}_n.$$ The null hypothesis is true when $\rho=0$, and the alternative hypothesis is true when $\rho\neq0$.

\begin{figure}[ht!]  %The figure environment for floating and captioning    
\centering % Centers the image
\includegraphics[width=1\textwidth, keepaspectratio]{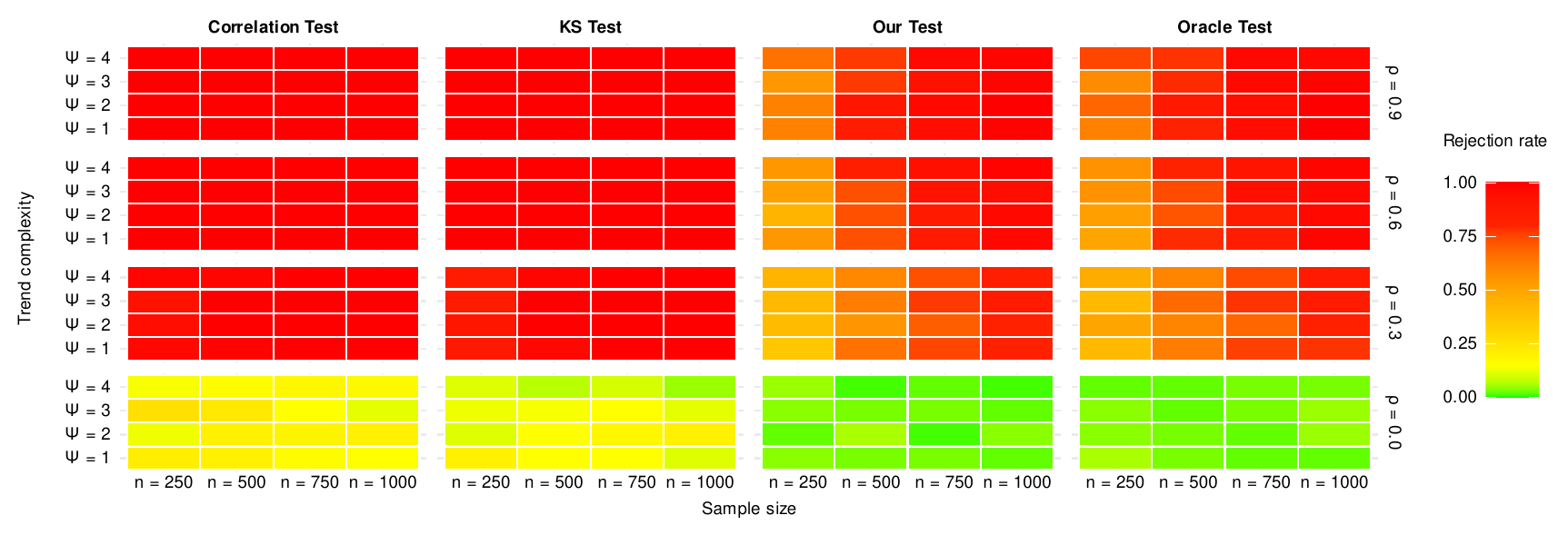}\caption{Our test holds the level even with fairly small sample sizes, and gains power as we increase the correlation and sample size. The other tests fail to hold the level.}
\label{figure:corrshock_indep_test_heatmap}
\end{figure}

\begin{figure}[ht!]  %The figure environment for floating and captioning    
\centering % Centers the image
\includegraphics[width=1\textwidth, keepaspectratio]{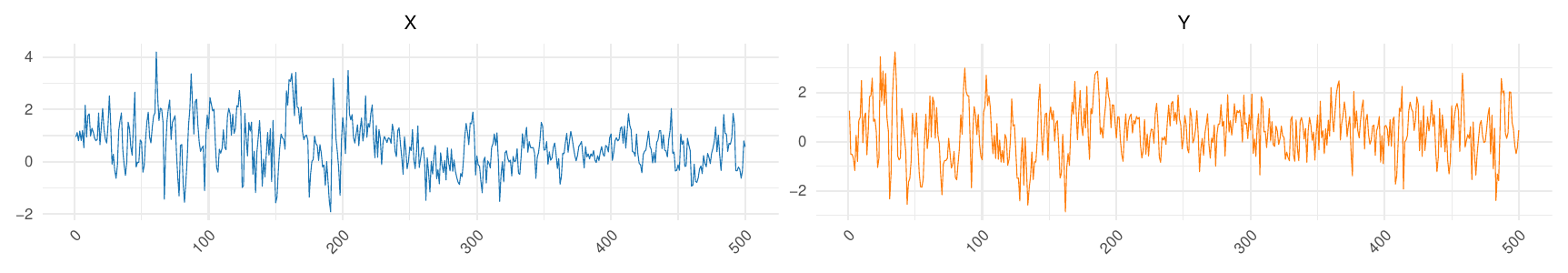}\caption{One realization from the null distribution with $\rho=0$, $\Psi=1$, $n=500$.}
\label{figure:XYtsplot_corrshock_n_500_k_1_corr_0_r_14}
\end{figure}

\begin{figure}[ht!]  %The figure environment for floating and captioning    
\centering % Centers the image
\includegraphics[width=1\textwidth, keepaspectratio]{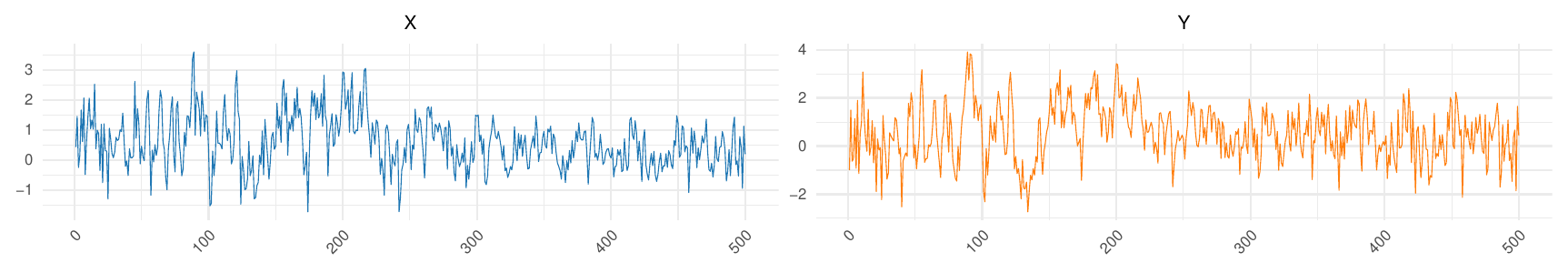}\caption{One realization from the alternative distribution with $\rho=0.9$, $\Psi=1$, $n=500$.}
\label{figure:XYtsplot_corrshock_n_500_k_1_corr_9_r_11}
\end{figure}

\begin{figure}[ht!]  %The figure environment for floating and captioning    
\centering % Centers the image
\includegraphics[width=1\textwidth, keepaspectratio]{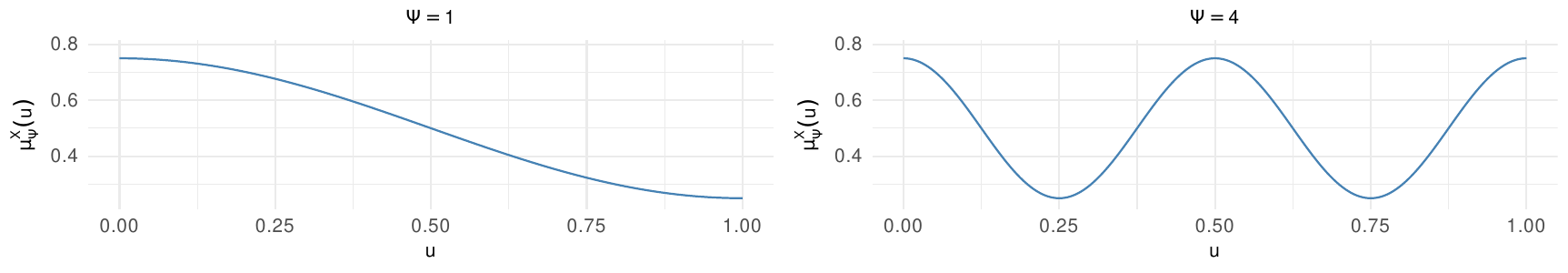}\caption{The time-varying mean function $\mu^X_{\Psi}(u)$ from \eqref{eqn:mu_x_trend_corrshock} at different complexities.}
\label{figure:mu_x_trend_function}
\end{figure}

\begin{figure}[ht!]  %The figure environment for floating and captioning    
\centering % Centers the image
\includegraphics[width=1\textwidth, keepaspectratio]{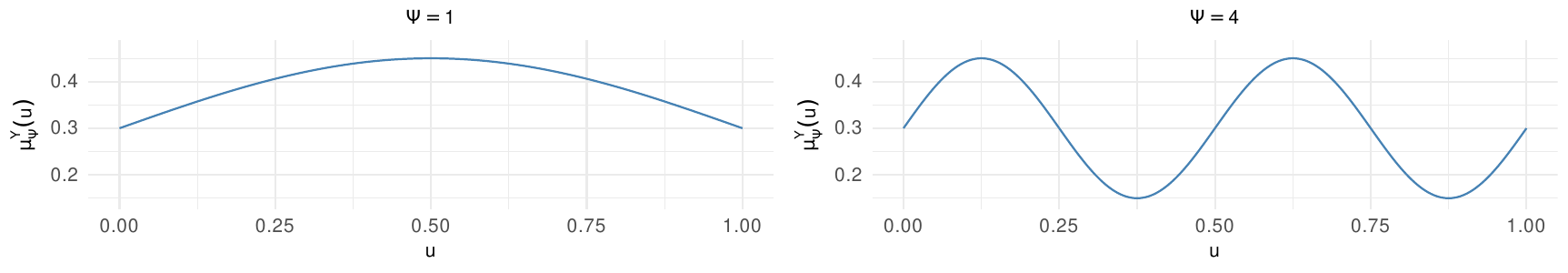}\caption{The time-varying mean function $\mu^Y_{\Psi}(u)$ from \eqref{eqn:mu_y_trend_corrshock} at different complexities.}
\label{figure:mu_y_trend_function}
\end{figure}

\subsection{Connections with Expected Conditional Covariance}\label{subsection:connections_with_ECCov}

To begin, let us briefly summarize the main ideas behind the univariate version of the original generalized covariance measure (GCM) test from \textcite{shah_gcm_2020}. For this paragraph, momentarily redefine $X$, $Y$ to be two random variables and $Z$ to be a random vector. The GCM test is based on the weak conditional independence criterion of \textcite{daudin_ci_1980}, which states that if $X \!\perp \!\!\! \perp Y \mid Z$, then $\mathbb{E}_P[\phi(X,Z)\varphi(Y,Z)]=0$ for all functions $\phi \in L^2_{X,Z}$ and $\varphi \in L^2_{Y,Z}$ such that $\mathbb{E}_P[\phi(X,Z)\mid Z]=0$ and $\mathbb{E}_P[\varphi(Y,Z)\mid Z]=0$.  Thus, under the null hypothesis of conditional independence, the expectation of the products of errors, $\mathbb{E}_P(\varepsilon  \xi)$, from the regressions $X=\phi(Z)+\varepsilon$ and $Y=\varphi(Z)+\xi$, or equivalently the expected conditional covariance, $\mathbb{E}_P[\mathrm{Cov}_P(X,Y|Z)]$, is equal to zero. This can be seen as a generalization of the fact that the partial correlation coefficient % defined as the correlation between the residuals of linear regressions of $X$ on $Z$ and $Y$ on $Z$ 
is equal to zero if and only if $X \!\perp \!\!\! \perp Y \mid Z$ when $(X,Y,Z)$ are jointly Gaussian. The GCM test is based on the normalized sum of the products of residuals from the nonlinear regressions of $X$ on $Z$ and $Y$ on $Z$.

Now, let us translate the weak conditional independence criterion of \textcite{daudin_ci_1980} into our setting. For some $t \in \mathcal{T}_n$ and $m=(i,j,a,b)\in\mathcal{D}_n$, if $X_{t,n,i,a}\!\perp \!\!\! \perp Y_{t,n,j,b} \mid \bm{Z}_{t,n}$, then $$\mathbb{E}_P[\phi(X_{t,n,i,a},\bm{Z}_{t,n}) \varphi(Y_{t,n,j,b},\bm{Z}_{t,n})]=0,$$ for all functions $\phi \in L^2_{X_{t,n,i,a},\bm{Z}_{t,n}}$ and $\varphi\in L^2_{Y_{t,n,j,b},\bm{Z}_{t,n}}$ such that $\mathbb{E}_P[\phi(X_{t,n,i,a},\bm{Z}_{t,n})\mid \bm{Z}_{t,n}]=0$ and $\mathbb{E}_P[\varphi(Y_{t,n,j,b},\bm{Z}_{t,n})\mid \bm{Z}_{t,n}]=0$. Hence, the expected conditional covariance $$\rho_{P,t,n,m}=\mathbb{E}_P[\mathrm{Cov}_P(X_{t,n,i,a},Y_{t,n,j,b}|\bm{Z}_{t,n})],$$ is always equal to zero under the null. Equivalently, the mean of the products of the errors, $\mathbb{E}_{P}(R_{P,t,n,m})$, will always be zero under the null. This can be seen as a generalization of the partial correlation coefficient being equal to zero under conditional independence in the linear-Gaussian time series context; see \textcite{rao_basu}.

Crucially, the expected conditional covariances $\rho_{P,t,n,m}$ can be zero at all times, even under alternatives in which the corresponding conditional dependencies always hold. Consequently, we can only hope to have power against alternatives in which the time-varying expected conditional covariances $\rho_{P,t,n,m}$ are non-zero for at least some times. Hence, our statistic is designed to detect non-zero covariances between the errors $\varepsilon_{P,t,n,i,a}$ and $\xi_{P,t,n,j,b}$.

\subsection{Full Sum Test Statistics}\label{subsection:alternative_test_statistics}

Consider the test statistic $$S_{n,p}^{\star}(\bm{\hat{R}}_{n})=\left|\left|\frac{1}{\sqrt{T_{n,L}}}\sum_{t\in\mathcal{T}_{n,L}}\bm{\hat{R}}_{t,n}\right|\right|_p,$$ based on the $\ell_p$-norm ($p\geq 2$) of the full sum of residual products. For example, we can use the test statistics $$ S_{n,\infty}^{\star}(\bm{\hat{R}}_{n})=  \left|\left|\frac{1}{\sqrt{T_{n,L}}} \sum_{t\in\mathcal{T}_{n,L}} \bm{\hat{R}}_{t,n} \right|\right|_{\infty}, \text{ }S_{n,2}^{\star}(\bm{\hat{R}}_{n})=\left|\left|\frac{1}{\sqrt{T_{n,L}}} \sum_{t\in\mathcal{T}_{n,L}} \bm{\hat{R}}_{t,n} \right|\right|_2.$$ Crucially, the full sum test statistic $S_{n,p}^{\star}(\bm{\hat{R}}_n)$ will not have power against alternatives in which the time-averages of the time-varying expected conditional covariances are close to zero (e.g., positive during the first half of times, and negative during the second half). On the other hand, the maximum partial sum test statistic $S_{n,p}(\bm{\hat{R}}_n)$ from Section 2.3 does have power against these alternatives. If the time-varying expected conditional covariances are suspected to consistently maintain the same sign (whether positive or negative), then users might be able to gain some power by using $S_{n,p}^{\star}(\bm{\hat{R}}_n)$, although we emphasize that $S_{n,p}(\bm{\hat{R}}_n)$ will also have power against these alternatives. However, in settings where we have little prior knowledge about the time-varying expected conditional covariances between the nonstationary processes under alternatives, then the maximum partial sum test statistic $S_{n,p}(\bm{\hat{R}}_n)$ should be used because it has power against a wider range of alternatives. For similar reasons, we recommend using $S_{n,p}(\bm{\hat{R}}_n)$ when conducting automated multiple conditional independence testing (e.g., for screening out irrelevant time series in a large database of possible forecasting signals).

It is perhaps most intuitive to frame the problem in the following way. Consider the time-varying partially linear model
$$\mathbb{E}_P(Y_{t,n,j,b}|X_{t,n,i,a},\bm{Z}_{t,n})=\beta_{P,t,n,m} X_{t,n,i,a}+h_{P,t,n,j,b}(\bm{Z}_{t,n}),$$
for some function $h_{P,t,n,j,b}(\cdot)$. When the time-varying conditional expectation $\mathbb{E}_P(Y_{t,n,j,b}|X_{t,n,i,a},\bm{Z}_{t,n})$ is assumed to have this time-varying partially linear form, the time-varying coefficient $\beta_{P,t,n,m}$ is equal to the expected conditional covariance of $X_{t,n,i,a}$ and $Y_{t,n,j,b}$ given $\bm{Z}_{t,n}$ divided by the expected conditional variance of $X_{t,n,i,a}$ given $\bm{Z}_{t,n}$; see \textcite{SemiparametricMinimaxRates,demystify_eccov_22} for more discussion. If domain knowledge suggests that the time-varying coefficients $(\beta_{P,t,n,m})_{m=(i,j,a,b)\in\mathcal{D}_n}$ consistently maintain the same sign over time $t\in\mathcal{T}_n$, then the full sum test statistic $S_{n,p}^{\star}(\bm{\hat{R}}_n)$ can be used to gain some power. Otherwise, if we cannot make this assumption, then use the maximum partial sum test statistic $S_{n,p}(\bm{\hat{R}}_n)$ because it has power against a broader range of alternatives.

\subsection{Simplifications Under Stationarity}\label{subsection:special_cases_stationary}

Throughout this paper, we have completely avoided the assumption of stationarity. However, it is worth explaining how things would simplify if we are willing to assume that the processes are stationary. Overall, the takeaway is that the original GCM test from \textcite{shah_gcm_2020} would require minimal modifications.

To begin, suppose we have $n\in\mathbb{N}$ observations of a stationary mixing time series, so that the regression functions are time-invariant. Further, suppose that the errors are iid. The statistical guarantees of many machine learning algorithms and statistical models, such as support vector machines \parencite{Steinwart_SVM_2009,Steinwart_2009,steinwart_2014}, random forests \parencite{rf_dep_proc,davis_rf_2020}, lasso \parencite{tewari_lasso_2020}, and high-dimensional vector autoregressive models \parencite{rate_optimal_var_tsay_2023}, have been studied in the context of stationary mixing time series with iid errors. Over the last decade, the literature on statistical learning theory for time series has been able to move beyond the restrictive assumptions of stationarity and mixing \parencite{Yu,tewari_lasso_2020,Karandikar2002,Alquier2014} (or asymptotic stationarity \parencite{Agarwal}) by describing nonstationarity in terms of discrepancy measures \parencite{mohri_learn_thry_ts_2014,mohri_learn_thry_ts_2015,mohri2017,Hanneke2019,mohri_discrep_learn_thry_ts_2020}. This literature has recently considered new notions of learnability for general non-iid stochastic processes \parencite{Dawid_Tewari_learnable_stoch_proc} and conditions under which learning from general non-iid stochastic processes is possible \parencite{Hanneke2021}. 

\textcite{wbwu_timevar_npreg} considered the setting in which the regression functions are time-varying and the errors are iid. Since the errors are iid, a multiplier bootstrap testing procedure can be justified by the Gaussian approximation from \textcite{chernozhukov_maxima_sums_2013} which was used by \textcite{shah_gcm_2020}. Hence, the resulting test would be very similar to the original GCM test for the iid setting from \textcite{shah_gcm_2020}. The main difference is that there can be time-delayed conditional dependencies in the time series setting.

Suppose that the observed processes are temporally dependent (e.g., some form of mixing) and stationary so that the regression functions are time-invariant as before, but the errors are also temporally dependent. The guarantees of the lasso and vector autoregressive models are fairly well-studied in this setting. \textcite{basu_2015} investigated high-dimensional vector autoregressive models with serially correlated errors. \textcite{gupta_2012,Xie_2018} studied the lasso with errors satisfying various weak dependence conditions. \textcite{Peng_2023,Xie_2017} studied the lasso with $\phi$-mixing and $\beta$-mixing errors, respectively. \textcite{wbwu_lasso_corrError} studied the lasso in the setting with temporally dependent errors by using the functional dependence measure of \textcite{wu_funct_dep_meas}.

In the serially correlated error setting, the key difference with the GCM test from \textcite{shah_gcm_2020} is that one must use a suitable Gaussian approximation result to justify a multiplier bootstrap-type testing procedure. See \textcite{xiaohui_clt_dep_2021} for a comprehensive overview of Gaussian approximations for dependent data. \textcite{chernozhukov_2019} investigated a block multiplier bootstrap under a $\beta$-mixing assumption, and \textcite{zhang_cheng_2014} explored a wild multiplier bootstrap under the functional dependence measure of \textcite{wu_funct_dep_meas}. Also, \textcite{zhang_wu_2017} discuss estimators for the long-run covariance matrix so that their Gaussian approximation for high-dimensional time series can be applied in practice. See \textcite{wbwu_hac_estimator_consistency,asymp_thry_stat_proc} for more discussion about long-run covariance matrix estimation for stationary time series.

One could also have time-invariant regression functions with errors that are nonstationary and temporally dependent. For instance, \textcite{xia_2024} studies the lasso with locally stationary errors. However, the statistical guarantees of other machine learning algorithms and statistical models have not been studied in this setting. If the process of error products is mean-nonstationary (i.e. time-varying expected conditional covariance) under alternatives, then the same test statistics from Section 2.3 can be used. Otherwise, if domain knowledge suggests that the time-varying expected conditional covariances usually maintain the same sign, then the test statistics from Section~\ref{subsection:alternative_test_statistics} can be used.

To recap, we considered how the assumption of stationarity would vastly simplify the problem. We find that the original GCM test for the iid setting from \textcite{shah_gcm_2020} can be adapted to the stationary time series setting by making the previously mentioned changes. In contrast, we consider the much more complicated setting in which the observed processes can be nonstationary and temporally dependent, the regression functions can vary over time, and the error processes can be nonstationary and temporally dependent. We emphasize that our dGCM test can be used with stationary processes and iid sequences, which are special cases of the general framework from Section 3.

\subsection{Piecewise Locally Stationary Processes}\label{subsection:piecewise_locally_stationary}

We briefly describe how to extend the Sieve-dGCM test from Section~\ref{appendix:CI_test_SIEVE} from locally stationary processes \parencite{dahlhaus1997,zhou_wu_2009_quantile_loc_stat,dahlhaus_locally_stationary,dahlhaus_richter_wu_locally_stationary_general} to a more general class of nonstationary processes known as piecewise locally stationary (PLS) processes introduced in \textcite{zhou_zhou_2013}. Specifically, the class of PLS processes generalizes the stochastic Lipschitz nonstationarity condition from Assumption~\ref{asmpt_stoch_lip_SIEVE} by allowing for finitely many breakpoints \parencite{zhou_zhou_2013,multiscale_jump_nsts,change_point_corr_nsts}. We emphasize that PLS processes are included in the even more general class of nonstationary processes from Section 3 with the total variation-type nonstationarity condition from Assumption 3.6.

The main idea is to identify the breakpoints, fit a separate sieve model on each locally stationary segment, and run Algorithm 1 on all the residuals. If the breakpoints are known exactly, then the same arguments can be used to show that the sieve time-varying regression estimators achieve the required convergence rates (i.e. within each locally stationary segment). If the breakpoints must be identified, then our arguments must be extended to account for this. As far as we know, \textcite{multiscale_jump_nsts} is the most relevant work on identifying breakpoints for PLS processes. We leave the full details for future work.

\subsection{Cyclostationary Processes}\label{subsection:cyclostationary}

The general triangular array framework from Section 3 also allows for nonstationary processes that exhibit some form of repetition over time, such as periodic stationary processes or cyclostationary processes \parencite{bennett_1958,parzen_1979,bloomfield_1994,gardner_1994,cyclo_half_century_2006,napolitano_2016}, which are not necessarily locally stationary. The theoretical justification of the dGCM test from Theorem 3.1 requires that we improve our estimates of the time-varying regression functions as $n$ grows. See Remark 2.1 in \textcite{mult_str_wu_2022} and the preceding discussion about time-varying regression with periodic stationary or cyclostationary processes. Also, see Section 2.5.1 of \textcite{wbwu_sga_2024} for a discussion of how strong Gaussian approximations for nonstationary nonlinear processes with causal representations as in Sections 3.1, 3.3, 3.4 can be used with cyclostationary processes. Ideally, we would like to be able to handle even more complex forms of nonstationarity than cyclostationarity. See \textcite{cyclo_half_century_2006,generalized_cyclo_2016} for generalizations of this concept.

\subsection{Additional Tests for Locally Stationary Processes}\label{subsection:local_testing}

We discuss three conditional independence tests for locally stationary processes that we did not pursue in this paper. Crucially, the test statistics below require local long-run covariance estimation. Most local long-run covariance estimators use kernel smoothing, and therefore require selecting bandwidths. Unfortunately, test statistics that use kernel smoothing can be very sensitive to the choice of the bandwidths, which can be hard to select in practice. Inspired by the success of bandwidth-free approaches in other areas of time series analysis \parencite{lobato_2001,shao2010_self_normalized_ts,bfi_shao_2013,shao2015_self_normalized_ts_review,xiaofeng_sample_splitting_bfi}, we designed the dGCM test so that it does not require local long-run covariance estimation and therefore avoids kernel smoothing. 

Recall the notation for the locally stationary setting from Section~\ref{appendix:CI_test_SIEVE}. To begin, let us translate the \say{weak} conditional independence criterion of \textcite{daudin_ci_1980} into the locally stationary setting as follows. For some $n\in\mathbb{N}$, $u\in [0,1]$, $(i,j,a,b)\in\mathcal{D}_n$, $t\in\mathcal{T}_n$, if $$\tilde{X}_{\lfloor un\rfloor,n,i,a}(u)\!\perp \!\!\! \perp \tilde{Y}_{\lfloor un\rfloor,n,j,b}(u) \mid \bm{\tilde{Z}}_{\lfloor un\rfloor,n}(u),$$ then $$\mathbb{E}_P[\phi(\tilde{X}_{\lfloor un\rfloor,n,i,a}(u),\bm{\tilde{Z}}_{\lfloor un\rfloor,n}(u)) \varphi(\tilde{Y}_{\lfloor un\rfloor,n,j,b}(u),\bm{\tilde{Z}}_{\lfloor un\rfloor,n}(u))]=0,$$ for all functions $$\phi \in L^2_{\tilde{X}_{\lfloor un\rfloor,n,i,a}(u),\bm{\tilde{Z}}_{\lfloor un\rfloor,n}(u)},\text{ } \varphi\in L^2_{\tilde{Y}_{\lfloor un\rfloor,n,j,b}(u),\bm{\tilde{Z}}_{\lfloor un\rfloor,n}(u)},$$ such that \begin{align*}
\mathbb{E}_P[\phi(\tilde{X}_{\lfloor un\rfloor,n,i,a}(u),\bm{\tilde{Z}}_{\lfloor un\rfloor,n}(u))\mid \bm{\tilde{Z}}_{\lfloor un\rfloor,n}(u)]&=0,\\ \mathbb{E}_P[\varphi(\tilde{Y}_{\lfloor un\rfloor,n,j,b}(u),\bm{\tilde{Z}}_{\lfloor un\rfloor,n}(u))\mid \bm{\tilde{Z}}_{\lfloor un\rfloor,n}(u)]&=0.    
\end{align*} Hence, the corresponding local expected conditional covariance $$\rho_{P,n,m}(u)=\mathbb{E}_P[\mathrm{Cov}_P(\tilde{X}_{\lfloor un\rfloor,n,i,a}(u),\tilde{Y}_{\lfloor un\rfloor,n,j,b}(u)|\bm{\tilde{Z}}_{\lfloor un\rfloor,n}(u))],$$ is equal to zero for $m=(i,j,a,b)\in \mathcal{D}_n$.

% Note that the global null hypothesis~\eqref{eqn:global_null_hypoth_all_times_SIEVE} implies the global null hypothesis from Section 2.2. We will reject~\eqref{eqn:global_null_hypoth_all_times_SIEVE} if and only if we reject the global null hypothesis from Section 2.2, and this induced test will have uniformly asymptotic level. By the same arguments of Lemma 1 from \textcite{runge_ci_rvec_2023}, this induced test will have uniformly asymptotic level if the original test does. Similar to the discussion in Section 2.2, there are four alternative hypotheses that can be used depending on whether the conditional independence relationships are time-invariant or dimension/time-offset index-invariant. 
% Also, note that $(\tilde{X}_{t,n,i,a}(u))_{t\in\mathbb{Z}}$, $(\tilde{Y}_{t,n,j,b}(u))_{t\in\mathbb{Z}}$, $(\bm{\tilde{Z}}_{t,n}(u))_{t\in\mathbb{Z}}$ are all stationary time series for each fixed $u\in [0,1]$, so we may replace the time $t$ for the input sequence $t$ by any other three times $t_1,t_2,t_3\in\mathbb{Z}$. 

First, consider the global null hypothesis of conditional independence 
\begin{equation} \tilde{X}_{\lfloor un\rfloor,n,i,a}(u)\!\perp \!\!\! \perp \tilde{Y}_{\lfloor un\rfloor,n,j,b}(u) \mid \bm{\tilde{Z}}_{\lfloor un\rfloor,n}(u) \text{ for all } u\in [0,1], \text{ for all } (i,j,a,b)\in\mathcal{D}_n.\label{eqn:global_null_hypoth_ALL_times_LSTS}\end{equation}
In the univariate setting, $\mathcal{D}_n$ simply consists of one dimension/time-offset tuple as in Section~\ref{subsection:cond_indep_null_hypoth_for_nsts}. Also, note that this hypothesis can be extended to the group of time series setting as discussed in Section~\ref{subsection:cond_indep_null_hypoth_for_nsts}. To test for the null hypothesis~\eqref{eqn:global_null_hypoth_ALL_times_LSTS}, we could, for example, use the test statistic   
$$\underset{u\in [0,1]}{\sup} \text{ } \left| \left|\frac{1}{\sqrt{T_n }} \sum_{t=\mathbb{T}_n^{-}}^{\lfloor u n\rfloor} (\bm{\hat{\Sigma}}_{t,n}^{\bm{R}})^{-1/2} \bm{\hat{R}}_{t,n} \right|\right|_p,$$ based on some $\ell_p$ norm $(p\geq 2)$ of the studentized partial sum process, where $\bm{\hat{\Sigma}}_{t,n}^{\bm{R}}$ is an estimate of the local long-run covariance matrix at time $t$. The theoretical guarantees for the test based on this test statistic would utilize the recent results from \textcite{mies_random_mult_sga_2024} about strong Gaussian approximations with random multipliers.

Second, it is possible to develop a test for the null hypothesis  \begin{equation}
   \tilde{X}_{\lfloor un\rfloor,n,i,a}(u)\!\perp \!\!\! \perp \tilde{Y}_{\lfloor un\rfloor,n,j,b}(u) \mid \bm{\tilde{Z}}_{\lfloor un\rfloor,n}(u) \text{ for all } (i,j,a,b)\in\mathcal{D}_n, \label{eqn:null_hypoth_ONE_time_LSTS}
\end{equation} for a \textit{particular} rescaled time $u\in [0,1]$ (i.e. instead of for \textit{all} $u\in [0,1]$ as in~\eqref{eqn:global_null_hypoth_ALL_times_LSTS}) by using, for example, the test statistic
$$\underset{m=(i,j,a,b)\in \mathcal{D}_n}{\max} \text{ } \left|\frac{1}{\sqrt{T_n h_{n,m}}} \sum_{t\in\mathcal{T}_n} K\left(\frac{t/n-u}{h_{n,m}}\right)\hat{R}_{t,n,m} \right| / \hat{\sigma}_{n,m}^R(u),$$
for some bandwidths $h_{n,m}\xrightarrow[]{} 0$ and local long-run variance estimates $(\hat{\sigma}_{n,m}^R(u))^2$. The main idea for this local conditional independence test is that since $\mathbb{E}_P(\tilde{R}_{P,\lfloor un\rfloor,n,m}(u))=0$ under the null and the process of error products is \say{approximately stationary} over short periods of time, we can expect that the means of $R_{P,t,n,m}=\tilde{R}_{P,t,n,m}(t/n)$ for rescaled times $t/n$ near $u$ are also close to zero under the null. This statement can be made rigorous by using the tools for nonlinear locally stationary processes developed by \textcite{dahlhaus_richter_wu_locally_stationary_general}.

Third, it is also possible to simultaneously test whether conditional independence 
\begin{equation} \tilde{X}_{\lfloor un\rfloor,n,i,a}(u)\!\perp \!\!\! \perp \tilde{Y}_{\lfloor un\rfloor,n,j,b}(u) \mid \bm{\tilde{Z}}_{\lfloor un\rfloor,n}(u) \text{ for all } (i,j,a,b)\in\mathcal{D}_n, \label{eqn:null_hypoth_SIMULT_time_LSTS}\end{equation} holds at \textit{all} rescaled times $u\in [0,1]$ (i.e. instead of for a \textit{particular} $u\in [0,1]$ as in~\eqref{eqn:null_hypoth_ONE_time_LSTS}). This can be done by creating simultaneous confidence bands (i.e. over time) for expected conditional covariance curves $(\rho_{P,n,m}(u))_{u\in [0,1]}$ for each $m=(i,j,a,b)\in\mathcal{D}_n$. Depending on whether or not estimates of the local long-run variances $(\hat{\sigma}_{n,m}^R(u))^2$ are used, these simultaneous confidence bands will have time-varying or time-invariant widths, respectively. The main idea is that the local null hypothesis of conditional independence at rescaled time $u\in [0,1]$ for some dimension/time-offset tuple $m=(i,j,a,b)\in\mathcal{D}_n$ can be rejected if zero is not included in the corresponding confidence interval for the local expected conditional covariance $\rho_{P,n,m}(u)$. This can be done using similar arguments as \textcite{weichi_wu_network}, which focuses on inferring time-varying correlation curves. However, due to the problem of post-selection inference \parencite{arun_psi_review}, this would require either stronger assumptions (e.g., Donsker-type), data decomposition techniques (e.g., splitting, fission, or thinning) for nonstationary time series, or two independent realizations of the same process. 

An approach for inferring expected conditional covariance curves would have a wide range of applications outside of testing for conditional independence, since this functional frequently appears in the causal inference literature \parencite{edward_kennedy_review2022,HigherOrderInfluenceFunctions2008,SemiparametricMinimaxRates,higher_order_inference_treatment2011,MinimaxEstimationFunctional2017,crossfitting2018}. We suspect that similar approaches can be used to infer curves based on other functionals of interest in causal inference. Hence, this line of work would be of significant interest to the emerging field of time series causal inference \parencite{reconstruct_causal_relations_ts_runge,causalinf_temp_runge,ci_ts_runge,causal_assoc_runge,causal_network_runge,cd_groups_runge_2023}. Lastly, we note that it may be possible to extend the tests discussed here to the piecewise locally stationary setting, though we leave the details for future work.

\subsection{Weakening the Assumptions on the Error Processes}\label{subsection:weaker_dependence_asmpt_errors}

In Assumption 3.5, we assume that there are distribution-uniform upper bounds on the $L^{\infty}(P)$ norms and $L^{\infty}(P)$ functional dependence measures of the error processes. We use this assumption to show inequality~\eqref{eqn:PE_ineq2} in the proof of Theorem 3.1. Afterward, we use the time-uniform convergence rates for the time-varying regression estimators to show Step 1.2 in the proof of Theorem 3.1. It is possible to weaken the assumptions imposed on the error processes by making stronger assumptions about the time-varying regression estimators, or more complicated assumptions about the terms in~\eqref{eqn:PE_ineq2}. Instead, we opt for simpler assumptions on the errors and estimation errors for the sake of transparency. Note that the Sieve-dGCM test from Section~\ref{appendix:CI_test_SIEVE} performs well even when the error processes violate Assumption 3.5, at least in the settings we considered for our simulations in Section 4. To satisfy Assumption 3.5, we can use truncated Gaussian error processes, for example.

\subsection{Variable Selection for Forecasting in Unstable Environments}\label{subsection:variable_selection}
% A central problem in statistics and machine learning is variable selection. 
Conditional independence tests offer a principled approach to variable selection. In the context of forecasting, the goal is to identify a minimal subset $S\subseteq \{1,...,p\}$ out of $p$ signals (including relevant lags) such that, for all times $t$, the forecasting target $Y_{t+h}$ at horizon $h$ is conditionally independent of the other signals $(X_{t}^i)_{i \not\in S}$ given $(X_{t}^i)_{i\in S}$. As discussed in \textcite{causal_assoc_runge}, one can construct parsimonious forecasting models by using conditional independence tests for nonlinear time series to detect time-delayed conditional dependencies while using the Benjamini–Hochberg procedure to control the false discovery rate (FDR) \parencite{Benjamini_Hochberg}. For applications to macroeconomic forecasting, see \textcite{copula_cit, cond_distr_cit}. We contribute to this literature by providing a test flexible enough to be used for identifying relevant forecasting signals in unstable environments, for which we are unaware of past work.

\subsection{Causal Discovery for Nonstationary Time Series}\label{subsection:note_cd_nsts}

The discovery of time-delayed causal relationships (see Figure~\ref{figure:unrolled_time_series_graph}) from observational time series is an important problem in numerous scientific domains. Conditional independence tests for nonlinear time series are a core component of many constraint-based and hybrid causal discovery algorithms; see \textcite{runge_earth_ts}. Crucially, several additional assumptions are needed for valid causal discovery, e.g., Markov and faithfulness.

Over the last several years, causal discovery for nonstationary time series has become an increasingly active area of research \parencite{biwei_huang_cd_nsts}. We emphasize that the tests used in the causal discovery algorithm must be appropriately tailored to the characteristics of the data. For instance, a constraint-based causal discovery algorithm may produce incorrect conclusions about the causal structure of a process if the independence and conditional independence tests fail to account for temporal dependence and nonstationarity. Our work feeds naturally into this research area.

%%%%%%%%%%%%% uncomment \iffalse and \fi no room for figure
%\iffalse
\begin{figure}[ht!]
    \centering
    \begin{tikzpicture}[
        processnode/.style={circle, draw=black, fill=pink!100, thick, minimum size=12mm},
        arrow/.style={-Stealth, thick},
        grayarrow/.style={-Stealth, thick, gray},
        node distance=15mm and 20mm
        ]

        % Unrolled Time Series Graph
        \node[processnode] (X1) at (0, 4) {$X_{t-1}$};
        \node[processnode] (X2) [right=of X1] {$X_{t}$};
        \node[processnode] (X3) [right=of X2] {$X_{t+1}$};

        \node[processnode, thick, font=\bfseries] (Z1) at (0, 2) {$Z_{t-1}$};
        \node[processnode, thick, font=\bfseries] (Z2) [right=of Z1] {$Z_{t}$};
        \node[processnode, thick, font=\bfseries] (Z3) [right=of Z2] {$Z_{t+1}$};

        \node[processnode] (Y1) at (0, 0) {$Y_{t-1}$};
        \node[processnode] (Y2) [right=of Y1] {$Y_{t}$};
        \node[processnode] (Y3) [right=of Y2] {$Y_{t+1}$};

        % Arrows indicating causation
        \draw[arrow] (X1) -- (X2);
        \draw[arrow] (X2) -- (X3);

        \draw[arrow] (Y1) -- (Y2);
        \draw[arrow] (Y2) -- (Y3);

        \draw[arrow] (Z1) -- (Z2);
        \draw[arrow] (Z2) -- (Z3);

        \draw[arrow] (Z2) -- (Y3);
        \draw[arrow] (Z2) -- (X2);
        
        \draw[arrow] (Z1) -- (Y2);
        \draw[arrow] (Z1) -- (X1);
        
        \draw[arrow] (Z3) -- (X3);
        
        % Gray arrows for continuation in time
        \draw[grayarrow] (X3) -- ++(1.5, 0);
        \draw[grayarrow] (Y3) -- ++(1.5, 0);
        \draw[grayarrow] (Z3) -- ++(1.5, 0);
        \draw[grayarrow] (Z3) -- ++(1.5, -1);

        \draw[grayarrow] (-1.5, 4) -- ++(X1);
        \draw[grayarrow] (-1.5, 0) -- ++(Y1);
        \draw[grayarrow] (-1.5, 1) -- ++(Y1);
        \draw[grayarrow] (-1.5, 2) -- ++(Z1);

        % Labels
        %\node[above,font=\large\bfseries] at (current bounding box.north) {Causal Time Series Graph};
        
    \end{tikzpicture} 
    \caption{Causal graph depicting the time-delayed causal relationships among the stochastic processes $X=(X_t)_{t\in\mathbb{Z}}$, $Y=(Y_t)_{t\in\mathbb{Z}}$, and $Z=(Z_t)_{t\in\mathbb{Z}}$. The causal graph shows that $Z$ is a common cause of both $X$ and $Y$, directly influencing $X$ in the same time period and affecting $Y$ with a one time step delay. In this example, the causal graphical structure remains fixed over time, although the causal effects themselves may vary over time.} \label{figure:unrolled_time_series_graph}
\end{figure} 
%\fi
%%%%%%%%%%%%%%%%%

A common practice in time series analysis is to difference a time series until the augmented Dickey-Fuller (ADF) test says it is stationary. One may presume that a similar approach can be taken in the context of causal discovery for time series. That is, applying a causal discovery algorithm for stationary time series to a transformed (e.g., $d$-times differenced) time series in the process of identifying a causal graph for the originally nonstationary time series. Unfortunately, as discussed in \textcite{spirtes_var_example}, even differencing once can induce spurious correlations among the time series. Consequently, the resultant causal graph for the original time series can be wildly incorrect.

In contrast, causal discovery algorithms based on conditional independence tests for nonstationary time series (e.g., our dGCM test) can identify more interpretable causal graphs because the time series need not be stationary. Instead, we can consider \say{locally stationary transformations} of the original time series, such as log growth rates, percent changes, or single differences. In other words, users of dGCM need not assume that the transformed time series is stationary over the entire time period, only that it is approximately stationary over short time windows. In practice, this means that causal graphs identified by dGCM-based causal discovery algorithms can be easier to interpret, because they can be constructed for locally stationary transformations rather than stationary transformations (e.g., $d$-times differenced time series). We emphasize that constraint-based causal discovery algorithms for nonstationary nonlinear time series can use the dGCM test for conditional independence and the test for independence from Section~\ref{subsection:bonus_indep_test}.

\subsection{Granger Causality}\label{subsection:granger_causality}

Throughout this paper, we have discussed causality as in \textcite{pearl} and causal discovery as in \textcite{spirtes_cps}. However, most economists are more familiar with the concept of Granger causality, particularly in the stationary linear time series setting. Therefore, it may be helpful to review different notions of Granger causality in the nonstationary nonlinear time series setting. To do so, we provide a highly condensed summary of Sections 2 and 4 in \textcite{shojaie_fox_granger_review}.

The original definition of Granger causality from \textcite{Granger1969} is about prediction. Informally, a process $X$ is said to be Granger noncausal for another process $Y$ if, for all times $t$, the variance of the error from the optimal prediction of $Y_t$ based on all relevant information up to time $t-1$ is not reduced by including the history of $X$ up to time $t-1$. See Section 2 in \textcite{shojaie_fox_granger_review} for the exact definition and the stringent conditions under which this predictive definition corresponds to genuine causality as in \textcite{pearl}. While this original definition does not assume linear dynamics, much of the following methodology revolves around the identification of coefficients in linear vector autoregressive (VAR) models with $p$ time series \parencite{Granger1980,Lutkepohl2005,Basu2015}.

Another definition of Granger causality, referred to as \textit{strong} Granger causality \parencite{florens}, is stated in terms of conditional independence relationships among stochastic processes. Let $(X^i)_{i\in [p]}$ be $p$ signals used to predict the target $Y$. The process $X^i$ is said to be (strongly) Granger noncausal for $Y$ if, for all times $t$, $Y_t$ is conditionally independent of the history of the signal $X^i$ up to time $t-1$ given the history of the other signals $(X^j)_{j\in [p]\backslash\{i\}}$ up to time $t-1$. See Definition 2 in \textcite{shojaie_fox_granger_review} for the exact definition and the rest of Section 4 therein for more discussion.

Notably, \textcite{eichler} introduced a comprehensive graphical modeling framework for time series based on strong Granger causality, which can be detected using conditional independence tests for nonlinear time series \parencite{assess_granger,su_white_char_ci_test,copula_cit,song_wei_2021,proj_cit_time_series}. In a similar vein, our proposed conditional independence test can be used to detect strong Granger causality for nonlinear time series with time-varying dynamics. This can be incorporated into graphical modeling frameworks for nonstationary nonlinear time series, analogous to \textcite{rao_basu}. Also, we note that \textcite{local_granger_nsts} developed a methodology for inferring so-called \say{local Granger causality}, which is a frequency domain characterization of time-evolving relationships between locally stationary time series.

There are various techniques for assessing nonlinear Granger causality that do not use conditional independence testing. For instance, the neural Granger causality method from \textcite{neural_granger} extracts Granger causal structures by using sparsity-inducing penalties on the weights of structured multilayer perceptrons (MLPs) and recurrent neural networks (RNNs). Additionally, there is an influential strand of literature connecting Granger causality and directed information theory \parencite{granger_directed_info_relationship,directed_info_graphs}. See Section 4 of \textcite{shojaie_fox_granger_review} for more discussion of nonlinear Granger causality.

\subsection{All Test Results for the Real Data Application}\label{subsection:all_results_for_real_data}

The following table displays the p-value and adjusted p-value from the Benjamini-Hochberg procedure \parencite{Benjamini_Hochberg} of each hypothesis test from Section 5.

\sisetup{
  round-mode          = places,
  round-precision     = 3,
  detect-weight       = true,
  detect-family       = true
}

\begin{table}[p] % [ht!]
    \centering
%%% knitr 
\resizebox{\textwidth}{!}{
\begin{tabular}{lll}
\toprule
\textbf{Null Hypothesis} & \textbf{P-value} & \textbf{Adjusted P-value}\\
\midrule
$\texttt{S\&P(t)} \!\perp \!\!\! \perp \texttt{FTSE(t)} \text{ for all times } t$ & \SI{6e-4}{} & \num{0.0048}\\
$\texttt{S\&P(t)} \!\perp \!\!\! \perp \texttt{HangSeng(t)} \text{ for all times } t$ & \num{0.0186} & \num{0.0318857142857143}\\
$\texttt{S\&P(t)} \!\perp \!\!\! \perp \texttt{Nikkei(t)} \text{ for all times } t$ & \num{0.0556} & \num{0.0741333333333333}\\
$\texttt{FTSE(t)} \!\perp \!\!\! \perp \texttt{HangSeng(t)} \text{ for all times } t$ & \num{0.0018} & \num{0.00864}\\
$\texttt{FTSE(t)} \!\perp \!\!\! \perp \texttt{Nikkei(t)} \text{ for all times } t$ & \num{0.0388} & \num{0.0547764705882353}\\
$\texttt{HangSeng(t)} \!\perp \!\!\! \perp \texttt{Nikkei(t)} \text{ for all times } t$ & \num{0.0114} & \num{0.0228}\\
$\texttt{S\&P(t)} \!\perp \!\!\! \perp \texttt{FTSE(t)} \mid \texttt{HangSeng(t)} \text{ for all times } t$ & \SI{4e-4}{} & \num{0.0048}\\
$\texttt{S\&P(t)} \!\perp \!\!\! \perp \texttt{FTSE(t)} \mid \texttt{Nikkei(t)} \text{ for all times } t$ & \SI{2e-4}{} & \num{0.0048}\\
$\texttt{S\&P(t)} \!\perp \!\!\! \perp \texttt{HangSeng(t)} \mid \texttt{FTSE(t)} \text{ for all times } t$ & \num{0.7968} & \num{0.7968}\\
$\texttt{S\&P(t)} \!\perp \!\!\! \perp \texttt{HangSeng(t)} \mid \texttt{Nikkei(t)} \text{ for all times } t$ & \num{0.1302} & \num{0.135860869565217}\\
$\texttt{S\&P(t)} \!\perp \!\!\! \perp \texttt{Nikkei(t)} \mid \texttt{FTSE(t)} \text{ for all times } t$ & \num{0.0904} & \num{0.103314285714286}\\
$\texttt{S\&P(t)} \!\perp \!\!\! \perp \texttt{Nikkei(t)} \mid \texttt{HangSeng(t)} \text{ for all times } t$ & \num{0.083} & \num{0.0996}\\
$\texttt{FTSE(t)} \!\perp \!\!\! \perp \texttt{HangSeng(t)} \mid \texttt{S\&P(t)} \text{ for all times } t$ & \num{0.0022} & \num{0.0088}\\
$\texttt{FTSE(t)} \!\perp \!\!\! \perp \texttt{HangSeng(t)} \mid \texttt{Nikkei(t)} \text{ for all times } t$ & \num{0.0014} & \num{0.0084}\\
$\texttt{FTSE(t)} \!\perp \!\!\! \perp \texttt{Nikkei(t)} \mid \texttt{S\&P(t)} \text{ for all times } t$ & \num{0.0294} & \num{0.0441}\\
$\texttt{FTSE(t)} \!\perp \!\!\! \perp \texttt{Nikkei(t)} \mid \texttt{HangSeng(t)} \text{ for all times } t$ & \num{0.0622} & \num{0.0785684210526316}\\
$\texttt{HangSeng(t)} \!\perp \!\!\! \perp \texttt{Nikkei(t)} \mid \texttt{S\&P(t)} \text{ for all times } t$ & \num{0.0064} & \num{0.01536}\\
$\texttt{HangSeng(t)} \!\perp \!\!\! \perp \texttt{Nikkei(t)} \mid \texttt{FTSE(t)} \text{ for all times } t$ & \num{0.0054} & \num{0.0144}\\
$\texttt{FTSE(t)} \!\perp \!\!\! \perp \texttt{Nikkei(t)} \mid \texttt{S\&P(t-1)} \text{ for all times } t$ & \num{0.1072} & \num{0.116945454545455}\\
$\texttt{FTSE(t)} \!\perp \!\!\! \perp \texttt{HangSeng(t)} \mid \texttt{S\&P(t-1)} \text{ for all times } t$ & \num{0.0036} & \num{0.0114}\\
$\texttt{Nikkei(t)} \!\perp \!\!\! \perp \texttt{HangSeng(t)} \mid \texttt{S\&P(t-1)} \text{ for all times } t$ & \num{0.016} & \num{0.0295384615384615}\\
$\texttt{FTSE(t)} \!\perp \!\!\! \perp \texttt{S\&P(t-1)} \text{ for all times } t$ & \num{0.0074} & \num{0.0161454545454545}\\
$\texttt{Nikkei(t)} \!\perp \!\!\! \perp \texttt{S\&P(t-1)} \text{ for all times } t$ & \num{0.0038} & \num{0.0114}\\
$\texttt{HangSeng(t)} \!\perp \!\!\! \perp \texttt{S\&P(t-1)} \text{ for all times } t$ & \num{0.0234} & \num{0.03744}\\
\bottomrule
\end{tabular}}
%%% knitr
\label{tab:knitr_adjusted_pvalues}
\end{table}

%%%

\pagebreak 

\printbibliography[segment=\therefsegment,
                   title={References}]

%\printbibliography[heading=subbibliography,title={References}]

\end{refsegment}

\end{document}